\documentclass[10pt,twocolumn]{article}
\usepackage[margin=0.75in,columnsep=0.28in]{geometry}
\usepackage{amsmath,amssymb}
\usepackage{booktabs}
\usepackage{longtable}
\usepackage{array}
\usepackage{xcolor}
\usepackage[numbers,sort&compress]{natbib}
\usepackage[colorlinks=true,linkcolor=blue!45!black,citecolor=blue!45!black,urlcolor=blue!45!black]{hyperref}
\usepackage[T1]{fontenc}
\usepackage[expansion=false]{microtype}
\usepackage{graphicx}
\usepackage[font=small,labelfont=bf]{caption}
\usepackage{authblk}
\usepackage{enumitem}
\usepackage{placeins}

\usepackage{float}
\usepackage{balance}
\usepackage{braket}

\renewcommand{\emph}[1]{#1}

\setkeys{Gin}{width=\linewidth,keepaspectratio}

\definecolor{deeppink}{RGB}{255,20,147}
\newcommand{\code}[1]{\texttt{#1}}
\newcommand{\todo}[1]{}

\newcommand{\eh}{E_{\mathrm{h}}}

\title{A Hardware-Trained Generative Quantum Eigensolver for Sample-Based Quantum Diagonalization of Organotin Hydroxides up to 70 Qubits}

\author[1]{Kunal Kumar\thanks{Corresponding author: kunalkumar@ibm.com}}
\author[2,4]{Kimberlee Keithley}
\author[6]{Toshiya Iwamori}
\author[2,4]{Shu Kanno}
\author[5]{Ikko Hamamura}
\author[2]{Takao Kobayashi}
\author[6]{Akira Koseki}
\author[2,4]{Qi Gao\thanks{Corresponding author: qi.gao.mb@mcgc.com}}
\author[3,4]{Naoki Yamamoto}
\author[1]{Hiroshi Horii}
\author[1]{Yuri Kobayashi}
\author[7]{Borja Peropadre}
\author[8]{Kouhei Nakaji}
\author[1]{Toru Imai}
\author[6]{Takayuki Osogami}

\affil[1]{IBM Quantum, IBM Japan, Tokyo 103-8510, Japan}
\affil[2]{Mitsubishi Chemical Corporation, Science \& Innovation Center,
1000 Kamoshida-cho, Aoba-ku, Yokohama 227-8502, Japan}
\affil[3]{Department of Applied Physics and Physico-Informatics,
Keio University, Hiyoshi 3-14-1, Kohoku, Yokohama 223-8522, Japan}
\affil[4]{Quantum Computing Center, Keio University, Hiyoshi 3-14-1,
Kohoku, Yokohama 223-8522, Japan}
\affil[5]{NVIDIA, ATT EAST 12F, 2-11-7 Akasaka, Minato-ku, Tokyo 107-0052,
Japan}
\affil[6]{IBM Research, IBM Japan, Tokyo 103-8510, Japan}
\affil[7]{IBM Quantum, IBM Research Cambridge, Cambridge, Massachusetts, USA}
\affil[8]{NVIDIA, Santa Clara, California, USA}

\date{}

\begin{document}
\maketitle

\begin{abstract}
\noindent
Quantum hardware can provide feedback for training circuit generators beyond the reach of exact classical simulation.
We train generative quantum eigensolver (GQE) policies, transformers conditioned on molecular descriptors, to generate the state-preparation circuits for sample-based quantum diagonalization (SQD).
Training runs first on a noiseless simulator and then with the IBM Heron processor \code{ibm\_kobe} inside the loop, at 46 and 70 qubits ($1.2\times10^{8}$ shots for the four main training runs). A classical CCSD(T) reference energy of the training molecules enters only an admission gate and two auxiliary reward terms.
During hardware training, a reward penalty prevents collapse to circuits that sample only the Hartree--Fock determinant.
Policies trained on transition-metal complexes are applied without retraining to Sn(OH)$_3$-Me and its hydrates, minimal models of organotin photoresists for extreme-ultraviolet lithography, from 32 to 70 qubits.
After one round of hardware sampling, followed by classical ranking and recovery of the cached counts with the molecular Hamiltonian and no further QPU calls, 27 of 28 runs at 46--70 qubits are within 15 mHa of the density-matrix renormalization group (DMRG) reference (median 4.5 mHa), and in all eight runs with an untrained control the trained policy is more accurate.
At 56--70 qubits the selected circuits use 34--72 times fewer two-qubit gates than the local unitary cluster Jastrow (LUCJ) circuit transpiled for the same device.
These results demonstrate that circuit generators can be trained with quantum hardware at 70 qubits.
\end{abstract}

\section{Introduction}
\label{sec:intro}

Strongly correlated electrons, as in transition-metal complexes and organometallic compounds, are a main benchmark for quantum computational chemistry~\cite{mcardle2020quantum}. 
Exact classical diagonalization becomes impossible when the active space grows, because its cost increases exponentially.
Density-matrix renormalization group (DMRG) methods~\cite{white1992dmrg} work well for low-dimensional systems, but the growth of entanglement limits them in complex active spaces.
Quantum computers are therefore expected to help beyond the classically solvable regime.
Early work reduced molecular problems to a few qubits and ran them on small superconducting devices~\cite{bravyi2017tapering,kandala2017hardware}, and processors now have more than a hundred qubits and run circuits beyond exact classical simulation~\cite{kim2023evidence}.
Present quantum processors are noisy and allow only shallow circuits.
Quantum-selected configuration interaction (QSCI)~\cite{kanno2023qsci} and sample-based quantum diagonalization (SQD)~\cite{robledomoreno2024chemistry} suit this situation: the quantum processor only samples important electronic configurations from a prepared state, and a classical supercomputer diagonalizes the Hamiltonian in the subspace that these configurations define.
This hybrid scheme has made active spaces of several tens of qubits accessible on superconducting processors~\cite{robledomoreno2024chemistry}, and it has been extended to molecular vibrations~\cite{kumar2024vibrational,ollitrault2020quantum}. Recent work has also explored broken-symmetry reference states for SQD and reported near-DMRG accuracy~\cite{kumar2026brokensymmetry}.

The quality of an SQD calculation depends on the circuit that prepares the state. Fixed ans{\"a}tze such as the local unitary cluster Jastrow (LUCJ) ansatz~\cite{motta2023lucj} often give deep circuits, which must then be adapted to the connectivity of the device. The generative quantum eigensolver (GQE) offers another route: an autoregressive model, first formulated as a transformer (``GPT-QE''), generates a sequence of operations from a predefined operator pool, so the circuit is learnt and not designed by hand~\cite{nakaji2024gqe}. Transformer-based GQE has been used to numerically design circuits for QSCI, with a substantial reduction in two-qubit gates compared with Trotterized baselines, for active spaces of up to 32 qubits~\cite{kemmoku2026generative}. 
A GQE and QSCI pipeline has also been applied to tin photoresist models of 14 to 44 qubits, including methyltin trihydroxide, with hardware runs on an IQM processor for smaller tin systems at 14 and 22 qubits~\cite{elgammal2026scaling}.

Training one model on many molecules (amortization) would let the same model propose circuits for a new molecule, without designing a new ansatz for each active space.
Training such a model requires an energy estimate for every circuit it proposes.
In this work, the noiseless training stage uses molecules of up to 32 qubits, and exact classical simulation becomes intractable from about 46 qubits (Section~\ref{sec:training}); for the 46- and 70-qubit targets we obtain the energy feedback from the quantum processor.
As a result, the policy is trained on hardware-measured (noisy) outcomes, and circuits are scored according to their performance under the device noise and transpilation constraints.
Here we extend amortized circuit design to transition-metal active spaces of 46--70 qubits and train the policy directly on superconducting hardware.

Sustained semiconductor scaling demands the transition of lithographic patterning from deep ultraviolet toward extreme ultraviolet lithography (EUVL).
In this context, organotin materials have emerged as leading photoresist candidates owing to the high absorption cross-section of tin at the EUV excitation energy of 92~eV \cite{ma2022organotin,du2024tinoxo,haitjema2017patterning}.
Consequently, diverse organometallic and hybrid tin clusters have been synthesized, including butyltin oxo hydroxo complexes \cite{frederick2019thermal}, sodium-templated alkyltin Keggin clusters \cite{diulus2020effect,saha2017alkyltin}, and organotin carboxylates \cite{sharps2017implications}.
Photolysis of the Sn--C bond is believed to drive the EUV chemistry of these materials, and its description requires multireference methods~\cite{chalabala2018}. Such systems therefore need careful treatment when the size of the problem is chosen for a quantum algorithm.
Here, we employ the tin-organometallic family as held-out targets for an amortized generative policy trained on benchmark transition-metal complexes, using Sn(OH)$_3$-Me (methyltin trihydroxide), the smallest model of a monoorganotin hydroxide, and its mono- and di-hydrates.

To make policy training feasible under device noise, we use a two-stage curriculum. Stage 1 uses a noiseless simulator and small molecules, while Stage 2 continues the training on the IBM processor \code{ibm\_kobe} using transition-metal complexes with 46- and 70-qubit active spaces. The resulting policies are then evaluated zero-shot on the held-out Sn(OH)$_3$-Me targets introduced above. This progression from easier simulated problems to larger hardware-based problems is intended to stabilize training under device noise~\cite{patel2024curriculum}. In Stage 2, we compare two rewards: one incorporating a Hartree--Fock penalty, which prevents the policy from producing only the reference determinant, and one tuned jointly on a noisy simulator (Section~\ref{sec:pipelines}). Our evidence that the policy transfers to new chemistry is qualitative (Section~\ref{sec:discussion}).

The trained policies give energies close to DMRG with short circuits.
For each target, the policy generates ten circuits, each circuit is sampled once on \code{ibm\_kobe}, and the circuit with the lowest energy computed from its measured shots is used for configuration recovery.
At 46 to 70 qubits, 27 of 28 runs are within 15~mHa of the DMRG reference (median 4.5~mHa), and in all eight runs with an untrained control the trained policy is more accurate (Section~\ref{sec:zero-shot}).
At 56 to 70 qubits the selected circuits use 64 to 160 two-qubit gates.
This is 34 to 72 times fewer than the LUCJ circuit, which needs 4{,}248 to 6{,}952 two-qubit gates in an offline estimate on an idealized heavy-hex model of the device (\code{ibm\_heron\_r2} basis, Qiskit optimization level 1); the LUCJ circuits actually submitted to \code{ibm\_kobe} were larger, so the reduction is a lower bound (Section~\ref{sec:lucj-comparison}).
Sampling ten circuits takes about 11 times the QPU time per job of sampling one LUCJ circuit (Section~\ref{sec:cost}).

The primary contributions of this work are as follows:
\begin{enumerate}[leftmargin=1.8em,label=(\roman*)]
  \item We develop amortized GQE policies trained with a two-stage curriculum (a noiseless-simulator warm-up followed by continuation on \code{ibm\_kobe}) for 46- and 70-qubit active spaces, with a Hartree--Fock penalty in the hardware-stage reward (Section~\ref{sec:pipelines}), and compare them with a jointly tuned reward (Section~\ref{sec:three-model-inference}).
  \item We perform a zero-shot evaluation across the eleven active-space tiers of the Sn(OH)$_3$-Me family (32--70 qubits) against DMRG references, comparing five distinct circuit selection and pooling strategies, with matched untrained controls.
  \item We present a circuit-level analysis (depth, valid-shot yield, spin-resolved excitation hierarchy and operator statistics) of the sampled quantum circuits (Sections~\ref{sec:lucj-comparison} and~\ref{sec:analysis-selection}, Appendix~\ref{sec:appendix-chemistry}).
  \item We report the quantum and classical cost of training and inference: shots, QPU time and classical wall time (Section~\ref{sec:cost}). To our knowledge, no GQE policy has been trained on quantum hardware at this size before, so these costs are a first reference point for future studies.
  \item We open-source the related code, the generated sequence tokens, and the analyses of their excitation content, its correlation with the spin sectors, and the correctness of the populated orbitals.
\end{enumerate}

The remainder of the paper is organized as follows. Section~\ref{sec:methods} describes the method: the pipeline, sample-based quantum diagonalization, the policy and its architecture, the molecules, the training procedure, the hyperparameters and the computing resources. Section~\ref{sec:results} reports the results: the two training stages (Sections~\ref{sec:stage1-results} and~\ref{sec:stage2-results}), the zero-shot inference on unseen molecules (Section~\ref{sec:zero-shot}), the circuit cost against LUCJ (Section~\ref{sec:lucj-comparison}), the comparison of the two rewards (Section~\ref{sec:three-model-inference}), the divergence from the LUCJ circuit (Section~\ref{sec:divergence}), the comparison with DMRG (Section~\ref{sec:refs}), the determinants of recovery accuracy (Section~\ref{sec:analysis-selection}) and the computational cost (Section~\ref{sec:cost}). Section~\ref{sec:discussion} discusses the results, and Section~\ref{sec:conclusion} concludes.

\section{Methods}
\label{sec:methods}

\subsection{Pipeline overview}
\label{sec:methods-pipeline}
\label{sec:pipeline-overview}

\begin{figure*}[t]
\centering
\includegraphics[width=\textwidth]{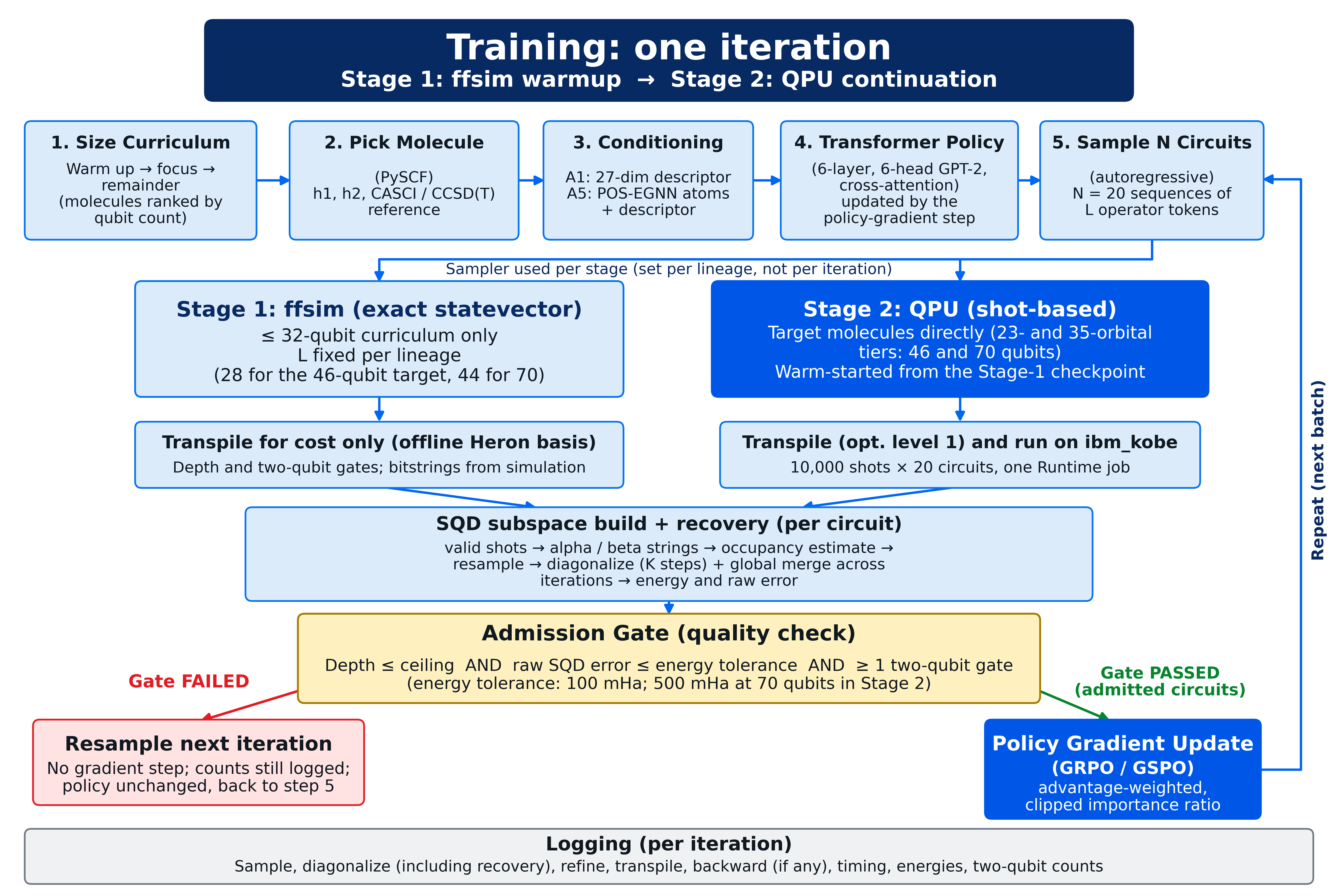}
\caption{Training control flow for one iteration. The size curriculum picks a molecule; its conditioning (A1 or A5) drives the transformer policy, which samples $N=20$ circuits of $L$ operator tokens each. In Stage~1 the circuits are evaluated with an exact statevector simulator (curriculum of at most 32 qubits; the transpilation only measures depth and two-qubit-gate count); in Stage~2 they are transpiled and run on \texttt{ibm\_kobe} (10{,}000 shots for each of 20 circuits in one Runtime job) on the 23- and 35-orbital target molecules, initialized from the full Stage-1 policy checkpoint (every weight is carried over; Section~\ref{sec:training}). The SQD subspace build with self-consistent recovery and global merge gives each circuit's energy and raw error; an admission gate then checks the best circuit of the batch (depth within the ceiling, raw error within the energy tolerance, at least one two-qubit gate). Admitted iterations drive a GRPO/GSPO policy-gradient update over all 20 circuits; for rejected iterations the counts are logged but nothing is back-propagated and the next iteration resamples.}
\label{fig:flow-training}
\end{figure*}

\begin{figure*}[t]
\centering
\includegraphics[width=\textwidth]{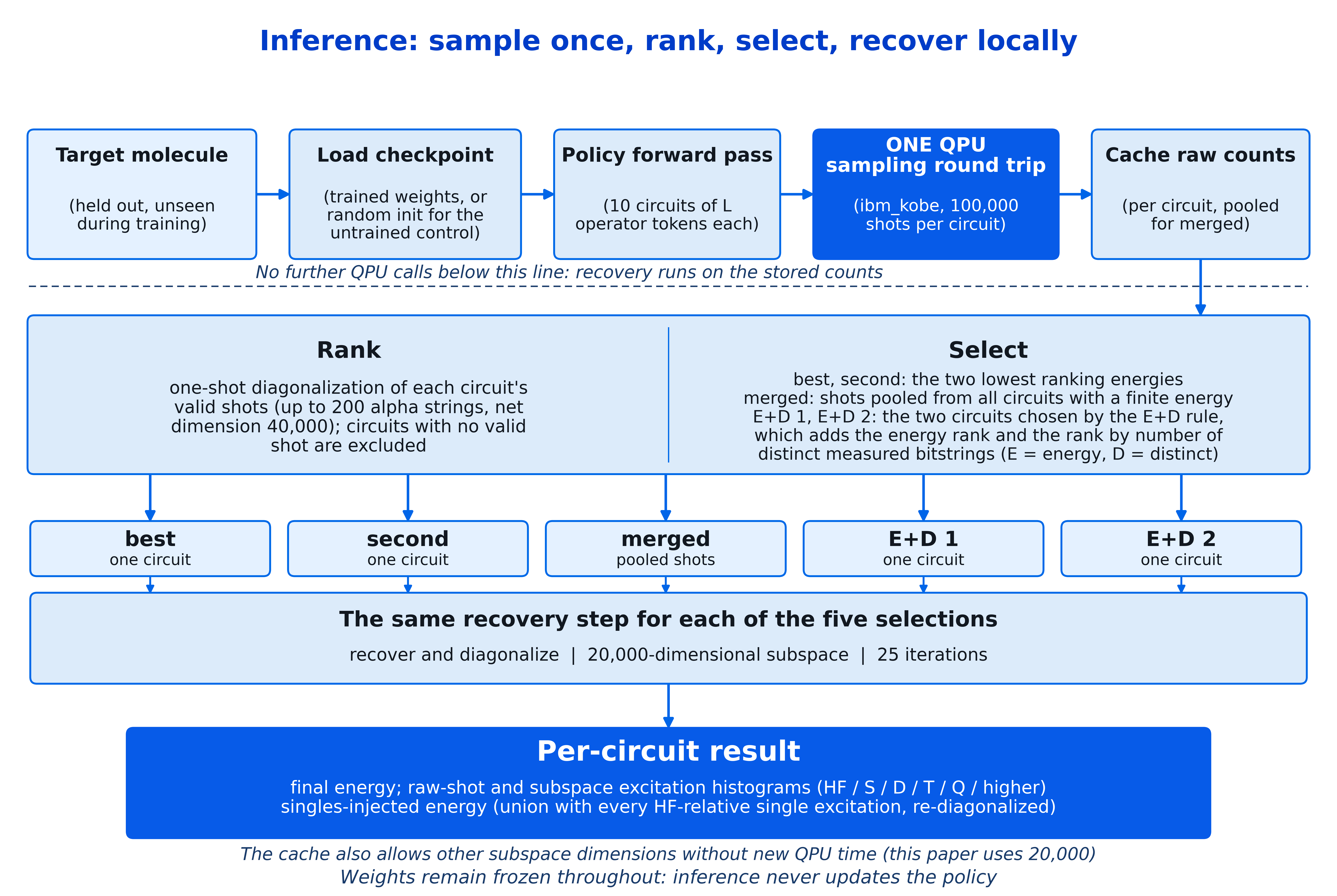}
\caption{Inference control flow. A frozen policy (trained, or randomly initialized for the untrained control) emits 10 circuits for an unseen molecule; they are sampled once on \texttt{ibm\_kobe} (100{,}000 shots each) and the raw counts are cached. Below the dashed line everything runs locally on the cached counts: the circuits are ranked by a one-shot diagonalization (up to 200 alpha strings, net dimension 40{,}000; circuits without a valid shot are excluded); five results are selected: best and second (the two lowest ranking energies), merged (the shots pooled from all circuits with a finite energy), and E+D~1 and E+D~2 (the two circuits chosen by the E+D rule, which combines the rank by energy, E, with the rank by the number of distinct measured bitstrings, D); the same recovery step is then applied to each of the five, with $D=20{,}000$ strings per spin sector and 25 iterations. The result per circuit is the energy, the raw-shot and subspace excitation histograms and the singles-injected energy.}
\label{fig:flow-inference}
\end{figure*}

Figures~\ref{fig:flow-training} and~\ref{fig:flow-inference} summarize the
full training and inference control flow,
elaborated in the remainder of this section. Training proceeds per
molecule-selection iteration: the size curriculum (Section~\ref{sec:training})
picks a molecule, its descriptor conditions the policy
(Section~\ref{sec:architecture}, Fig.~\ref{fig:encoder-arch}), the policy emits a batch of candidate gate
sequences, and---depending on which stage is active for that lineage---the
resulting circuits are either simulated via statevectors (Stage 1,
Section~\ref{sec:training}) or executed on physical superconducting hardware (Stage 2). Every
sampled circuit is transpiled, its subspace is built via recovery and global
merge (Section~\ref{sec:recovery}), and the resulting raw energy error is checked,
together with the circuit depth and non-triviality criteria, against a hard admission gate;
only circuits that pass the gate contribute a policy-gradient update
(Section~\ref{sec:gqe-formalism}). Inference reuses the identical policy forward
pass and recovery machinery with frozen weights, with one key addition
relative to training: a single hardware sampling round trip's raw
counts are cached, enabling subsequent ranking, selection, and subspace recovery
to execute locally without invoking additional quantum processor queries.

\subsection{Sample-based quantum diagonalization}
\label{sec:sqd-formalism}

Let an ansatz circuit $U(\boldsymbol{\theta})$ act on $n_q = 2 n_o$ qubits,
where $n_o$ is the number of active spatial orbitals (Jordan--Wigner-mapped,
one qubit per spin-orbital, spin-separated so that an $n_o$-orbital active
space on $(n_\alpha, n_\beta)$ electrons occupies $n_q$ qubits in total).
We write $\ket{\psi_{\boldsymbol\theta}} = U(\boldsymbol\theta)\ket{\mathrm{HF}}$ for the state prepared by the circuit.
Measuring $\ket{\psi_{\boldsymbol\theta}}$ in the computational
basis yields a set of bitstrings, each splittable into an $\alpha$-string and
a $\beta$-string of Hamming weight $n_\alpha$ and $n_\beta$ respectively. Let
$\mathcal{A}, \mathcal{B}$ denote the sets of distinct $\alpha$- and
$\beta$-strings observed over $N_{\mathrm{shots}}$ measurements, with
empirical counts $c(a,b)$ for each observed pair $(a,b)\in\mathcal{A}\times
\mathcal{B}$; the observed determinants form the set $S_{\mathrm{raw}}$.

A classical configuration-interaction subspace $S$ is constructed from the samples, but is not restricted to $S_{\mathrm{raw}}$. In our recovery procedure (Section~\ref{sec:recovery}), the observed strings are repaired, the Hartree--Fock determinant and carryover strings are added, and the determinant set is taken as the product $S_{\mathrm{rec}}=\mathcal A'\times\mathcal B'$, which in general contains determinants that were not directly measured. For a circuit's own recovered subspace we set $S=S_{\mathrm{rec}}$, while the pooled result uses a merged set $S_{\mathrm{merge}}$.

The molecular Hamiltonian $\hat H$ is projected onto $S$ and diagonalized exactly. Here $\ket{v}$ is a trial vector in the span of the determinants of $S$; it is not the state $\ket{\psi_{\boldsymbol\theta}}$ prepared by the circuit, and the minimum can lie below $\braket{\psi_{\boldsymbol\theta}|\hat H|\psi_{\boldsymbol\theta}}$:
\begin{equation}
  E_S = \min_{\ket{v} \in \operatorname{span}(S)}
        \frac{\braket{v| \hat{H} | v}}{\braket{v|v}}.
  \label{eq:subspace-diag}
\end{equation}
The subspace $S$ comes from the samples of one circuit $U_{\mathbf g}$ (Section~\ref{sec:gqe-formalism}), so $E_S$ is the SQD energy of that circuit; we also write it $E_{\mathrm{SQD}}(\mathbf g)$. We report the signed deviation $\Delta E = E_{\mathrm{SQD}} - E_{\mathrm{DMRG}}$ from the DMRG reference energy (in mHa); a positive value means that the SQD energy lies above the reference. Because the DMRG energy is itself an approximation, the sign is informative. All deviations in this study are positive. Because $S$ is built from sampled determinants rather than an a priori
truncation, the quality of $E_S$ depends both on the circuit's ability to
place amplitude on determinants near the true ground state and on how $S$ is
selected and enlarged from the raw sample (Section~\ref{sec:recovery}).

\subsection{Amortized generative policy}
\label{sec:gqe-formalism}

Rather than fixing $U(\boldsymbol\theta)$ to a hand-designed form, a policy
$\pi_\phi$ generates the circuit itself, token by token. At each of $L$ autoregressive decoding steps, $\pi_\phi$ samples a gate token $g_t$ from a problem-specific operator pool $\mathcal{P}$ (Section~\ref{sec:architecture}), conditioned on the tokens emitted so far and on a molecule-specific descriptor $\mathbf{d}$. Here $L$ is the number of gate tokens per circuit: $L=28$ for the policies trained for 46 qubits and $L=44$ for those trained for 70 qubits. The sampling is
defined as follows:
\begin{equation}
  g_t \sim \pi_\phi(\,\cdot \mid g_{<t}, \mathbf{d}\,), \qquad t = 1,\dots,L,
  \label{eq:autoregressive}
\end{equation}
yielding a full gate sequence $\mathbf{g} = (g_1,\dots,g_L)$ and hence a
circuit $U_{\mathbf g}$. A batch of $N_{\mathrm{samples}}$ sequences is drawn
per training iteration; each is scored by its resulting SQD energy
$E_{\mathrm{SQD}}(\mathbf g)$ (Eq.~\ref{eq:subspace-diag}), and the policy is updated to
favor sequences with lower energy via a group-relative policy-gradient
objective (GRPO/GSPO; \cite{shao2024deepseekmath,schulman2017ppo}): with the batch energies $E_i = E_{\mathrm{SQD}}(\mathbf g_i)$, $i=1,\dots,N_{\mathrm{samples}}$,
\begin{equation}
  \begin{aligned}
  A_i &= \frac{\bar E - E_i}{\operatorname{std}(\{E_j\}) + \varepsilon},\\
  \mathcal{L}(\phi) &= -\min\!\Big(r_i(\phi)\,A_i,\;
    \operatorname{clip}\big(r_i(\phi), 1-\epsilon_{\mathrm{lo}}, 1+\epsilon_{\mathrm{hi}}\big)\,A_i\Big),
  \end{aligned}
  \label{eq:grpo}
\end{equation}
where $\bar E$ is the batch mean energy, $r_i(\phi)$ is the
(token- or sequence-level) importance ratio between the current and the
behavior policy, and the lowest-energy sequence in the batch additionally
receives a direct imitation term. $A_i$ is undefined (and the ratio term
dropped) when the batch has zero energy variance, which the training logs
record as a `reward-degenerate' iteration.

In each training iteration, the policy samples $N=20$ circuits and we compute a raw SQD energy $E_i$ for each. The admission gate is evaluated once per iteration, on the best circuit (lowest $E_i$): its transpiled depth must be within the ceiling, its raw SQD error within the energy tolerance, and it must contain at least one two-qubit gate (ruling out the trivial empty-circuit optimum). If these criteria are met, we compute advantages over all 20 circuits in the batch and apply one GRPO/GSPO policy-gradient update. If the gate fails, the iteration is logged but no gradient is back-propagated and the next iteration resamples.

\subsection{Recovery and global-merge subspace construction}
\label{sec:recovery}

Sampled bitstrings obtained from physical quantum processors are typically sparse and corrupted by device noise, precluding their direct use as $S$ in Eq.~\ref{eq:subspace-diag}.
Two complementary subspace-construction mechanisms are employed to resolve this limitation.

\paragraph{Self-consistent configuration recovery.} Given raw sampled
counts $c(a,b)$, an iterative recovery loop (following the
configuration-recovery procedure of the Qiskit SQD addon \cite{robledomoreno2024chemistry,javadiabhari2024qiskit}; code labels in Table~\ref{tab:glossary}) alternates
between (i) estimating a running per-orbital occupancy vector
$\boldsymbol{\mu}^{(k)}$ for each spin from the current best determinant set, (ii)
repairing the observed bitstrings, in both the $\alpha$ and the $\beta$ half, against the occupancy of each spin and post-selecting those whose halves have the correct electron numbers $n_\alpha$ and $n_\beta$ (Hamming weight), (iii)
subsampling down to a fixed target number $D$ of strings per spin sector (the recovery subspace size)---where the
Hartree--Fock determinant and carryover determinants from the prior step
are enforced, with the remainder drawn from the recovered distribution; for the closed-shell molecules of this work the repaired $\alpha$ and $\beta$ halves are first folded into one list of distinct strings, used for both spin sectors---and (iv)
diagonalizing the determinant subspace $\mathcal A_k\times\mathcal B_k$ with $\lvert\mathcal A_k\rvert=\lvert\mathcal B_k\rvert=D$, whose net dimension is $D^2$ (Eq.~\ref{eq:subspace-diag}), on GPU. Here $D$ counts strings per spin sector: $D=200$ at the ranking stage (40{,}000 determinants) and $D=20{,}000$ at the final recovery ($4\times10^{8}$ determinants, capped by the number of strings that exist). The lowest-energy determinant subspace
over $K$ recovery iterations is retained, with each step
seeding the subsequent occupancy estimate and carryover list.
In a variant called K-batch SQD~\cite{robledomoreno2024chemistry}, $B$ independent subspaces are drawn at each step from the same recovered distribution and only the one with the lowest energy is kept; this costs $B$ diagonalizations per step and gives a steadier occupancy estimate for the next step.

\paragraph{Global merge.} Across \emph{training} iterations (as opposed to
within one circuit's own recovery loop), two global-refinement policies
determine how much of the recovery history is retained and merged with the
current batch's local result: a \emph{bounded} mode keeps the union of the
previous retained state and the current local state, capped at a fixed
refine dimension, while a cumulative mode keeps the union of the
previous retained state, the current local state, \emph{and} every
per-circuit state sampled this iteration, capped only by an overall
cumulative dimension. The merged state does not enter the training reward. The reward of each circuit is its own recovered energy, computed from that circuit's samples alone (Eq.~\ref{eq:subspace-diag}), before any merging. The merge is applied afterwards to the batch, and its energy is recorded as the best energy of the iteration, which is reported but not used as a reward or in the admission gate (which tests the best circuit's own raw error). At inference, the merge pools the shots of several circuits into the reported "merged" result (Section~\ref{sec:zero-shot}).

\subsection{Model architecture}
\label{sec:architecture}

\begin{figure}[!t]
\centering
\includegraphics[width=\columnwidth]{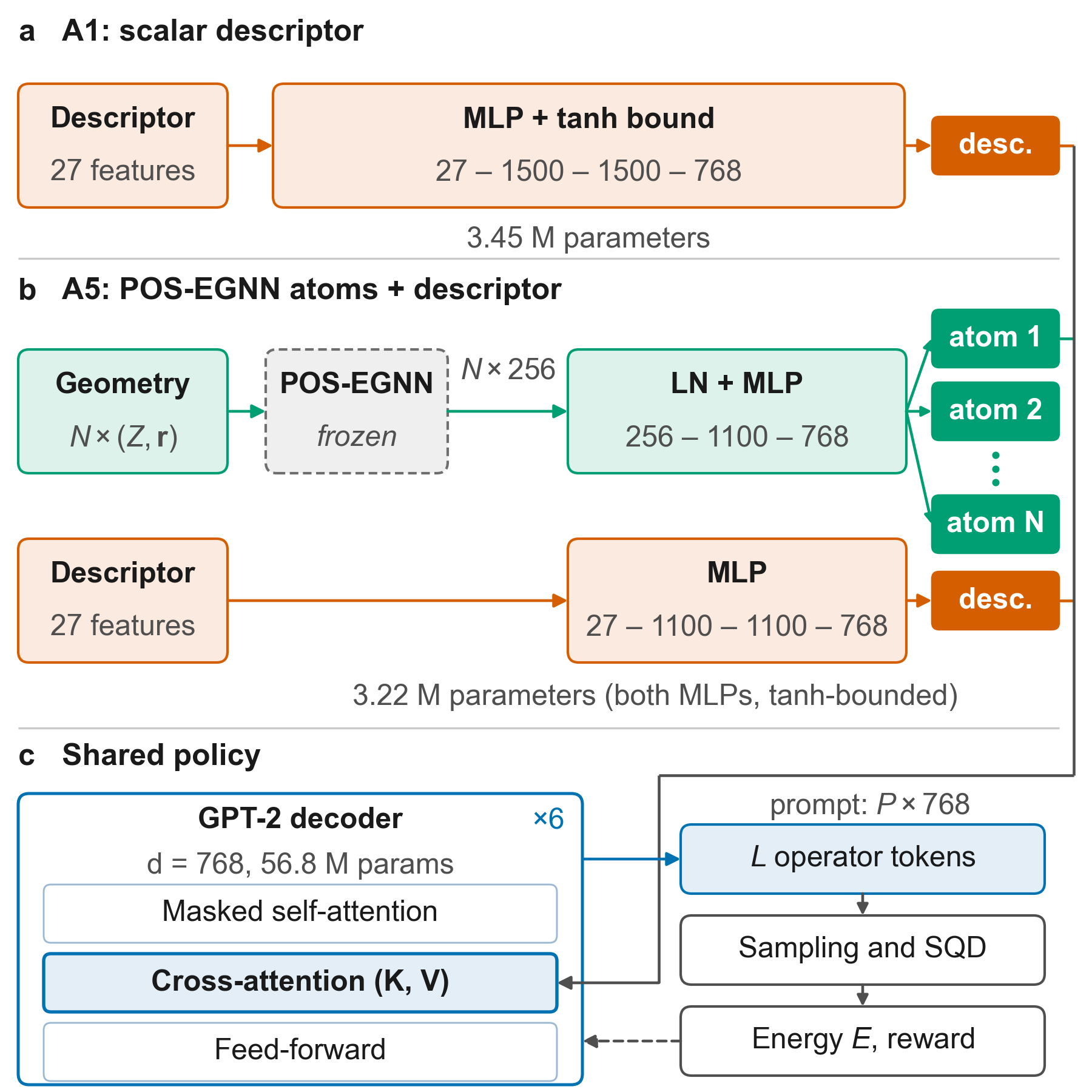}
\caption{Conditioning encoders and shared policy. \textbf{(a) A1.} The 27-dimensional active-space descriptor is mapped by a three-layer MLP (27--1500--1500--768, GELU, 3.45~M parameters) to a single 768-dimensional prompt token; the output is bounded element-wise as $3\tanh(x/3)$. \textbf{(b) A5.} The $N$ atoms of the molecule (atomic numbers $Z$ and positions $\mathbf r$) are embedded by the pre-trained POS-EGNN \cite{ibm2025posegnn,deng2023chgnet} into $N\times256$ invariant per-atom features. This network is frozen: its output is computed once per geometry and stored. The features are layer-normalised and mapped by a two-layer MLP (256--1100--768) to $N$ atom tokens. Independently, the descriptor (the A1 features, with the HOMO--LUMO gap and the mean and standard deviation of the orbital energies recomputed from the eigenvalues of the active-space one-body Hamiltonian) is mapped by a separate MLP (27--1100--1100--768) to one further token. The $N+1$ tokens are bounded as in A1 and kept as separate prompt tokens; the two MLPs have 3.22~M parameters in total. \textbf{(c) Shared policy.} Both arms feed the same GPT-2 decoder (6 layers, 6 heads, embedding dimension 768, $\approx$56.8~M parameters). In every layer, masked self-attention over the gate tokens generated so far is followed by a cross-attention block whose keys and values (K, V) are the prompt tokens ($P=1$ for A1, $P=N+1$ for A5) and a feed-forward block. The decoder samples $L$ operator tokens ($L=28$ for the 46-qubit and 44 for the 70-qubit lineages), which are assembled into a circuit, sampled and processed by SQD; the resulting energy defines the reward. The dashed arrow marks the policy-gradient update of the decoder and of the MLP encoders. Vermillion: descriptor stream; green: atom stream; blue: policy; grey dashed: frozen network.}
\label{fig:encoder-arch}
\end{figure}

The policy $\pi_\phi$ is a GPT-2-style autoregressive transformer decoder with 6 layers, 6 attention heads and about 56.8 million parameters. The decoder follows the GPT-2 architecture~\cite{radford2019gpt2} that GPT-QE uses for GQE~\cite{nakaji2024gqe}; the cross-attention to a molecule prompt is our addition. We did not tune the model size. Its cross-attention layers let the decoder attend to a molecule-specific prompt of variable length, instead of a single conditioning vector. Its position embeddings cover the $L+1$ positions of the gate-token sequence (a start token and $L$ gate tokens), where $L$ is defined in Section~\ref{sec:gqe-formalism}. The prompt enters only through cross-attention and carries no position embedding. The decoder is shared by the two conditioning schemes of this work, called A1 and A5. They differ only in how the prompt is built from the molecule; both are described in detail below and shown in Figure~\ref{fig:encoder-arch}.

\paragraph{A1: scalar cross-attention conditioning.} A fixed 27-dimensional
hand-built active-space descriptor $\mathbf{d}\in\mathbb{R}^{27}$ (per-system
scalars: orbital/electron counts, HF/CASCI-adjacent energetics, and related
active-space statistics) is mapped by a three-layer MLP (hidden width 1500,
GELU activations, 3.45~M parameters) to a single prompt token of the decoder's
embedding dimension $E=768$, with the output bounded element-wise by a
scaled hyperbolic tangent (bound $b=3$):
\begin{equation}
  \mathbf{p}_{\mathrm{A1}} = b\,\tanh\!\big(\mathrm{MLP}_{27\to E}(\mathbf d)/b\big) \in \mathbb{R}^{1\times E}.
\end{equation}
The bound caps the scale of the cross-attention keys and values, which would
otherwise grow without limit as the encoder weights train. This one-token prompt
is computed once per molecule and cached and cross-attended at every one
of the $L$ autoregressive decoding steps.

\paragraph{A5: POS-EGNN structural + scalar cross-attention conditioning.}
POS-EGNN (position-based equivariant graph neural network) is a model released
by IBM Research within its Foundation Models for Materials collection
\cite{ibm2025posegnn}. It was pre-trained as a machine-learning potential, predicting
energies, forces and stresses, on about 1.4 million structures of the Materials Project
trajectory data set MPtrj \cite{deng2023chgnet}, and its invariant per-atom features can be
extracted for downstream tasks. We use it as a frozen feature extractor: for a molecule
with $N$ atoms and geometry $\{(Z_i,\mathbf r_i)\}$ it yields 256-dimensional invariant
embeddings $\mathbf e_i\in\mathbb{R}^{256}$, $i=1,\dots,N$. Because the geometry is fixed per molecule,
these embeddings are computed once, in a separate environment (POS-EGNN pins a different PyTorch
version than the training environment), and stored; they are neither recomputed nor updated during training.
The embeddings are layer-normalized and projected by a two-layer MLP (256--1100--768) to $N$ prompt
tokens, and the descriptor, independently projected with separate weights by a three-layer MLP
(27--1100--1100--768), gives one additional token. The descriptor $\mathbf d_{\mathrm{A5}}$ contains the same 27 features
as in A1 except that the three orbital-energy entries are recomputed from the eigenvalues of the active-space one-body
Hamiltonian, which repairs a defect of the original entries for systems loaded from integral files. The two MLPs have
3.22~M parameters in total.
\begin{equation}
\begin{aligned}
  \mathbf{p}_{\mathrm{A5}} &= b\,\tanh\!\big([\,\mathbf q_1;\dots;\mathbf q_N;\,\mathbf q_d\,]/b\big)
     \in \mathbb{R}^{(N+1)\times E},\\
  \mathbf q_i &= \mathrm{MLP}_{256\to E}\big(\mathrm{LN}(\mathbf e_i)\big),\qquad
  \mathbf q_d = \mathrm{MLP}_{27\to E}(\mathbf d_{\mathrm{A5}}),
\end{aligned}
\end{equation}
i.e.\ a prompt of $N{+}1$ tokens, kept as separate tokens
(rather than pooled into one vector) so that the decoder's
cross-attention, not a fixed pooling operator, determines which atoms
matter for a given decoding step. Both token streams pass through the same
bound $b=3$ as in A1, so neither stream can dominate cross-attention through an
unbounded activation scale. The prompt tokens carry no position or token-type embedding: the atom tokens are an unordered set, and the descriptor token differs from them only through its separate projection.

\subsubsection{Data flow through the policy}
 
\paragraph{Input $\to$ output, training.} For a given molecule, both arms'
encoders and the shared decoder are run forward together: build the prompt
($\mathbf p_{\mathrm{A1}}$ or $\mathbf p_{\mathrm{A5}}$) from the molecule's
descriptor (and geometry, for A5) $\to$ cache it $\to$
autoregressively sample $N_{\mathrm{samples}}$ gate-token sequences of
length $L$, cross-attending the cached prompt at every step $\to$ assemble
each sequence into a circuit via the operator pool $\mathcal P$ $\to$
sample/simulate each circuit (Stage 1 or Stage 2,
Section~\ref{sec:training}) $\to$ build and recover a subspace per circuit
(Section~\ref{sec:recovery}) $\to$ admission gate $\to$ for admitted
iterations (those whose best circuit passes), backpropagate the GRPO/GSPO loss (Eq.~\ref{eq:grpo}) using all $N=20$ circuits in the batch through
\emph{both} the decoder and the active encoder (A1's MLP or A5's two MLPs),
since the prompt is kept attached to its autograd graph rather than
detached. Both the decoder and the conditioning encoder are therefore
trained jointly, end to end, from the SQD energy signal alone.

\paragraph{Input $\to$ output, inference.} Identical forward path through
the encoder and decoder, with every weight frozen (no backward pass, no
optimizer step): molecule descriptor (and geometry, for A5) $\to$ prompt
$\to$ sampled circuits $\to$ one QPU sampling round trip (cached,
Fig.~\ref{fig:flow-inference}) $\to$ ranking and selection of the best, second,
merged and E+D circuits $\to$ recovery/diagonalization at the reported
subspace size $\to$ final energy and excitation histograms. The ten circuits are ranked by a one-shot diagonalization of their valid shots; \emph{best} and \emph{second} are the two circuits with the lowest ranking energy, \emph{merged} pools the shots of all circuits with a finite ranking energy, and \emph{E+D~1} and \emph{E+D~2} are the two circuits that combine the rank by energy (E) with the rank by the number of distinct measured bitstrings (D). An \emph{untrained} control uses the same path with randomly initialized decoder and encoder weights instead of a checkpoint, so that the difference to a trained policy shows what the training contributed beyond the architecture and the ranking and recovery steps. We ran eight such controls, on three molecules (SnOH3-me L, SnOH3-me-h2o M and SnOH3-me LL).

The operator pool $\mathcal{P}$ comprises local
two-orbital gates (Givens rotations, which have single-excitation character, and number--number phase gates, each acting on
one spatial-orbital pair, i.e.\ four qubits including both spin sectors)
together with full-coverage operators (four fixed orbital rotations, two
diagonal-Coulomb layers and one LUCJ layer) that touch every qubit
in the active space. For a molecule with $n_{\mathrm{orb}}$ spatial orbitals, the pool has $3\,n_{\mathrm{orb}}+5$ operators: 74 for the 23-orbital (46-qubit) and 110 for the 35-orbital (70-qubit) molecules. The vocabulary of the transformer is fixed at 110 for all molecules, with the remaining token ids acting as identity padding when $3\,n_{\mathrm{orb}}+5<110$ (token map in Table~\ref{tab:token-map}).

\begin{table}[t]
\centering
\caption{Token map of the operator pool (\code{FfsimLUCJPool}) for a molecule with $n=n_{\mathrm{orb}}$ spatial orbitals. Token ids above $3n+4$ are identity padding to reach the fixed vocabulary size 110.}
\label{tab:token-map}
\small
\begin{tabular}{@{}p{2.8cm}p{6.5cm}@{}}
\toprule
Token id(s) & Operator \\
\midrule
0 & identity \\
1 & LUCJ layer \\
2--5 & orbital rotations (four fixed seeds) \\
6--$2n+3$ & Givens rotations on neighbouring pairs \\
& $(p,p{+}1)$ (two angles per pair) \\
$2n+4$, $2n+5$ & diagonal-Coulomb layers (two time values) \\
$2n+6$--$3n+4$ & number--number phase gates on neighbouring \\
& pairs $(p,p{+}1)$ (one gate per pair) \\
\midrule
Total & $3n+5$ operators \\
\bottomrule
\end{tabular}
\end{table}

\subsection{Molecules and active spaces}
\label{sec:molecules}

\paragraph{Training curriculum.} A size-ordered curriculum of small and
medium molecules (H$_2$, LiH, H$_2$O, BeH$_2$, NH$_3$, CH$_4$, N$_2$,
CH$_2$O, HCN, C$_2$H$_4$, CO$_2$) together with three transition-metal
complexes --- Fe(CN)$_6$, Co(NH$_3$)$_6$, and ferrocene --- at a small,
$\leq$16-orbital active-space tier for Stage-1 pretraining, and at a
23-orbital (46-qubit) or 35-orbital (70-qubit) tier for the target of
Stage-2 continuation (Section~\ref{sec:training}). The molecules are ordered by qubit count, which sets the length of the generated sequence and the size of the operator pool and the cost of the simulation; we did not order them by the difficulty of sampling their ground states.

\paragraph{Zero-shot/held-out target.} The SnOH3-me family (and its
mono- and di-hydrate adducts) is used as a held-out target,
at active-space tiers S/M/L/LL spanning 24--50 electrons and up to
35 orbitals (70 qubits), with DMRG references available at every tier
(Table~\ref{tab:dmrg-reference}).

\begin{figure}[H]
\centering
\includegraphics[width=\columnwidth]{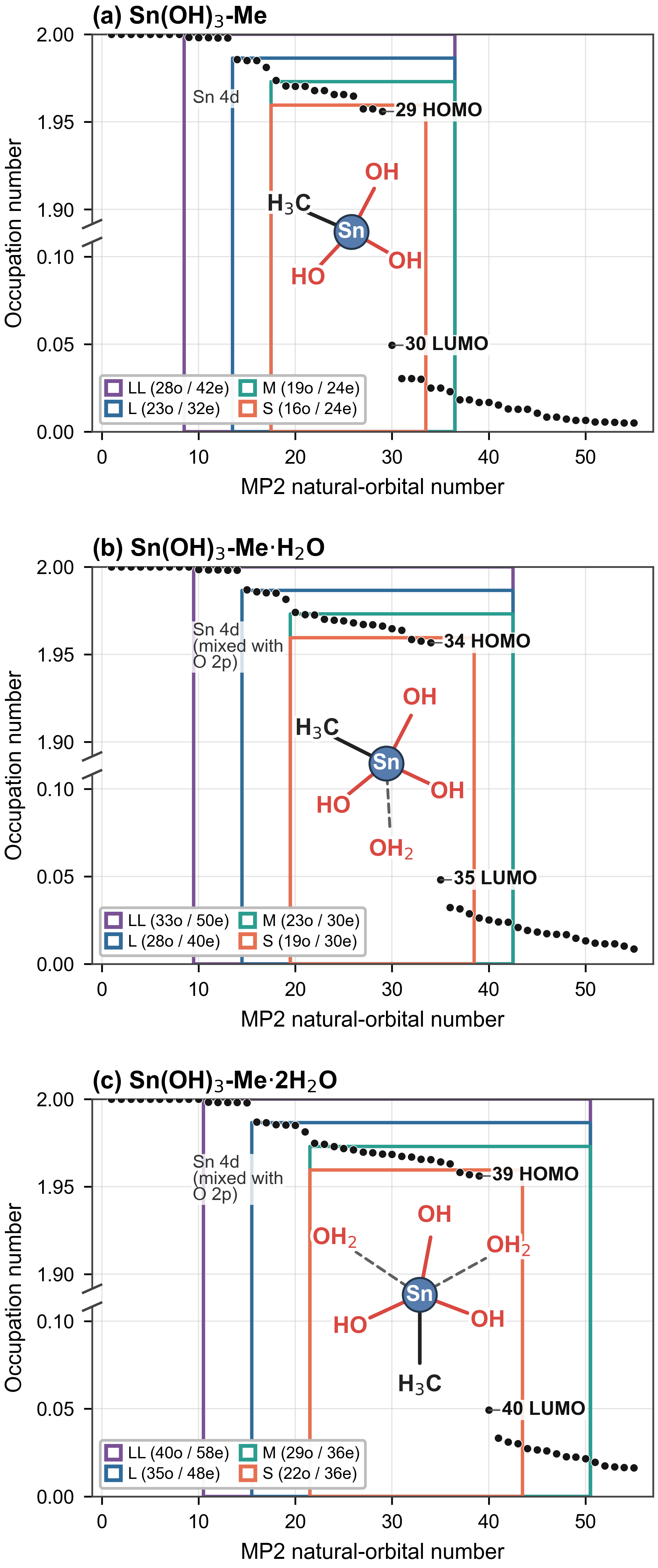}
\caption{MP2 natural-orbital occupation numbers (MP2/def2-SVP, frozen core) of (a) Sn(OH)$_3$-Me, (b) Sn(OH)$_3$-Me$\cdot$H$_2$O and (c) Sn(OH)$_3$-Me$\cdot$2H$_2$O, with the S, M, L and LL active spaces boxed (orbitals, electrons; the natural-orbital ranges are in Table~\ref{tab:active-spaces}). The orbitals labelled Sn~4d are mixed with O~2p. The occupation axis is compressed between 0.1 and 1.9. The sketch in each panel is the structure of the molecule (Sn blue, O red, C black; dashed: water bound to tin). In the optimized geometry used for the calculations the first water of the dihydrate is coordinated to tin (Sn--O 2.4~\AA{}) and the second lies 3.5~\AA{} from tin, hydrogen-bonded to two hydroxyl groups.}
\label{fig:no-occupation}
\end{figure}

\paragraph{Held-out molecules and how their active spaces were chosen.} The held-out targets are Sn(OH)$_3$-Me and its mono- and di-hydrate adducts, Sn(OH)$_3$-Me$\cdot$H$_2$O and Sn(OH)$_3$-Me$\cdot$2H$_2$O. Sn(OH)$_3$-Me is the smallest model of a monoorganotin hydroxide, the Sn--C and Sn--O motif of the organotin oxo-hydroxo clusters discussed in the introduction (Section~\ref{sec:intro}); in the optimized structures one water is coordinated to tin in each hydrate (Sn--O 2.45~\AA{} in the monohydrate, 2.41~\AA{} in the dihydrate; the Sn--OH distances are 2.0~\AA{}) and is hydrogen-bonded to two hydroxyl groups, while the second water of the dihydrate lies in the second shell (Sn--O 3.53~\AA{}) and is hydrogen-bonded to two hydroxyl groups, so that the water changes the environment of tin while keeping the chemistry the same, and together the three molecules give a ladder of active spaces from 32 to 70 qubits with DMRG references at every system size (Table~\ref{tab:dmrg-reference}). The structures were optimized and shown, by vibrational-frequency analysis, not to be saddle points; the MP2 natural orbitals were computed with Gaussian~16 \cite{gaussian16} at the frozen-core MP2 level \cite{moller1934} in the def2-SVP basis \cite{weigend2005}, and the active spaces are built on these natural orbitals \cite{lowdin1955}, not on Hartree--Fock canonical orbitals. The natural orbitals are ordered by occupation number (Figure~\ref{fig:no-occupation}). An active space is a contiguous range of them: it starts at the lowest occupied natural orbital of the range, runs through the HOMO and includes the first virtual natural orbitals. Eleven tiers were defined (Table~\ref{tab:active-spaces}): S, M, L and LL for Sn(OH)$_3$-Me and for the monohydrate, and S, M and L for the dihydrate (there is no LL tier for the dihydrate). S and M share their lowest orbital (occupation 1.974--1.975) and differ in the number of virtual orbitals; L and LL add occupied orbitals whose occupation is closer to 2 (lowest orbital of L: 1.986--1.987; of LL: 1.998), and only LL reaches down to the orbitals labelled Sn~4d, which mix with O~2p (Figure~\ref{fig:no-occupation}). The occupation-number threshold that ends the virtual range is not well defined; the tiers use the first 4 (S) or 7 to 11 (M, L, LL, depending on the molecule) virtual natural orbitals. Whether the Sn~4d orbitals must be in the active space is not settled by this work. The natural orbitals themselves are shown in Figures~\ref{fig:no-orbitals-snoh3me}, \ref{fig:no-orbitals-h2o} and~\ref{fig:no-orbitals-2h2o} and discussed in Appendix~\ref{sec:appendix-orbitals}. Natural orbitals were preferred to selections by Hartree--Fock orbital energy because they are more localized on bonds and atoms and so have more distinct characters (an assessment made when the active spaces were chosen; the two selections were not benchmarked against each other here).

\subsection{Training procedure}
\label{sec:training}

Training proceeds in two stages per (architecture, qubit-count-target) pair.

\paragraph{Stage 1 --- noiseless-simulator warmup.} The policy is trained from scratch, that is, from random weights, with an exact-statevector (\code{ffsim}) sampler on the 
$\leq$32-qubit curriculum (Section~\ref{sec:molecules}), using
$L=28$ gate tokens per circuit for the 46-qubit-target lineage and $L=44$
for the 70-qubit-target lineage (gate count cannot change once a lineage is
fixed, since it resizes the transformer's position-embedding table). In both stages every generated circuit acts on the Hartree--Fock reference state, so the policy only chooses the operators applied to it. The policy's vocabulary is 110 operators from the first iteration, the size of the pool of the 35-orbital targets, so that the output head and token embedding keep their shape in Stage~2.
Stage~1 runs for a fixed 150 iterations, without early stopping. This stage is run once per architecture
(A1 or A5) and per circuit-length lineage ($L=28$ or $44$), not per target molecule, since
the exact-statevector sampler's memory footprint scales as $\binom{n_o}{
n_\alpha}\binom{n_o}{n_\beta}$ and becomes intractable above $n_o\approx 23$
--- the curriculum therefore never includes the actual 46-/70-qubit target
molecules in Stage~1. This Stage~1 is shared by the Stage-2 configurations described below.

\paragraph{Stage 2 --- physical hardware continuation.} The Stage-1 checkpoint
initialises the policy for 150 further iterations on the target molecules
(the 23- or 35-orbital tier of Fe(CN)$_6$/Co(NH$_3$)$_6$/ferrocene) with a
shot-based sampler executing on the IBM quantum processing unit (QPU) \code{ibm\_kobe} (Heron
generation), rather than the statevector
simulator, since Stage-1's qubit-count memory ceiling does not constrain shot-based
hardware sampling. Circuits were submitted at 10{,}000 shots per circuit, batched
20 circuits per Qiskit Runtime job, in standard job execution mode
rather than a persistent Runtime session to prevent prolonged device locking
on shared multi-user allocations. The Stage-2 reward adds the Hartree--Fock penalty (Section~\ref{sec:reward-tuning}).

\paragraph{What the warm start carries over.} The vocabulary is 110 in both stages, so every policy tensor keeps its shape: the transformer layers, the position embedding, the cross-attention, the token embedding, the output head and the prompt encoder are all carried from Stage~1 into Stage~2. The price of the larger vocabulary is that operator ids above the $3\,n_{\mathrm{orb}}+5$ real operators of a molecule's pool act as identity padding; on a 16-orbital Stage-1 molecule, $3\cdot16+5=53$ ids are real and 57 of 110 ids are padding (52\%). Inference restores the whole checkpoint, including the prompt encoder. The Stage-1 and Stage-2 histories are single runs of 150 iterations each (Appendix~\ref{sec:appendix-training-curves}).

\subsubsection{The two reward configurations}
\label{sec:pipelines}

Stage 2 is run with two different rewards, which define the two training configurations (Table~\ref{tab:pipelines}). Everything else is the same: the architecture, the optimiser (learning rate $5\times10^{-6}$, gradient-norm clipping at 8.0), the size curriculum, the seed (1234), the vocabulary of 110 operators, the fixed budget of 150 iterations per stage without early stopping, and the sampler and recovery settings. Both configurations are trained for the four lineages (A1 and A5 conditioning, each for 46 and for 70 qubits; Section~\ref{sec:architecture}).

\begin{description}[leftmargin=1.4em,style=nextline,itemsep=3pt]
  \item[\textbf{Hartree--Fock penalty} (the main configuration).] The Stage-2 reward contains a penalty $\lambda_{\mathrm{HF}}=0.05$ on the share of right-spin-sector shots that is the Hartree--Fock determinant. It lowers the reward of circuits whose valid shots are mostly the reference determinant, which a policy without this term converges to on the hardware (Appendix~\ref{sec:appendix-chemistry}). The final checkpoint is used at inference. The Stage-1 policy is shared with the second configuration.
  \item[\textbf{Tuned reward} (a comparison).] The Stage-2 reward weights were chosen by a joint search on a noisy 24-qubit simulator (Section~\ref{sec:reward-tuning}): $\lambda_{\mathrm{HF}}=0.07$, a spin-balance term $\lambda_{\mathrm{spin}}=0.77$, an entropy bonus $c_H=0.26$, $\lambda_{\mathrm{exc}}=3.9$ and $\lambda_{\mathrm{cov}}=7.2$. For A1-70q the policy used at inference is the one saved at iteration 129, because the run ended by a segmentation fault at iteration 137.
\end{description}

The Hartree--Fock penalty is the main configuration because it is the simpler reward and, on the 43 test cases that both have, it gives the more reliable result at inference (Section~\ref{sec:three-model-inference}). A third variant without any penalty was completed for one lineage only and is used in the training comparisons of the appendices, not at inference.

\begin{table}[!t]
\centering
\caption{The two training configurations (Section~\ref{sec:pipelines}); everything not listed is identical. The meaning of each reward weight is in Table~\ref{tab:t2-reward}. A variant without a penalty ($\lambda_{\mathrm{HF}}=0$, $\lambda_{\mathrm{exc}}=2.0$, $\lambda_{\mathrm{cov}}=10.0$) was completed for A1-46q only (the other lineages were stopped after 17 or 18 iterations).}
\label{tab:pipelines}
\small
\renewcommand{\arraystretch}{1.25}
\setlength{\tabcolsep}{4pt}
\begin{tabular}{@{}p{0.34\linewidth}p{0.28\linewidth}p{0.3\linewidth}@{}}
\toprule
 & \textbf{Hartree--Fock penalty} (main) & \textbf{Tuned reward} \\
\midrule
Policy vocabulary, Stage 1 / Stage 2 & 110 / 110 & 110 / 110 \\
Iterations, Stage 1 / Stage 2 & 150 / 150 & 150 / 150 \\
$\lambda_{\mathrm{HF}}$ & 0.05 & 0.07 \\
$\lambda_{\mathrm{spin}}$, $c_H$ & 0, 0 & 0.77, 0.26 \\
$\lambda_{\mathrm{exc}}$, $\lambda_{\mathrm{cov}}$ & 2.0, 10.0 & 3.9, 7.2 \\
Checkpoint used at inference & final & final (A1-70q: iteration 129) \\
\bottomrule
\end{tabular}
\end{table}

\subsection{Hyperparameters and tuning}
\label{sec:hyperparams}

\paragraph{Settings.} Table~\ref{tab:hyperparams} lists the settings of the four lineages, Table~\ref{tab:t2-reward} the Stage-2 reward weights of the two configurations (Table~\ref{tab:pipelines}), and Table~\ref{tab:infer-hyperparams} the settings used at inference. Appendix~\ref{sec:appendix-glossary} defines every setting by its name in the code (Table~\ref{tab:glossary}), and Appendix~\ref{sec:appendix-tuning} describes the preliminary runs that fixed them, including the depth sweep that set the admission gate. The optimiser, the size curriculum, the seed and the recovery settings are the same for all runs.

\subsubsection{The reward and its weights}
\label{sec:reward-tuning}

\paragraph{The adjusted energy.} For each of the $N=20$ circuits sampled at an iteration, the policy is trained on an adjusted energy
\begin{equation}
\begin{aligned}
E_i^{\mathrm{adj}} \;=\; E_i \;&+\; g_i\,\bigl(\lambda_{\mathrm{exc}}\,x^{\mathrm{exc}}_i + \lambda_{\mathrm{cov}}\,x^{\mathrm{cov}}_i\bigr)
\;+\; \lambda_{\mathrm{over}}\,x^{\mathrm{over}}_i \\
&+\; \lambda_{\mathrm{HF}}\,x^{\mathrm{HF}}_i \;+\; \lambda_{\mathrm{spin}}\,x^{\mathrm{spin}}_i ,
\end{aligned}
\label{eq:adjusted-energy}
\end{equation}
where $E_i$ is the SQD energy of the circuit (Hartree) and $g_i\in\{0,1\}$ is one when $E_i$ is within the auxiliary-term threshold (50~mHa) of the reference energy, distinct from the gradient-admission threshold of 100 or 500~mHa (Table~\ref{tab:hyperparams}), and zero otherwise. The
group-relative objective then subtracts the batch mean and divides by the batch standard deviation of $E^{\mathrm{adj}}$, so only the differences between circuits of one batch
matter, not the absolute scale. Table~\ref{tab:t2-reward} lists the five weights and the entropy coefficient. The terms are:
\begin{itemize}[leftmargin=1.5em,itemsep=2pt]
  \item $x^{\mathrm{exc}}_i$: the share of sampled determinants with excitation rank above 3 relative to the Hartree--Fock determinant, from the raw counts. It favours a wavefunction dominated by singles, doubles and triples. Gated, because for a circuit far from the reference energy the rank distribution is not informative.
  \item $x^{\mathrm{cov}}_i$: the share of qubits that the transpiled circuit leaves idle. It pushes the policy to use the qubits of the molecule. Gated.
  \item $x^{\mathrm{over}}_i=\log_{10}(d_i/d_{\max})$ when the transpiled depth $d_i$ exceeds the ceiling $d_{\max}$ (Table~\ref{tab:hyperparams}), zero otherwise. It is the one depth term that is not gated, because on a hard molecule no circuit reaches the gate.
  \item $x^{\mathrm{HF}}_i$: the share of right-spin-sector shots (those with the target alpha and beta electron counts) that are exactly the Hartree--Fock determinant. It penalizes a circuit that leaves the reference state unchanged, which the policy converged to in the first Stage-2 run. Ungated.
  \item $x^{\mathrm{spin}}_i=|w_\alpha-w_\beta|$, where $w_\alpha$ ($w_\beta$) is the share of right-spin-sector shots with an excitation in the alpha (beta) channel only. It is zero for a balanced state. Ungated.
  \item The entropy coefficient $c_H$ acts on the loss and not on the reward, $\mathcal{L}\leftarrow\mathcal{L}-c_H\,\overline{H}$, with $\overline{H}$ the mean token entropy of the sampled sequences. It counters the collapse of the policy onto one sequence.
\end{itemize}
The reasons for the Hartree--Fock, spin and entropy terms are given in Appendix~\ref{sec:appendix-tuning}.
The three optional penalties on a proxy gate count, on the diagonalisation subspace size and on the transpiled depth (Table~\ref{tab:glossary}) are zero throughout.

\paragraph{Reference energy in training.} Each training molecule is associated with an approximate classical reference energy (CCSD(T) for the Stage-2 molecules). This reference enters only through the admission gate and two auxiliary reward terms (Section~\ref{sec:training}); the policy-gradient update depends on it only through these binary decisions, and not through backpropagation of the reference energy itself. At 70 qubits the admission gate passed 99--100\% of iterations once the tolerance was loosened (Section~\ref{sec:stage2-results}), so in that regime it constrained the updates only weakly. We do not estimate an uncertainty of the reference energies. For systems where a suitable classical reference is unavailable, a different gating strategy would be required; we do not test such a reference-free protocol here. Inference uses no reference energy.

\paragraph{How the weights were chosen.} The depth and coverage weights were fixed in preliminary runs, where the depth ceiling was the dominant lever of the admission gate. The Hartree--Fock weight was set to 0.05 in a test on a 24-qubit simulator with readout noise: it lowered the Hartree--Fock share of the valid shots without raising the raw error, whereas larger values removed that share only by moving the policy onto a single excited determinant. Because the five weights interact, we then searched them jointly (42 configurations, two seeds each) on a noisy 24-qubit simulator, and took the tuned reward from the configurations that met our criteria on raw error, Hartree--Fock share, policy diversity and circuit content. Two seeds cannot separate the passing configurations, so the search supports a region of the weight space and not one point in it, and it does not show that the tuned weights improve the hardware result, which is why the tuned reward is a comparison and the Hartree--Fock penalty is the main configuration (Section~\ref{sec:three-model-inference}). The search, its criteria and its limits are given in Appendix~\ref{sec:appendix-tuning}.

\paragraph{Comparison protocol.} To compare the two rewards at inference, we ran the same zero-shot protocol with the policies of each configuration: the final policy of each run (for A1-70q with the tuned reward, the policy saved at iteration 129 of 137), the same lineages and molecules, one batch of ten circuits sampled at 100{,}000 shots, ranking at net dimension 40{,}000, and the same recoveries. A variant without any penalty is not evaluated, because three of its four Stage-2 runs were stopped after at most 18 iterations.

The remaining optimizer and curriculum settings (the learning rate, the gradient-norm clipping threshold, the fraction of iterations spent on the largest molecules, and the fraction of molecules counted as ``largest''; exact definitions and code labels are in Table~\ref{tab:glossary}, Appendix~\ref{sec:appendix-glossary}) were tuned to varying degrees of rigor, distinguished in Table~\ref{tab:tuning-narrative}. The fraction of molecules counted as ``largest'' was set by a controlled pair of runs with a clear, measured winner. The gradient-clipping threshold was recalibrated by measuring the policy's gradient-norm distribution rather than by a grid sweep. The learning rate's existing default was validated against, and beat, one higher alternative, and was not derived from a wider search. The fraction of iterations spent on the largest molecules was changed together with the fraction of molecules counted as ``largest'' and was never isolated in its own test.

\begin{table*}[h]
\centering
\caption{Final hyperparameters, per lineage. All four lineages
share the same optimizer settings (base learning rate $5{\times}10^{-6}$,
encoder LR multiplier 1.0, weight decay 0.01, grad clip 8.0) and the same
size-curriculum schedule (warmup/focus/remainder fractions 0.25/0.30/0.45,
top-40\% molecules by qubit count treated as the curriculum's ``large set'');
only the values shown here differ across lineages. The recovery size $D$ is the number of strings per spin sector kept during training (same quantity as at inference time, Table~\ref{tab:infer-hyperparams}); the resulting determinant subspace contains up to $D^2$ determinants (Section~\ref{sec:recovery}). Code labels for every row are in Table~\ref{tab:glossary}.}
\label{tab:hyperparams}
\footnotesize
\begin{tabular}{@{}lcccc@{}}
\toprule
Setting (code labels: Table~\ref{tab:glossary}) & A1-46q & A1-70q & A5-46q & A5-70q \\
\midrule
Conditioning & scalar27 & scalar27 & POS-EGNN+s27 & POS-EGNN+s27 \\
Gate tokens per circuit, $L$ & 28 & 44 & 28 & 44 \\
Policy params (M) & 56.79 & 56.83 & 56.79 & 56.83 \\
Encoder params (M) & 3.45 & 3.45 & 3.22 & 3.22 \\
Depth ceiling (Stage 1) & 10000 & 20000 & 10000 & 20000 \\
Depth ceiling (Stage 2) & 10000 & 50000 & 10000 & 50000 \\
Coverage penalty weight $\lambda_{\mathrm{cov}}$ & 10.0 & 10.0 & 10.0 & 10.0 \\
Depth-overshoot penalty weight $\lambda_{\mathrm{over}}$ & 2.0 & 2.0 & 2.0 & 2.0 \\
Excitation penalty weight $\lambda_{\mathrm{exc}}$ & 2.0 & 2.0 & 2.0 & 2.0 \\
Energy admission tolerance, mHa (Stage 1) & 100 & 100 & 100 & 100 \\
Energy admission tolerance, mHa (Stage 2) & 100 & 500 & 100 & 500 \\
Recovery size $D$ (strings per spin; Stage 1) & 20000 & 20000 & 20000 & 20000 \\
Recovery size $D$ (strings per spin; Stage 2) & 40000 & 40000 & 40000 & 40000 \\
Shots per circuit (Stage 2) & 10000 & 10000 & 10000 & 10000 \\
Stage-2 backend & ibm\_kobe & ibm\_kobe & ibm\_kobe & ibm\_kobe \\
\bottomrule
\end{tabular}
\end{table*}

\begin{table*}[ht]
\centering
\caption{Reward and loss weights of the Stage-2 runs; Section~\ref{sec:reward-tuning} defines every term, and Appendix~\ref{sec:appendix-tuning} describes how the weights were tuned.
The first five terms add to the per-circuit reward energy $E_i^{\mathrm{adj}}$ (Hartree); the entropy coefficient enters the loss. A ``gated'' term is applied only when the
circuit's energy is within the auxiliary-term threshold (50~mHa; Table~\ref{tab:glossary}), distinct from the gradient-admission threshold of 100 or 500~mHa (Table~\ref{tab:hyperparams}), of the reference energy; the other terms always apply. ``Reference'' is the setting of the preliminary runs and of
the runs without a penalty, ``HF penalty'' changes only $\lambda_{\mathrm{HF}}$, and ``Tuned'' is the configuration selected by the joint search of
Section~\ref{sec:reward-tuning} (rounded to the values submitted). The search range is the interval explored. The optional penalties on a proxy gate count, on the diagonalisation subspace size and on the transpiled depth (Table~\ref{tab:glossary}) are zero in every run. The code label of each weight is the entry in the ``Flag'' column; all labels are defined in Table~\ref{tab:glossary}.}
\label{tab:t2-reward}
\small
\renewcommand{\arraystretch}{1.25}
\setlength{\tabcolsep}{4pt}
\begin{tabular}{@{}>{\raggedright\arraybackslash}p{0.5in}>{\raggedright\arraybackslash}p{1.3in}>{\raggedright\arraybackslash}p{1.85in}>{\centering\arraybackslash}p{0.5in}>{\centering\arraybackslash}p{0.6in}>{\centering\arraybackslash}p{0.5in}>{\centering\arraybackslash}p{0.5in}>{\centering\arraybackslash}p{0.5in}@{}}
\toprule
\textbf{Term} & \textbf{Flag} & \textbf{Quantity $x_i$ of circuit $i$} & \textbf{Gated} & \textbf{Range} & \textbf{Ref.} & \textbf{HF pen.} & \textbf{Tuned} \\
\midrule
$\lambda_{\mathrm{exc}}$ & \texttt{cost-\allowbreak{}lambda-\allowbreak{}excitation} & weight of excitation ranks $>3$ among the sampled determinants & yes & 0--4 & 2.0 & 2.0 & 3.9 \\
$\lambda_{\mathrm{cov}}$ & \texttt{cost-\allowbreak{}lambda-\allowbreak{}coverage} & idle-qubit fraction of the transpiled circuit & yes & 0--20 & 10.0 & 10.0 & 7.2 \\
$\lambda_{\mathrm{over}}$ & \texttt{cost-\allowbreak{}lambda-\allowbreak{}overshoot} & $\log_{10}(d_i/d_{\max})$ when the depth $d_i$ exceeds the ceiling $d_{\max}$ & no & fixed & 2.0 & 2.0 & 2.0 \\
$\lambda_{\mathrm{HF}}$ & \texttt{cost-\allowbreak{}lambda-\allowbreak{}hf} & share of right-spin-sector shots that are the Hartree--Fock determinant & no & 0--0.3 & 0 & 0.05 & 0.07 \\
$\lambda_{\mathrm{spin}}$ & \texttt{cost-\allowbreak{}lambda-\allowbreak{}spin} & $|$alpha-only $-$ beta-only$|$ share of right-spin-sector shots & no & 0--1 & 0 & 0 & 0.77 \\
$c_H$ & \texttt{entropy-\allowbreak{}coef} & mean token entropy of the sampled sequences (nats), subtracted from the loss & --- & 0--0.3 & 0 & 0 & 0.26 \\
\bottomrule
\end{tabular}
\end{table*}

\begin{table*}[ht]
\centering
\caption{Inference-time hyperparameters (zero-shot campaign on unseen
molecules, Section~\ref{sec:zero-shot}), unique to inference and not
already listed in Table~\ref{tab:hyperparams}. All values are
uniform across every lineage and every target molecule in this
campaign -- unlike training (Table~\ref{tab:hyperparams}), nothing here
varies by architecture or qubit-count tier. The recovery size $D$ is the
same quantity as in training (Table~\ref{tab:hyperparams}, Section~\ref{sec:recovery}): the
final-recovery number of strings per spin sector, which can be changed with zero
further QPU calls once a lineage's circuits are sampled and cached
(Fig.~\ref{fig:flow-inference}). The ranking cap (distinct from
the training recovery size, although the code flag has the same name) is a separate, smaller cap used
only to rank the 10 sampled circuits against each other before the
final recovery -- not swept, intentionally cheap: 200 strings per spin sector (net dimension 40{,}000) in the fresh-shot campaign. The E+D rule uses the same ranking.}
\label{tab:infer-hyperparams}
\footnotesize
\begin{tabular}{@{}ll@{}}
\toprule
Parameter (code labels in Table~\ref{tab:glossary}) & Value \\
\midrule
Shots per circuit & 100000 \\
Recovery size $D$ (strings per spin sector; final recovery; the cache allows other values) & 20000 \\
Final recovery iterations $K$ & 25 \\
Ranking cap (pre-recovery ranking only; fresh-shot campaign) & 200 strings per spin sector (net dimension 40{,}000) \\

Depth limit for diagonalisation (70q lineages, matches Stage-2 training ceiling) & 50000 \\
\bottomrule
\end{tabular}
\end{table*}

The full per-parameter tuning narrative (what was varied, what the
sweep/ablation found, and the setting this motivated) is reported in
Table~\ref{tab:tuning-narrative}, Appendix~\ref{sec:appendix-training-detail}.
\subsection{Computing resources}
\label{sec:computing-resources}

All classical computation (policy training, circuit generation, and the SQD ranking and recovery) ran on the \emph{roquo} cluster at the RIKEN Center for Computational Science (R-CCS), a GPU supercomputer coupled to IBM quantum processors through the JHPC-quantum project. It has 135 nodes, each with 144 Arm (Grace) cores, about 1.7~TB of memory and four NVIDIA GB200 GPUs. Jobs are scheduled with Slurm; a typical job used a quarter of a node (one GPU, 36 cores), and the recoveries used two GPUs. The quantum circuits were executed on \code{ibm\_kobe} through IBM's Qiskit Runtime service.

\section{Results}
\label{sec:results}

\subsection{Stage-1 (simulator) training}
\label{sec:stage1-results}

The Stage-1 policy is the first stage of both training configurations (Section~\ref{sec:training}); all numbers in this subsection are from it. It ran for a fixed 150 iterations per lineage.

\begin{figure*}[!t]
\centering
\includegraphics[width=\linewidth]{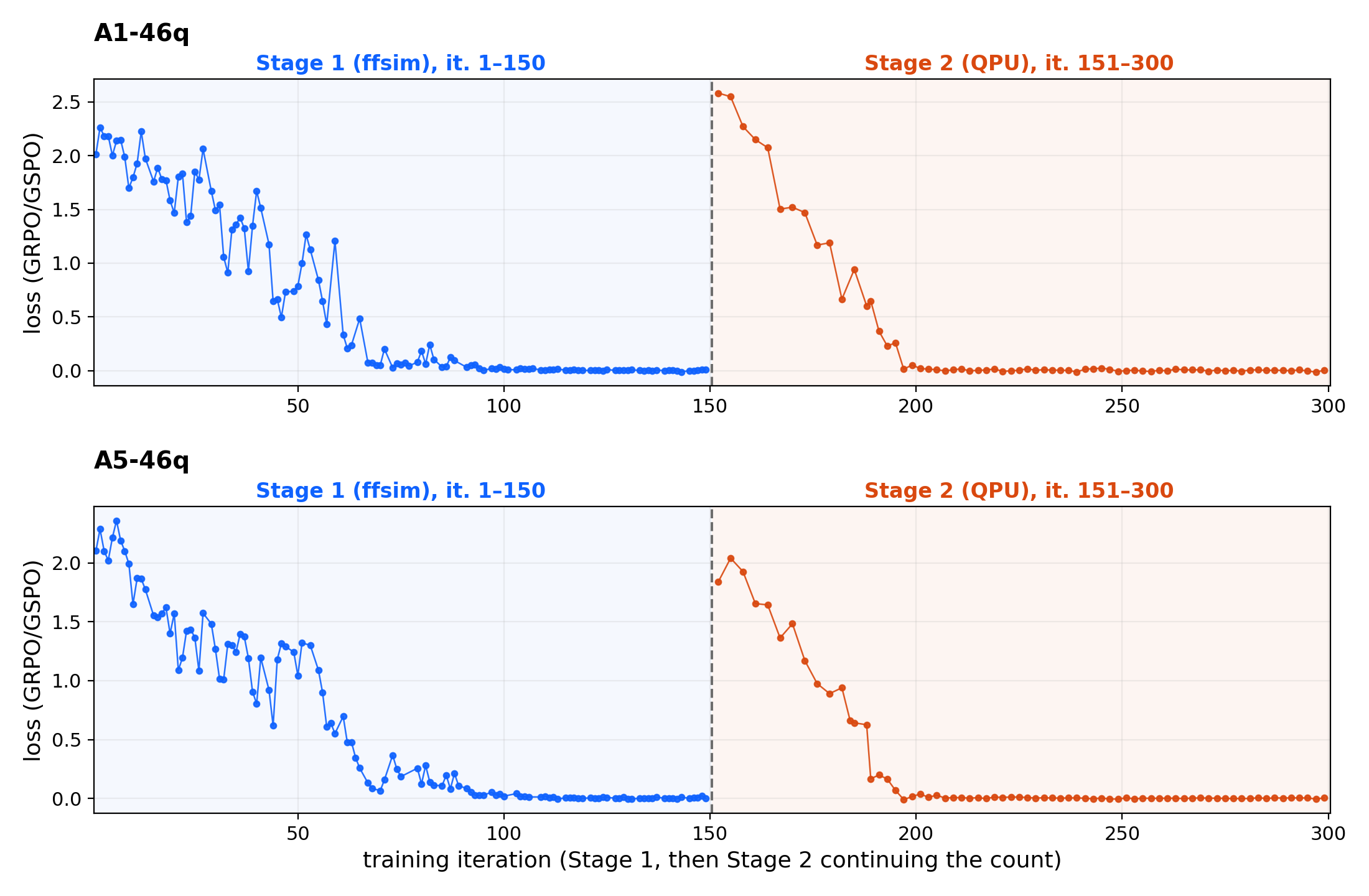}
\caption{Training loss (GRPO/GSPO) of the 46-qubit lineages, A1-46q (top) and A5-46q (bottom): Stage~1 (ffsim simulator, iterations 1--150, left of the dashed line) followed by Stage~2 with the Hartree--Fock penalty (\code{ibm\_kobe} QPU, iterations 151--300, right of the line), with the iteration count continuing across the two stages. Points are admitted iterations only; an iteration that fails the admission gate triggers no update and has no loss. The 70-qubit lineages are in Figure~\ref{fig:stage-loss-70q} (Appendix~\ref{sec:appendix-training-curves}).}
\label{fig:stage-loss-46q}
\end{figure*}

Loss on admitted iterations (Figure~\ref{fig:stage-loss-46q} for the 46-qubit lineages,
Figure~\ref{fig:stage-loss-70q} in Appendix~\ref{sec:appendix-training-curves}
for the 70-qubit ones) starts at 2.2--2.5 (mean of the first three iterations) for all four lineages and falls below 0.1 after 58 (A1-46q), 60 (A5-46q), 92 (A1-70q) and 90 (A5-70q) iterations, ending at 0.002--0.010. The raw energy error of the same iterations
(Figures~\ref{fig:stage-rawerr-46q} and~\ref{fig:stage-rawerr-70q}) does not
decay monotonically because the size curriculum cycles through the
training molecules: it returns to small values on the small
warm-up molecules and rises to tens of mHa on the larger ones in every
cycle. Averaged over the admitted iterations of the last 30, the error against the reference is 36.1, 28.1, 27.6 and 21.1~mHa for A1-46q, A5-46q, A1-70q and A5-70q (Table~\ref{tab:t2-compare}). These errors are against a coupled-cluster or exact reference on the training molecules, which are not the inference molecules, so they bound the accuracy of the warm start but do not predict the zero-shot error (Section~\ref{sec:three-model-inference}). The admission gate passes in 82--86\% of the Stage-1 iterations (Figure~\ref{fig:gate-pass}, Appendix~\ref{sec:appendix-training-curves}).

The depth of the best of the 20 circuits sampled in each iteration
(Figures~\ref{fig:stage-depth-46q} and~\ref{fig:stage-twoq-46q}; 70-qubit lineages in Figures~\ref{fig:stage-depth-70q} and~\ref{fig:stage-twoq-70q}, Appendix~\ref{sec:appendix-training-curves}) \emph{grows} during Stage~1: its median over the first ten iterations is 436--1158 and over the last ten 3997 (A1-46q), 6634 (A5-46q), 8384 (A1-70q) and 10{,}454 (A5-70q), as the policy learns to put more gates on the 16-orbital simulator molecules, where depth is not penalised by a device. The Stage-2 values (right of the dashed line in the same figures) are 50--150, but the two stages use different molecules and samplers and are not a like-for-like comparison (Section~\ref{sec:discussion}).

\begin{figure*}[!t]
\centering
\includegraphics[width=\linewidth]{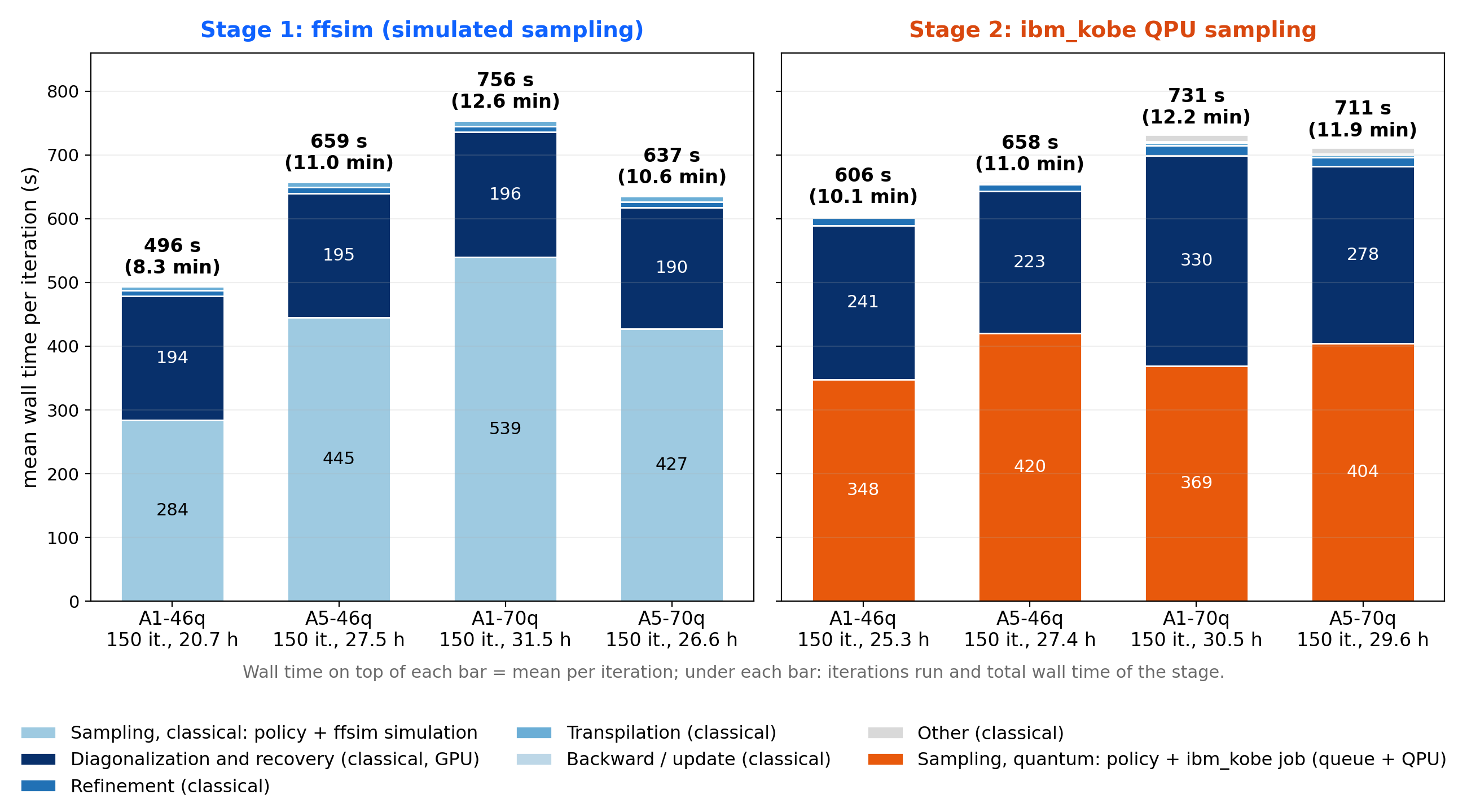}
\caption{Mean wall time per training iteration by pipeline step, for Stage~1 (left) and Stage~2 with the Hartree--Fock penalty (right) and the four lineages. The sampling step (policy generation of the gate sequences, then an \code{ffsim} simulation in Stage~1 or an IBM Runtime job on \code{ibm\_kobe}, including queue time and QPU execution, in Stage~2) is the only step that involves the quantum computer (orange); every other step is classical (blue shades): diagonalization and recovery on the GPUs, refinement, transpilation, and the backward pass. Bold numbers on top are the mean wall time per iteration; below each bar are the iterations run and the total wall time of the stage. One A5-46q, A1-70q and A5-70q iteration whose sampling exceeded one hour (device-queue stalls) is excluded from the Stage-2 means.}
\label{fig:stage-timing}
\end{figure*}

Sampling dominates Stage-1 wall time (Figure~\ref{fig:stage-timing}, left; Table~\ref{tab:t2-timing}):
284--539~s per iteration out of 496--756~s in total (57--71\%), even though Stage~1 uses an exact-statevector
\code{ffsim} sampler rather than hardware: the cost is the
policy's autoregressive generation of $L$ gate tokens for
20 circuits per iteration and their simulation, not a hardware round-trip, and it is largest for the 70-qubit lineages (427--539~s) whose sequences have 44 gates. Diagonalization (including recovery) is the second-largest cost at
190--196~s; refine, transpile, and backward are each under 15~s. A Stage-1 run takes 20.7--31.5~h.

\begin{figure*}[!t]
\centering
\includegraphics[width=\linewidth]{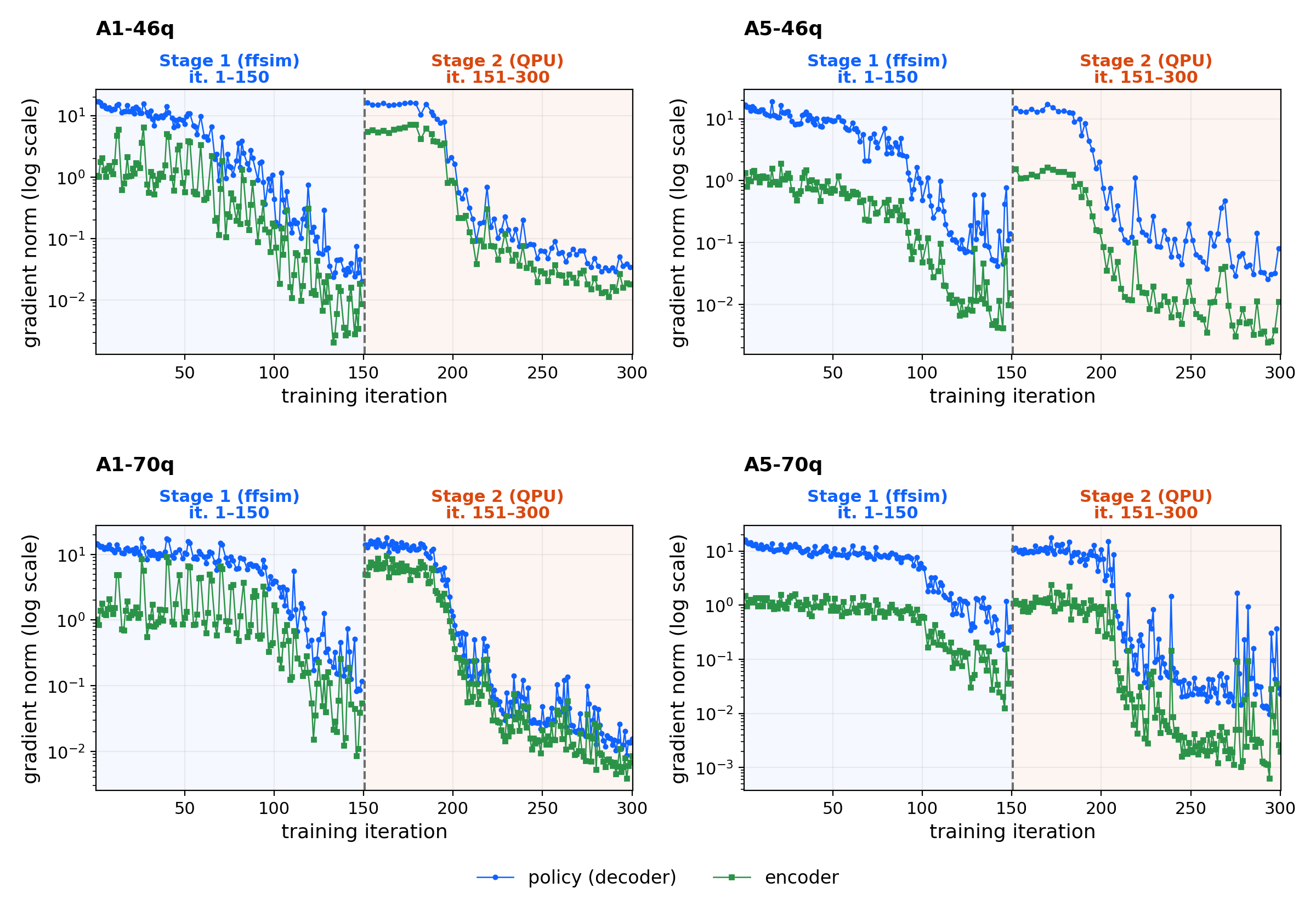}
\caption{Gradient norm of the policy transformer (decoder) and of the prompt encoder on admitted iterations, for the four lineages (log scale). In each panel Stage~1 (ffsim) is left and Stage~2 (\code{ibm\_kobe} QPU, HF penalty) right of the dashed line, with the iteration count continuing across the stages.}
\label{fig:stage-gradnorm}
\end{figure*}

The gradient-norm breakdown by model component
(Figure~\ref{fig:stage-gradnorm}, left of each dashed line) shows the policy transformer's
gradient norm exceeding the prompt encoder's, by a median factor of
7.3--10.8 across all four
lineages in Stage~1 -- consistent with the parameter-count
asymmetry between the two components (a 56.8M-parameter policy
backbone vs.\ a 3.2--3.4M-parameter prompt encoder, Section~\ref{sec:architecture}).
Both norms decay as training converges and the encoder's never reaches the policy's. This is the
closest per-component analogue this architecture logs to a separate
``encoder loss / decoder loss'' decomposition, since both components are
optimized jointly against a single scalar GRPO/GSPO objective
(Section~\ref{sec:gqe-formalism}).

\subsection{Stage-2 (hardware) training with the Hartree--Fock penalty}
\label{sec:stage2-results}

On hardware, all four lineages run for 150 iterations and settle on shallow circuits, but the sampled circuits narrow to a small family. The loss on admitted iterations (Figures~\ref{fig:stage-loss-46q} and~\ref{fig:stage-loss-70q}) restarts at 1.9--2.9 and falls below 0.1 after 18 (A1-46q and A5-46q), 48 (A1-70q) and 51 (A5-70q) iterations, ending at 0.001--0.004. The raw error on the hardware targets (Figures~\ref{fig:stage-rawerr-46q} and~\ref{fig:stage-rawerr-70q}) is nearly flat, with a mean over the stage of 46.2 and 46.5~mHa for the 46-qubit lineages and 300.6 and 294.1~mHa for the 70-qubit lineages (Table~\ref{tab:t2-compare}).

The admission gate passes 100\% and 99\% of the iterations for A1-70q and A5-70q, and 46\% and 47\% for A1-46q and A5-46q (Figure~\ref{fig:gate-pass}). The depth of the best of the 20 circuits is low from the start and stays so; its median over the last ten iterations is 56 (A1-46q), 54 (A5-46q), 76 (A1-70q) and 152 (A5-70q), against 44--66 in the first ten (Figures~\ref{fig:stage-depth-46q}--\ref{fig:stage-twoq-70q}).

The chemistry of the sampled circuits (Appendix~\ref{sec:appendix-chemistry}) shows the narrowing. By iteration 50--55 the effective vocabulary of the policy is about one token. In the best circuit, 6--21\% of the shots are in the right spin sector, and 0--4\% of those are the Hartree--Fock determinant (60--76\% in Stage~1). The 46-qubit circuits are strongly beta-only, with a beta-only minus alpha-only weight difference of 0.56 (A1-46q) and 0.67 (A5-46q). The tuned reward keeps an effective vocabulary of 21--27 (Table~\ref{tab:t2-compare}). Per-iteration times and billed QPU time are given in Tables~\ref{tab:t2-timing} and~\ref{tab:t2-qpu} (Appendix~\ref{sec:appendix-training-detail}) and the total cost in Section~\ref{sec:cost}. Each result is from one seed per lineage.

\subsection{Zero-shot inference on unseen molecules}
\label{sec:zero-shot}

\begin{figure*}[!tp]
\centering
\includegraphics[width=\textwidth]{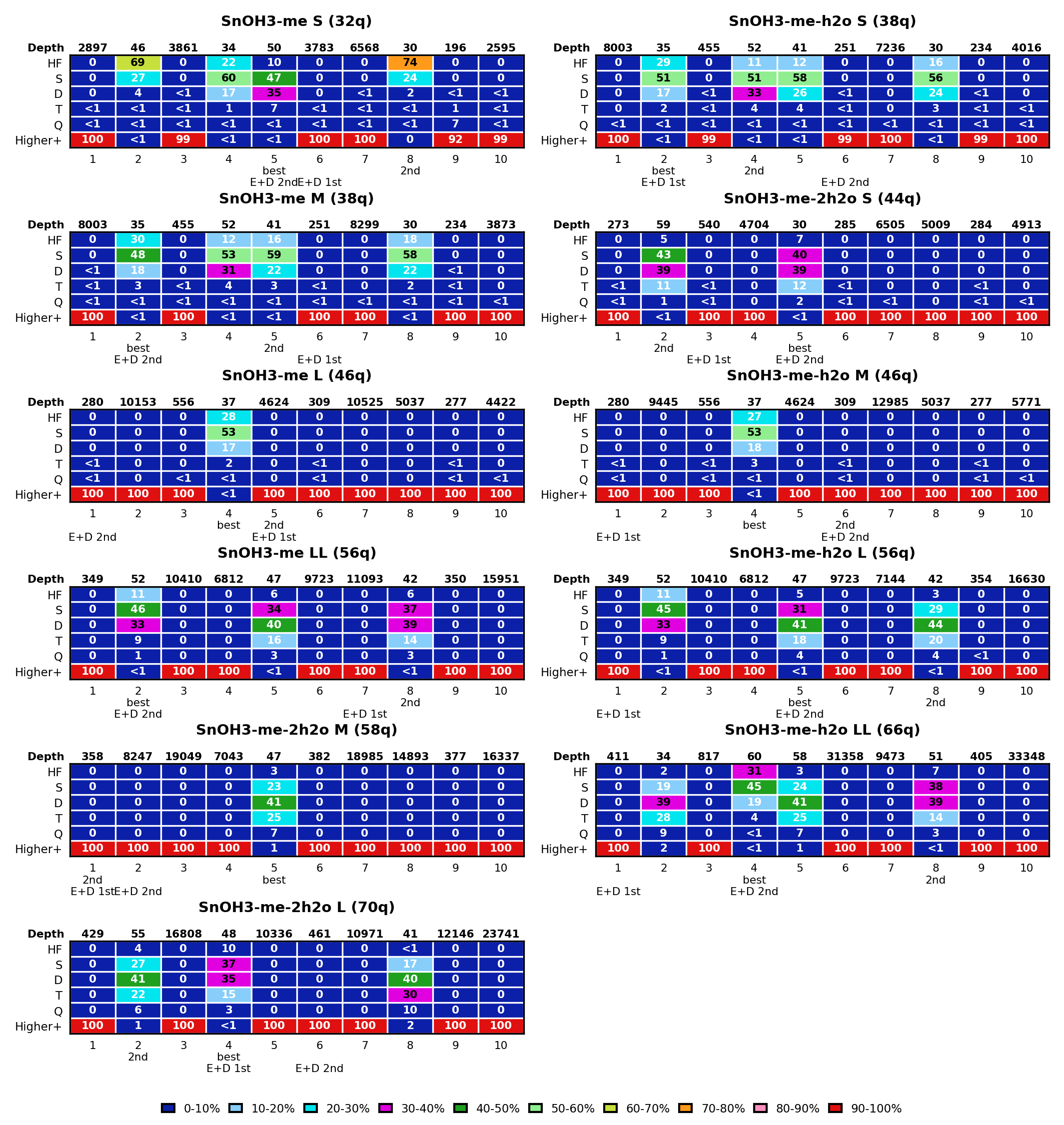}
\caption{Raw-shot excitation composition (HF, S, D, T, Q, higher; percent of 100{,}000 shots) of the ten circuits sampled by the A1 policy trained at 46 qubits with the Hartree--Fock penalty, for each of the eleven compounds, ordered from the smallest to the largest active space. Numbers above each column are the circuit depth as-submitted to \code{ibm\_kobe}; ``best'' and ``2nd'' mark the two circuits with the lowest ranking energy, and ``E+D 1st'' and ``E+D 2nd'' the two circuits chosen by the combined energy and distinct-bitstring rank (a circuit can carry two labels). In every compound the best circuit puts at most 1.2\% of its shots in ``higher'' and is among the four shallowest of the ten (the shallowest in four of the eleven panels); the second circuit does so in eight of the eleven panels, and in the SnOH3-me L, SnOH3-me-h2o M and SnOH3-me-2h2o M panels it has essentially all its weight in ``higher''. Most of the other circuits are one to three orders of magnitude deeper with all their weight in ``higher''.}
\label{fig:collage-a1-46q}
\end{figure*}

\begin{table*}[!t]
\centering
\caption{Zero-shot inference comprehensive summary, trained policies only, every lineage$\times$molecule combination. Each table entry is energy (E$_h$)/error vs.\ DMRG (mHa). Depth/2q are the depth and two-qubit-gate count of the best circuit as transpiled and submitted to \code{ibm\_kobe} (live transpilation at Qiskit optimization level 1); the offline estimate on the \code{ibm\_heron\_r2} basis is not used. The distance of the generated circuits from the idealised LUCJ circuit is in Section~\ref{sec:divergence}. Matched untrained controls are in Appendix~\ref{sec:appendix-inference-detail}. All values come from the fresh-shot re-ranking campaign: every run's ten sampled circuits were sampled on \code{ibm\_kobe} with 100{,}000 new shots each (no error mitigation), and the raw counts and transpiled circuits were stored. Circuits are ranked by a one-shot diagonalization of the electron-number-conserving shots in a subspace of up to 200 most-measured alpha strings (net CI dimension 40{,}000); circuits with no valid shot have no ranking energy and are excluded. ``Best''/``second'' are the two lowest-ranking-energy circuits, ``merged'' pools the shots of every circuit with a finite ranking energy, and ``E+D 1''/``E+D 2'' are the two circuits with the lowest sum of two ranks, the rank of the ranking energy (the same 40{,}000-determinant ranking) and the rank by number of distinct measured bitstrings, recovered on the same fresh shots. Every recovery uses $D=20{,}000$ strings per spin sector and 25 iterations. Errors are relative to the DMRG reference.}
\label{tab:zeroshot-master}
\footnotesize
\setlength{\tabcolsep}{3pt}
\begin{tabular}{@{}llrrrrrr@{}}
\toprule
Molecule (tier, qubits) & Arch.\ (lineage) & Best E$_h$/mHa & Second E$_h$/mHa & Merged E$_h$/mHa & E+D 1/2 (mHa) & Depth & 2q \\
\midrule
SnOH3-me S (32q) & A1 (A1-46q) & $-$479.193266/0.12 & $-$479.192672/0.71 & $-$479.193374/0.01 & 0.01/0.12 & 50 & 56 \\
SnOH3-me S (32q) & A1 (A1-70q) & $-$479.193164/0.22 & $-$479.193270/0.11 & $-$479.193374/0.01 & 0.01/0.01 & 41 & 68 \\
SnOH3-me S (32q) & A5 (A5-46q) & $-$479.192727/0.65 & $-$479.193081/0.30 & $-$479.193374/0.01 & 0.30/0.01 & 31 & 36 \\
SnOH3-me S (32q) & A5 (A5-70q) & $-$479.192924/0.46 & $-$479.193134/0.25 & $-$479.193374/0.01 & 0.01/0.46 & 69 & 60 \\
\addlinespace
SnOH3-me-h2o S (38q) & A1 (A1-46q) & $-$555.108538/0.29 & $-$555.108684/0.15 & $-$555.108826/0.01 & 0.29/0.01 & 35 & 52 \\
SnOH3-me-h2o S (38q) & A1 (A1-70q) & $-$555.108766/0.07 & $-$555.108751/0.08 & $-$555.108826/0.01 & 0.01/0.01 & 57 & 92 \\
SnOH3-me-h2o S (38q) & A5 (A5-46q) & $-$555.108713/0.12 & $-$555.108826/0.01 & $-$555.108826/0.01 & 0.12/0.01 & 34 & 40 \\
SnOH3-me-h2o S (38q) & A5 (A5-70q) & $-$555.108740/0.09 & $-$555.108764/0.07 & $-$555.108826/0.01 & 0.09/0.19 & 110 & 84 \\
\addlinespace
SnOH3-me M (38q) & A1 (A1-46q) & $-$479.292683/9.6 & $-$479.296591/5.7 & $-$479.286434/15.8 & 152.0/9.6 & 35 & 52 \\
SnOH3-me M (38q) & A1 (A1-70q) & $-$479.298463/3.8 & $-$479.295672/6.6 & $-$479.282569/19.7 & 144.4/139.1 & 57 & 92 \\
SnOH3-me M (38q) & A5 (A5-46q) & $-$479.285058/17.2 & $-$479.282828/19.5 & $-$479.271416/30.9 & 143.9/130.6 & 32 & 52 \\
SnOH3-me M (38q) & A5 (A5-70q) & $-$479.288699/13.6 & $-$479.152486/149.8 & $-$479.265913/36.4 & 131.8/149.1 & 152 & 104 \\
\addlinespace
SnOH3-me-2h2o S (44q) & A1 (A1-46q) & $-$631.083103/0.21 & $-$631.083239/0.07 & $-$631.083269/0.04 & 0.04/0.21 & 30 & 60 \\
SnOH3-me-2h2o S (44q) & A1 (A1-70q) & $-$631.083257/0.06 & $-$631.083084/0.23 & $-$631.083269/0.04 & 0.06/0.04 & 60 & 100 \\
SnOH3-me-2h2o S (44q) & A5 (A5-46q) & $-$631.083185/0.13 & $-$631.083207/0.11 & $-$631.083267/0.04 & 0.11/0.06 & 49 & 64 \\
SnOH3-me-2h2o S (44q) & A5 (A5-70q) & $-$631.083092/0.22 & $-$631.083044/0.27 & $-$631.083269/0.04 & 0.22/0.27 & 168 & 104 \\
\addlinespace
SnOH3-me L (46q) & A1 (A1-46q) & $-$479.299926/14.3 & $-$479.121483/192.7 & $-$479.258786/55.4 & 192.7/198.1 & 37 & 64 \\
SnOH3-me L (46q) & A1 (A1-70q) & $-$479.306409/7.8 & $-$479.116356/197.9 & $-$479.272504/41.7 & 193.1/198.6 & 69 & 100 \\
SnOH3-me L (46q) & A5 (A5-46q) & $-$479.301748/12.5 & $-$479.306572/7.6 & $-$479.284378/29.8 & 199.6/12.5 & 55 & 80 \\
SnOH3-me L (46q) & A5 (A5-70q) & $-$479.308601/5.6 & $-$479.297326/16.9 & $-$479.284631/29.6 & 199.1/5.6 & 144 & 124 \\
\addlinespace
SnOH3-me-h2o M (46q) & A1 (A1-46q) & $-$555.123748/1.4 & $-$555.102442/22.7 & $-$555.118018/7.1 & 22.7/22.7 & 37 & 64 \\
SnOH3-me-h2o M (46q) & A1 (A1-70q) & $-$555.124450/0.71 & $-$555.102207/23.0 & $-$555.118773/6.4 & 23.0/22.6 & 69 & 100 \\
SnOH3-me-h2o M (46q) & A5 (A5-46q) & $-$555.123672/1.5 & $-$555.123151/2.0 & $-$555.123131/2.0 & 1.4/2.0 & 58 & 64 \\
SnOH3-me-h2o M (46q) & A5 (A5-70q) & $-$555.123764/1.4 & $-$555.123482/1.7 & $-$555.124246/0.91 & 1.4/1.7 & 108 & 100 \\
\addlinespace
SnOH3-me LL (56q) & A1 (A1-46q) & $-$479.300183/14.9 & $-$479.299292/15.8 & $-$479.286822/28.2 & 203.8/14.9 & 52 & 80 \\
SnOH3-me LL (56q) & A1 (A1-70q) & $-$479.304625/10.4 & $-$479.112860/202.2 & $-$479.279821/35.2 & 202.2/202.1 & 62 & 124 \\
SnOH3-me LL (56q) & A5 (A5-70q) & $-$479.301896/13.2 & $-$479.111752/203.3 & $-$479.261236/53.8 & 203.3/204.3 & 79 & 104 \\
\addlinespace
SnOH3-me-h2o L (56q) & A1 (A1-46q) & $-$555.132208/2.8 & $-$555.132421/2.6 & $-$555.129520/5.5 & 33.1/2.8 & 47 & 80 \\
SnOH3-me-h2o L (56q) & A1 (A1-70q) & $-$555.133193/1.9 & $-$555.132552/2.5 & $-$555.127221/7.8 & 33.1/33.0 & 65 & 124 \\
SnOH3-me-h2o L (56q) & A5 (A5-46q) & $-$555.131417/3.6 & $-$555.132660/2.4 & $-$555.129768/5.3 & 33.1/3.6 & 53 & 64 \\
SnOH3-me-h2o L (56q) & A5 (A5-70q) & $-$555.132405/2.7 & $-$555.132408/2.6 & $-$555.130714/4.3 & 2.7/33.1 & 44 & 76 \\
\addlinespace
SnOH3-me-2h2o M (58q) & A1 (A1-46q) & $-$631.112084/6.5 & $-$631.077480/41.1 & $-$631.095105/23.5 & 41.1/41.1 & 47 & 88 \\
SnOH3-me-2h2o M (58q) & A1 (A1-70q) & $-$631.114317/4.3 & $-$631.113058/5.6 & $-$631.106177/12.4 & 41.1/41.1 & 65 & 132 \\
SnOH3-me-2h2o M (58q) & A5 (A5-46q) & $-$631.114080/4.5 & $-$631.113620/5.0 & $-$631.105052/13.6 & 41.1/41.1 & 45 & 72 \\
SnOH3-me-2h2o M (58q) & A5 (A5-70q) & $-$631.113705/4.9 & $-$631.111954/6.7 & $-$631.110745/7.9 & 41.1/41.1 & 62 & 68 \\
\addlinespace
SnOH3-me-h2o LL (66q) & A1 (A1-46q) & $-$555.135328/2.1 & $-$555.133503/3.9 & $-$555.131333/6.1 & 35.4/2.1 & 60 & 92 \\
SnOH3-me-h2o LL (66q) & A1 (A1-70q) & $-$555.134846/2.6 & $-$555.102002/35.4 & $-$555.131599/5.8 & 2.6/35.4 & 51 & 156 \\
SnOH3-me-h2o LL (66q) & A5 (A5-46q) & $-$555.133877/3.5 & $-$555.134389/3.0 & $-$555.130850/6.5 & 35.4/3.5 & 44 & 84 \\
SnOH3-me-h2o LL (66q) & A5 (A5-70q) & $-$555.135231/2.2 & $-$555.134346/3.1 & $-$555.133797/3.6 & 3.1/2.2 & 57 & 84 \\
\addlinespace
SnOH3-me-2h2o L (70q) & A1 (A1-46q) & $-$631.123614/8.6 & $-$631.123272/9.0 & $-$631.119423/12.8 & 8.6/54.8 & 48 & 100 \\
SnOH3-me-2h2o L (70q) & A1 (A1-70q) & $-$631.124552/7.7 & $-$631.077480/54.8 & $-$631.099821/32.4 & 54.8/7.7 & 60 & 160 \\
SnOH3-me-2h2o L (70q) & A5 (A5-46q) & $-$631.124857/7.4 & $-$631.123201/9.0 & $-$631.122669/9.6 & 54.8/7.2 & 44 & 96 \\
SnOH3-me-2h2o L (70q) & A5 (A5-70q) & $-$631.120256/12.0 & $-$631.124322/7.9 & $-$631.110701/21.5 & 12.0/54.8 & 134 & 128 \\
\bottomrule
\end{tabular}
\end{table*}

Every lineage's \emph{trained} policy of the HF-penalty recipe (A1/A5, 46q-trained
and 70q-trained) was run zero-shot on the unseen molecules of the Sn(OH)$_3$-Me family --
the small, medium, large and extra-large active-space tiers of the bare molecule and its
mono- and di-hydrate adducts, none of which appear in any lineage's own
training curriculum. All four lineages are run on all eleven tiers (44 runs, all measured on \code{ibm\_kobe}; 43 of the Hartree--Fock-penalty runs have a result in the table, see below). Table~\ref{tab:zeroshot-master} is the comprehensive summary,
trained policies only; the matched untrained controls (eight runs on three molecules; the controls are randomly initialized and so do not depend on the training recipe) and the full
per-molecule detail -- excitation-level composition,
circuit-cost/coverage breakdowns, per-stage timing -- are in
Appendix~\ref{sec:appendix-inference-detail}.
Section~\ref{sec:crossnorb} explains why some lineage/molecule
combinations are a matched-norb case and others a cross-norb
vocab-mismatch ablation (the lineage's own training norb differs from
the target's). Figures~\ref{fig:collage-a1-46q} and~\ref{fig:collage-a1-70q} show, for the A1 policy trained at 46 and at 70 qubits, the raw excitation composition and as-submitted depth of all ten sampled circuits of every compound; they make visible that the selected circuits are the ones with weight at low excitation ranks and small depth, which Section~\ref{sec:analysis-selection} quantifies.

\textbf{Summary of the HF-penalty runs.}
Of the 44 runs, 43 have a recovered best circuit; the other (SnOH3-me LL with A5-46q) produced valid shots for one circuit only, so the ranking found no best and second pair (the selections best, second, merged and E+D are defined in Section~\ref{sec:methods-pipeline} and Figure~\ref{fig:flow-inference}).
Sampling the ten circuits again with a different seed gave eight circuits with valid shots but a recovered error of about 200~mHa (Appendix~\ref{sec:appendix-repeat}), so this test case is not reported.
For the tuned reward, the one run with a single usable circuit (SnOH3-me-2h2o S with A5-70q) was sampled again with a different seed, which gave five usable circuits and a best-circuit error of 0.
07~mHa; this second draw is the one reported (Appendix~\ref{sec:appendix-repeat}).
The median error of the best circuit against DMRG is 2.8~mHa; 13 runs are within 1~mHa, 28 within 5~mHa and 42 within 15~mHa.
The eight untrained controls have a median error of 106~mHa (one within 15~mHa, at the 56-qubit tier, where the untrained policy happened to pick a depth-89 circuit).
The comparison is paired by molecule and lineage and uses the same sampling, ranking and recovery.
The depth of the selected circuit is 30--168 (median 55) and its two-qubit-gate count 36--160 (median 84), against a median depth of 12{,}300 for the untrained controls' selected circuits.
The error grows with the size of the active space, as expected for a fixed recovery size of 20{,}000 strings per spin sector; the exceptions at small size are the SnOH3-me M (38q) and SnOH3-me L (46q) tiers (best-circuit errors 3.8--17.2 and 5.6--14.3~mHa), where the non-hydrate molecule is harder than its size suggests.
At the three S tiers, the recovery cap of 20{,}000 strings exceeds the number of $\alpha$ strings (1,820--7,315; Table~\ref{tab:active-spaces}), so the cap does not restrict the strings that can be retained, although the basis still depends on the sampled configurations. The near-exact merged energies at these tiers therefore give limited evidence that the circuits preferentially sample important configurations. Larger spaces, where the cap is smaller than the full $\alpha$-string space, are a more demanding test. For SnOH3-me M (38 qubits), the cap covers at most 39.7\% of the 50,388 $\alpha$ strings, which may contribute to its larger errors.

The matched untrained controls are compared with the trained policies in Appendix~\ref{sec:appendix-inference-detail} (Table~\ref{tab:crossnorb-detail}).
Paired by lineage, the trained policy is more accurate in all eight comparisons.
At SnOH3-me L the four controls give 188--200~mHa against 5.6--14.3~mHa for the four trained policies, and at SnOH3-me-h2o M the three controls give 22.6--22.9~mHa against 0.71--1.5~mHa.
At SnOH3-me LL the margin is small (10.4 against 12.3~mHa for A1-70q), and the control lies, inside the 10.4--14.9~mHa range of the three trained policies, so the size of the separation depends on the molecule.

\begin{table*}[!t]
\centering
\caption{Lowest recovered SQD energy vs.\ DMRG reference for every SnOH3-me-family target, with the signed deviation $\Delta E = E_{\mathrm{SQD}} - E_{\mathrm{DMRG}}$ (all positive), taken over all lineages, policies and result types (best, second, merged, E+D 1/2) in Tables~\ref{tab:zeroshot-detail} and \ref{tab:crossnorb-detail}; the lineage and type that achieved it are noted. DMRG references are from Table~\ref{tab:dmrg-reference} (maximum bond dimension 4000, 17 sweeps). Selecting the minimum over many results is a best-of-$k$ statistic, not a single pre-declared run.}
\label{tab:dmrg-vs-sqd}
\small
\setlength{\tabcolsep}{3pt}
\begin{tabular}{@{}lrrrll@{}}
\toprule
Target & DMRG ($\eh$) & Best SQD ($\eh$) & $\Delta E$ (mHa) & Lineage & Type \\
\midrule
SnOH3-me S (32q) & $-$479.193382 & $-$479.193374 & <0.01 & A1-46q trained & merged \\
SnOH3-me-h2o S (38q) & $-$555.108833 & $-$555.108826 & <0.01 & A1-46q trained & merged \\
SnOH3-me M (38q) & $-$479.302278 & $-$479.298463 & 3.8 & A1-70q trained & best \\
SnOH3-me-2h2o S (44q) & $-$631.083312 & $-$631.083269 & 0.04 & A1-46q trained & merged \\
SnOH3-me L (46q) & $-$479.314216 & $-$479.308601 & 5.6 & A5-70q trained & best \\
SnOH3-me-h2o M (46q) & $-$555.125158 & $-$555.124450 & 0.71 & A1-70q trained & best \\
SnOH3-me LL (56q) & $-$479.315057 & $-$479.304625 & 10.4 & A1-70q trained & best \\
SnOH3-me-h2o L (56q) & $-$555.135058 & $-$555.133193 & 1.9 & A1-70q trained & best \\
SnOH3-me-2h2o M (58q) & $-$631.118616 & $-$631.114317 & 4.3 & A1-70q trained & best \\
SnOH3-me-h2o LL (66q) & $-$555.137396 & $-$555.135328 & 2.1 & A1-46q trained & best \\
SnOH3-me-2h2o L (70q) & $-$631.132238 & $-$631.125022 & 7.2 & A5-46q trained & E+D 2 \\
\bottomrule
\end{tabular}
\end{table*}

\subsection{Circuit cost: trained policy vs.\ pure LUCJ ansatz}
\label{sec:lucj-comparison}
The transpilation costs and the hardware results of the LUCJ benchmark are collected in Appendix~\ref{sec:appendix-lucj}.

The relevant LUCJ comparator is the circuit that was benchmarked on hardware: the spin-balanced LUCJ operator of the \code{ffsim} library \cite{sung2026ffsim} built from CCSD amplitudes, with one ansatz repetition (the shallowest variant). Depth is the total circuit depth over all gates and measurements (Qiskit's \code{QuantumCircuit.depth()}). For the generated circuits, depths and two-qubit-gate counts are those of the circuits as transpiled at Qiskit optimization level 1 \cite{javadiabhari2024qiskit} and submitted to \code{ibm\_kobe}. For LUCJ, we report an offline estimate: the same circuit transpiled at optimization level 1 for the \code{ibm\_heron\_r2} basis gates on an idealized 193-qubit heavy-hex lattice, the model used during training. The LUCJ circuits submitted to \code{ibm\_kobe} were transpiled live and are larger (for example depth 8{,}262 and 14{,}116 two-qubit gates at 70 qubits, against 536 and 6{,}952 in the estimate), so the reductions below are lower bounds. The LUCJ circuit has depth 356 and 3040 two-qubit gates at 46 qubits, and depth 428--536 and 4248--6952 two-qubit gates at 56--70 qubits (Tables~\ref{tab:lucj-hw} and~\ref{tab:lucj-hw-raw}).

\paragraph{Selected circuits.} At 56--70 qubits the best circuits of the 19 HF-penalty runs (SnOH3-me LL, SnOH3-me-h2o L, SnOH3-me-2h2o M, SnOH3-me-h2o LL and SnOH3-me-2h2o L, with the lineages that were run) have depth 44--134 and 64--160 two-qubit gates, i.e.\ 4--12$\times$ shallower and 34--72$\times$ fewer two-qubit gates than the LUCJ circuit of the same molecule (428--536 deep with 4{,}248--6{,}952 two-qubit gates). At the 46-qubit hydrate (SnOH3-me-h2o M) the four trained lineages' best circuits have depth 37--108 and 64--100 two-qubit gates against 356 and 3040 for LUCJ. No run selected a very deep circuit: the largest selected depth is 168. The shallowest selected circuits are not always the most accurate: the deepest of the 56--70-qubit selections (A5-70q at SnOH3-me-2h2o L, depth 134) has an error of 12.0~mHa, and the A5-70q SnOH3-me LL circuit (depth 79) 13.2~mHa; the two largest errors of the 19 runs (14.9 and 13.2~mHa) are at SnOH3-me LL.

\paragraph{Typical circuits.} The selected circuits are not typical of what the policy samples. Over all ten circuits of each HF-penalty run at 56--70 qubits, the median as-submitted depth is between 58 and 12{,}424 (median over the 20 runs, 5{,}960; 20 rather than 19 because the run without a recovered best circuit has sampled circuits), and only 1--9 of the ten circuits per run are shallower than the LUCJ circuit; the A1-70q runs are the deepest (6{,}515--12{,}424) and the A5-70q runs the shallowest (58--6{,}733). The cost advantage therefore depends on selecting the right circuit among the ten; the selection (lowest ranking energy) is made after sampling, so the advantage belongs to the sample-and-select procedure and not to every circuit the policy emits. A smaller-scale version of this comparison (16 qubits, N$_2$/CH$_4$/H$_2$O/benzene) previously found the trained policy's circuits 60--180$\times$ shallower in two-qubit-gate count at matched accuracy. The LUCJ depths and gate counts in Table~\ref{tab:lucj-hw-raw} are the offline estimate described in Section~\ref{sec:lucj-comparison}, not the transpilation of the submitted LUCJ job, which was larger.

The LUCJ circuit was also run on hardware with the pipeline of the other campaigns of this study: 100{,}000 shots on \code{ibm\_kobe}, a 25-iteration GPU recovery with $D=20{,}000$ strings per spin sector, and no error mitigation, dynamical decoupling or twirling. Unlike the trained GQE policy, LUCJ is a single fixed ansatz and not a sampled distribution: there is one circuit per molecule, sampled once and recovered once, so no best/second-best selection applies. Table~\ref{tab:lucj-hw} reports all completed results across all 9
target molecules in the experimental campaign; Table~\ref{tab:lucj-hw-raw}
reports the corresponding raw-shot excitation distributions evaluated across the
100{,}000 physical shots (following the same convention as
Table~\ref{tab:raw-shot-excitation}) alongside transpiled circuit metrics.

The LUCJ results show a trend across the nine target molecules. At the three smallest active spaces (SnOH3-me S at 32 qubits, SnOH3-me-h2o S at 38 qubits and SnOH3-me-2h2o S at 44 qubits), LUCJ reaches the DMRG reference energy to within 0.04~mHa, as do the GQE lineages; there the recovery cap of 20{,}000 strings already covers the whole $\alpha$-string space (Table~\ref{tab:active-spaces}). From 46 qubits onward, the raw error of LUCJ, that is, the error of the energy recovered from the sampled bitstrings, ranges from 22.7 to 204.1~mHa; it is 22.7--54.8~mHa for all targets except SnOH3-me LL (56 qubits), an outlier at 204.1~mHa. We also report the error after \emph{singles injection}: the single excitations of the Hartree--Fock determinant are added to the recovered set of configurations, and the Hamiltonian is diagonalized again in the enlarged basis (Appendix~\ref{sec:appendix-inference-detail}). This lowers all errors from 46 qubits onward to 5.2--12.6~mHa, except SnOH3-me LL, which falls from 204.1 to 41.2~mHa. At SnOH3-me-h2o M (46 qubits), the one target where LUCJ and the untrained GQE policies can be compared on identical footing, the singles-injected error of LUCJ (5.2~mHa) is close to that of the three untrained GQE lineages (4.9--5.1~mHa, Table~\ref{tab:crossnorb-detail}), although the LUCJ circuit is much deeper (depth 356 against 37--108 for the trained GQE circuits). The best circuits selected by the ranking for the four trained Hartree--Fock-penalty lineages reach 0.71--1.5~mHa at this target, with as-submitted depths of 37--108 and 64--100 two-qubit gates, so at SnOH3-me-h2o M the trained policy is better than LUCJ in accuracy and in circuit cost. This rests on one molecule and one fresh hardware sample per circuit. The GQE circuit was selected from ten by ranking energy, not by comparison with DMRG, whereas LUCJ is a single circuit with no such selection step.

\subsection{Hartree--Fock penalty and tuned reward}
\label{sec:three-model-inference}

We compare the two training configurations at inference, with the protocol of Section~\ref{sec:reward-tuning}, on the molecule--lineage pairs for which both have a best and a second circuit. A lineage is one trained policy (A1 or A5, trained at 46 or 70 qubits), and each molecule--lineage pair is one \emph{test case}: that policy run once on one of the eleven Sn(OH)$_3$-Me active spaces, giving $4\times11=44$ test cases per configuration. The per-test-case values are in Appendix~\ref{sec:appendix-three-way}, and the per-run results of each configuration are in the reports described in Appendix~\ref{sec:appendix-availability}.
Table~\ref{tab:three-way} gives the 43 of 44 test cases for which both have a best and a second circuit; for one Hartree--Fock-penalty run no pair exists (Appendix~\ref{sec:appendix-repeat}). All entries are medians over these test cases.

\begin{table*}[!t]
\centering
\caption{The two training configurations on the 43 test cases (molecule--lineage pairs) for which both have a best and a second circuit (of 44; the Hartree--Fock-penalty run of SnOH3-me LL with A5-46q has none, see Appendix~\ref{sec:appendix-repeat}). Medians over test cases, by lineage (last block: all cells). Deviation columns (mHa, $\Delta E = E_{\mathrm{SQD}} - E_{\mathrm{DMRG}}$): ``best $\Delta E$'' and ``2nd $\Delta E$'' are the deviations of the recovered energy of the best and second circuit, and ``min(best,2nd)'' is the smaller of the two. ``depth'' is the as-submitted depth of the best circuit. The remaining columns give the shot composition of the best circuit, in percent: ``valid'' is the share of all shots with the correct electron numbers in each spin sector; the next columns are shares of these valid shots. ``HF'' is the share equal to the Hartree--Fock reference determinant, ``HL'' the share that contains a HOMO$\to$LUMO excitation (the highest occupied to the lowest unoccupied orbital), and ``$\alpha$ only'', ``$\beta$ only'' and ``both'' the shares that excite only the spin-up channel, only the spin-down channel, or both.}
\label{tab:three-way}
\small
\setlength{\tabcolsep}{4pt}
\begin{tabular}{@{}llrrrrrrrrrrr@{}}
\toprule
Lineage & Model set & test cases & best $\Delta E$ & 2nd $\Delta E$ & min(best,2nd) & depth & valid & HF & HL & $\alpha$ only & $\beta$ only & both \\
\midrule
A1-46q & HF penalty & 11 & 2.85 & 5.69 & 2.64 & 47 & 5.0 & 53.5 & 0.8 & 11.8 & 18.6 & 3.8 \\
 & tuned reward & 11 & 2.30 & 22.35 & 2.30 & 40 & 6.0 & 26.5 & 0.2 & 23.8 & 19.6 & 6.5 \\
\midrule
A5-46q & HF penalty & 10 & 3.58 & 2.70 & 2.70 & 44 & 10.6 & 54.7 & 1.6 & 4.7 & 22.4 & 2.9 \\
 & tuned reward & 10 & 2.37 & 3.03 & 2.31 & 49 & 10.8 & 39.9 & 0.6 & 8.4 & 14.8 & 7.0 \\
\midrule
A1-70q & HF penalty & 11 & 2.55 & 6.61 & 2.55 & 60 & 7.3 & 49.3 & 2.5 & 7.6 & 26.4 & 4.4 \\
 & tuned reward & 11 & 2.55 & 22.90 & 2.55 & 63 & 2.9 & 21.1 & 0.8 & 22.8 & 5.0 & 14.3 \\
\midrule
A5-70q & HF penalty & 11 & 2.65 & 3.05 & 2.65 & 108 & 7.5 & 38.6 & 2.3 & 13.8 & 16.7 & 6.5 \\
 & tuned reward & 11 & 2.62 & 4.46 & 2.62 & 61 & 2.5 & 17.0 & 0.4 & 9.6 & 23.2 & 22.9 \\
\midrule
all & HF penalty & 43 & 2.85 & 5.00 & 2.64 & 55 & 8.4 & 49.3 & 1.8 & 10.3 & 21.7 & 4.4 \\
 & tuned reward & 43 & 2.40 & 7.33 & 2.39 & 56 & 4.3 & 25.0 & 0.7 & 14.4 & 19.0 & 14.3 \\
\bottomrule
\end{tabular}
\end{table*}

The two sets are close, and which one is better depends on the statistic. The tuned reward has the lower median error of the best circuit (2.4~mHa against 2.8~mHa for the Hartree--Fock penalty), of the better of the best and second circuit (2.4 against 2.6~mHa), and is the lower of the two in 26 of the 43 test cases. The Hartree--Fock penalty has the lower mean (4.8~mHa against 10.5) and the more reliable tail: none of its best circuits is above 30~mHa, against three for the tuned reward (41, 41 and 198~mHa, at SnOH3-me-2h2o M with the 46-qubit lineages and at SnOH3-me L with A5-70q). The two sets are within 5~mHa in 28 and 24 of the 43 test cases, and within 1~mHa in 13 and 12. The second circuit is less accurate for the tuned reward (median 7.3~mHa against 5.0~mHa), and the selected circuits have the same depth (median 55 and 56).

The circuits of the two sets differ in the chemistry of their shots. The best circuits of the Hartree--Fock penalty have the larger share of shots in the right spin sector (8.0\% against 4.4\%), a larger Hartree--Fock share of the valid shots (50\% against 24\%), fewer alpha-only shots (10.7\% against 14.5\%) and a smaller both-spin share (4.4\% against 14.8\%), with similar beta-only shares (21\% and 19\%).

\subsection{Divergence of the generated circuits from the idealised LUCJ circuit}
\label{sec:divergence}

The depth and error comparisons above do not say how \emph{different} a generated circuit is, as a physical object, from the idealised circuit it is meant to replace. We therefore compare each generated circuit $U_g$ (the product of the pool operators selected by the token sequence) with the LUCJ circuit $U_{\mathrm{LUCJ}}$ of the same molecule (the spin-balanced LUCJ operator of \code{ffsim} \cite{sung2026ffsim} built from CCSD amplitudes, one ansatz repetition, as in the hardware baseline), in an exact, noiseless calculation with state vectors in the full $N$-electron space. This is possible for the three tiers of at most 44 qubits whose determinant space fits in memory (SnOH3-me S, SnOH3-me-h2o S and SnOH3-me-2h2o S; 16, 19 and 22 orbitals); all tiers of 46 qubits and above cannot be computed exactly: their full state vectors need 1 to 4 terabytes at 46 qubits, 22 and 155 terabytes at 56 qubits and $3\times10^{9}$ gigabytes at 70 qubits (16 bytes per amplitude). SnOH3-me M (38 qubits, $2.5\times10^{9}$ determinants, a 41-gigabyte state vector) was not computed. The comparison is therefore limited to the three tiers that were computed, and the larger tiers are compared with LUCJ through circuit cost and recovered energy (Sections~\ref{sec:lucj-comparison} and~\ref{sec:refs}). All circuits of both training configurations on these tiers are included: 12 runs of ten circuits each for each of the two configurations (Table~\ref{tab:divergence-summary}); the tuned-reward run at SnOH3-me-2h2o S with A5-70q is the repeated draw of Appendix~\ref{sec:appendix-repeat}.

\paragraph{Measures.} With $|\psi_g\rangle=U_g|\mathrm{HF}\rangle$ and $|\phi\rangle=U_{\mathrm{LUCJ}}|\mathrm{HF}\rangle$ and the determinant distributions $p=|\psi_g|^2$, $q=|\phi|^2$ we report (i) the state fidelity $F=|\langle\psi_g|\phi\rangle|^2$ \cite{jozsa1994fidelity,nielsen2010}; (ii) the Hellinger distance $H=\sqrt{1-\mathrm{BC}}$ \cite{hellinger1909} with the Bhattacharyya coefficient $\mathrm{BC}=\sum_x\sqrt{p_xq_x}$ \cite{bhattacharyya1943}, which is 0 for identical and 1 for disjoint distributions; (iii) the Jensen--Shannon divergence in bits \cite{lin1991js}, bounded by 1; and (iv) $d_{\mathrm{HF}}=\sqrt{2-2|\langle\psi_g|\phi\rangle|}$, the distance between the two states up to a global phase, which is a lower bound of the phase-minimised operator distance $\min_\alpha\|U_g-e^{i\alpha}U_{\mathrm{LUCJ}}\|$ because the two operators are compared on one input state.

We also report two operator-level quantities, both without phase minimisation. First, for the best and second circuits of SnOH3-me S we estimate the operator norm $\|U_g-U_{\mathrm{LUCJ}}\|$ (maximum 2) by a matrix-free Lanczos iteration \cite{lanczos1950} on $(U_g-U_{\mathrm{LUCJ}})^\dagger(U_g-U_{\mathrm{LUCJ}})$ with a stopping criterion of top-Ritz residual below $10^{-4}$; 12 of 16 circuits met this criterion, while four reached the 60-step cap without convergence. Second, we estimate the normalised Hilbert--Schmidt distance as $\|U_g-U_{\mathrm{LUCJ}}\|_F/\sqrt{D}$ (with $D$ the Hilbert-space dimension) using a 3-vector Hutchinson trace estimate. This quantity is 0 for identical unitaries and $\sqrt2$ for unrelated Haar-random unitaries.

\paragraph{Result.} Table~\ref{tab:divergence-summary} gives the medians; the per-circuit values are in Appendix~\ref{sec:appendix-divergence} and in the released reports. The circuits that the ranking selects (best and second best) have high fidelity with the LUCJ state in the Hartree--Fock-penalty set without exception: all 24 selected circuits have $F>0.95$ (median $F=0.998$, Hellinger distance $0.035$, $d_{\mathrm{HF}}=0.05$). However, the LUCJ state itself has fidelity 0.9976--0.9997 with the Hartree--Fock determinant, so the unchanged reference would also score comparably high; the selected circuits exceed the Hartree--Fock fidelity by at most 0.0004. For the tuned reward 19 of the 24 selected circuits are near the LUCJ state ($F>0.5$; 17 have $F>0.95$): 11 of the 12 best circuits, but only 8 of the 12 second circuits. The other five selected circuits are orthogonal to it (fidelity 0), among them one best circuit. The circuits that are not selected are mostly the opposite: their median fidelity is $0$ and their median Hellinger distance $1.0$ (essentially orthogonal states, $d_{\mathrm{HF}}\approx\sqrt2$), although about half of them (46 of 96 for the Hartree--Fock penalty) are also near the LUCJ state. Divergence is related to depth: 98\% of the circuits of the Hartree--Fock penalty that are shallower than 200 are near the LUCJ state, against 36\% of the deeper ones (95\% against 27\% for the tuned reward). Closeness to the LUCJ state does not translate into a lower recovered error for the second circuit: on the three tiers where the divergence is computed, its median error is 0.13~mHa for the Hartree--Fock penalty, where all second circuits are near the LUCJ state, and 0.07~mHa for the tuned reward, where four of twelve are not.

\begin{table}[!tp]
\centering
\caption{Ideal (noiseless) distance of the generated circuits from the idealised LUCJ circuit (Section~\ref{sec:divergence}), medians over the circuits of the three small tiers (SnOH3-me S, SnOH3-me-h2o S, SnOH3-me-2h2o S; ten circuits per run; 12 runs for each of the two configurations). $F$: state fidelity; $H$: Hellinger distance between the determinant distributions; $d_{\mathrm{HF}}$: distance of the two states up to a global phase (a lower bound of the operator distance); ``near'': circuits with $F>0.5$. Selected = best and second circuit of the ranking. The per-circuit values are in Appendix~\ref{sec:appendix-divergence}.}
\label{tab:divergence-summary}
\footnotesize
\renewcommand{\arraystretch}{1.15}
\setlength{\tabcolsep}{3pt}
\begin{tabular}{@{}llrrrrrr@{}}
\toprule
Set & Circuits & $n$ & $F$ & $H$ & $d_{\mathrm{HF}}$ & near & depth \\
\midrule
HF penalty & selected & 24 & 0.998 & 0.035 & 0.049 & 24/24 & 57 \\
 & not selected & 96 & 0.000 & 0.999 & 1.413 & 46/96 & 3372 \\
\midrule
tuned reward & selected & 24 & 0.998 & 0.035 & 0.049 & 19/24 & 61 \\
 & not selected & 96 & 0.000 & 1.000 & 1.414 & 28/96 & 3830 \\
\bottomrule
\end{tabular}
\end{table}

\paragraph{Operator-level distances are not used.} The operator-level quantities above are reported without phase minimisation, whereas $d_{\mathrm{HF}}$ minimises over a global phase on the Hartree--Fock input state. Because the global phase of \code{ffsim}'s LUCJ circuit is arbitrary, the unphased operator-level distances can substantially overstate the discrepancy. For the selected circuits, the unphased state distance $\sqrt{2-2\mathrm{Re}\langle\psi_g|\phi\rangle}$ ranges from 0.17 to 0.59, while $d_{\mathrm{HF}}$ ranges from 0.05 to 0.22; for one tuned circuit it is 2.00 while $d_{\mathrm{HF}}$ is 0.05. We therefore do not use the operator-level values in the analysis; a phase-minimised operator distance was not computed here.

\paragraph{What the ideal state does not show.} The reference state is almost the Hartree--Fock determinant (99.8--99.97\% of its probability, because the CCSD amplitudes are small), so a high fidelity to it mostly says that the generated state is also close to the Hartree--Fock determinant (99.5\%, 100\% and 98\% for the best A1-46q circuit at the three tiers). The hardware does not sample that state: the measured shots of the same circuits are 7--29\% Hartree--Fock, 40--51\% single and 17--39\% double excitations (Table~\ref{tab:raw-shot-excitation}), so device noise, not the ideal circuit, produces the excitation content that the recovery step exploits. The divergence is thus a statement about the ideal circuits, not about what the device returns; the two should be read together.
Noiseless sampling with the matrix-product-state (MPS) simulator of cuQuantum~\cite{cuquantum} is consistent with this reading (Appendix~\ref{sec:appendix-noiseless-simulation}).
The simulation used 8-token checkpoints, so its circuits differ from those run on \code{ibm\_kobe}.
Without noise, circuits without a global orbital-rotation operator (\code{orb\_rot} in Table~\ref{tab:glossary}) returned the Hartree--Fock determinant in at least 90\% of their shots, and 4--7 circuits per test case returned only that determinant in $10^5$ shots.
Circuits containing a global \code{orb\_rot} operator instead produced many distinct configurations.

\paragraph{Relation to the recovered error.}
Among the recovered circuits of the three small tiers, the fidelity is only weakly related to the recovered error (Spearman correlation $0.31$ for the HF-penalty set), because every error on these tiers is within a fraction of a mHa (median 0.1~mHa both for circuits near the LUCJ state and for those far from it).

As noted in Section~\ref{sec:zero-shot}, the recovery cap exceeds the full alpha string dimension at these tiers.
In the noiseless MPS sampling of Appendix~\ref{sec:appendix-noiseless-simulation}, the merged recovered bases contained more than 99.5\,\% of all alpha strings.
The best individual circuits recovered to 0.25--0.51~mHa and had broad alpha-string coverage.
The comparison cannot say whether closeness to LUCJ helps at the larger tiers, where it cannot be computed.

\FloatBarrier
\subsection{Comparison against DMRG}
\label{sec:refs}

Table~\ref{tab:dmrg-vs-sqd} gives, for each SnOH3-me-family target, the lowest recovered energy over every lineage,
policy and result type in the fresh-shot campaign, with the DMRG reference (Table~\ref{tab:dmrg-reference}) and the
signed deviation $\Delta E$; the lineage and result type that achieved it are noted. This is a best-of-$k$ statistic over up to
eight lineage/policy combinations and five result types, so it is more favourable than any single pre-declared run;
per-lineage values are in Tables~\ref{tab:zeroshot-detail} and~\ref{tab:crossnorb-detail}.

The three small active spaces (SnOH3-me S (32q), SnOH3-me-h2o S (38q), SnOH3-me-2h2o S (44q)) are recovered to within 0.1~mHa of DMRG by the best
result of each (0.01, 0.01 and 0.04~mHa). On the larger tiers the lowest errors are 0.71~mHa (SnOH3-me-h2o M (46q)), 2.1~mHa (SnOH3-me-h2o LL (66q)), 1.9~mHa (SnOH3-me-h2o L (56q)), 3.8~mHa (SnOH3-me M (38q)), 4.3~mHa (SnOH3-me-2h2o M (58q)), 5.6~mHa (SnOH3-me L (46q)), 7.2~mHa (SnOH3-me-2h2o L (70q)) and 10.4~mHa (SnOH3-me LL (56q)). The SnOH3-me (non-hydrate) series is the hardest of the family for the policy: its 38-, 46- and 56-qubit tiers have the largest errors, and the SnOH3-me M (38q) tier, which has fewer qubits than the 46-qubit hydrate, is 3.8--17.2~mHa from DMRG where the hydrate is within 1.5~mHa.

The choice of circuit matters more than shot noise. Selecting the two circuits with the lowest
one-shot ranking energy at net CI dimension 40{,}000 gives errors of 0.71--1.5~mHa for Sn(OH)3-h2o M for the best circuit of each trained lineage, whereas the E+D circuits (Section~\ref{sec:methods-pipeline}), recovered on the same fresh shots, remain at 22.6--23.0~mHa in two of the four lineages (A1-46q and A1-70q), as do all untrained controls (22.6--22.9~mHa). Each circuit was sampled once, so this campaign cannot separate the choice of circuit from shot noise. The best circuits have $5.3\times10^{3}$--$2.4\times10^{4}$ valid shots of 100{,}000 at these test cases; the E+D picks that stay near 23~mHa have 13--2457 valid shots, but an E+D pick of A5-46q with 1992 valid shots reaches 1.4~mHa, so the valid-shot count alone does not separate them. We did not run the ablation that
would separate valid-shot count from ranking energy or depth, so we do not attribute the improvement to either.

\subsection{What distinguishes a good recovery: valid shots and selection}
\label{sec:analysis-selection}
The stored per-circuit data (released with the paper, see Appendix~\ref{sec:appendix-data}) allow a few observational checks. They are not controlled
experiments: valid-shot count, circuit depth and connectivity are correlated with one another, and the runs are few.
We use the 94 distinct individual-circuit recoveries of the HF-penalty trained runs at 44 qubits and above (best, second and the two E+D picks, duplicates removed). 

\paragraph{Valid shots.} The recovered error falls steeply with the number of valid shots
of the recovered circuit: median 41.1~mHa for circuits with fewer than 100 valid shots (25 recoveries; 4 below 25~mHa), 41.1~mHa for 100--1000
(11; 3 below 25~mHa), 5.3~mHa for 1000--10{,}000 (42; 39 below 25~mHa) and 1.8~mHa above 10{,}000 (16; all below 25~mHa). Within a single molecule the
Spearman correlation between $\log_{10}$ valid shots and error is negative at 7 of the 8 molecules with enough recoveries ($-0.65$ to $-0.89$); the one positive value ($+0.57$) occurs at SnOH3-me-2h2o S, where every
recovery is within a fraction of a mHa.

\paragraph{Ranking energy does not order recoveries.} Of the 43 HF-penalty runs in which both the best and the second best circuit were recovered, the best circuit
(lowest ranking energy) recovered to an energy no higher than the second in 28, i.e.\ 65\%. The ranking energy is a useful screen (it discards
circuits with too few valid shots) but its order among the leading circuits is only weakly informative about the recovered order.

\FloatBarrier
\subsection{Computational cost: training and inference}
\label{sec:cost}
This section reports the quantum and classical resources of the training and inference runs. Wall-clock figures are Slurm elapsed times, shot counts are circuits times shots per circuit, and QPU time is IBM's billed quantum time per job. The same numbers, with the per-run breakdown, are in Appendix~\ref{sec:appendix-data} and in the released HTML. The \code{ibm\_kobe} calibration in force during the experiments (T$_1$, T$_2$, readout, single- and two-qubit gate errors) is given in Section~\ref{sec:kobe-calibration}, Tables~\ref{tab:kobe-calibration-daily} and~\ref{tab:kobe-calibration-jobs}.

\subsubsection{Training}
Stage 1 runs on an exact-statevector simulator and uses no QPU time. Stage 2 samples on \code{ibm\_kobe} at 10{,}000 shots per circuit,
20 circuits per iteration batched in one Qiskit Runtime job, so each iteration is 200{,}000 shots and one job. For the Hartree--Fock-penalty configuration the four lineages ran 150 iterations of Stage 1 and 150 of Stage 2 each, i.e.\ 600 Stage-2 iterations,
12{,}000 circuits and 120{,}000{,}000 shots; Stage 1 took 106.3~h and Stage 2 112.8~h of wall time in total (the lineages ran concurrently). IBM billed 648 QPU minutes (10.8~h) for the 600 jobs, 65~s per job on average: 59.8 and 59.4~s per job for the 46-qubit lineages and 69.9 and 70.0~s for the 70-qubit lineages (Table~\ref{tab:t2-qpu}), so the 70-qubit iterations cost 1.2 times as much QPU time as the 46-qubit ones. The tuned reward costs 1175 QPU minutes (120~s per job) because its 70-qubit circuits are deeper. Table~\ref{tab:cost-summary} gives the quantum resources and Table~\ref{tab:cost-appendix} (Appendix~\ref{sec:appendix-cost}) the budget per lineage. Iterations lost to crashed Runtime jobs are not counted, so the shot totals are lower bounds.

\begin{table*}[!t]
\centering
\caption{Quantum resources on \code{ibm\_kobe}: circuits, shots, Qiskit Runtime jobs and IBM-billed QPU time (minutes) from the account's own job records. A training job is one iteration (20 circuits at 10{,}000 shots); an inference job is one run (10 circuits at 100{,}000 shots). Stage~1 of the training is a simulation and uses no QPU time. The inference rows are the sampling jobs of the 44 runs of each configuration; the untrained controls are the 8 control runs, and the LUCJ baseline one run per molecule at each of two transpilation levels. Training without a penalty is the A1-46q run and the stopped runs of the other lineages.}
\label{tab:cost-summary}
\small
\setlength{\tabcolsep}{5pt}
\begin{tabular}{@{}lrrrrr@{}}
\toprule
Phase & Circuits & Shots & Jobs & QPU time (min) & mean QPU s/job \\
\midrule
HF penalty, training Stage 2, 46-qubit models & 6{,}000 & 60{,}000{,}000 & 300 & 298 & 60 \\
HF penalty, training Stage 2, 70-qubit models & 6{,}000 & 60{,}000{,}000 & 300 & 350 & 70 \\
HF penalty, training total & 12{,}000 & 120{,}000{,}000 & 600 & 648 & 65 \\
HF penalty, inference (44 runs) & 440 & 44{,}000{,}000 & 44 & 315 & 429 \\
\midrule
Tuned reward, training Stage 2 & 11{,}740 & 117{,}400{,}000 & 587 & 1175 & 120 \\
Tuned reward, inference (44 runs) & 440 & 44{,}000{,}000 & 44 & 366 & 500 \\
\midrule
No penalty, training Stage 2 & 3{,}700 & 37{,}000{,}000 & 185 & 205 & 67 \\
Untrained controls, inference (8 runs) & 80 & 8{,}000{,}000 & 8 & 85 & 639 \\
LUCJ baseline, inference & 18 & 1{,}800{,}000 & 18 & 11 & 38 \\
\bottomrule
\end{tabular}
\end{table*}

The mean iteration time by pipeline step is in Figure~\ref{fig:stage-timing} (Table~\ref{tab:t2-timing}, Appendix~\ref{sec:appendix-training-detail}). In Stage 1 the circuit sampling (exact simulation) accounts for the largest share
(57--71\% of an iteration); in Stage 2 of the HF-penalty recipe the sampling share is
47--61\% and includes the IBM device queue, with the SQD diagonalization/recovery taking most of the remainder.
The refinement, transpilation and backward steps together are under 5\% of an iteration in every lineage.

\begin{table*}[!t]
\centering
\caption{Inference timing and cost of the two training configurations (one zero-shot run = ten circuits sampled once on \code{ibm\_kobe} at 100{,}000 shots, ranked, then up to five recoveries). Wall times are Slurm elapsed times of the finished jobs on the roquo cluster; ``jobs'' counts finished jobs, failed, cancelled and timed-out jobs are listed separately and not counted. Recovery jobs use two GPUs each. QPU time is the time IBM billed for the sampling jobs (job records); the untrained controls are not included. Recoveries still running when the table was made are listed in the row ``Still running'' and are not counted.}
\label{tab:inference-timing}
\small
\renewcommand{\arraystretch}{1.15}
\setlength{\tabcolsep}{4pt}
\begin{tabular}{@{}llrrrr@{}}
\toprule
Set & Stage & jobs & wall (h) & mean per job (min) & GPU-hours \\
\midrule
HF penalty & Sampling and transpilation (IBM round trip) & 44 & 128.1 & 175 & 128.1 \\
 & Ranking diagonalisation & 44 & 1.0 & 1 & 2.0 \\
 & Recovery: best / second & 86 & 196.5 & 137 & 392.9 \\
 & Recovery: E+D 1 / 2 & 48 & 306.9 & 384 & 613.8 \\
 & Recovery: merged & 43 & 191.5 & 267 & 383.0 \\
 & All recoveries & 177 & 694.9 & 236 & 1389.8 \\
 & QPU time billed by IBM (sampling) & \multicolumn{4}{l}{44/44 jobs: 18{,}877 s (315 min), 429 s per job} \\
 & Not counted: failed, cancelled or timed out & 1 & 0.00 & & \\
\midrule
tuned reward & Sampling and transpilation (IBM round trip) & 44 & 191.5 & 261 & 191.5 \\
 & Ranking diagonalisation & 44 & 1.2 & 2 & 2.4 \\
 & Recovery: best / second & 86 & 254.1 & 177 & 508.3 \\
 & Recovery: E+D 1 / 2 & 50 & 332.9 & 399 & 665.8 \\
 & Recovery: merged & 43 & 228.2 & 318 & 456.3 \\
 & All recoveries & 179 & 815.2 & 273 & 1630.4 \\
 & QPU time billed by IBM (sampling) & \multicolumn{4}{l}{44/44 jobs: 21{,}980 s (366 min), 500 s per job} \\
\bottomrule
\end{tabular}
\end{table*}

\subsubsection{Inference}
Inference runs once on hardware per run (a batch of 10 circuits at 100{,}000 shots each, one Runtime job) and then entirely on classical hardware. Each of the two training configurations has 44 runs (440 circuits, $4.4\times10^{7}$ shots, 44 Runtime jobs); the billed QPU time and the classical wall time are in Table~\ref{tab:inference-timing}. The LUCJ baseline used 18 Runtime jobs (1{,}800{,}000 shots, 11~QPU minutes; one per molecule at each of the two transpilation levels) and the eight untrained controls 8 jobs (Table~\ref{tab:cost-summary}).

Table~\ref{tab:inference-timing} gives the inference timing and the billed QPU time of the two configurations side by side, stage by stage.

In both configurations the classical recovery dominates the wall time, far more than the hardware round trip (mostly IBM queue time and local transpilation) or the ranking (Table~\ref{tab:inference-timing}); merged recoveries are the most expensive per job because they pool up to ten times the shots. Failed, cancelled and timed-out jobs are not counted in the table, and the per-run times are in the released reports.


\FloatBarrier
\section{Discussion}
\label{sec:discussion}

The two rewards give results that are equivalent in accuracy at the resolution of this study (Section~\ref{sec:three-model-inference}): on the 43 test cases that both have, the median error of the best circuit is 2.8~mHa for the Hartree--Fock penalty and 2.4~mHa for the tuned reward, with one training seed per lineage.
We centre the paper on the Hartree--Fock penalty because it is the simpler reward, with one added term, and because it has no best-circuit error above 30~mHa, against three for the tuned reward.
The more diverse policy of the tuned reward (an effective vocabulary of 21--27 tokens against one; Appendix~\ref{sec:appendix-chemistry}) does not translate into more accurate circuits on unseen molecules in this setting.
At inference, the penalty does not remove the Hartree--Fock weight of the shots as it does in training; it changes the balance between the spin channels (Section~\ref{sec:three-model-inference}).
We have not tested whether these differences cause the differences in accuracy. What both configurations share, and what the comparison with the untrained controls supports, is that a trained policy recovers well below an untrained one at the tiers where untrained policies fail (a median of 106~mHa for the eight untrained controls against 2.8~mHa for the 43 runs of the Hartree--Fock penalty).

The HF-penalty policy has collapsed onto about one operator sequence by iteration 50--55 of Stage~2, and its 46-qubit circuits excite one spin channel only (Section~\ref{sec:stage2-results}). Both statements describe the policy as trained, and the zero-shot circuits are drawn from it with the molecule-conditioned input, so they are not a fixed circuit; but ten samples are far from ten independent circuits (Section~\ref{sec:zero-shot} shows that some runs' ten circuits are nearly identical). The practical consequence is that a large share of the accuracy comes from the ranking and recovery steps that follow the sampling, and from a modest number of well-yielding circuits, not from a rich learnt distribution. The tuned reward gives a slightly lower median error (2.4 against 2.8~mHa on the 43 paired test cases) but a worse tail, at the price of deeper 70-qubit circuits and 1.8 times the QPU time per job. Whether a more diverse policy would help at larger scale, or with more circuits per run, is open question.

The dominant lever found during hyperparameter tuning was the circuit-depth
admission ceiling, not the magnitude of the coverage/overshoot reward
terms: once the ceiling is set generously relative to the transpiled
depths produced by a given circuit length $L$ and qubit count, a wide range of
reward-weight combinations learn comparably, whereas an overly tight
ceiling produces sustained zero-gradient stalls regardless of reward
shaping. The admission gate's role as a hard constraint on
which samples ever reach the policy-gradient update is a more consequential
design choice, for this architecture and reward formulation, than the
relative weighting of the soft reward terms. The pass rate of the Stage-2 gate differs by qubit count and not by conditioning scheme (100\% and 99\% at 70 qubits against 46\% and 47\% at 46 qubits, Section~\ref{sec:stage2-results}); because A1 and A5 agree, we attribute it to the energy-admission tolerance of the 46-qubit lineage and not to either architecture.

A related, more acute version of the same structural issue appeared at the
70-qubit hardware tier in the preliminary runs: the energy admission tolerance that worked at
46 qubits produced zero admitted iterations over many consecutive
hardware iterations, a genuine deadlock rather than
slow convergence, since without an admitted iteration no gradient exists. Loosening the tolerance by a fixed
factor broke the deadlock, and the 70-qubit lineages of the final runs, which start from that loosened tolerance, admit 99--100\% of their iterations (Table~\ref{tab:t2-compare}). The 46-qubit lineages never needed the loosening, yet it is they that show the weaker admission outcome:
both settle near a 46--47\% pass rate (Figure~\ref{fig:gate-pass}), two independent conditioning schemes giving the same
outcome, which points at the 46-qubit lineage's \emph{energy tolerance itself} being comparatively
tight for its own curriculum, not at either architecture. It is the mirror image of the 70-qubit
finding: there the tolerance blocked learning until loosened; here learning proceeds, but roughly half of all hardware circuits fail the gate throughout training. We did not test a different tolerance for the 46-qubit lineages.

The depth and two-qubit-gate curves cannot be read as evidence of transfer. The best circuit of Stage~2 is shallow from the first iterations (Section~\ref{sec:stage2-results}), whereas Stage~1 ends with depths of several thousand (Section~\ref{sec:stage1-results}), but the two stages differ in the molecules (the Stage-1 curriculum against the 23- and 35-orbital targets), the sampler, the best-of-20 selection by energy and the depth ceiling, so the change cannot be attributed to the carried policy. We did not establish that the Stage-1 warm start improves Stage~2 over a random initialization under the same budget. A short test on a small molecule (ethylene) on a noisy simulator showed no clear difference between warm-started and randomly initialized policies; however, this test was not designed as a controlled comparison of the hardware stage, so no quantitative conclusion about warm-start benefits can be drawn from it.

The divergence measures (Section~\ref{sec:divergence}) are computed for the ideal, noiseless circuit on the Hartree--Fock input, and only for the three tiers of at most 44 qubits. They support three statements. First, the circuits that the ranking selects have a fidelity above 0.95 with the LUCJ state, but no higher than that of the Hartree--Fock determinant itself (0.9976--0.9997), so this fidelity does not show that the circuits reproduce the correlated part of LUCJ. Second, the deep circuits that the policy also samples are orthogonal to it (fidelity 0, Hellinger distance 1), which matches their absence of valid shots on the device; the same circuits carry the diagonal-Coulomb layers that dominate their depth. Third, closeness is not implied by shallowness: in the SnOH3-me S run of the A1-46q policy, a circuit of depth 46 has fidelity 0.954 and one of depth 196 has fidelity 0, whereas the circuits of depth 30--57 reach 0.993--0.998. Operator-level distances are reported without phase minimisation and are not used in the analysis (Section~\ref{sec:divergence}). The accuracy of the method is therefore established by the recovered energies against DMRG, and not by the closeness to LUCJ.

We looked for two trends and found neither clearly. (i) \emph{Does a policy predict better on active spaces close to the size it was trained for?} For the HF-penalty set, the median error of the selected circuit at the spaces of at most 46 qubits is 0.85 and 1.07~mHa for the A1-46q and A5-46q policies, and 0.46 and 0.93~mHa for the A1-70q and A5-70q policies; at 56--70 qubits it is 6.5 and 4.1~mHa (46-qubit policies) against 4.3 and 4.9~mHa (70-qubit policies). The 70-qubit policy is better in 5 of 6 test cases (A1) and 5 of 6 (A5) at the small spaces and in 4 of 5 (A1) and 2 of 4 (A5) at the large ones, so the larger policy is slightly better throughout, including on the spaces the smaller policy was trained for, and the gap does not grow with the size of the target. The medians over all eleven spaces (2.6--3.6~mHa) do not separate the four lineages with one seed each, so we find no specialisation of a policy to the size it was trained for; the largest errors go with few usable circuits in the draw (Appendix~\ref{sec:appendix-repeat}), an observation on one draw per test case and not a test. (ii) \emph{Does the heavier A5 encoder generate deeper circuits?} No. Over all 220 sampled circuits per encoder, the A1 circuits are deeper (median as-submitted depth 3{,}427 against 528 for A5; mean 5{,}693 against 3{,}550). This is carried by the A1-70q policy, whose circuits are the deepest of the four lineages (median 5{,}581, 65\% deeper than 1{,}000), whereas A5-70q has the shallowest and the highest valid-shot yield. The extra structural information of the A5 encoder is therefore not paid for in circuit depth, and it does not give a consistently lower error either (median over all spaces 3.6 and 2.7~mHa for A5, 2.9 and 2.6~mHa for A1). All comparisons rest on one training seed per lineage.

The depth advantage of Section~\ref{sec:lucj-comparison} is paid for with shots. A fresh-shot run samples ten circuits at 100{,}000 shots each ($10^6$ shots, one Runtime job) to obtain one or two circuits for recovery, and ranks all ten by a diagonalization; eight or nine of the ten are discarded, and in most runs the typical sampled circuit is deeper than LUCJ. A conventional LUCJ-based SQD run uses one fixed circuit and 100{,}000 shots, needs no training or selection step, uses 11 to 13 times less QPU time (38~s per job against 429 and 500~s per job for the Hartree--Fock-penalty and tuned-reward runs, Table~\ref{tab:inference-timing}), and its depth can be tuned directly through the number of ansatz repetitions and the transpiler optimization level. Whether the policy is worth its additional shots and its one-time training (Table~\ref{tab:cost-summary}) depends on the size of the active space. At the three smallest spaces (32, 38 and 44 qubits) the recovery subspace cap alone reached the DMRG answer from the LUCJ samples, as it did from every GQE circuit, so there the additional shots and the policy buy nothing. From 46 qubits up, LUCJ-SQD's raw errors are 22.7--204.1~mHa and its singles-injected errors 5.2--41.2~mHa (5.2--12.6~mHa except at SnOH3-me LL). The policy's accuracy gain over LUCJ is established only at SnOH3-me-h2o M, with one fresh sample per circuit; at the other tiers its errors (Table~\ref{tab:dmrg-vs-sqd}) are comparable to or only modestly better than LUCJ's singles-injected errors, and the benefit there is the shallower circuit (Section~\ref{sec:lucj-comparison}).

\FloatBarrier
\section{Conclusion}
\label{sec:conclusion}

We trained amortized GQE policies in two stages, a noiseless-simulator warm-up and 150 iterations with the IBM Heron processor \code{ibm\_kobe} in the training loop, at 46 and 70 qubits, with a Hartree--Fock penalty in the hardware reward. Applied without retraining to the eleven active spaces of the Sn(OH)$_3$-Me family, 32 to 70 qubits, the policies gave, after one round of hardware sampling, followed by classical ranking and recovery of the cached counts with the molecular Hamiltonian and no further QPU calls, a median error of 2.8~mHa against DMRG over the 43 trained runs (13 within 1~mHa, 42 within 15~mHa). At 46 to 70 qubits, 27 of 28 runs are within 15~mHa (median 4.5~mHa); the errors are 0.06--0.65~mHa at the 32--44-qubit spaces (3.8--17~mHa at SnOH3-me M, 38 qubits) and 0.7--1.5~mHa at the 46-qubit hydrate. The trained policy is more accurate than its untrained control in all eight runs that have one (median error of the controls 106~mHa). At 56--70 qubits the selected circuits used 34--72$\times$ fewer two-qubit gates and were 4--12$\times$ shallower than the LUCJ circuit (offline estimate; the submitted LUCJ circuits were larger), at the price of about 11 times the QPU time per job, because ten circuits are sampled instead of one. These results show that training a circuit generator with the quantum hardware in the loop is feasible at 70 qubits. The tuned reward gives a slightly lower median error, 2.4 against 2.8~mHa on the 43 test cases that both have (lower in 26), but a worse tail: three of its best circuits exceed 30~mHa, none does for the Hartree--Fock penalty. We therefore keep the simpler Hartree--Fock penalty as the main configuration, for its robustness in these single-seed runs.

The evidence has clear limits. The cost and accuracy advantages are those of the selected circuit, not of the typical sampled one; the ranking energy screens out circuits with too few valid shots but does not order the leading circuits; the trained policy collapses onto essentially one operator sequence in training, and its 46-qubit circuits excite one spin channel only; each circuit was sampled once and each configuration trained with one seed; training uses an approximate classical reference energy (CCSD(T) for the Stage-2 molecules) in the admission gate and in two reward terms, so the protocol applies to systems for which such a reference exists; the effect of model size was not studied; the evidence that the policy transfers to new chemical systems remain qualitative; the untrained controls cover three molecules; and no ablation separates valid-shot count, depth and connectivity as the cause of the gains, or tests whether idle qubits matter (Appendix~\ref{sec:appendix-inference-detail}). The four main training runs required $1.2\times10^{8}$ hardware shots and 648 QPU minutes. The code and the per-run results are released (Appendix~\ref{sec:appendix-availability}).

Several follow-up studies are suggested by these limits. Controlled hardware comparisons of warm-started and randomly initialized policies, multiple training seeds per lineage, and smaller or parameter-efficient transformers would help quantify how sensitive the results are to the training recipe and model size. Whether a more diverse policy improves performance at larger scale, or with more circuits per run, remains open. Because several runs rely on modest valid-shot yields, a natural next step is to combine the approach with error mitigation (not used here) and to order training molecules by a measure of how difficult their ground states are to sample. Finally, the cost accounting reported here can guide the choice of the number of circuits and shots when extending the approach to larger active spaces.

\section*{Acknowledgements}

We thank the RIKEN JHPC-quantum operations team
for providing access to the Fugaku supercomputers' IBM Quantum allocation
(\code{ibm\_kobe} based on the IBM Heron
processor) and for promptly flagging and help-
ing resolve a technical issue during the
hardware training campaign. Part
of this work was carried out on the Miyabi supercomputer system (Miyabi-G) of the Joint Center for Advanced High Performance Computing (JCA-
AHPC), through its ``Large-scale HPC Challenge'', a call-for-proposals program that gives exclusive use of the accelerated GPU nodes of Miyabi-G, the general-purpose CPU nodes of Miyabi-C, or both. All main-text training, inference and LUCJ results were obtained on \code{ibm\_kobe}. Additional preliminary experiments that are not part of the reported analyses were performed on \code{ibm\_kawasaki} managed by the University of Tokyo.
A part of this work was performed for Council for Science, Technology and Innovation (CSTI), Cross-ministerial Strategic Innovation Promotion Program (SIP), ``Promoting the application of advanced quantum technology platforms to social issues'' (Funding agency: QST).

We used Claude Sonnet 5.5 (Anthropic) to assist with drafting and editing the manuscript and the analysis code. The authors reviewed the output and are responsible for the content of this paper.

The sampled circuits, raw counts and result files behind every table are described in Appendix~\ref{sec:appendix-availability}.

\appendix
\onecolumn
\renewcommand{\thefigure}{A\arabic{figure}}\renewcommand{\thetable}{A\arabic{table}}
\setcounter{figure}{0}\setcounter{table}{0}
\FloatBarrier
\section{Classical reference benchmarks}
\label{sec:appendix-classical}

\begin{table}[h]
\centering
\caption{DMRG reference energies, SnOH3-me family, all active-space tiers (maximum bond dimension 4000, 17 sweeps, energy-convergence threshold $10^{-5}\,\eh$). The truncation error is the discarded weight of the last sweep; it is not an uncertainty estimate. No sweep-by-sweep energies or lower-bond-dimension calculations are reported, so differences below about $10^{-2}$~mHa should not be over-interpreted.}
\label{tab:dmrg-reference}
\begin{tabular}{@{}llrr@{}}
\toprule
Molecule & Tier (active space) & DMRG energy ($\eh$) & Truncation error \\
\midrule
SnOH3-me & S (16o, 24e) & $-479.193382$ & $2.87\times10^{-12}$ \\
SnOH3-me & M (19o, 24e) & $-479.302278$ & $4.73\times10^{-7}$ \\
SnOH3-me & L (23o, 32e) & $-479.314216$ & $1.45\times10^{-6}$ \\
SnOH3-me & LL (28o, 42e) & $-479.315057$ & $2.04\times10^{-7}$ \\
SnOH3-me-h2o & S (19o, 30e) & $-555.108833$ & $2.53\times10^{-12}$ \\
SnOH3-me-h2o & M (23o, 30e) & $-555.125158$ & $2.33\times10^{-9}$ \\
SnOH3-me-h2o & L (28o, 40e) & $-555.135058$ & $4.70\times10^{-9}$ \\
SnOH3-me-h2o & LL (33o, 50e) & $-555.137396$ & $1.64\times10^{-8}$ \\
SnOH3-me-2h2o & S (22o, 36e) & $-631.083312$ & $3.04\times10^{-12}$ \\
SnOH3-me-2h2o & M (29o, 36e) & $-631.118616$ & $2.80\times10^{-7}$ \\
SnOH3-me-2h2o & L (35o, 48e) & $-631.132238$ & $5.76\times10^{-7}$ \\
\bottomrule
\end{tabular}
\end{table}

\FloatBarrier
\section{MP2 natural orbitals and active spaces of the target molecules}
\label{sec:appendix-orbitals}
\begin{figure}[t]
\centering
\includegraphics[width=\linewidth]{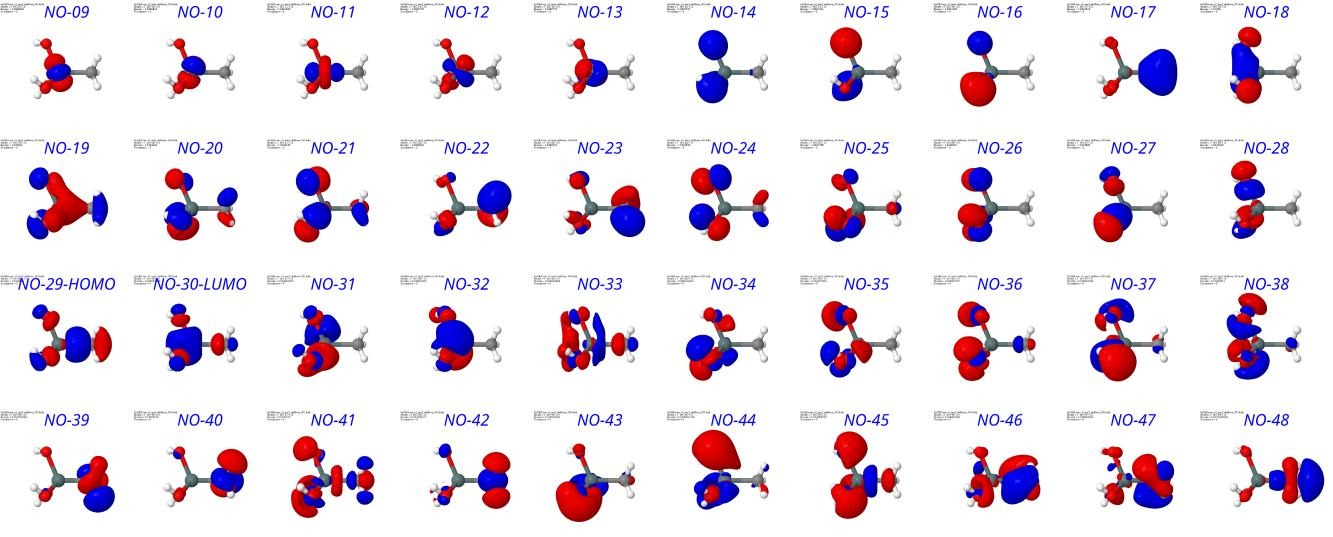}
\caption{MP2 natural orbitals NO-9 to NO-48 of Sn(OH)$_3$-Me (MP2/def2-SVP, frozen core, Gaussian~16; isosurfaces, the two colours are the two signs of the orbital). The orbitals are ordered by occupation number; the HOMO and LUMO are labelled. The active spaces of Table~\ref{tab:active-spaces} are ranges of these orbitals (Figure~\ref{fig:no-occupation}). The first eight natural orbitals are not shown.}
\label{fig:no-orbitals-snoh3me}
\end{figure}

\begin{figure}[t]
\centering
\includegraphics[width=\linewidth]{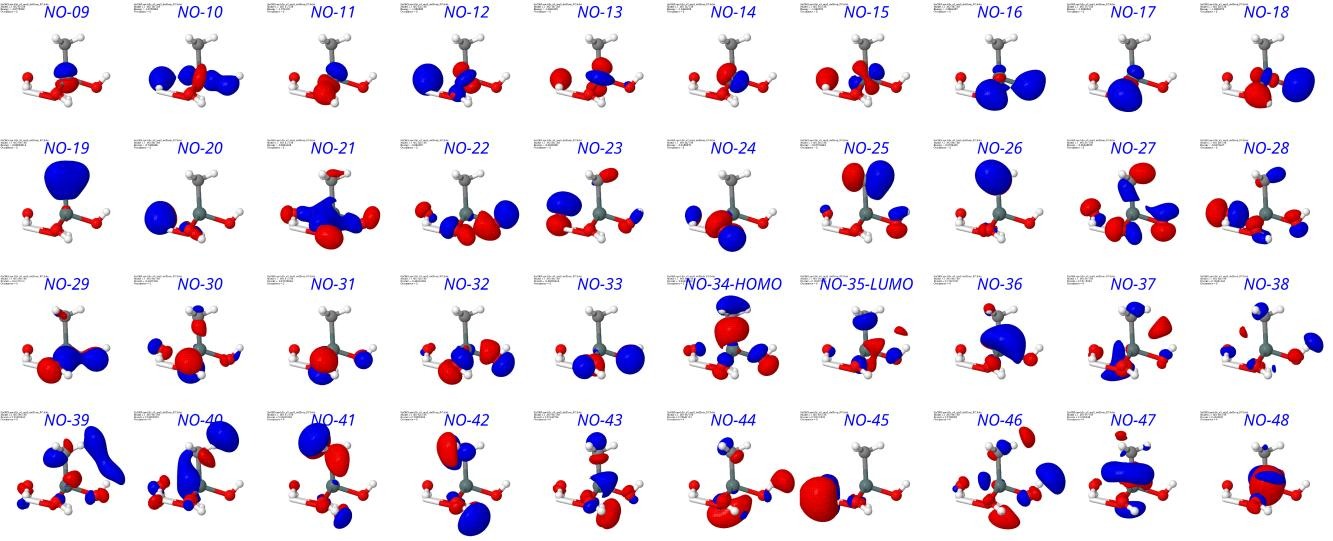}
\caption{MP2 natural orbitals NO-9 to NO-48 of Sn(OH)$_3$-Me$\cdot$H$_2$O (MP2/def2-SVP, frozen core, Gaussian~16; isosurfaces, the two colours are the two signs of the orbital). The orbitals are ordered by occupation number; the HOMO and LUMO are labelled. The active spaces of Table~\ref{tab:active-spaces} are ranges of these orbitals (Figure~\ref{fig:no-occupation}). The first eight natural orbitals are not shown.}
\label{fig:no-orbitals-h2o}
\end{figure}

\begin{figure}[t]
\centering
\includegraphics[width=\linewidth]{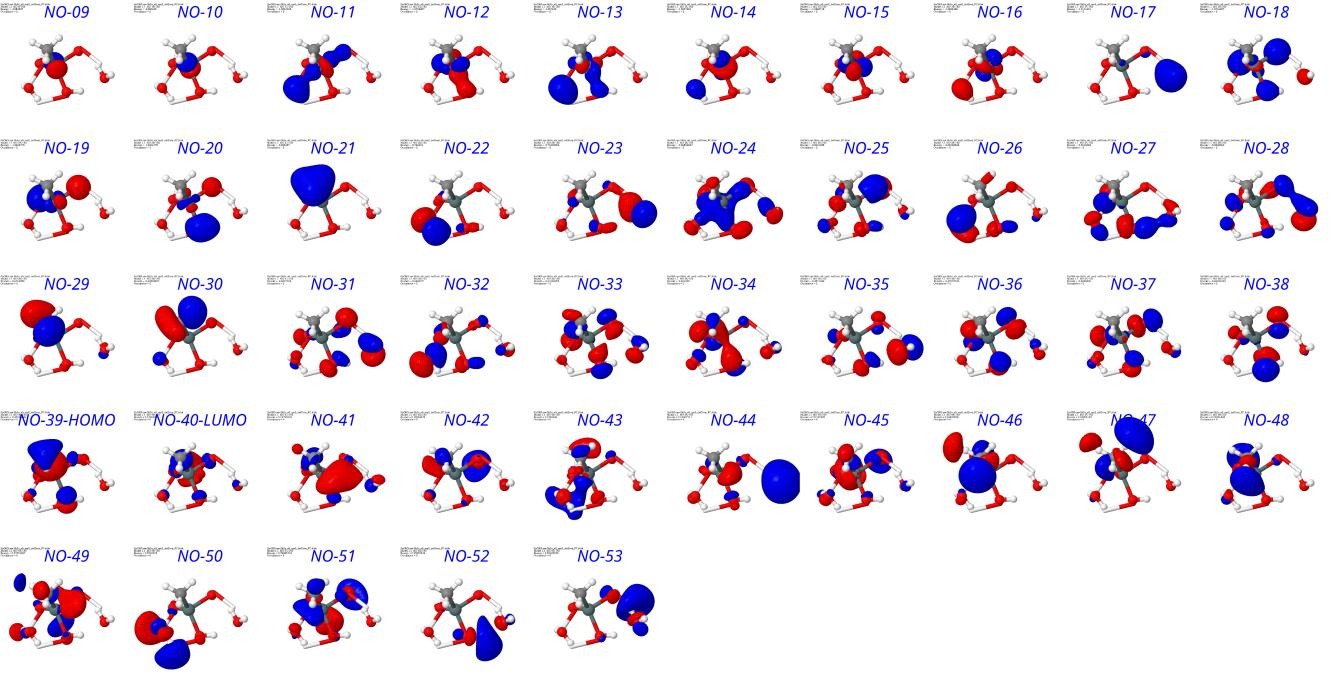}
\caption{MP2 natural orbitals NO-9 to NO-53 of Sn(OH)$_3$-Me$\cdot$2H$_2$O (MP2/def2-SVP, frozen core, Gaussian~16; isosurfaces, the two colours are the two signs of the orbital). The orbitals are ordered by occupation number; the HOMO and LUMO are labelled. The active spaces of Table~\ref{tab:active-spaces} are ranges of these orbitals (Figure~\ref{fig:no-occupation}). The first eight natural orbitals are not shown.}
\label{fig:no-orbitals-2h2o}
\end{figure}

This appendix documents the natural orbitals behind the active spaces of the held-out molecules (Section~\ref{sec:molecules}): the definition of the tiers (Table~\ref{tab:active-spaces}) and the orbitals of all three molecules (Figures~\ref{fig:no-orbitals-snoh3me}--\ref{fig:no-orbitals-2h2o}), followed by comments on the orbital shapes. The occupation numbers are those of the Gaussian~16 MP2 densities (for Sn(OH)$_3$-Me as stored with the natural orbitals, for the hydrates from the MP2 density and the def2-SVP overlap matrix)~\cite{gaussian16,moller1934,weigend2005}. The comments are our reading of the plotted isosurfaces and occupation numbers; they are qualitative and have not been checked by a population analysis.

\begin{table}[ht]
\centering
\caption{%
Active spaces built on the MP2 natural orbitals (NOs) of the three target molecules (MP2/def2-SVP, frozen core, Gaussian~16).
A tier is a contiguous range of NOs ordered by occupation number, from the lowest NO of the range through the HOMO and the first virtual NOs; orbitals and electrons give the active space, ``occupied'' and ``virtual'' the NOs of the range below and above the HOMO.
Alpha strings is the number of alpha-spin occupation strings, $\binom{n_\mathrm{orb}}{n_\mathrm{elec}/2}$, which equals the number of beta strings for these closed-shell active spaces.
The last column says whether the tier was run in this paper (all eleven tiers of 32 to 70 qubits were run for the four lineages of each configuration).}
\label{tab:active-spaces}
\small
\resizebox{\linewidth}{!}{%
\begin{tabular}{@{}llccrrrrcc@{}}
\toprule
Molecule & Tier & (orbitals, electrons) & NO range & Occupied & Virtual & Qubits & Alpha strings & HOMO/LUMO NO & Run here \\
\midrule
Sn(OH)$_3$-Me & LL & 28, 42 & 9--36 & 21 & 7 & 56 & 1,184,040 & 29 / 30 & yes \\
Sn(OH)$_3$-Me & L & 23, 32 & 14--36 & 16 & 7 & 46 & 245,157 & 29 / 30 & yes \\
Sn(OH)$_3$-Me & M & 19, 24 & 18--36 & 12 & 7 & 38 & 50,388 & 29 / 30 & yes \\
Sn(OH)$_3$-Me & S & 16, 24 & 18--33 & 12 & 4 & 32 & 1,820 & 29 / 30 & yes \\
\addlinespace
Sn(OH)$_3$-Me$\cdot$H$_2$O & LL & 33, 50 & 10--42 & 25 & 8 & 66 & 13,884,156 & 34 / 35 & yes \\
Sn(OH)$_3$-Me$\cdot$H$_2$O & L & 28, 40 & 15--42 & 20 & 8 & 56 & 3,108,105 & 34 / 35 & yes \\
Sn(OH)$_3$-Me$\cdot$H$_2$O & M & 23, 30 & 20--42 & 15 & 8 & 46 & 490,314 & 34 / 35 & yes \\
Sn(OH)$_3$-Me$\cdot$H$_2$O & S & 19, 30 & 20--38 & 15 & 4 & 38 & 3,876 & 34 / 35 & yes \\
\addlinespace
Sn(OH)$_3$-Me$\cdot$2H$_2$O & L & 35, 48 & 16--50 & 24 & 11 & 70 & 417,225,900 & 39 / 40 & yes \\
Sn(OH)$_3$-Me$\cdot$2H$_2$O & M & 29, 36 & 22--50 & 18 & 11 & 58 & 34,597,290 & 39 / 40 & yes \\
Sn(OH)$_3$-Me$\cdot$2H$_2$O & S & 22, 36 & 22--43 & 18 & 4 & 44 & 7,315 & 39 / 40 & yes \\
\bottomrule
\end{tabular}
}
\end{table}

\paragraph{Occupation numbers.} In all three molecules the natural-orbital occupations stay close to 2 or 0 (Figure~\ref{fig:no-occupation}): no orbital has an occupation between about 0.1 and 1.9. The largest deviations are at the frontier: the HOMO has an occupation of 1.956--1.957 and the LUMO of 0.048--0.049, in each molecule. The occupied orbitals fall off smoothly from 2.00 to the HOMO, so there is no gap in the occupied part that would define a natural boundary for an active space; the virtual orbitals drop from the LUMO (0.048--0.049) to 0.031--0.033 and decay slowly. The orbitals nearest the frontier, which are the most fractionally occupied, are the first that every tier includes. This is a limitation, not a finding: MP2 occupations stay close to 0 and 2 for weakly correlated systems and do not show whether strong static correlation exists. The comparison with DMRG validates the SQD energies within each selected active space, not the physical completeness of that space.

\paragraph{Orbitals near the frontier.} In Sn(OH)$_3$-Me the HOMO (NO-29) is concentrated along the Sn--CH$_3$ axis, with lobes of opposite sign on the tin and on the methyl side and small amplitude on the oxygens, which looks like a Sn--C $\sigma$-type orbital. The LUMO (NO-30) has a large lobe on the tin and hydroxyl side and a smaller lobe on the methyl carbon. In the hydrates the frontier orbitals are less symmetric because the water ligands break the three-fold arrangement of the hydroxyl groups. The HOMO of the monohydrate (NO-34) has its main lobe above tin towards the methyl group and small lobes on the hydroxyl oxygens, and its LUMO (NO-35) consists of small lobes spread over the hydroxyl region. In the dihydrate the HOMO (NO-39) has lobes on the methyl side and on one oxygen and the LUMO (NO-40) lobes on tin and neighbouring oxygens. If, as these shapes suggest, the Sn--C bond region contributes to the frontier orbitals, the active spaces are consistent with the Sn--C bond cleavage that matters for organotin photoresists, but this is a reading of the pictures, not a result.

\paragraph{Orbitals below the frontier and the Sn~4d orbitals.} The occupied natural orbitals at the start of the LL range (NO-9 to about NO-15 in Sn(OH)$_3$-Me, with the corresponding ranges in the hydrates) have compact, multi-lobed shapes around tin with small contributions from the oxygens, which is the mixing of Sn~4d with O~2p noted in Figure~\ref{fig:no-occupation}; their occupations are 1.998 or higher. These orbitals are only in the LL tier. The orbitals between them and the HOMO are mostly lone-pair-like on single hydroxyl or water oxygens, or localized along Sn--O bonds, and the hydrates add orbitals that extend over the water ligands.

\paragraph{Higher virtual orbitals.} The virtual orbitals above the LUMO have more nodes, and several of them are large diffuse lobes that extend outside the molecular frame. Such orbitals are probably correlating or diffuse functions rather than bonding orbitals. NO-44 and NO-45 of the monohydrate lie above its active spaces (they end at NO-42), so they are in no tier. NO-44 of the dihydrate is inside its M and L tiers (ending at NO-50) and outside S (ending at NO-43). Whether adding a few higher natural orbitals would improve the active spaces, as the selection of these spaces suggested, and whether the Sn~4d orbitals are needed, were not tested here.

\FloatBarrier
\section{Glossary of code-level hyperparameters and labels}
\label{sec:appendix-glossary}

The main text uses descriptive names for every setting; this appendix gives, for each, the flag or identifier it carries in the released code, its exact meaning, and the value used. Units: energies in mHa (10$^{-3}$ Hartree) unless stated; ``iteration'' is one policy update from a batch of sampled circuits; ``admitted'' means that the iteration passed the admission gate (the best circuit of the batch of 20) and so contributed a gradient step.

\begingroup
\small
\renewcommand{\arraystretch}{1.2}
\begin{longtable}{@{}>{\raggedright\arraybackslash}p{3.0cm}>{\raggedright\arraybackslash}p{3.6cm}>{\raggedright\arraybackslash}p{7.2cm}@{}}
\caption{Glossary. ``Name in the text'' is the descriptive name used in the main text; ``Code label'' is the flag or identifier in the released code.}\label{tab:glossary}\\
\toprule
Name in the text & Code label & Definition \\
\midrule
\endfirsthead
\toprule
Name in the text & Code label & Definition \\
\midrule
\endhead
\multicolumn{3}{l}{\emph{Policy and circuit}}\\
Number of gate tokens per circuit, $L$ & \code{ngates} & Number of operator tokens the policy generates for one circuit (28 for the 46-qubit lineages, 44 for the 70-qubit lineages). It fixes the size of the policy's position-embedding table and therefore cannot change between training stages. \\
\addlinespace
Operator pool; vocabulary & \code{FfsimLUCJPool} & The set of unitary operators that the tokens select. For a molecule with $n_{\mathrm{orb}}$ spatial orbitals it has $3\,n_{\mathrm{orb}}+5$ operators; token ids above that are identity padding so that one output layer serves every molecule. \\
Local excitation gates & \code{givens}, \code{num\_num} & Givens rotation between two neighbouring spatial orbitals (single-excitation character) and number--number (density--density) phase gate between two orbitals. \\
Global operators & \code{orb\_rot}, \code{diag\_coulomb}, \code{LUCJ} & Operators acting on every qubit: four fixed (seeded) orbital rotations, a diagonal-Coulomb layer with two time values, and one LUCJ layer (Table~\ref{tab:token-map}). \\
Molecule conditioning input & \code{set\_prompt} & Computes the molecule descriptor (and, for the A5 encoder, the molecular graph embedding) once and stores it, so that the decoder attends to it at every step. \\
\addlinespace
\multicolumn{3}{l}{\emph{Optimiser and curriculum}}\\
Learning rate & \code{base-lr} & Step size of the AdamW optimiser, $5\times10^{-6}$ for all lineages. \\
Prompt-encoder learning-rate multiplier & \code{encoder-lr-mult} & The prompt encoder's learning rate is the learning rate times this factor; 1.0 here. \\
Gradient-norm clipping threshold & \code{grad-clip} & Maximum $L_2$ norm of the combined policy and encoder gradient; larger gradients are rescaled to this norm before the update. 8.0 here. \\
Warm-up fraction & \code{warmup-frac} & Fraction of the iterations in the first phase of the three-phase size curriculum, in which the training molecules are visited in plain round-robin order (0.25). \\
Focus fraction & \code{focus-frac} & Fraction of the iterations in the second phase, which visits only the largest molecules, without noise (0.30 here). The remaining 0.45 form the third phase, with the same restricted set and Gaussian noise (standard deviation 0.01 Hartree) added to the reward. \\
Large-molecule fraction & \code{curriculum-top-frac} & Fraction of the training molecules, ordered by qubit count, that count as the ``largest'' set in the second and third phases (0.4 here; rounded up to at least one molecule). \\
\addlinespace
\multicolumn{3}{l}{\emph{Admission gate and reward}}\\
Depth ceiling & \code{depth-ceiling-max} & Largest transpiled circuit depth that is still admitted: a batch whose best circuit is deeper triggers no gradient update. 10{,}000 (46 qubits) and 20{,}000 (70 qubits) in Stage~1; 10{,}000 (46 qubits) and 50{,}000 (70 qubits) in Stage~2. \\
Energy admission tolerance (gradient-admission threshold) & \code{gate-energy-mha} & A batch is admitted only if the magnitude $|\Delta E|$ of the raw deviation of its best circuit from the reference energy is at most this value (mHa): 100 in Stage~1, and 100 (46 qubits) or 500 (70 qubits) in Stage~2. \\
Auxiliary-term threshold & \code{gate-loose-mha} & Accuracy threshold (50~mHa) inside which the gated reward terms (excitation and coverage penalties) apply; wider than chemical accuracy (1.6~mHa) because the training molecules are far from it. \\
Excitation penalty weight, $\lambda_{\mathrm{exc}}$ & \code{cost-lambda-excitation} & Weight of the penalty on the share of sampled determinants of excitation rank above three. \\
Coverage penalty weight, $\lambda_{\mathrm{cov}}$ & \code{cost-lambda-coverage} & Weight of the penalty on the fraction of qubits that carry no gate in the transpiled circuit. \\
Depth-overshoot penalty weight, $\lambda_{\mathrm{over}}$ & \code{cost-lambda-overshoot} & Weight of the penalty on $\log_{10}(d/d_{\max})$ when the depth $d$ exceeds the ceiling $d_{\max}$. \\
Hartree--Fock penalty weight, $\lambda_{\mathrm{HF}}$ & \code{cost-lambda-hf} & Weight of the penalty on the share of right-spin-sector shots that is the Hartree--Fock determinant. \\
Spin-balance penalty weight, $\lambda_{\mathrm{spin}}$ & \code{cost-lambda-spin} & Weight of the penalty on the imbalance between alpha-only and beta-only excitations. \\
Entropy coefficient, $c_H$ & \code{entropy-coef} & Weight of the mean token entropy of the sampled sequences, subtracted from the loss to keep the policy from collapsing. \\
Gate-count, subspace-size and real-depth weights & \code{cost-lambda}, \code{cost-lambda-dim}, \code{cost-lambda-real-depth} & Optional accuracy-gated penalties on a proxy gate count, on $\log_{10}$ of the diagonalisation subspace size, and on $\log_{10}$ of the backend-transpiled depth; all zero in every run reported here. \\
\addlinespace
\multicolumn{3}{l}{\emph{Subspace recovery and diagonalisation}}\\
Recovery size, $D$ & \code{max-dim} (training), \code{--subspace-dim} (inference), \code{subspace\_dim} & Target number of strings per spin sector kept in each configuration-recovery step; the resulting determinant subspace contains up to $D^2$ determinants (closed-shell product subspace). $D=20{,}000$ in the final recoveries; $D=40{,}000$ as the training cap in Stage~2. \\
Number of recovery iterations, $K$ & \code{recover-max-iterations}, \code{FINAL\_RECOVER\_ITERS} & Number of configuration-recovery iterations; 25 in the final recoveries. \\
Independent subspaces per step, $B$ & \code{num\_batches} & Number of independent subspaces drawn from the recovered distribution in one step, of which the lowest-energy one is kept. \\
Ranking cap & \code{--max-dim} (inference) & Smaller cap used only to rank the ten sampled circuits before the final recovery: 200 strings per spin sector (net dimension 40{,}000); the E+D rule of the trained policies uses the same ranking. Only the E+D picks of the eight untrained controls, sampled in the first campaign, come from the earlier ranking with a cap of 2{,}000 determinants (\code{RANKING\_PAIR\_CAP}). \\
Depth limit for diagonalisation & \code{--max-depth-for-diag} & Circuits deeper than this are not diagonalised at inference (50{,}000 for the 70-qubit lineages, matching the training ceiling). \\
\addlinespace
\multicolumn{3}{l}{\emph{Software and devices}}\\
Recovery routine & \code{recover\_and\_diagonalize}, \code{qiskit\_addon\_sqd} & Our GPU implementation of the configuration-recovery loop of SQD, built on the Qiskit SQD addon. \\
Local unitary cluster Jastrow ansatz & \code{UCJOpSpinBalanced}, \code{n\_reps} & ffsim's spin-balanced LUCJ operator built from CCSD amplitudes; \code{n\_reps} is the number of ansatz repetitions (1 in the LUCJ baseline). \\
Hardware backend & \code{ibm\_kobe}, \code{ibm\_kawasaki}, \code{ibm\_heron\_r2} & IBM superconducting backends (Heron generation). \code{ibm\_kobe} is the physical processor used for all reported sampling. \code{ibm\_kawasaki} was used only in preliminary runs not included in the reported analyses. \code{ibm\_heron\_r2} is an offline backend model whose coupling map is used for the per-iteration training transpilation; depths reported for training are on this model (not necessarily the circuits submitted). Inference depths are live-transpiled for \code{ibm\_kobe} and submitted. \\
\bottomrule
\end{longtable}
\endgroup

\section{Hyperparameter tuning: preliminary runs and reward search}
\label{sec:appendix-tuning}

This appendix describes the preliminary runs and the joint search behind the settings of Section~\ref{sec:hyperparams} and the tuned reward of Section~\ref{sec:reward-tuning}. The tables of the tuning (Tables~\ref{tab:tuning-narrative} and~\ref{tab:tuning-ranking}) are in Appendix~\ref{sec:appendix-training-detail}.

\paragraph{Settings.} Table~\ref{tab:hyperparams} lists the settings of the four lineages; the Stage-2 reward weights of the two configurations are in Table~\ref{tab:pipelines} and Section~\ref{sec:reward-tuning}, and Table~\ref{tab:glossary} defines every setting by its name in the code. The optimiser, the size curriculum, the seed and the recovery settings are the same for all runs.

Most of these settings were fixed in preliminary Stage-2 runs, which also determined the admission gate. Two findings from them are used here:

\begin{enumerate}[leftmargin=1.5em]
  \item \textbf{The depth ceiling is the dominant lever of the admission gate.}
  In a sweep over the depth ceiling and the coverage and overshoot weights, the ceiling decided whether training stalled with zero gradient: with a loose ceiling a large share of the sampled circuits cleared the admission gate and contributed a gradient update, and the particular coverage and overshoot weights mattered much less; a tight ceiling stalled regardless of the reward shaping.
  \item \textbf{The energy admission tolerance must be loosened for the 70-qubit targets.}
  With the tolerance that worked at 46 qubits, no circuit cleared the gate over many consecutive iterations at 70 qubits. Without an admitted iteration there is no gradient, so the policy distribution cannot change (the next stochastic batch can still differ), a deadlock and not slow learning. Loosening the tolerance by a factor of five (100 to 500~mHa) restored a non-trivial pass rate.
\end{enumerate}

\paragraph{Why the Hartree--Fock, spin and entropy terms were added.} In the first Stage-2 run on \code{ibm\_kobe}, the policy converged to a circuit with no gate able to change the Hartree--Fock determinant, and 94\% of its valid shots were the Hartree--Fock determinant itself; this collapse happens before the energy gate opens, which is why the Hartree--Fock term is ungated. With the Hartree--Fock penalty alone, the 46-qubit lineages settled on circuits whose valid shots were mostly beta-only (about 60\%, with almost no alpha-only weight), which motivated the spin term. Without the entropy term, the effective vocabulary of the policy, $\exp\overline{H}$, fell from about 100 to 1 in all lineages, which removes the exploration that the group-relative advantage needs.

\paragraph{How the weights were chosen.} Three stages, in order, each informed by the previous one.
\emph{(i) Preliminary runs and their depth sweeps (Table~\ref{tab:tuning-narrative}).} $\lambda_{\mathrm{cov}}=10$, $\lambda_{\mathrm{over}}=2$ and $\lambda_{\mathrm{exc}}=2$ were fixed in the
36-job sweep over the depth ceiling and the coverage and overshoot weights: the depth ceiling was the dominant lever, and several weight combinations performed comparably.
$\lambda_{\mathrm{exc}}$ was set based on an observation in an earlier zero-shot run of a Stage-1 checkpoint on Sn(OH)$_3$-Me L (accepted circuit with 97.5\% high-rank weight); the organotin targets were therefore not untouched when this weight was set, and $\lambda_{\mathrm{exc}}$ was not swept.
\emph{(ii) A one-term test of $\lambda_{\mathrm{HF}}$.} On a 24-qubit CPU test (C$_2$H$_4$, 12 orbitals, 3 seeds, 60 iterations, symmetric 5\% readout flips) we compared
$\lambda_{\mathrm{HF}}\in\{0,0.05,0.2,1.0\}$ and a noise-free control. At 0.05 the Hartree--Fock share of the valid shots fell from 57\% to 37\%, the number of HOMO--LUMO rotations per circuit
doubled and the raw error stayed within 1~mHa of the unpenalized run (10.7 against 11.8~mHa). At 0.2 and 1.0 the share fell to 1\% and 0\% but the raw error rose to 44 and 75~mHa: the
policy escaped Hartree--Fock by landing on one excited determinant, not on a correlated state. We therefore started the hardware runs with $\lambda_{\mathrm{HF}}=0.05$.
\emph{(iii) A joint search.} With the penalty on hardware, the policy still collapsed to one sequence, and in the 46-qubit lineages onto a spin-asymmetric one. A grid of the spin weight
($\lambda_{\mathrm{spin}}\in\{0,0.2,0.6\}$) and the entropy coefficient ($c_H\in\{0,0.05,0.2\}$) at fixed $\lambda_{\mathrm{HF}}=0.05$, with symmetric and with asymmetric readout noise, showed
that $c_H=0.2$ was the only setting that kept the effective vocabulary above one, that the spin weight changed nothing under symmetric noise (alpha and beta weights agreed to 0.001 in every arm), that
$\lambda_{\mathrm{spin}}=0.2$ barely reduced the imbalance under asymmetric noise (0.114 against 0.128), and that $\lambda_{\mathrm{spin}}=0.6$ removed it only by degenerating
(Hartree--Fock share 0.01 and raw error 62~mHa). Since the five weights interact (a larger $\lambda_{\mathrm{cov}}$ or $\lambda_{\mathrm{exc}}$ changes which circuits the entropy bonus
protects), we then searched them together: 40 configurations drawn as a Latin hypercube over $\lambda_{\mathrm{HF}}\in[0,0.3]$, $\lambda_{\mathrm{spin}}\in[0,1]$, $c_H\in[0,0.3]$,
$\lambda_{\mathrm{exc}}\in[0,4]$ and $\lambda_{\mathrm{cov}}\in[0,20]$, plus the two reference settings ($\lambda_{\mathrm{HF}}=0$ and 0.05, the rest at their reference values), two
seeds each, 80 iterations each, on the same 24-qubit test. This search used a noise model fitted to the \code{ibm\_kobe} counts of the first Stage-2 run: 3{,}000 circuits at 23 orbitals show that the shots
in the right spin sector fall from 49\% at depth 45--60 to 12\% at depth 60--100 and to 2--3\% above depth 300, equivalent to independent per-bit flip probabilities of 0.017, 0.07 and 0.3--0.45.
The simulator therefore flips each bit with probability $p(d)=0.5\,[1-\exp(-(d-45)/200)]$ (plus 1\% readout noise) from the transpiled depth $d$, which reproduces 0.017, 0.08 and 0.42 at depths
52, 80 and 400, and applies it to the alpha and beta registers separately. The model is crude: independent flips cannot reproduce which bitstrings the device returns.

\paragraph{Selection.} Each configuration was scored on the last 20 of its 80 iterations by the raw error of the best circuit, the share of the valid shots that is Hartree--Fock, the number of gates
that can change that determinant, the effective vocabulary and the depth. We required the raw error to be at most 20~mHa (the best unpenalized runs reach 11~mHa), the Hartree--Fock share at most 0.30,
an effective vocabulary of at least 3 tokens, and at least 1.5 gates per circuit able to change the determinant. Table~\ref{tab:tuning-ranking} lists all 42 configurations, 80 runs in all.
The hardware runs had to start before the search had finished, so the configuration used for them (number~3: $\lambda_{\mathrm{HF}}=0.07$, $\lambda_{\mathrm{spin}}=0.77$, $c_H=0.26$,
$\lambda_{\mathrm{exc}}=3.9$, $\lambda_{\mathrm{cov}}=7.2$) was chosen on the 26 configurations that had a finished run at that point, not on the final ranking.

Of the 42 configurations, five meet the criteria: numbers 24, 2, 16, 32 and 3, with mean raw errors of 11.9, 12.4, 12.5, 14.5 and 14.8~mHa. The selected configuration is the fifth of these and tenth
overall by raw error. Over its two seeds it gave 16.8 and 12.7~mHa, a Hartree--Fock share of 0.17--0.21, an effective vocabulary of 5.8--7.5 tokens, 2.9--3.8 Hartree--Fock-changing gates per circuit and
a depth of 63--68. The unpenalized reference (configuration~0) gave 104 and 11.1~mHa over its two seeds: one seed collapsed at depth 7{,}056 and the other onto the Hartree--Fock circuit, with 63\% of its valid
shots Hartree--Fock. $\lambda_{\mathrm{HF}}=0.05$ alone (configuration~1) gave 123 and 10.8~mHa. Two seeds per configuration cannot separate the five passing configurations from each other: their
errors differ by less than the difference between the two seeds of the selected one. The ranking therefore supports the choice of a region of the weight space (a Hartree--Fock weight of 0.04--0.15, a spin weight above 0.2,
an entropy coefficient above 0.1), not of one point in it. Configurations 7, 13 and 22 reach 11--13~mHa at shallow depth, but their policy has collapsed to 1.4--2.2 effective tokens, and configuration 37 (11.4~mHa, 3.5 tokens) fails on its Hartree--Fock share of 0.34; the five passing configurations keep 3.7--9.6 effective tokens.
The spin weight cannot be judged in this simulator, because its readout noise is the same for the alpha and beta registers in the joint search (the imbalance is below 0.002 in every row); its value (0.77) was therefore chosen heuristically for the hardware runs, because the hardware showed one-sided excitation, and it has not been tested against that failure.

\paragraph{Limits.} The search used one molecule, 24 qubits and 80 iterations on a noise model that is fitted to one hardware run, with two seeds per configuration. It shows which settings keep the policy
diverse and the raw error moderate under that model. It does not show that the selected weights improve the hardware result; the hardware runs with this setting are reported next to the runs with $\lambda_{\mathrm{HF}}=0.05$ and
without a penalty (Section~\ref{sec:results} and Appendix~\ref{sec:appendix-chemistry}).

\section{Training detail: hyperparameter tuning, comparison of the runs and timing}
\label{sec:appendix-training-detail}

This appendix collects the tables of the hyperparameter tuning (Appendix~\ref{sec:appendix-tuning}): the preliminary runs (Table~\ref{tab:tuning-narrative}), the comparison of the training runs (Table~\ref{tab:t2-compare}), the reward search (Table~\ref{tab:tuning-ranking}), and the training time and billed QPU time (Tables~\ref{tab:t2-timing} and~\ref{tab:t2-qpu}). It is referenced from Section~\ref{sec:hyperparams} and Section~\ref{sec:results}.

\begingroup
\small
\setlength{\tabcolsep}{4pt}
\renewcommand{\arraystretch}{1.15}
\begin{longtable}{@{}>{\raggedright\arraybackslash}p{2.1cm}
                  >{\raggedright\arraybackslash}p{4.3cm}
                  >{\raggedright\arraybackslash}p{4.3cm}
                  >{\raggedright\arraybackslash}p{2.3cm}@{}}
\caption{Hyperparameter tuning in the preliminary runs: what was varied, what the sweep or ablation found, and the setting this motivated for the final runs. The names are the descriptive names of the code labels defined in Table~\ref{tab:glossary}.}\label{tab:tuning-narrative}\\
\toprule
Setting & What was varied & What was found & Setting used \\
\midrule
Depth ceiling & Swept jointly with the two reward weights below (grid over ceiling $\times$ coverage $\times$ overshoot weight) on the $\leq$32q curriculum. &
  The ceiling, not the reward-weight combination, was the dominant lever for avoiding sustained zero-gradient stalls: a loose ceiling let a wide range of reward weights learn comparably; a tight one stalled regardless of reward shaping. &
  10000 (46q), 20000 (70q, Stage 1); loosened to 50000 for 70q Stage 2 once hardware depths ran higher than in simulation. \\
\addlinespace
 Coverage and overshoot weights & Same grid as above. & Secondary to the ceiling: multiple weight combinations performed comparably once the ceiling itself was set generously; no single combination was uniquely best across all swept ceilings. & 10.0 / 2.0, all four lineages, both stages. \\
\addlinespace
 Energy admission tolerance & Observed directly on the first Stage-2 attempts at the 70q target: with the Stage-1 value (100), the energy tolerance never admitted a single iteration across many consecutive hardware iterations. & A structural deadlock, not slow learning: with zero admitted iterations the policy cannot update, so the policy distribution cannot change (the next stochastic batch can still differ). Loosening by $5\times$ broke the deadlock immediately; both 70q lineages then admitted 99--100\% of their iterations for the rest of training. & 100 (46q, both stages; never hit this deadlock), 500 (70q, Stage 2 only). \\
\addlinespace
 Gate tokens per circuit & Fixed per (encoder, qubit-target) lineage before training begins, not swept within a lineage. & Cannot change across a warmstart transfer: it sizes the transformer's position-embedding table ($n_{\mathrm{positions}}=\mathrm{ngates}{+}1$), so Stage 1 and Stage 2 of one lineage must agree on it by construction. & 28 (46q lineage), 44 (70q lineage). \\
\addlinespace
 Recovery subspace size & Fixed at 20000 for every Stage-1 run (ffsim); raised to 40000 for every Stage-2 run (hardware). & Not independently swept here; chosen to give the hardware stage, which collects far fewer effective samples per circuit than ffsim's exact simulation, more room in the per-iteration subspace cap. & 20000 (Stage 1), 40000 (Stage 2), uniformly. \\
\addlinespace
 Fraction of molecules counted as largest & Controlled pair of runs on a 5-molecule subset, 0.2 (the original default) vs.\ 0.4, holding every other setting fixed. & 0.4 won on every hard molecule measured: co2 21.2 vs.\ 112.4~mHa, fe\_cn6 132.7 vs.\ 240.8~mHa, co\_nh3\_6 294.1 vs.\ 459.1~mHa (for ferrocene the difference was smaller: 173.5 vs.\ 181.1~mHa). A wider ``hard'' set (top 40\% of the molecule pool by qubit count, rather than top 20\%) gave the curriculum's focus phase more molecules to distribute attention across. & 0.4, all four lineages. \\
\addlinespace
 Fraction of iterations on the largest molecules & Changed alongside curriculum-top-frac adjusted in the same tuning round (0.5$\to$0.3), but \emph{not} isolated in its own A/B test the way curriculum-top-frac was. & The value 0.3 was carried over from the preceding tuning round on the rationale that a shorter focus phase leaves more of the run for the final consolidation phase; this rationale was not tested in isolation. & 0.3, all four lineages (a fixed choice, not swept). \\
\addlinespace
 Learning rate & Controlled pair of runs on a 5-molecule subset, the existing default $5{\times}10^{-6}$ vs.\ a $4\times$ higher $2{\times}10^{-5}$. & The higher rate reached a numerically higher early gate-pass rate (41.7\%, 10/24 admitted iterations) but produced the worst final accuracy of any hyperparameter test run in this comparison (fe\_cn6 312~mHa, ferrocene 478~mHa) -- confirming the pre-existing default was not improved on by this specific alternative, rather than establishing $5{\times}10^{-6}$ as independently optimal from a wider sweep. & $5{\times}10^{-6}$, all four lineages, both stages. \\
\addlinespace
 Gradient-norm clipping threshold & Not swept over a grid; recalibrated by measuring the policy's gradient-norm distribution across 24 training runs, after the originally-used value of 1.0 was found to fire on essentially every accepted step. & Raw gradient norms' 90th percentile measured at $\approx$8.2 across those runs; clip threshold set to 8.0 (its measured 90th percentile) so it acts as an occasional safety net rather than a de facto fixed step size -- confirmed to then fire on only 3.5\% of accepted steps (90/2561) in the recalibrated runs. This recalibration showed no measurable accuracy or gate-pass-rate benefit over no clipping at all (medians differed by only 1--3 percentage points across 8 seeds); it was adopted as zero-cost insurance against rare large updates, not because it measurably improved training. & 8.0, all four lineages, both stages. \\ \bottomrule \end{longtable}
\renewcommand{\arraystretch}{1.0}
\endgroup

\FloatBarrier
\begin{table*}[ht]
\centering
\caption{The training runs (single seed per run), by stage and lineage. $n$ is the number of iterations of the run; ``gate'' the share of iterations that passed the admission gate; ``error'' the mean raw SQD error (mHa) of the best circuit over the admitted iterations of the last 30 iterations (``last'') and of the whole run (``all''); ``depth'' the median depth of the best circuit as transpiled for the offline backend model \code{ibm\_heron\_r2} at Qiskit optimization level 1 (not the circuit submitted) and ``$>$1000'' the share of iterations whose best circuit is deeper than 1{,}000; ``eff.\ voc.'' the policy's effective vocabulary defined as $\exp(\bar H)$ with $\bar H$ the token entropy in nats over all 110 token ids (including padding); ``valid'' the share of shots in the right spin sector, ``HF'' the share of the valid shots that is the Hartree--Fock determinant, ``HL'' the share that is a HOMO$\to$LUMO excitation and ``$|\alpha-\beta|$'' the difference of the alpha-only and beta-only weights among the valid shots, all tracked over the last 30 iterations. Stage~1 is shared by the three Stage-2 variants. Runs that were stopped (Stage~2 without penalty, A5-46q and A5-70q) are shown with the iterations they have; the tuned run of A1-70q ended by a segmentation fault at iteration 137.}
\label{tab:t2-compare}
\scriptsize
\setlength{\tabcolsep}{3pt}
\begin{tabular}{@{}llrrrrrrrrrrrr@{}}
\toprule
Lineage & Run & $n$ & gate (\%) & error last & error all & depth & $>$1000 (\%) & eff.\ voc. & valid & HF & HL & $|\alpha-\beta|$ \\
\midrule
A1-46q & Stage 1 & 150 & 84 & 36.1 & 26.5 & 4394 & 92 & 1.0 & 1.00 & 0.76 & 0.16 & 0.000 \\
A1-46q & Stage 2, no penalty & 150 & 51 & 26.2 & 37.3 & 48 & 0 & 1.0 & 0.51 & 0.95 & 0.00 & 0.008 \\
A1-46q & Stage 2, HF penalty & 150 & 46 & 48.1 & 46.2 & 56 & 0 & 1.0 & 0.10 & 0.02 & 0.07 & 0.558 \\
A1-46q & Stage 2, tuned & 150 & 59 & 48.0 & 46.7 & 44 & 5 & 21.5 & 0.03 & 0.03 & 0.00 & 0.018 \\
\midrule
A5-46q & Stage 1 & 150 & 82 & 28.1 & 22.3 & 6008 & 95 & 1.0 & 1.00 & 0.70 & 0.15 & 0.000 \\
A5-46q & Stage 2, no penalty & 17 & 47 & 43.8 & 43.8 & 39 & 0 & 61.2 & 0.05 & 0.06 & 0.05 & 0.064 \\
A5-46q & Stage 2, HF penalty & 150 & 47 & 45.0 & 46.5 & 53 & 0 & 1.0 & 0.21 & 0.00 & 0.00 & 0.666 \\
A5-46q & Stage 2, tuned & 150 & 63 & 56.7 & 52.5 & 43 & 7 & 20.9 & 0.04 & 0.06 & 0.00 & 0.043 \\
\midrule
A1-70q & Stage 1 & 150 & 86 & 27.6 & 24.0 & 7008 & 93 & 1.1 & 1.00 & 0.72 & 0.14 & 0.000 \\
A1-70q & Stage 2, HF penalty & 150 & 100 & 287.9 & 300.6 & 76 & 0 & 1.0 & 0.07 & 0.01 & 0.03 & 0.045 \\
A1-70q & Stage 2, tuned & 137 & 99 & 284.7 & 286.6 & 79 & 39 & 26.1 & 0.02 & 0.00 & 0.00 & 0.014 \\
\midrule
A5-70q & Stage 1 & 150 & 86 & 21.1 & 20.9 & 6832 & 91 & 1.2 & 1.00 & 0.60 & 0.20 & 0.000 \\
A5-70q & Stage 2, no penalty & 18 & 94 & 348.4 & 348.4 & 66 & 6 & 56.5 & 0.02 & 0.02 & 0.00 & 0.020 \\
A5-70q & Stage 2, HF penalty & 150 & 99 & 276.0 & 294.1 & 152 & 1 & 1.0 & 0.06 & 0.04 & 0.15 & 0.073 \\
A5-70q & Stage 2, tuned & 150 & 100 & 269.6 & 285.4 & 79 & 39 & 27.0 & 0.02 & 0.01 & 0.00 & 0.015 \\
\bottomrule
\end{tabular}
\end{table*}

\begin{table*}[ht]
\centering
\caption{Joint reward search on the noisy 24-qubit simulator: all 42 configurations, ranked by the mean raw error of the best circuit over the last 20 of 80 iterations (mean over the finished seeds, two for all but configuration 34, which has one, 80 runs in all). Configurations 0 and 1 are the two references (no Hartree--Fock penalty, and $\lambda_{\mathrm{HF}}=0.05$ alone, the other weights at their reference values); 2--41 are Latin-hypercube draws over the ranges of Table~\ref{tab:t2-reward}. Columns: the five weights and the entropy coefficient $c_H$; the share of the valid (right-spin-sector) shots that is the Hartree--Fock determinant; the policy's effective vocabulary; the number of gates per circuit that can change the Hartree--Fock determinant; the share of valid shots on HOMO$\to$LUMO determinants; the share of all shots that is valid; the median depth. A check mark means: raw error $\le20$~mHa, Hartree--Fock share $\le0.30$, effective vocabulary $\ge3$ and $\ge1.5$ gates able to change the determinant (5 of 42). The configuration used for the hardware runs (3) is in bold. The alpha/beta imbalance is below 0.002 in every row, because the noise of this simulator is the same for both registers.}
\label{tab:tuning-ranking}
\scriptsize
\setlength{\tabcolsep}{3pt}
\begin{tabular}{@{}rrrrrrrrrrrrrrl@{}}
\toprule
rank & cfg & $\lambda_{\mathrm{HF}}$ & $\lambda_{\mathrm{spin}}$ & $c_H$ & $\lambda_{\mathrm{exc}}$ & $\lambda_{\mathrm{cov}}$ & error (mHa) & HF share & eff.\ voc. & HF-changing & HOMO$\to$LUMO & valid & depth & ok \\
\midrule
1 & 7 & 0.01 & 0.86 & 0.14 & 3.2 & 10.7 & 11.0 & 0.52 & 1.4 & 1.07 & 0.17 & 0.31 & 66 &  \\
2 & 37 & 0.06 & 0.90 & 0.18 & 3.0 & 19.1 & 11.4 & 0.34 & 3.5 & 1.82 & 0.20 & 0.26 & 70 &  \\
3 & 24 & 0.07 & 0.26 & 0.23 & 3.2 & 7.7 & 11.9 & 0.28 & 3.8 & 2.28 & 0.22 & 0.24 & 70 & \checkmark \\
4 & 22 & 0.05 & 0.38 & 0.18 & 0.3 & 2.8 & 12.3 & 0.21 & 2.2 & 1.33 & 0.06 & 0.13 & 99 &  \\
5 & 2 & 0.15 & 0.72 & 0.19 & 0.2 & 0.8 & 12.4 & 0.10 & 3.7 & 3.49 & 0.08 & 0.11 & 104 & \checkmark \\
6 & 16 & 0.04 & 0.68 & 0.28 & 2.3 & 13.7 & 12.5 & 0.23 & 6.3 & 2.74 & 0.13 & 0.18 & 69 & \checkmark \\
7 & 19 & 0.02 & 0.21 & 0.11 & 0.8 & 17.5 & 12.7 & 0.39 & 1.1 & 0.53 & 0.08 & 0.18 & 87 &  \\
8 & 13 & 0.02 & 0.25 & 0.16 & 3.7 & 16.0 & 12.9 & 0.37 & 1.9 & 2.14 & 0.26 & 0.38 & 63 &  \\
9 & 32 & 0.04 & 0.53 & 0.29 & 1.2 & 2.3 & 14.5 & 0.20 & 9.6 & 3.95 & 0.06 & 0.14 & 76 & \checkmark \\
\textbf{10} & \textbf{3} & \textbf{0.07} & \textbf{0.77} & \textbf{0.26} & \textbf{3.9} & \textbf{7.2} & \textbf{14.8} & \textbf{0.19} & \textbf{6.7} & \textbf{3.36} & \textbf{0.17} & \textbf{0.22} & \textbf{66} & \textbf{\checkmark} \\
11 & 29 & 0.14 & 0.67 & 0.03 & 0.5 & 14.4 & 28.0 & 0.10 & 1.0 & 1.50 & 0.44 & 0.27 & 88 &  \\
12 & 17 & 0.25 & 0.09 & 0.21 & 1.5 & 15.4 & 29.4 & 0.03 & 3.6 & 3.71 & 0.31 & 0.16 & 84 &  \\
13 & 12 & 0.12 & 0.34 & 0.24 & 4.0 & 4.3 & 38.1 & 0.04 & 5.9 & 3.69 & 0.11 & 0.10 & 97 &  \\
14 & 21 & 0.18 & 0.28 & 0.03 & 1.7 & 15.7 & 44.1 & 0.02 & 1.0 & 3.50 & 0.78 & 0.48 & 63 &  \\
15 & 0 & 0.00 & 0.00 & 0.00 & 2.0 & 10.0 & 57.6 & 0.31 & 1.0 & 4.00 & 0.10 & 0.23 & 3560 &  \\
16 & 26 & 0.10 & 0.99 & 0.11 & 2.0 & 9.2 & 57.7 & 0.07 & 1.1 & 4.25 & 0.39 & 0.31 & 2715 &  \\
17 & 33 & 0.22 & 0.32 & 0.04 & 0.9 & 14.5 & 66.5 & 0.02 & 1.0 & 3.50 & 0.11 & 0.09 & 1478 &  \\
18 & 1 & 0.05 & 0.00 & 0.00 & 2.0 & 10.0 & 66.8 & 0.18 & 1.0 & 3.50 & 0.13 & 0.14 & 2221 &  \\
19 & 36 & 0.18 & 0.15 & 0.20 & 1.1 & 5.8 & 68.2 & 0.03 & 5.8 & 2.85 & 0.13 & 0.10 & 1952 &  \\
20 & 39 & 0.26 & 0.41 & 0.17 & 1.5 & 13.3 & 69.5 & 0.01 & 2.7 & 4.07 & 0.13 & 0.09 & 1918 &  \\
21 & 20 & 0.27 & 0.02 & 0.07 & 2.6 & 17.4 & 73.2 & 0.01 & 1.0 & 5.67 & 0.06 & 0.08 & 3752 &  \\
22 & 6 & 0.28 & 0.19 & 0.10 & 1.0 & 19.8 & 73.3 & 0.01 & 1.1 & 6.46 & 0.38 & 0.21 & 549 &  \\
23 & 5 & 0.27 & 0.48 & 0.15 & 2.7 & 11.0 & 86.9 & 0.02 & 2.4 & 5.39 & 0.29 & 0.24 & 3775 &  \\
24 & 23 & 0.09 & 0.15 & 0.10 & 2.1 & 10.3 & 89.1 & 0.16 & 1.1 & 2.68 & 0.01 & 0.10 & 3578 &  \\
25 & 8 & 0.10 & 0.91 & 0.05 & 0.9 & 8.7 & 89.3 & 0.16 & 1.0 & 1.50 & 0.06 & 0.11 & 2370 &  \\
26 & 11 & 0.30 & 0.64 & 0.08 & 3.3 & 1.6 & 90.2 & 0.01 & 1.0 & 4.51 & 0.07 & 0.09 & 2050 &  \\
27 & 9 & 0.14 & 0.81 & 0.27 & 2.2 & 16.9 & 95.6 & 0.00 & 7.4 & 7.98 & 0.00 & 0.05 & 6364 &  \\
28 & 10 & 0.07 & 0.45 & 0.02 & 2.5 & 12.6 & 100.6 & 0.23 & 1.0 & 3.00 & 0.08 & 0.14 & 1737 &  \\
29 & 34 & 0.19 & 0.12 & 0.22 & 1.8 & 1.1 & 104.7 & 0.00 & 5.5 & 4.19 & 0.02 & 0.06 & 2420 &  \\
30 & 25 & 0.13 & 0.74 & 0.21 & 2.9 & 3.3 & 109.2 & 0.01 & 6.6 & 5.45 & 0.02 & 0.06 & 1414 &  \\
31 & 41 & 0.17 & 0.56 & 0.15 & 3.5 & 7.0 & 116.7 & 0.00 & 2.8 & 2.54 & 0.00 & 0.05 & 4786 &  \\
32 & 30 & 0.21 & 0.59 & 0.29 & 2.0 & 18.4 & 118.0 & 0.00 & 10.2 & 5.08 & 0.00 & 0.05 & 2318 &  \\
33 & 15 & 0.23 & 0.38 & 0.26 & 1.8 & 12.3 & 119.8 & 0.01 & 10.7 & 5.04 & 0.00 & 0.05 & 4318 &  \\
34 & 40 & 0.00 & 0.61 & 0.00 & 3.4 & 0.4 & 122.6 & 0.00 & 1.0 & 3.01 & 0.00 & 0.05 & 5265 &  \\
35 & 27 & 0.20 & 0.83 & 0.05 & 3.0 & 6.4 & 125.9 & 0.00 & 1.0 & 3.50 & 0.23 & 0.11 & 296 &  \\
36 & 31 & 0.23 & 0.06 & 0.24 & 1.3 & 8.2 & 135.5 & 0.00 & 7.2 & 5.38 & 0.00 & 0.05 & 3157 &  \\
37 & 38 & 0.24 & 0.51 & 0.14 & 0.4 & 18.6 & 138.2 & 0.00 & 2.3 & 3.96 & 0.00 & 0.05 & 5354 &  \\
38 & 14 & 0.29 & 0.94 & 0.01 & 3.8 & 3.6 & 139.4 & 0.00 & 1.0 & 7.50 & 0.00 & 0.05 & 6970 &  \\
39 & 35 & 0.09 & 0.78 & 0.12 & 0.4 & 5.1 & 139.5 & 0.00 & 1.4 & 2.55 & 0.00 & 0.05 & 7718 &  \\
40 & 28 & 0.20 & 0.96 & 0.25 & 2.8 & 11.6 & 146.4 & 0.00 & 7.1 & 4.09 & 0.00 & 0.05 & 3961 &  \\
41 & 18 & 0.16 & 0.47 & 0.06 & 0.1 & 9.5 & 155.6 & 0.00 & 1.0 & 2.99 & 0.00 & 0.05 & 4039 &  \\
42 & 4 & 0.12 & 0.04 & 0.08 & 0.6 & 4.5 & 159.0 & 0.00 & 1.0 & 3.23 & 0.00 & 0.05 & 2786 &  \\
\bottomrule
\end{tabular}
\end{table*}

\begin{table}[t]
\centering
\caption{Time per iteration (seconds, mean over the iterations of each run) by pipeline step, Stage~1 on the simulator and Stage~2 on \code{ibm\_kobe}; the last column is the analysed elapsed time of the run with the stalled Stage-2 iterations removed. In the Stage-2 rows, iterations whose circuit sampling took longer than one hour (device-queue stalls caused by Runtime jobs left queued by cancelled Slurm jobs) are excluded from the means and from the analysed elapsed time, and counted in the column ``stalled''; Stage-1 iterations are all kept, including the ferrocene iterations, whose simulation can take up to an hour. The sampling column of Stage~2 includes the IBM device queue. $n$ is the number of iterations the run has.}
\label{tab:t2-timing}
\scriptsize
\resizebox{\linewidth}{!}{%
\begin{tabular}{@{}llrrrrrrrrr@{}}
\toprule
Run & Lineage & $n$ & wall & sampling & diagonalize & refine & transpile & backward & stalled & analysed (h) \\
\midrule
Stage 1 & A1-46q & 150 & 496 & 284 & 194 & 9 & 6 & 1.2 & 0 & 20.7 \\
Stage 1 & A5-46q & 150 & 659 & 445 & 195 & 10 & 8 & 1.3 & 0 & 27.5 \\
Stage 1 & A1-70q & 150 & 756 & 539 & 196 & 10 & 8 & 1.2 & 0 & 31.5 \\
Stage 1 & A5-70q & 150 & 637 & 427 & 190 & 9 & 8 & 1.1 & 0 & 26.6 \\
\midrule
Stage 2, no penalty & A1-46q & 150 & 494 & 265 & 215 & 11 & 1 & 0.8 & 1 & 20.5 \\
Stage 2, no penalty & A5-46q & 17 & 474 & 212 & 239 & 13 & 4 & 0.8 & 0 & 2.2 \\
Stage 2, no penalty & A5-70q & 18 & 621 & 267 & 284 & 15 & 16 & 1.5 & 0 & 3.1 \\
\midrule
Stage 2, HF penalty & A1-46q & 150 & 606 & 348 & 241 & 13 & 2 & 0.8 & 0 & 25.3 \\
Stage 2, HF penalty & A5-46q & 150 & 606 & 368 & 223 & 11 & 1 & 0.7 & 1 & 25.1 \\
Stage 2, HF penalty & A1-70q & 150 & 677 & 315 & 330 & 16 & 5 & 1.5 & 1 & 28.0 \\
Stage 2, HF penalty & A5-70q & 150 & 654 & 348 & 278 & 14 & 4 & 1.3 & 1 & 27.1 \\
\midrule
Stage 2, tuned & A1-46q & 150 & 791 & 525 & 241 & 14 & 5 & 0.8 & 0 & 33.0 \\
Stage 2, tuned & A5-46q & 150 & 790 & 527 & 237 & 15 & 5 & 0.9 & 0 & 32.9 \\
Stage 2, tuned & A1-70q & 137 & 938 & 563 & 298 & 20 & 23 & 1.4 & 1 & 35.4 \\
Stage 2, tuned & A5-70q & 150 & 919 & 546 & 294 & 20 & 22 & 1.4 & 1 & 38.1 \\
\bottomrule
\end{tabular}}
\end{table}

\paragraph{The training runs.} Tables~\ref{tab:t2-compare} and~\ref{tab:t2-timing} compare the Stage-1 run with the three Stage-2 variants, which differ only in the reward. \emph{No penalty} uses the excitation, coverage and overshoot terms only. \emph{HF penalty}, the main configuration, adds the Hartree--Fock term ($\lambda_{\mathrm{HF}}=0.05$). \emph{Tuned} uses the jointly tuned reward of Table~\ref{tab:t2-reward}. Stage~1 is shared by the three variants (one Stage-1 run per lineage). The no-penalty Stage-2 run finished only for A1-46q; the other three lineages were stopped after 17 or 18 iterations (A1-70q never started its first hardware batch) so that the device could be used for the tuned runs. The tuned runs reached 150 iterations, except A1-70q, which ended by a segmentation fault in the transpiler at iteration 137; its latest saved policy is that of iteration 129, and no later checkpoint exists.

\paragraph{Why the wall times differ.} The per-iteration times of Table~\ref{tab:t2-timing} differ between runs for reasons that we can separate only in part.
(i) \emph{Stage-1 iterations are slower at 70 qubits} (539 and 427~s of sampling against 284 and 445~s at 46 qubits for A1 and A5), because the policy generates $L=44$ gate tokens per circuit instead of 28 and the simulator applies them; the longest single sampling step was 57~min (the ferrocene iteration of each 14-molecule cycle).
(ii) \emph{In Stage~2, the sampling column is the time to get a batch of 20 circuits back from \code{ibm\_kobe}}, which includes the device queue, not the QPU time. It grows from
the no-penalty run (212 to 267~s) to the HF-penalty runs (315 to 368~s) and the tuned runs (523 to 569~s). Two effects add up. The load on the queue was not constant: between one and eight of our own Stage-2 jobs were submitted at the
same time, on top of other tenants' jobs, and the sampling time of the same lineage is not reproducible across these periods; we did not control for this. The circuits themselves also differ. A collapsed policy emits the same short circuit
twenty times, whereas the tuned policy, which does not collapse, emits twenty different ones and, at 70 qubits, many deep ones. The billed QPU time per job, which excludes the queue, shows this: it is 59 to 60~s for the 46-qubit
HF-penalty runs and 83 to 85~s for the tuned ones, and 70~s against 159 to 167~s at 70 qubits (Table~\ref{tab:t2-qpu}). Diagonalization, which does not depend on the device, is similar in all Stage-2 regimes (215 to 330~s).
(iii) \emph{The stalled iterations} (Table~\ref{tab:t2-timing}, last column) are excluded from the Stage-2 means. There are six, one in each of six runs. Four (no-penalty A1-46q iteration 102, and the first iteration of three HF-penalty runs) are attributed to leftover Runtime jobs, which we did not verify run by run; two (tuned 70-qubit, iteration 128) are unexplained. One iteration of the no-penalty A1-46q run (8.8~h of sampling) waited
behind Qiskit Runtime jobs that cancelled Slurm jobs had left queued on \code{ibm\_kobe}, and the first iteration of each HF-penalty run of A5-46q, A1-70q and A5-70q (2.3 to 2.4~h) waited behind the same leftover jobs from the runs we had just replaced. Two were not:
iteration 128 of both 70-qubit tuned runs (1.8~h each) fell into a period of about three hours in which \code{ibm\_kobe} started none of our jobs. We do not know the reason; the device reported itself operational throughout.

\begin{table}[t]
\centering
\caption{IBM-billed QPU time of the Stage-2 runs, from the account's job records (one Runtime job per iteration, 20 circuits of 10{,}000 shots). ``Jobs'' counts the iterations that were submitted; ``mean'' is the billed seconds
per job. ``Stopped'': cancelled after 17 or 18 iterations; ``crashed'': A1-70q tuned ended by a segmentation fault in the transpiler at iteration 137. The 1372 jobs total 121{,}661~s (2{,}028~min, 33.8~h). These jobs are listed
separately in Table~\ref{tab:cost-summary}.}
\label{tab:t2-qpu}
\scriptsize
\begin{tabular}{@{}llrrr@{}}
\toprule
Regime & Lineage & Jobs & QPU time (min) & mean (s/job) \\
\midrule
No penalty & A1-46q & 150 & 146.7 & 58.7 \\
No penalty & A5-46q (stopped) & 17 & 21.2 & 74.9 \\
No penalty & A5-70q (stopped) & 18 & 37.2 & 124.2 \\
\midrule
HF penalty & A1-46q & 150 & 149.4 & 59.8 \\
HF penalty & A5-46q & 150 & 148.5 & 59.4 \\
HF penalty & A1-70q & 150 & 174.8 & 69.9 \\
HF penalty & A5-70q & 150 & 175.1 & 70.0 \\
\midrule
Tuned & A1-46q & 150 & 211.3 & 84.5 \\
Tuned & A5-46q & 150 & 206.3 & 82.5 \\
Tuned & A1-70q (crashed) & 137 & 365.7 & 160.1 \\
Tuned & A5-70q & 150 & 391.5 & 156.6 \\
\bottomrule
\end{tabular}
\end{table}

The chemistry that each regime produced is in Appendix~\ref{sec:appendix-chemistry}. The comparison of the energies in Table~\ref{tab:t2-compare} shows no regime that is better than the others on hardware,
and the trade-offs between them are discussed there.

\FloatBarrier
\section{Chemistry learnt during training}
\label{sec:appendix-chemistry}

This appendix shows what the policy generated in terms of the two quantities that connect a circuit to the chemistry of a closed-shell molecule, using the per-iteration tracking recorded during training.
\emph{HOMO--LUMO content.} The Hartree--Fock determinant fills the lowest orbitals. A rotation between two neighbouring occupied orbitals, or two neighbouring virtual ones, leaves it unchanged, whereas a rotation between the highest
occupied orbital (HOMO) and the lowest virtual one (LUMO) is the only neighbouring rotation that moves an electron from an occupied to a virtual orbital. We therefore count the HOMO--LUMO rotations per circuit, the share of
the right-spin-sector shots that are exactly a HOMO$\to$LUMO excitation, and the share that is the Hartree--Fock determinant itself.
\emph{Alpha/beta balance.} For a closed-shell molecule the alpha and beta electrons are equivalent, so excitations that move only an alpha electron (alpha-only) and only a beta electron (beta-only) should have equal weight. A large
imbalance means that the circuit, or the noise acting on it, favours one spin channel.
The right-spin-sector shots (called valid shots) are those with the right number of alpha and beta electrons; all other shots (wrong electron number, or an electron moved between the two channels) are noise, because every operator
in the pool conserves both numbers. Unless stated otherwise, the values reported below are means over the last 30 iterations (or over all iterations where noted), computed from per-iteration batch means over the $N=20$ circuits. The HF, alpha-only, beta-only and both-channel shares form a partition of the valid shots, while the HOMO$\to$LUMO share overlaps them.

\begin{figure}[t]
\centering
\includegraphics[width=\linewidth]{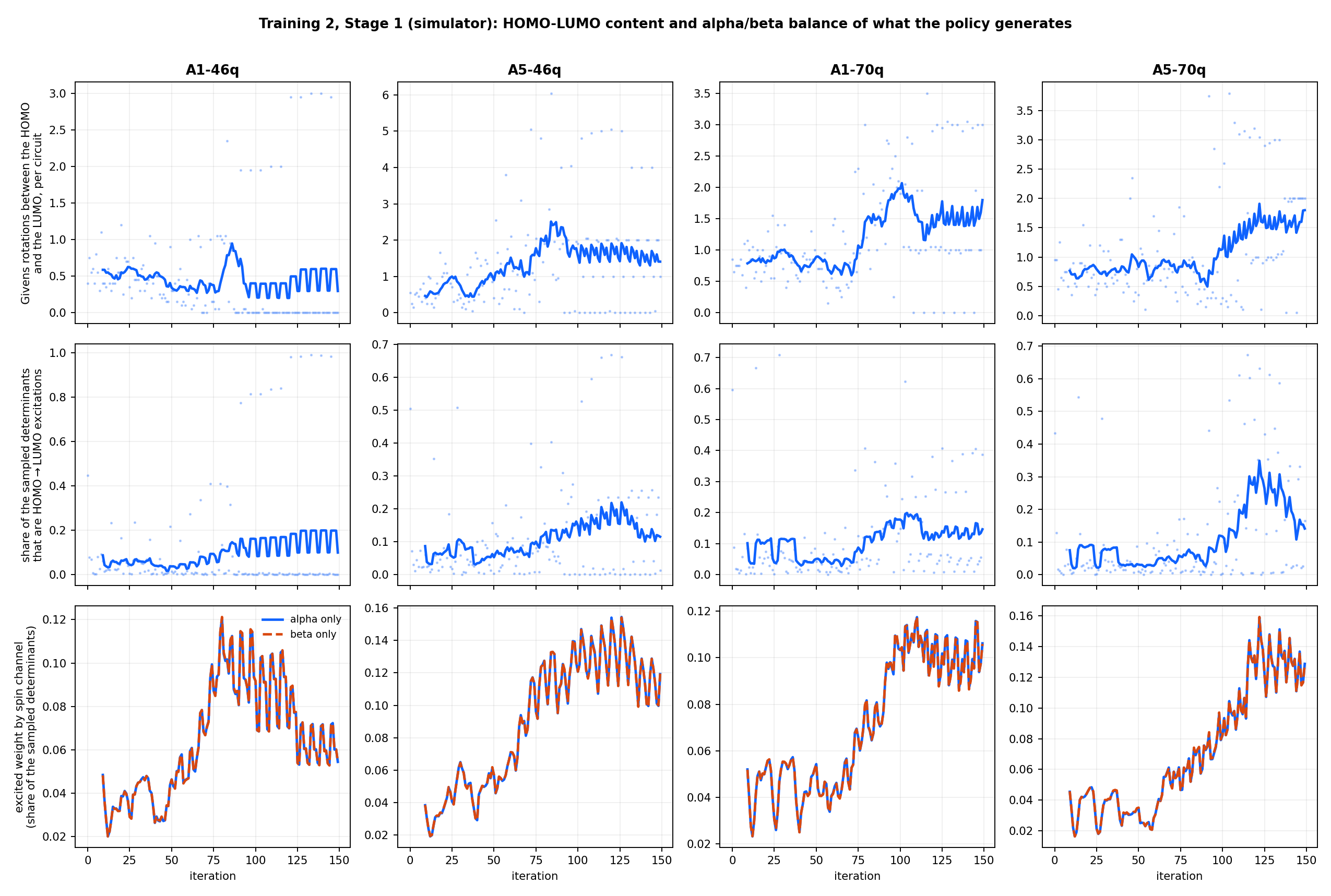}
\caption{Stage~1 (simulator), per lineage (columns). Top: HOMO--LUMO rotations per circuit. Middle: share of the sampled determinants that are HOMO$\to$LUMO excitations. Bottom: weight of the alpha-only (solid) and
beta-only (dashed) excitations. Dots are single iterations, lines are rolling means over 10 iterations. On the simulator all shots are in the right spin sector.}
\label{fig:chem-stage1}
\end{figure}

\begin{figure}[t]
\centering
\includegraphics[width=0.62\linewidth]{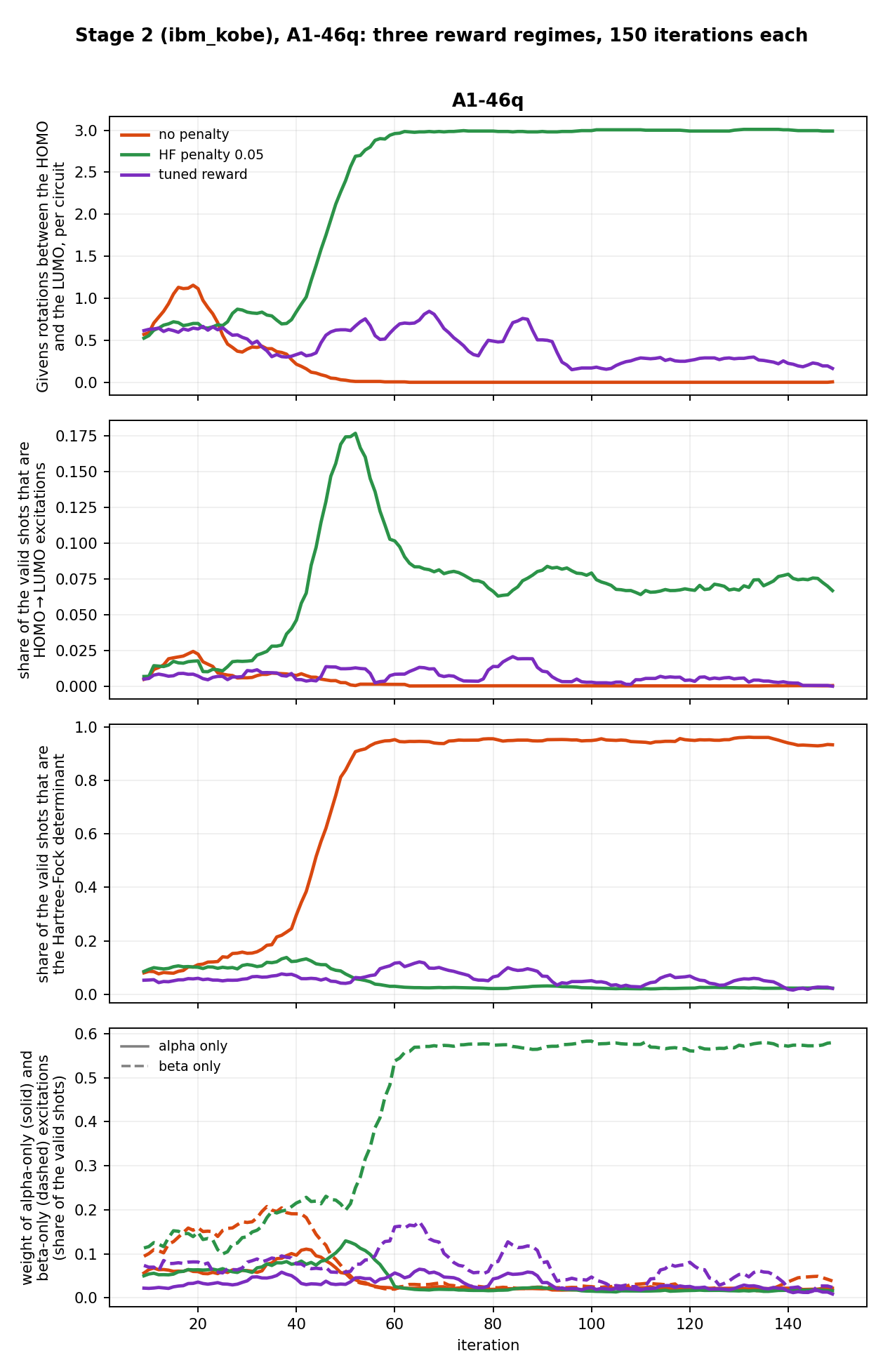}
\caption{Stage~2 (\code{ibm\_kobe}), A1-46q, the only lineage with all three reward regimes complete (150 iterations each): no penalty (orange), HF penalty 0.05 (green) and tuned reward (purple, 150 iterations).
Rows: HOMO--LUMO rotations per circuit; share of the valid shots that are HOMO$\to$LUMO excitations; share of the valid shots that are the Hartree--Fock determinant; weight of alpha-only (solid) and beta-only (dashed) excitations among the valid shots.
Rolling means over 10 iterations. A closed-shell molecule should give equal alpha-only and beta-only weight.}
\label{fig:chem-stage2-a1-46q}
\end{figure}

\begin{figure}[t]
\centering
\includegraphics[width=0.62\linewidth]{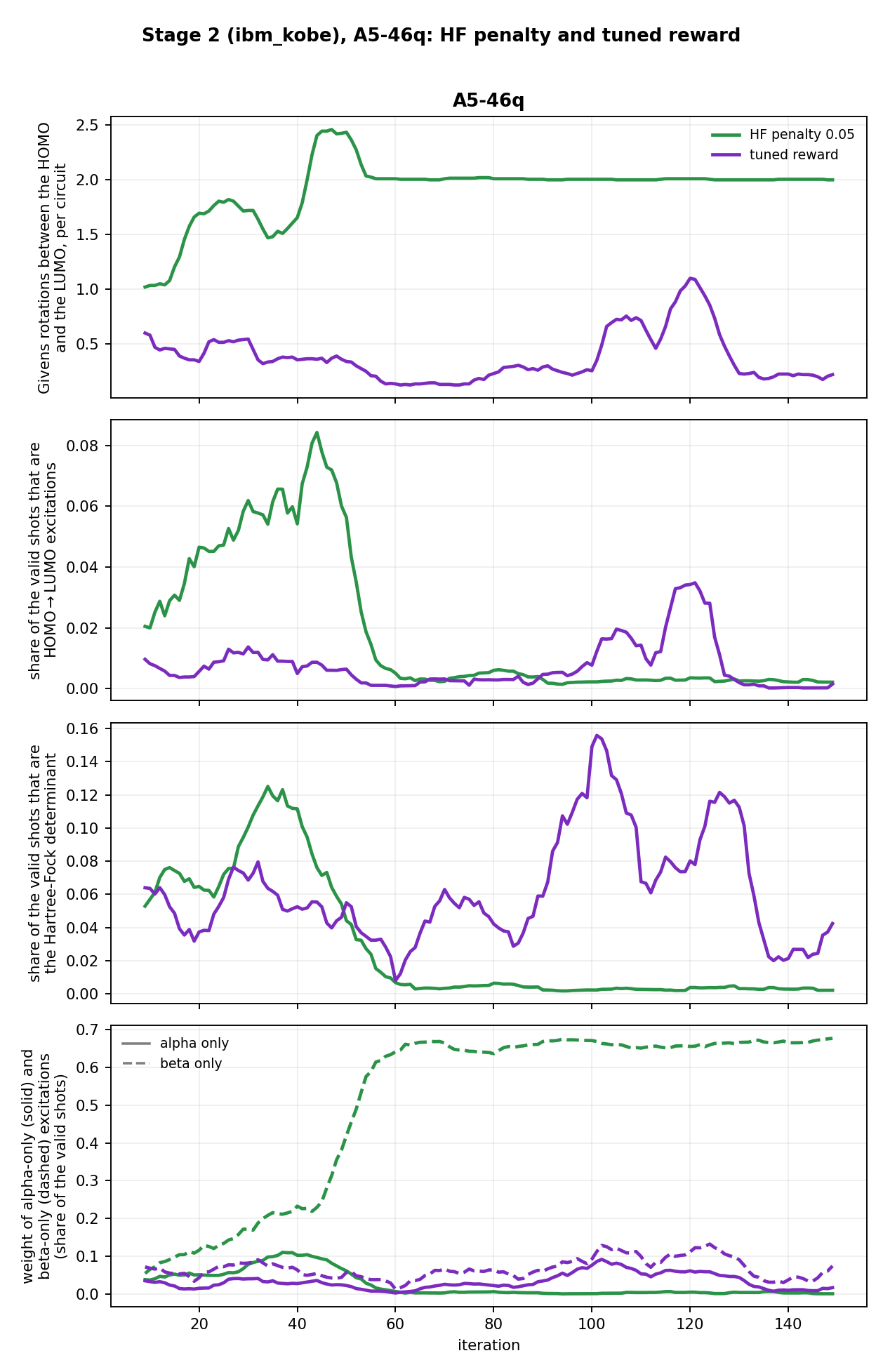}
\caption{As Figure~\ref{fig:chem-stage2-a1-46q} for A5-46q, HF penalty 0.05 (green, 150 iterations) and tuned reward (purple, 150 iterations). The no-penalty run of this lineage was stopped after 17 iterations and is not shown.}
\label{fig:chem-stage2-a5-46q}
\end{figure}

\begin{figure}[t]
\centering
\includegraphics[width=\linewidth]{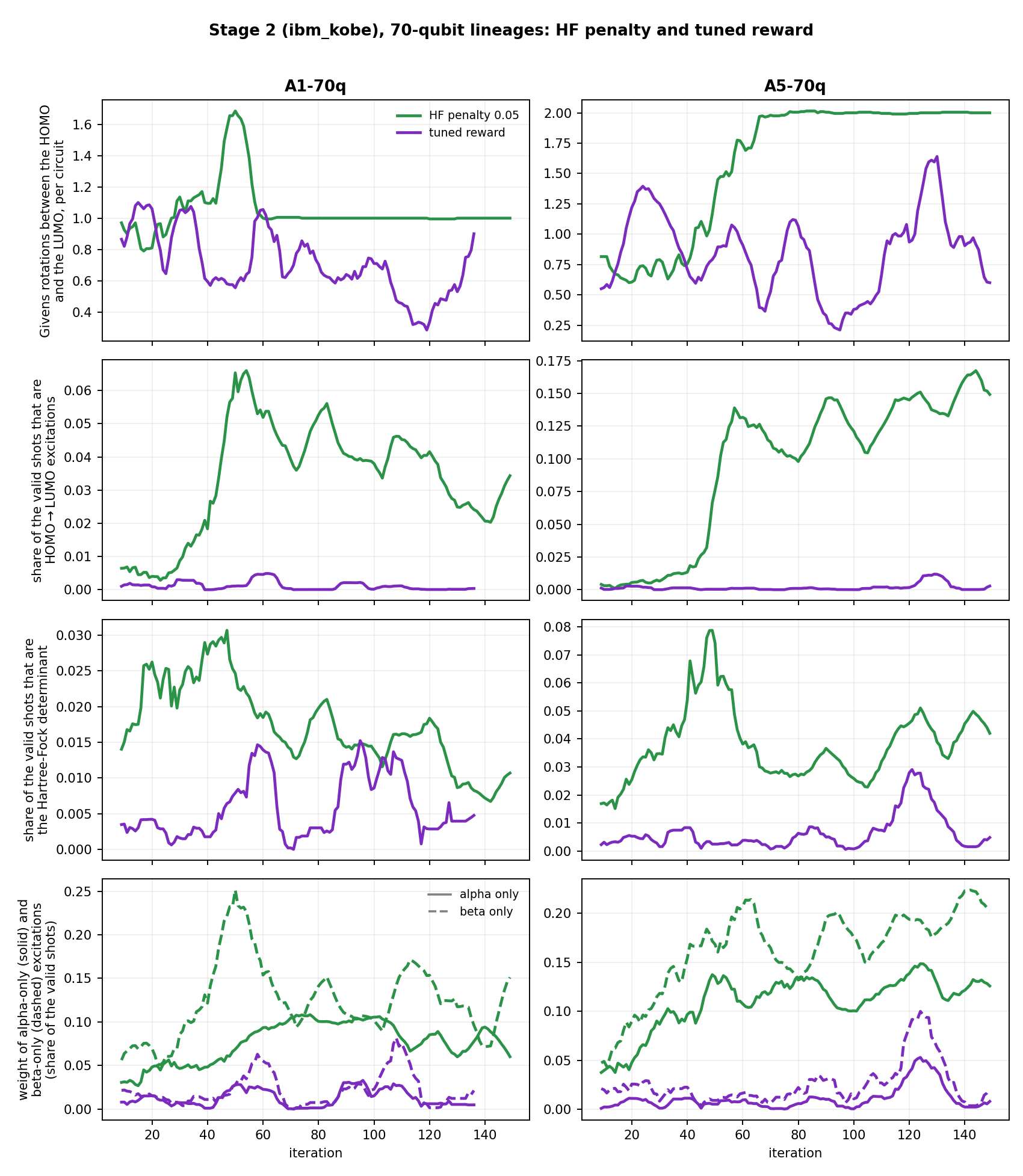}
\caption{As Figure~\ref{fig:chem-stage2-a1-46q} for the 70-qubit lineages A1-70q (left) and A5-70q (right), HF penalty 0.05 (green, 150 iterations) and tuned reward (purple; 137 iterations for A1-70q, 150 for A5-70q). The no-penalty runs of these lineages were
stopped after at most 18 iterations (A1-70q never started) and are not shown.}
\label{fig:chem-stage2-70q}
\end{figure}

\paragraph{Stage 1 (Figure~\ref{fig:chem-stage1}).} On the simulator the circuits keep a small excitation content and the two spin channels stay exactly equal (alpha-only and beta-only both 0.06 to 0.13 over the last 30 iterations; the simulator and all gates are
spin-symmetric, so this is expected). The policy generates 0.5 (A1-46q) to 1.6 HOMO--LUMO rotations per circuit, 14 to 20\% of the sampled weight is on HOMO$\to$LUMO determinants and 60 to 76\% is the Hartree--Fock determinant. The HOMO--LUMO content
rises during Stage~1 in all lineages except A1-46q, where it falls after iteration 80 before the policy collapses onto one sequence.

\paragraph{Stage 2 (Figures~\ref{fig:chem-stage2-a1-46q}, \ref{fig:chem-stage2-a5-46q} and~\ref{fig:chem-stage2-70q}).}
Table~\ref{tab:chem-summary} gives the last 30 iterations of each run.
\begin{table}[t]
\centering
\caption{Chemistry of the last 30 iterations of the Stage-2 runs. ``Valid'': share of all shots in the right spin sector. The other columns are shares of the valid shots: Hartree--Fock determinant (HF), HOMO$\to$LUMO excitation (HL),
alpha-only (a) and beta-only (b) excitation, and excitation in both channels at once. $b/a$ is the ratio of beta-only to alpha-only weight: for example, alpha-only weights 0.016 and 0.003 correspond to about 16 and 7 valid shots per circuit (at 10000 shots), so ratios such as 36 and 207 are uncertain by a large factor. HOMO--LUMO rotations (rot.) are per circuit. The tuned runs are complete (150 iterations) except A1-70q (137); the no-penalty run is complete for A1-46q only.}
\label{tab:chem-summary}
\scriptsize
\begin{tabular}{@{}llrrrrrrrrr@{}}
\toprule
Lineage & Regime & valid & HF & HL & a & b & both & $b/a$ & HL rot. \\
\midrule
A1-46q & no penalty & 0.51 & 0.95 & 0.00 & 0.022 & 0.030 & 0.00 & 1.4 & 0.00 \\
A1-46q & HF penalty & 0.10 & 0.02 & 0.07 & 0.016 & 0.574 & 0.39 & 36 & 3.00 \\
A1-46q & tuned & 0.03 & 0.03 & 0.00 & 0.017 & 0.035 & 0.92 & 2.1 & 0.23 \\
\midrule
A5-46q & HF penalty & 0.21 & 0.00 & 0.00 & 0.003 & 0.669 & 0.32 & 207 & 2.00 \\
A5-46q & tuned & 0.04 & 0.06 & 0.00 & 0.025 & 0.068 & 0.85 & 2.7 & 0.25 \\
\midrule
A1-70q & HF penalty & 0.07 & 0.01 & 0.03 & 0.072 & 0.117 & 0.80 & 1.6 & 1.00 \\
A1-70q & tuned & 0.02 & 0.01 & 0.00 & 0.024 & 0.040 & 0.93 & 1.7 & 0.57 \\
\midrule
A5-70q & HF penalty & 0.06 & 0.04 & 0.15 & 0.127 & 0.200 & 0.63 & 1.6 & 2.00 \\
A5-70q & tuned & 0.02 & 0.00 & 0.00 & 0.008 & 0.023 & 0.96 & 2.9 & 1.06 \\
\bottomrule
\end{tabular}
\end{table}

\paragraph{The three regimes on A1-46q, compared with Stage 1.} A1-46q is the one lineage for which all three Stage-2 regimes ran 150 iterations, so it is the clean comparison. Table~\ref{tab:chem-a1-46q} puts them next to the Stage-1 simulator run that all three started from. Two cautions apply to reading it.
First, the Stage-1 shares are shares of all shots, because on the simulator every shot is valid, whereas the Stage-2 shares are shares of the valid shots, which are 51, 10 and 2.7\% of all shots for the three regimes; the Stage-2 columns therefore describe few shots, and the tuned regime's the fewest.
Second, Stage 1 and Stage 2 use different molecules (14 small and medium molecules against the three transition-metal complexes), so the comparison is of the same quantities, not of the same circuits.

\begin{table}[t]
\centering
\caption{A1-46q: the chemistry of what the policy generates in the last 30 iterations of the simulator stage and of each of the three hardware regimes. Shares are of all shots in Stage 1 (all valid) and of the valid shots in Stage 2. ``HL'' is the HOMO$\to$LUMO excitation, ``a'' and ``b'' the alpha-only and beta-only excitation, ``both'' an excitation in both channels at once, ``rot.'' the HOMO--LUMO rotations per circuit, ``HF gates'' the gates per circuit that can change the Hartree--Fock determinant, ``ident.'' the share of identity tokens and ``depth'' the median depth of the best circuit.}
\label{tab:chem-a1-46q}
\scriptsize
\begin{tabular}{@{}lrrrrrrrrrr@{}}
\toprule
Run & valid & HF & HL & a & b & both & HL rot. & HF gates & ident. & depth \\
\midrule
Stage 1, simulator & 1.00 & 0.76 & 0.17 & 0.062 & 0.062 & 0.11 & 0.50 & 2.5 & 0.60 & 3478 \\
\midrule
Stage 2, no penalty & 0.51 & 0.95 & 0.00 & 0.022 & 0.030 & 0.00 & 0.00 & 0.0 & 0.36 & 48 \\
Stage 2, HF penalty & 0.10 & 0.02 & 0.07 & 0.016 & 0.574 & 0.39 & 3.00 & 3.0 & 0.25 & 56 \\
Stage 2, tuned & 0.03 & 0.03 & 0.00 & 0.017 & 0.035 & 0.92 & 0.23 & 1.8 & 0.34 & 48 \\
\bottomrule
\end{tabular}
\end{table}

\emph{No penalty.} Stage 1 left 76\% of the weight on the Hartree--Fock determinant and 17\% on HOMO$\to$LUMO, with two or three gates per circuit able to change that determinant. The unpenalized Stage 2 pushes this to its limit: no gate can change the determinant (0.0 against 2.5 per circuit),
no HOMO--LUMO rotation remains (0.50 to 0.00), and 95\% of the valid shots are the Hartree--Fock determinant. Of the three regimes it keeps the Stage-1 spin symmetry (0.022 and 0.030), because it excites almost nothing.
\emph{HF penalty.} It reverses the collapse onto Hartree--Fock (2\% of the valid shots) and has more HOMO--LUMO rotations than Stage 1 (3.0 against 0.5) and a similar share of HOMO$\to$LUMO weight (7\% against 17\%), but breaks the Stage-1 spin symmetry, with 57\% of the valid shots beta-only and 2\% alpha-only.
\emph{Tuned.} It removes the Hartree--Fock circuit (3\% of the valid shots) and keeps the alpha and beta weights within a factor of 2.1 of each other (0.017 and 0.035), closer to the Stage-1 symmetry than the HF penalty, but carries almost no HOMO$\to$LUMO weight (0.3\% against 17\%) and fewer HOMO--LUMO rotations than Stage 1 (0.23 against 0.50). Most of its valid shots are excited in both channels at once (92\%), against 11\% in Stage 1.

\paragraph{The three regimes on the other lineages.} The same comparison for A5-46q and the two 70-qubit lineages leaves out the no-penalty regime, whose runs were stopped after 17 or 18 iterations (A1-70q has none). For the HF penalty and the tuned reward the picture of A1-46q repeats.
The HF penalty gives HOMO--LUMO rotation counts of 2.0 (A5-46q, A5-70q) and 1.0 (A1-70q) per circuit and beta-only weights of 0.67 (A5-46q), 0.12 (A1-70q) and 0.20 (A5-70q) against alpha-only weights of 0.003, 0.072 and 0.127. The tuned reward gives 0.25 to 1.06 HOMO--LUMO rotations, a beta-only to alpha-only ratio of 2.7 (A5-46q), 3.1 (A1-70q) and 1.8 (A5-70q),
and HOMO$\to$LUMO weight of at most 0.5\%, with 0.4 to 6\% of the valid shots on the Hartree--Fock determinant. Stage 1 of these lineages has 1.5 to 1.6 HOMO--LUMO rotations per circuit and 14 to 20\% of the weight on HOMO$\to$LUMO determinants; neither Stage-2 regime reaches the second figure except the HF penalty at A5-70q (15\%).

\paragraph{Trade-offs.} Each regime gives up something that another keeps.
\begin{itemize}[leftmargin=1.5em,itemsep=2pt]
  \item \emph{No penalty: usable shots, no HOMO$\to$LUMO excitation.} The only regime with a high share of valid shots (0.51 in A1-46q), because its circuit does nothing to the Hartree--Fock determinant: 95\% of the valid shots are that determinant
  and none is a HOMO$\to$LUMO excitation. The shots are clean because there is nothing to disturb, and the recovered energy then comes from the configuration-recovery step, not from the circuit.
  \item \emph{HF penalty: chemistry content, but spin-asymmetric and a collapsed policy.} It removes the Hartree--Fock circuit (HF share 0.00 to 0.04) and has the most HOMO$\to$LUMO weight (0.07 in A1-46q, 0.15 in A5-70q), but the 46-qubit circuits
  excite the beta channel almost alone (beta-only to alpha-only weight 36 in A1-46q and 207 in A5-46q), which a closed-shell molecule should not do, and only 7 to 21\% of the shots are valid. The policy still collapses onto one
  sequence (effective vocabulary 1.0 in all lineages), so the penalty changes which circuit is repeated, not whether one is.
  \item \emph{Tuned reward: balance and diversity, but a low HOMO$\to$LUMO content and few valid shots.} It keeps the policy from collapsing (21 to 27 effective tokens in the last 30 iterations, against 1.0), keeps the alpha and beta weights within a factor of 1.7
  to 3.3 of each other, and removes the Hartree--Fock circuit (HF share 0.00 to 0.06). In return the HOMO$\to$LUMO weight is at most 0.005, the number of HOMO--LUMO rotations is 0.23 and 0.25 in the 46-qubit tuned runs (against 3.0 and 2.0 for the HF penalty), and only 2 to 4\% of the shots are valid. The valid shots are mostly excited in both channels at once (0.85 to 0.97), and with so few of them we cannot tell designed double excitations from noise that happens to land in the right sector.
  At 70 qubits it also emits very deep circuits in 35 to 44\% of the iterations (Table~\ref{tab:t2-compare}); for circuits deeper than 1{,}000, the median valid-shot share is $10^{-5}$ and the maximum is 0.6\% (214 circuits of the HF-penalty set).
  \item \emph{Cost.} The tuned regime is the slowest per iteration and uses the most QPU time per job (Tables~\ref{tab:t2-timing} and~\ref{tab:t2-qpu}), as discussed above.
  \item \emph{Energy.} None of this shows up as a better energy on hardware. For Co(NH$_3$)$_6$, the molecule whose iterations pass the gate in every run, the mean raw error over the last 45 admitted iterations is 26.2, 47.1 and 33.9~mHa for A1-46q (no penalty, HF penalty, tuned), 46.1 and 36.2~mHa for A5-46q (HF penalty, tuned; the no-penalty run was stopped), 225.5 and 224.1~mHa for A1-70q (HF penalty, tuned) and 225.2 and 217.5~mHa for A5-70q (HF penalty, tuned).
  These are single runs and the differences are of the size of the run-to-run noise.
\end{itemize}
In short, the no-penalty conclusions rest on A1-46q, the only lineage for which that regime completed 150 Stage-2 iterations. The penalties and the tuned reward change what the policy generates in the way they were designed to (no Hartree--Fock-only circuit, no collapse, balanced spin channels) but, within the noise of this device, they have not yet produced
more HOMO$\to$LUMO weight or lower energies than the plain reward. All statements rest on one run per regime and lineage.
\FloatBarrier

\section{Additional training curves}
\label{sec:appendix-training-curves}
The curves below complete the main-text Figures~\ref{fig:stage-loss-46q}, \ref{fig:stage-gradnorm} and~\ref{fig:stage-timing}. All use the run histories of the shared Stage~1 and the HF-penalty Stage~2; the training histories of all runs are part of the released data.

\begin{figure}[!ht]
\centering
\includegraphics[width=0.9\linewidth]{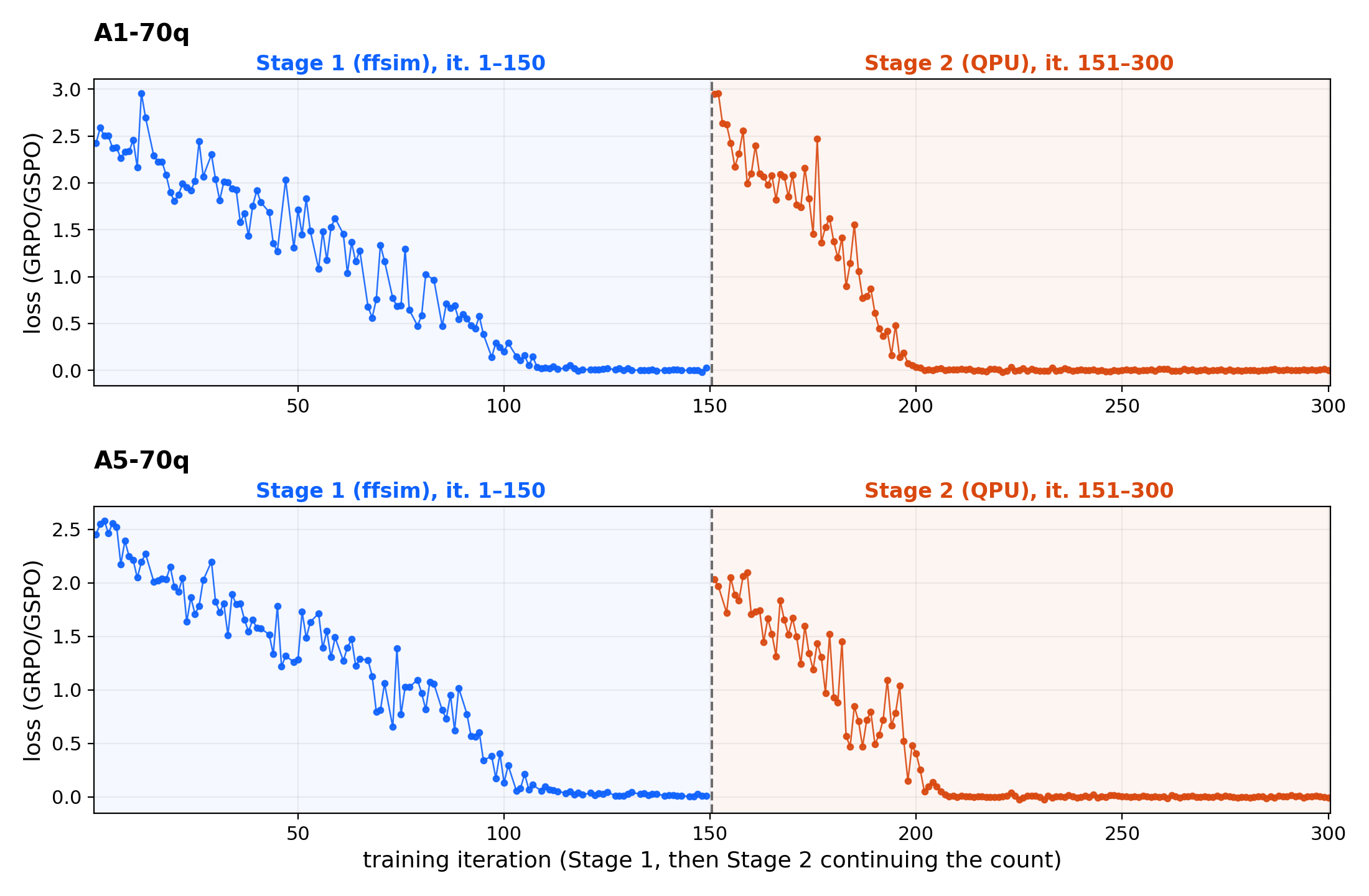}
\caption{Training loss (GRPO/GSPO) of the 70-qubit lineages, A1-70q (top) and A5-70q (bottom), on admitted iterations. Stage~1 is the shared Stage~1 (ffsim simulator, iterations 1--150, left of the dashed line) and Stage~2 the HF-penalty run (\code{ibm\_kobe} QPU, iterations 151--300, right of it).}
\label{fig:stage-loss-70q}
\end{figure}

\begin{figure}[t]
\centering
\includegraphics[width=\linewidth]{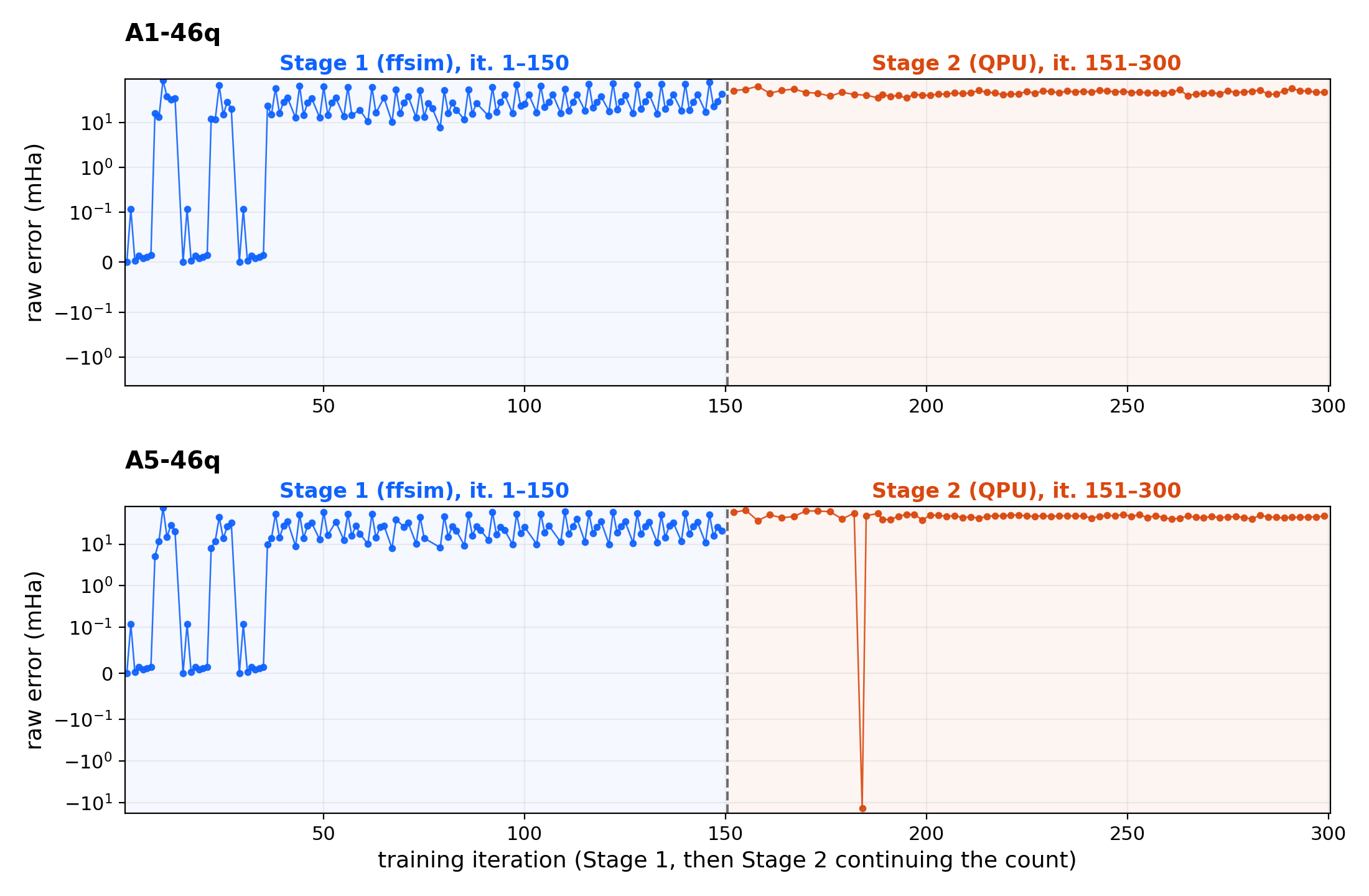}
\caption{Raw SQD energy error (mHa; symmetric logarithmic axis) of the 46-qubit lineages on admitted iterations. Stage~1 (ffsim simulator, iterations 1--150, shared Stage~1) is left and Stage~2 (\code{ibm\_kobe} QPU, HF penalty, iterations 151--300) right of the dashed line. In Stage~1 the error cycles with the size curriculum (small warm-up molecules near zero, larger ones tens of mHa). Negative values are iterations whose recovered energy lies below the CCSD(T) reference, which is not variational.}
\label{fig:stage-rawerr-46q}
\end{figure}

\begin{figure}[t]
\centering
\includegraphics[width=\linewidth]{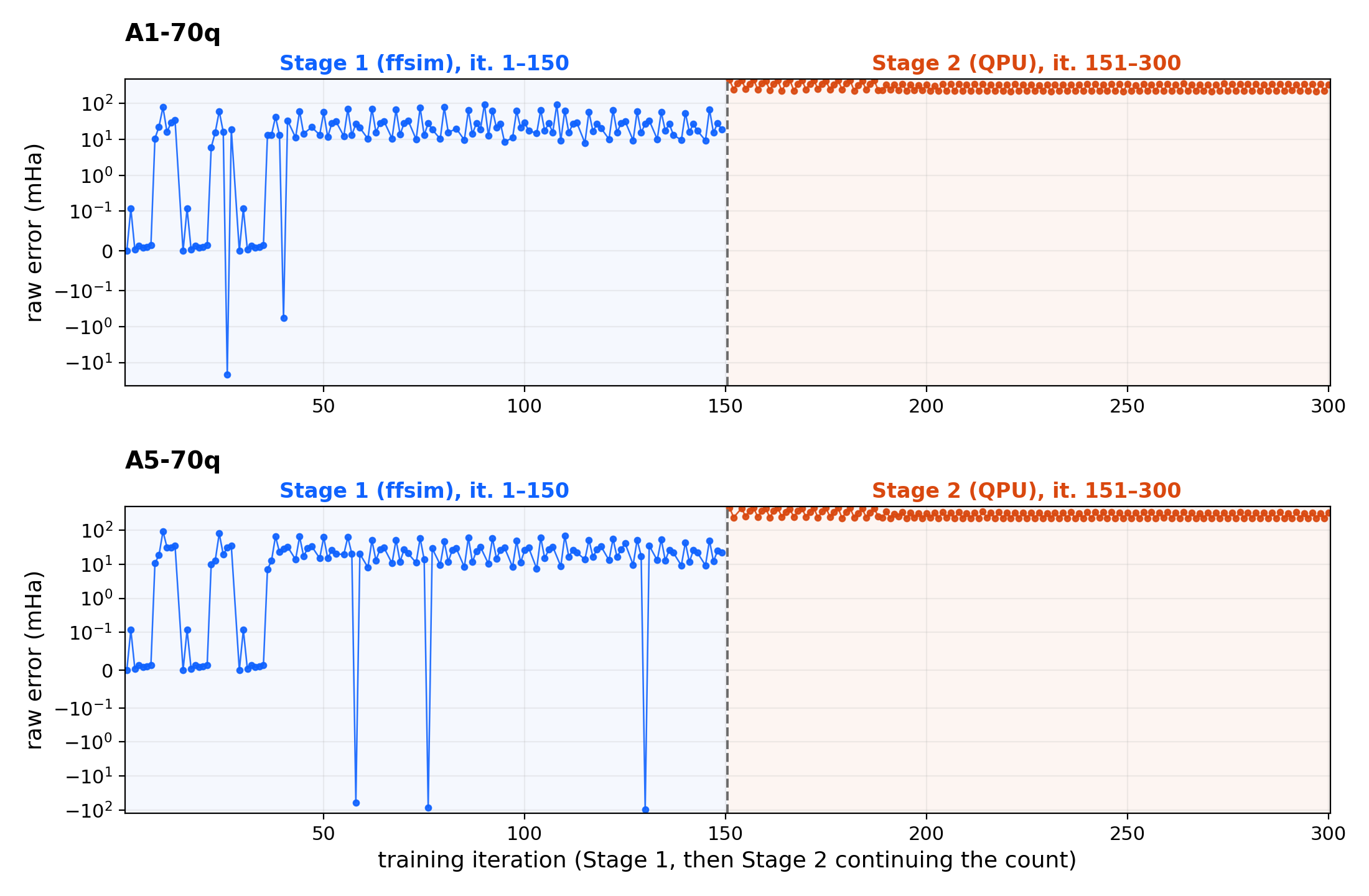}
\caption{As Figure~\ref{fig:stage-rawerr-46q} for the 70-qubit lineages. Stage~1 (ffsim simulator, iterations 1--150, shared Stage~1) is left and Stage~2 (\code{ibm\_kobe} QPU, HF penalty, iterations 151--300) right of the dashed line.}
\label{fig:stage-rawerr-70q}
\end{figure}

\begin{figure}[t]
\centering
\includegraphics[width=\linewidth]{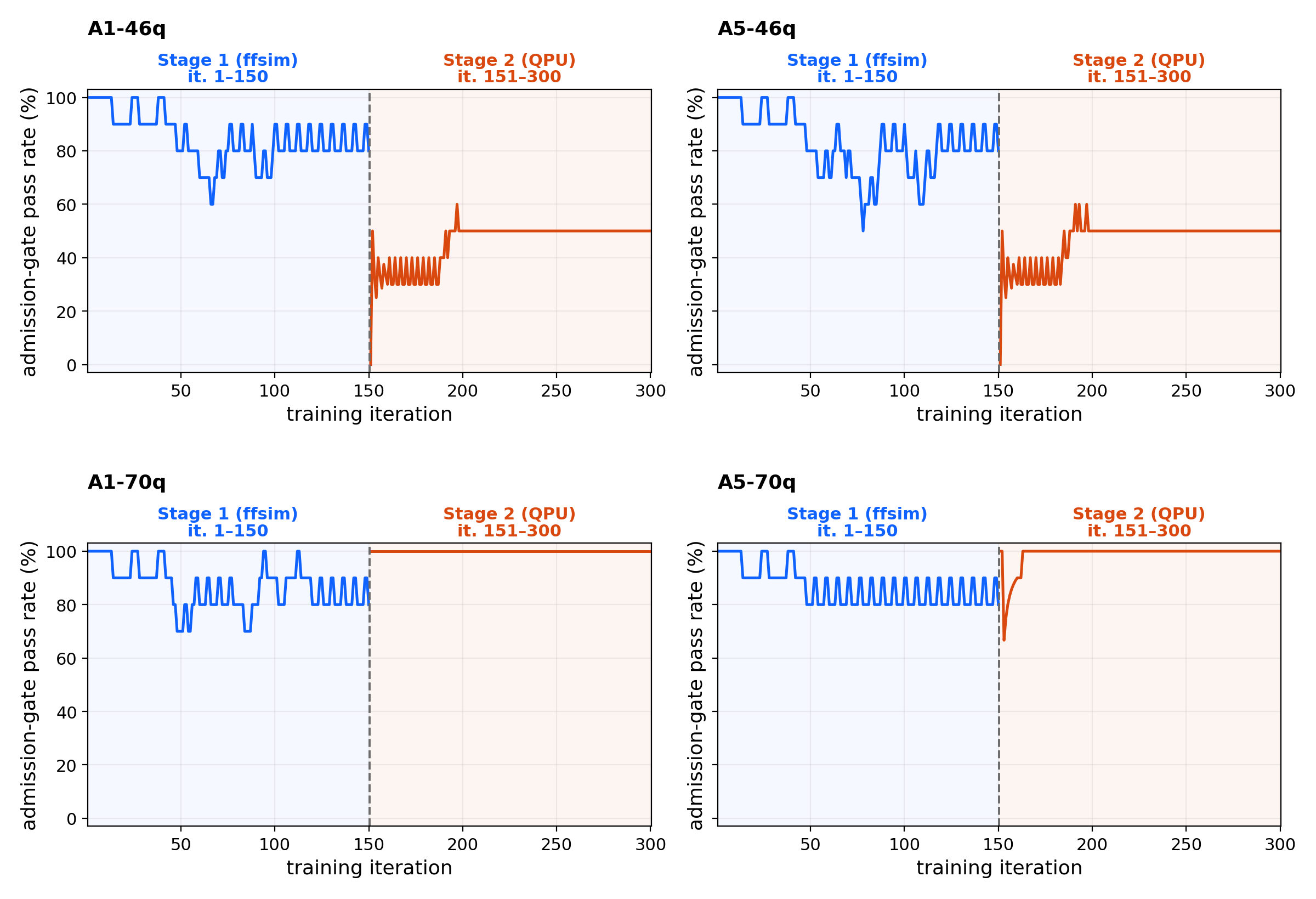}
\caption{Rolling admission-gate pass rate (window of 10 iterations) of the four lineages. Stage~1 (ffsim simulator, shared Stage~1) is left and Stage~2 (\code{ibm\_kobe} QPU, HF penalty) right of the dashed line; the iteration count continues across the stages.}
\label{fig:gate-pass}
\end{figure}

\begin{figure}[t]
\centering
\includegraphics[width=\linewidth]{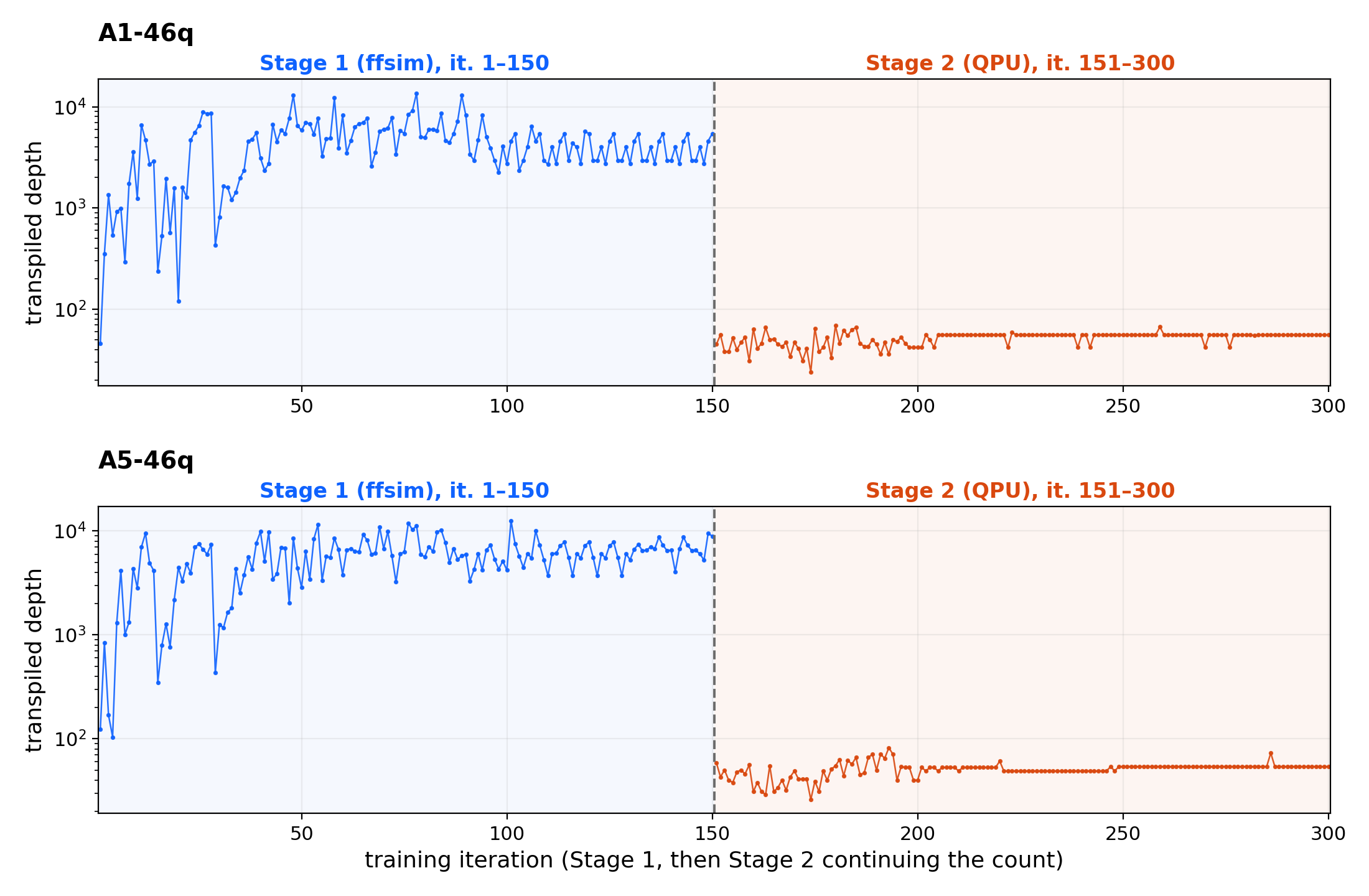}
\caption{Transpiled depth of the best circuit (lowest energy) of the 20 sampled in every iteration, admitted or not (logarithmic axis; transpiled at optimisation level~1 for the \code{ibm\_heron\_r2} coupling map). 46-qubit lineages. Stage~1 (ffsim simulator, iterations 1--150, shared Stage~1) is left and Stage~2 (\code{ibm\_kobe} QPU, HF penalty, iterations 151--300) right of the dashed line.}
\label{fig:stage-depth-46q}
\end{figure}

\begin{figure}[t]
\centering
\includegraphics[width=\linewidth]{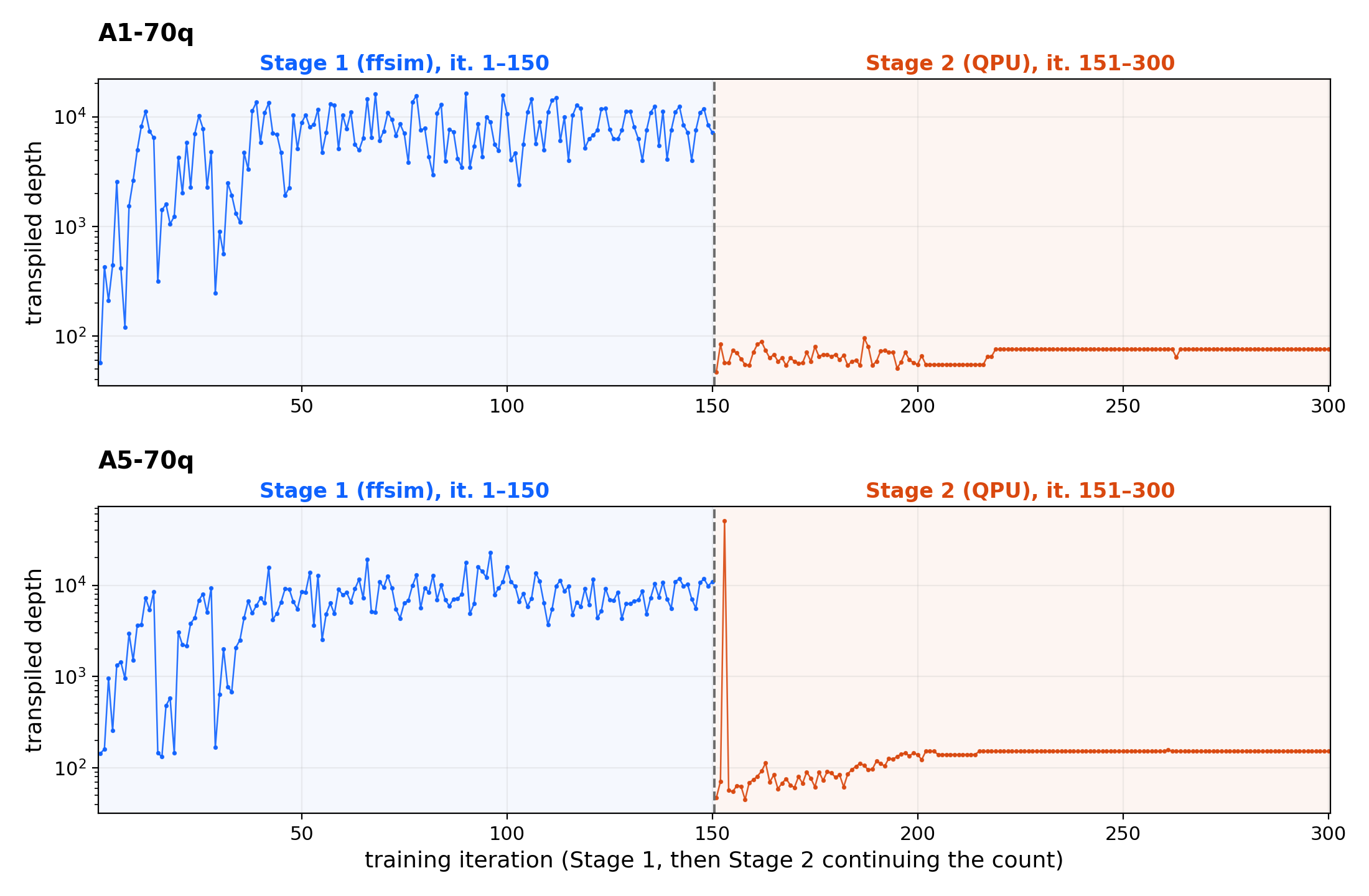}
\caption{Transpiled depth of the best circuit (lowest energy) of the 20 sampled in every iteration, admitted or not (logarithmic axis; transpiled at optimisation level~1 for the \code{ibm\_heron\_r2} coupling map). 70-qubit lineages. Stage~1 (ffsim simulator, iterations 1--150, shared Stage~1) is left and Stage~2 (\code{ibm\_kobe} QPU, HF penalty, iterations 151--300) right of the dashed line.}
\label{fig:stage-depth-70q}
\end{figure}

\begin{figure}[t]
\centering
\includegraphics[width=\linewidth]{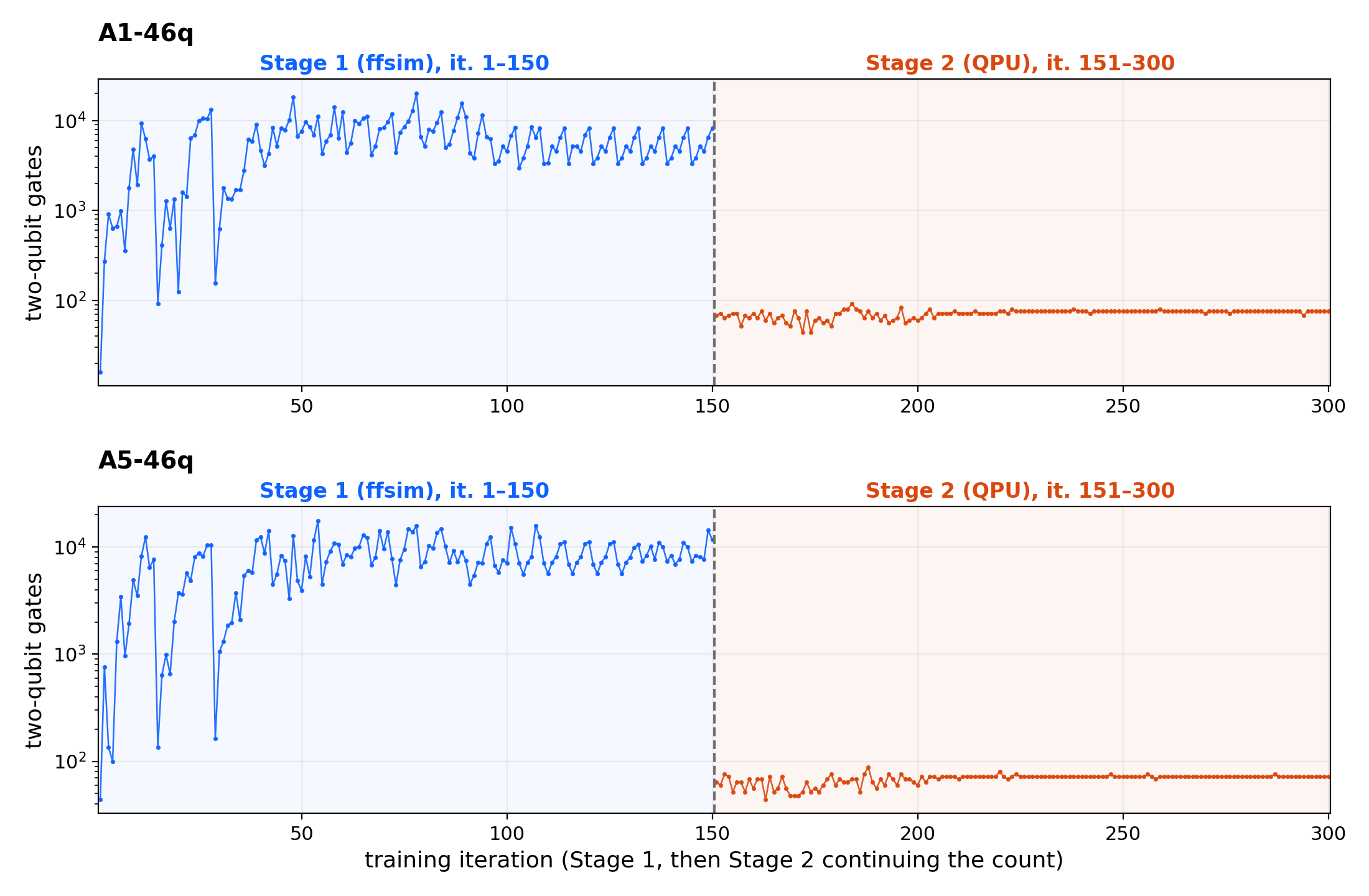}
\caption{Two-qubit-gate count of the best circuit (lowest energy) of the 20 sampled in every iteration, admitted or not (logarithmic axis). 46-qubit lineages. Stage~1 (ffsim simulator, iterations 1--150, shared Stage~1) is left and Stage~2 (\code{ibm\_kobe} QPU, HF penalty, iterations 151--300) right of the dashed line.}
\label{fig:stage-twoq-46q}
\end{figure}

\begin{figure}[t]
\centering
\includegraphics[width=\linewidth]{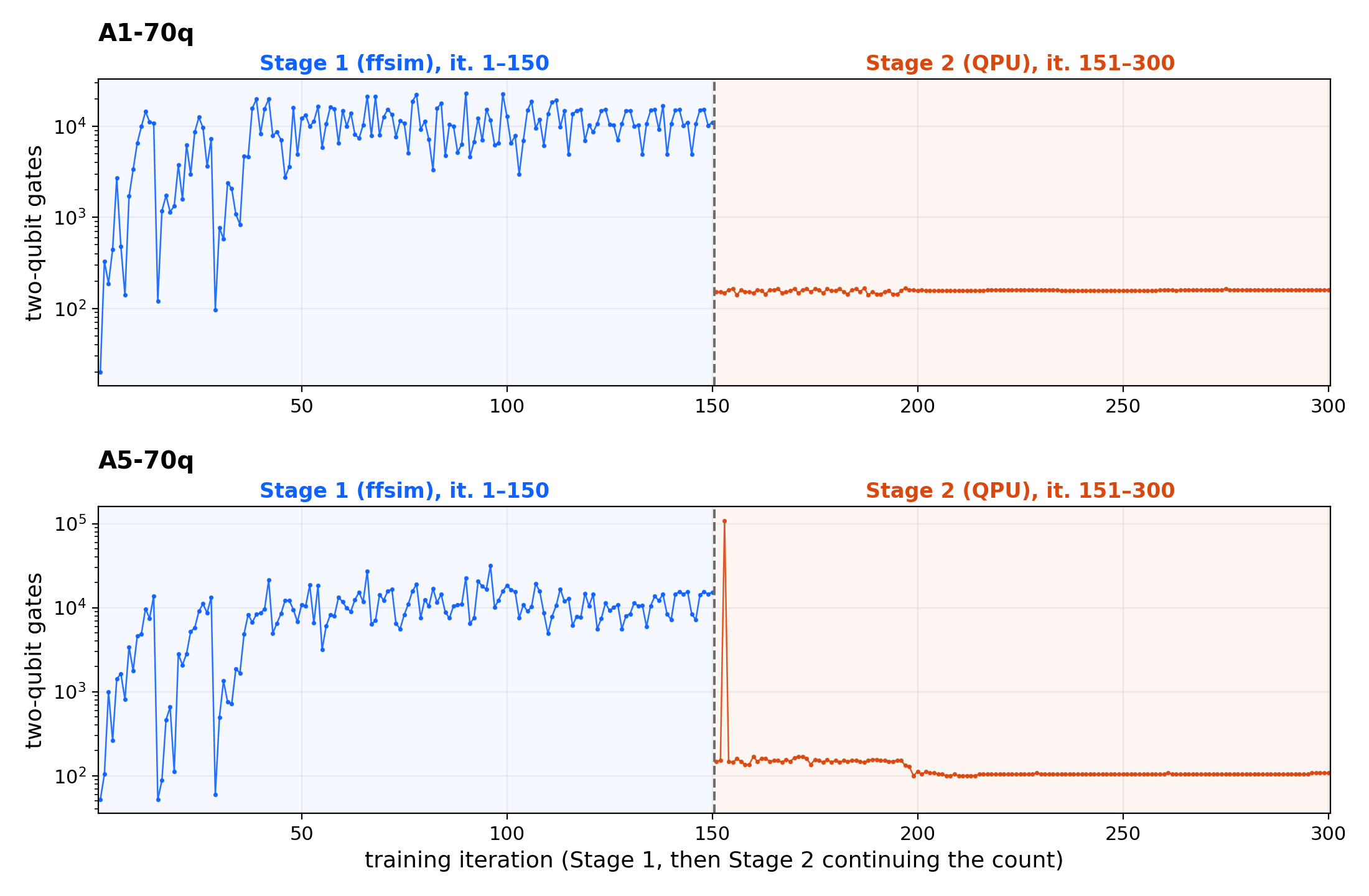}
\caption{Two-qubit-gate count of the best circuit (lowest energy) of the 20 sampled in every iteration, admitted or not (logarithmic axis). 70-qubit lineages. Stage~1 (ffsim simulator, iterations 1--150, shared Stage~1) is left and Stage~2 (\code{ibm\_kobe} QPU, HF penalty, iterations 151--300) right of the dashed line.}
\label{fig:stage-twoq-70q}
\end{figure}

\FloatBarrier
\section{Comparison of the two training configurations, per test case}
\label{sec:appendix-three-way}

Best and second circuit of every test case (molecule--lineage pair) in which both training configurations have both circuits recovered (43 of 44 test cases): recovered energy $E$ ($\eh$), signed deviation $\Delta E$ from the DMRG reference (mHa), as-submitted depth of the best circuit, and the spin-resolved shot analysis of the best circuit. ``valid'' is the share of all shots with the correct electron numbers in each spin sector; ``HF'' is the share of the valid shots equal to the Hartree--Fock determinant, ``HL'' the share containing a HOMO$\to$LUMO excitation (highest occupied to lowest unoccupied orbital), and $\alpha$/$\beta$/both the shares that excite only the spin-up channel, only the spin-down channel, or both. Two repeated energies in the table ($-631.077480$ and $-555.102002$) are Hartree--Fock determinant energies computed from the integrals, not converged SQD or DMRG results.

{\scriptsize
\setlength{\tabcolsep}{2.5pt}
\begin{longtable}{@{}lllrrrrrrrrr@{}}
\toprule
Target & Lineage & Set & best $E$ ($\eh$) & $\Delta E$ (mHa) & 2nd $E$ ($\eh$) & $\Delta E$ (mHa) & depth & valid (\%) & HF (\%) & HL (\%) & $\alpha$/$\beta$/both (\%) \\
\midrule
\endfirsthead
Target & Lineage & Set & best $E$ ($\eh$) & $\Delta E$ (mHa) & 2nd $E$ ($\eh$) & $\Delta E$ (mHa) & depth & valid (\%) & HF (\%) & HL (\%) & $\alpha$/$\beta$/both (\%) \\
\midrule
\endhead
S 32q & A1-46q & HF penalty & -479.193266 & 0.12 & -479.192672 & 0.71 & 50 & 2.7 & 58.1 & 0.8 & 24.8/10.9/6.3 \\
 &  & tuned reward & -479.193089 & 0.29 & -479.193374 & 0.01 & 27 & 6.4 & 56.5 & 0.0 & 35.1/5.0/3.4 \\
\addlinespace
S 32q & A1-70q & HF penalty & -479.193164 & 0.22 & -479.193270 & 0.11 & 41 & 21.3 & 12.6 & 0.0 & 0.6/82.4/4.4 \\
 &  & tuned reward & -479.193259 & 0.12 & -479.193255 & 0.13 & 34 & 5.8 & 21.1 & 28.6 & 4.2/63.2/11.5 \\
\addlinespace
S 32q & A5-46q & HF penalty & -479.192727 & 0.65 & -479.193081 & 0.30 & 31 & 40.9 & 94.6 & 1.4 & 3.3/2.0/0.1 \\
 &  & tuned reward & -479.193374 & 0.01 & -479.193374 & 0.01 & 2770 & 0.0 (9) & 0.0 & 0.0 & 0.0/0.0/100.0 \\
\addlinespace
S 32q & A5-70q & HF penalty & -479.192924 & 0.46 & -479.193134 & 0.25 & 69 & 33.1 & 31.0 & 51.9 & 1.8/63.1/4.0 \\
 &  & tuned reward & -479.191604 & 1.78 & -479.193374 & 0.01 & 92 & 3.0 & 17.0 & 4.4 & 5.3/54.8/22.9 \\
\addlinespace
h2o S 38q & A1-46q & HF penalty & -555.108538 & 0.29 & -555.108684 & 0.15 & 35 & 9.0 & 79.9 & 0.0 & 7.3/11.5/1.3 \\
 &  & tuned reward & -555.108769 & 0.06 & -555.108664 & 0.17 & 40 & 4.3 & 40.7 & 3.1 & 6.6/46.2/6.5 \\
\addlinespace
h2o S 38q & A1-70q & HF penalty & -555.108766 & 0.07 & -555.108751 & 0.08 & 57 & 14.7 & 76.7 & 0.5 & 20.6/2.1/0.6 \\
 &  & tuned reward & -555.108816 & 0.02 & -555.108680 & 0.15 & 68 & 2.9 & 35.5 & 0.8 & 6.0/48.1/10.4 \\
\addlinespace
h2o S 38q & A5-46q & HF penalty & -555.108713 & 0.12 & -555.108826 & 0.01 & 34 & 40.4 & 93.1 & 0.0 & 0.6/6.2/0.1 \\
 &  & tuned reward & -555.108740 & 0.09 & -555.108540 & 0.29 & 64 & 2.8 & 64.8 & 0.4 & 8.7/23.3/3.1 \\
\addlinespace
h2o S 38q & A5-70q & HF penalty & -555.108740 & 0.09 & -555.108764 & 0.07 & 110 & 12.9 & 61.5 & 6.6 & 21.7/12.4/4.4 \\
 &  & tuned reward & -555.108752 & 0.08 & -555.108824 & 0.01 & 45 & 22.4 & 3.5 & 0.1 & 0.0/95.0/1.4 \\
\addlinespace
M 38q & A1-46q & HF penalty & -479.292683 & 9.60 & -479.296591 & 5.69 & 35 & 12.0 & 64.2 & 11.1 & 11.8/20.2/3.8 \\
 &  & tuned reward & -479.297136 & 5.14 & -479.290408 & 11.87 & 40 & 8.4 & 48.9 & 8.5 & 23.2/18.2/9.7 \\
\addlinespace
M 38q & A1-70q & HF penalty & -479.298463 & 3.82 & -479.295672 & 6.61 & 57 & 14.8 & 52.1 & 9.1 & 35.2/7.5/5.2 \\
 &  & tuned reward & -479.295479 & 6.80 & -479.295776 & 6.50 & 68 & 12.5 & 2.5 & 32.8 & 8.8/20.3/68.4 \\
\addlinespace
M 38q & A5-46q & HF penalty & -479.285058 & 17.22 & -479.282828 & 19.45 & 32 & 20.7 & 3.8 & 4.0 & 0.1/91.5/4.6 \\
 &  & tuned reward & -479.290816 & 11.46 & -479.288843 & 13.43 & 49 & 21.7 & 7.8 & 0.4 & 0.3/89.5/2.4 \\
\addlinespace
M 38q & A5-70q & HF penalty & -479.288699 & 13.58 & -479.152486 & 149.79 & 152 & 7.0 & 45.0 & 0.4 & 11.6/33.9/9.5 \\
 &  & tuned reward & -479.292726 & 9.55 & -479.150765 & 151.51 & 56 & 4.0 & 25.0 & 0.4 & 32.6/19.0/23.4 \\
\addlinespace
2h2o S 44q & A1-46q & HF penalty & -631.083103 & 0.21 & -631.083239 & 0.07 & 30 & 1.5 & 64.1 & 2.7 & 8.0/25.3/2.6 \\
 &  & tuned reward & -631.083257 & 0.05 & -631.083235 & 0.08 & 59 & 5.7 & 25.3 & 1.0 & 31.8/19.6/23.2 \\
\addlinespace
2h2o S 44q & A1-70q & HF penalty & -631.083257 & 0.06 & -631.083084 & 0.23 & 60 & 1.9 & 61.7 & 2.9 & 7.6/26.4/4.3 \\
 &  & tuned reward & -631.083269 & 0.04 & -631.083269 & 0.04 & 298 & 0.0 (7) & 0.0 & 0.0 & 0.0/0.0/100.0 \\
\addlinespace
2h2o S 44q & A5-46q & HF penalty & -631.083185 & 0.13 & -631.083207 & 0.11 & 49 & 11.9 & 41.4 & 34.3 & 53.3/2.2/3.1 \\
 &  & tuned reward & -631.083122 & 0.19 & -631.083109 & 0.20 & 33 & 14.1 & 45.6 & 0.7 & 38.3/8.2/7.8 \\
\addlinespace
2h2o S 44q & A5-70q & HF penalty & -631.083092 & 0.22 & -631.083044 & 0.27 & 168 & 14.9 & 17.1 & 29.0 & 0.6/79.7/2.6 \\
 &  & tuned reward & -631.083237 & 0.07 & -631.083255 & 0.06 & 66 & 2.5 & 75.7 & 0.4 & 3.9/19.4/1.0 \\
\addlinespace
L 46q & A1-46q & HF penalty & -479.299926 & 14.29 & -479.121483 & 192.73 & 37 & 11.1 & 53.5 & 0.8 & 1.6/43.7/1.3 \\
 &  & tuned reward & -479.301231 & 12.98 & -479.115176 & 199.04 & 36 & 19.0 & 14.7 & 0.2 & 0.7/80.7/4.0 \\
\addlinespace
L 46q & A1-70q & HF penalty & -479.306409 & 7.81 & -479.116356 & 197.86 & 69 & 9.0 & 59.2 & 0.5 & 3.5/35.3/2.0 \\
 &  & tuned reward & -479.300283 & 13.93 & -479.116713 & 197.50 & 69 & 8.1 & 29.1 & 0.0 & 64.6/2.0/4.3 \\
\addlinespace
L 46q & A5-46q & HF penalty & -479.301748 & 12.47 & -479.306572 & 7.64 & 55 & 2.4 & 74.1 & 1.8 & 2.2/23.1/0.6 \\
 &  & tuned reward & -479.301200 & 13.02 & -479.120938 & 193.28 & 41 & 13.7 & 13.7 & 22.6 & 32.0/16.4/37.8 \\
\addlinespace
L 46q & A5-70q & HF penalty & -479.308601 & 5.62 & -479.297326 & 16.89 & 144 & 4.3 & 38.6 & 0.8 & 44.8/8.0/8.7 \\
 &  & tuned reward & -479.115938 & 198.28 & -479.115299 & 198.92 & 17042 & 0.0 (47) & 0.0 & 0.0 & 0.0/0.0/100.0 \\
\addlinespace
h2o M 46q & A1-46q & HF penalty & -555.123748 & 1.41 & -555.102442 & 22.72 & 37 & 10.3 & 52.9 & 0.7 & 3.0/41.8/2.4 \\
 &  & tuned reward & -555.123722 & 1.44 & -555.102808 & 22.35 & 36 & 22.6 & 8.5 & 0.1 & 0.4/87.6/3.5 \\
\addlinespace
h2o M 46q & A1-70q & HF penalty & -555.124450 & 0.71 & -555.102207 & 22.95 & 69 & 10.2 & 49.3 & 1.6 & 3.2/44.3/3.1 \\
 &  & tuned reward & -555.124065 & 1.09 & -555.102254 & 22.90 & 69 & 7.7 & 29.0 & 0.4 & 64.1/1.9/5.0 \\
\addlinespace
h2o M 46q & A5-46q & HF penalty & -555.123672 & 1.49 & -555.123151 & 2.01 & 58 & 23.9 & 23.4 & 1.5 & 0.9/72.8/2.8 \\
 &  & tuned reward & -555.122758 & 2.40 & -555.122769 & 2.39 & 45 & 9.5 & 66.1 & 0.0 & 14.6/13.1/6.1 \\
\addlinespace
h2o M 46q & A5-70q & HF penalty & -555.123764 & 1.39 & -555.123482 & 1.68 & 108 & 5.3 & 57.9 & 2.2 & 13.8/21.8/6.5 \\
 &  & tuned reward & -555.124431 & 0.73 & -555.124133 & 1.02 & 44 & 1.8 & 22.2 & 0.9 & 10.9/47.2/19.7 \\
\addlinespace
LL 56q & A1-46q & HF penalty & -479.300183 & 14.87 & -479.299292 & 15.76 & 52 & 3.8 & 41.6 & 1.0 & 8.7/41.7/7.9 \\
 &  & tuned reward & -479.305726 & 9.33 & -479.109979 & 205.08 & 52 & 2.2 & 26.5 & 0.7 & 29.2/20.5/23.8 \\
\addlinespace
LL 56q & A1-70q & HF penalty & -479.304625 & 10.43 & -479.112860 & 202.20 & 62 & 1.5 & 38.0 & 0.8 & 2.2/56.3/3.6 \\
 &  & tuned reward & -479.306343 & 8.71 & -479.110787 & 204.27 & 59 & 0.4 & 19.2 & 0.7 & 60.5/5.0/15.3 \\
\addlinespace
LL 56q & A5-70q & HF penalty & -479.301896 & 13.16 & -479.111752 & 203.30 & 79 & 1.8 & 24.0 & 2.3 & 30.5/16.7/28.9 \\
 &  & tuned reward & -479.302562 & 12.50 & -479.297283 & 17.77 & 63 & 14.9 & 56.5 & 3.7 & 14.4/23.2/5.8 \\
\addlinespace
h2o L 56q & A1-46q & HF penalty & -555.132208 & 2.85 & -555.132421 & 2.64 & 47 & 5.0 & 15.0 & 1.0 & 28.8/18.6/37.6 \\
 &  & tuned reward & -555.132762 & 2.30 & -555.102002 & 33.06 & 52 & 2.4 & 26.3 & 0.0 & 29.3/21.4/23.0 \\
\addlinespace
h2o L 56q & A1-70q & HF penalty & -555.133193 & 1.86 & -555.132552 & 2.51 & 65 & 7.3 & 47.9 & 2.5 & 39.6/6.7/5.8 \\
 &  & tuned reward & -555.133432 & 1.63 & -555.102199 & 32.86 & 59 & 0.9 & 16.3 & 1.1 & 67.6/2.9/13.2 \\
\addlinespace
h2o L 56q & A5-46q & HF penalty & -555.131417 & 3.64 & -555.132660 & 2.40 & 53 & 8.4 & 72.3 & 4.9 & 12.4/13.1/2.2 \\
 &  & tuned reward & -555.132718 & 2.34 & -555.132834 & 2.22 & 49 & 1.7 & 43.5 & 1.1 & 5.5/44.8/6.2 \\
\addlinespace
h2o L 56q & A5-70q & HF penalty & -555.132405 & 2.65 & -555.132408 & 2.65 & 44 & 14.2 & 6.7 & 44.7 & 14.6/24.2/54.5 \\
 &  & tuned reward & -555.132949 & 2.11 & -555.133484 & 1.57 & 60 & 2.1 & 7.8 & 14.3 & 9.6/35.7/46.9 \\
\addlinespace
2h2o M 58q & A1-46q & HF penalty & -631.112084 & 6.53 & -631.077480 & 41.14 & 47 & 2.3 & 18.4 & 4.8 & 39.3/11.9/30.4 \\
 &  & tuned reward & -631.077480 & 41.14 & -631.077480 & 41.14 & 10985 & 0.1 & 0.0 & 0.0 & 0.0/0.0/100.0 \\
\addlinespace
2h2o M 58q & A1-70q & HF penalty & -631.114317 & 4.30 & -631.113058 & 5.56 & 65 & 5.6 & 50.9 & 2.5 & 38.6/5.9/4.7 \\
 &  & tuned reward & -631.113121 & 5.49 & -631.077480 & 41.14 & 49 & 2.1 & 13.1 & 1.0 & 62.3/4.0/20.5 \\
\addlinespace
2h2o M 58q & A5-46q & HF penalty & -631.114080 & 4.54 & -631.113620 & 5.00 & 45 & 7.1 & 7.4 & 14.4 & 6.1/45.7/40.7 \\
 &  & tuned reward & -631.077480 & 41.14 & -631.077480 & 41.14 & 368 & 0.6 & 0.0 & 0.0 & 0.0/0.0/100.0 \\
\addlinespace
2h2o M 58q & A5-70q & HF penalty & -631.113705 & 4.91 & -631.111954 & 6.66 & 62 & 19.8 & 59.2 & 2.2 & 31.0/6.5/3.4 \\
 &  & tuned reward & -631.113566 & 5.05 & -631.077480 & 41.14 & 56 & 6.6 & 22.1 & 0.0 & 54.1/5.9/17.9 \\
\addlinespace
h2o LL 66q & A1-46q & HF penalty & -555.135328 & 2.07 & -555.133503 & 3.89 & 60 & 12.1 & 80.1 & 0.0 & 14.6/4.4/0.8 \\
 &  & tuned reward & -555.135415 & 1.98 & -555.133657 & 3.74 & 34 & 8.1 & 56.1 & 6.5 & 31.2/8.1/4.6 \\
\addlinespace
h2o LL 66q & A1-70q & HF penalty & -555.134846 & 2.55 & -555.102002 & 35.39 & 51 & 2.2 & 8.2 & 3.2 & 1.5/70.4/19.8 \\
 &  & tuned reward & -555.134843 & 2.55 & -555.134225 & 3.17 & 58 & 4.4 & 37.0 & 17.4 & 16.1/32.6/14.3 \\
\addlinespace
h2o LL 66q & A5-46q & HF penalty & -555.133877 & 3.52 & -555.134389 & 3.01 & 44 & 9.3 & 46.5 & 0.1 & 8.5/37.9/7.0 \\
 &  & tuned reward & -555.135580 & 1.82 & -555.133728 & 3.67 & 49 & 12.1 & 80.5 & 4.5 & 8.1/10.4/1.0 \\
\addlinespace
h2o LL 66q & A5-70q & HF penalty & -555.135231 & 2.17 & -555.134346 & 3.05 & 57 & 7.5 & 79.9 & 1.5 & 11.2/6.8/2.1 \\
 &  & tuned reward & -555.134772 & 2.62 & -555.132932 & 4.46 & 69 & 2.5 & 8.1 & 9.7 & 2.3/66.7/22.9 \\
\addlinespace
2h2o L 70q & A1-46q & HF penalty & -631.123614 & 8.62 & -631.123272 & 8.97 & 48 & 3.7 & 48.4 & 0.0 & 38.7/7.2/5.6 \\
 &  & tuned reward & -631.124462 & 7.78 & -631.077480 & 54.76 & 48 & 6.0 & 59.5 & 0.0 & 23.8/11.4/5.4 \\
\addlinespace
2h2o L 70q & A1-70q & HF penalty & -631.124552 & 7.69 & -631.077480 & 54.76 & 60 & 0.4 & 14.5 & 3.1 & 19.1/25.2/41.2 \\
 &  & tuned reward & -631.122201 & 10.04 & -631.077480 & 54.76 & 63 & 0.4 & 31.7 & 0.7 & 22.8/23.5/22.1 \\
\addlinespace
2h2o L 70q & A5-46q & HF penalty & -631.124857 & 7.38 & -631.123201 & 9.04 & 44 & 5.7 & 63.0 & 0.0 & 10.3/21.7/5.0 \\
 &  & tuned reward & -631.126224 & 6.01 & -631.124909 & 7.33 & 40 & 14.1 & 36.2 & 10.8 & 30.4/18.0/15.5 \\
\addlinespace
2h2o L 70q & A5-70q & HF penalty & -631.120256 & 11.98 & -631.124322 & 7.92 & 134 & 4.9 & 0.5 & 49.4 & 6.1/6.2/87.1 \\
 &  & tuned reward & -631.124065 & 8.17 & -631.124165 & 8.07 & 61 & 2.5 & 16.0 & 0.1 & 41.8/10.7/31.6 \\
\addlinespace
\bottomrule
\end{longtable}}

\section{Per-circuit divergence from the idealised LUCJ circuit}
\label{sec:appendix-divergence}

Fidelity, Hellinger distance, Jensen--Shannon divergence, $d_{\mathrm{HF}}$ and (where computed) the operator-norm lower bound of every circuit of the three small tiers, for the two training configurations; definitions and references are in Section~\ref{sec:divergence}. The tuned run of SnOH3-me-2h2o S with A5-70q is the repeated draw of Appendix~\ref{sec:appendix-repeat}. Larger tiers cannot be computed exactly, for the memory reason given there.

\begin{longtable}{@{}lllrrrrrr@{}}
\caption{Per-circuit ideal divergence from the LUCJ circuit, HF-penalty set (Section~\ref{sec:divergence}). $F$, $H$, JS (bits) and $d_{\mathrm{HF}}$ as in Table~\ref{tab:divergence-summary}. ``op.\ norm*'' is a lower-bound estimate of $\|U_g-U_{\mathrm{LUCJ}}\|$ (maximum 2) computed without phase minimisation, and only for the eight selected SnOH3-me S circuits of each set; 2.00 indicates a non-converged value (Lanczos stopped at the step cap). $F=0.000$ means $F<10^{-5}$, and $H$ or JS values above 0.998 are rounded to 1.000. Depth is as-submitted. ``sel.'': best or second circuit.}\label{tab:divergence-circuit_analysis_hf}\\
\toprule
Run & \# & sel. & depth & $F$ & $H$ & JS & $d_{\mathrm{HF}}$ & op.\ norm* \\
\midrule
\endfirsthead
\toprule
Run & \# & sel. & depth & $F$ & $H$ & JS & $d_{\mathrm{HF}}$ & op.\ norm* \\
\midrule
\endhead
SnOH3-me S (32q), A1-46q & 1 &  & 2897 & 0.000 & 0.999 & 1.000 & 1.414 & -- \\
 & 2 &  & 46 & 0.954 & 0.152 & 0.024 & 0.216 & -- \\
 & 3 &  & 3861 & 0.000 & 1.000 & 1.000 & 1.414 & -- \\
 & 4 &  & 34 & 0.998 & 0.035 & 0.001 & 0.049 & -- \\
 & 5 & best & 50 & 0.993 & 0.059 & 0.004 & 0.084 & 1.71 \\
 & 6 &  & 3783 & 0.000 & 1.000 & 1.000 & 1.414 & -- \\
 & 7 &  & 6568 & 0.000 & 1.000 & 1.000 & 1.414 & -- \\
 & 8 & 2nd & 30 & 0.998 & 0.035 & 0.001 & 0.049 & 1.33 \\
 & 9 &  & 196 & 0.000 & 1.000 & 1.000 & 1.414 & -- \\
 & 10 &  & 2595 & 0.954 & 0.152 & 0.024 & 0.215 & -- \\
\addlinespace
SnOH3-me S (32q), A1-70q & 1 &  & 6798 & 0.000 & 1.000 & 1.000 & 1.414 & -- \\
 & 2 &  & 2698 & 0.923 & 0.197 & 0.040 & 0.279 & -- \\
 & 3 &  & 3356 & 0.000 & 1.000 & 1.000 & 1.414 & -- \\
 & 4 & best & 41 & 0.998 & 0.035 & 0.001 & 0.049 & 1.79 \\
 & 5 & 2nd & 57 & 0.993 & 0.059 & 0.004 & 0.084 & 1.94 \\
 & 6 &  & 8368 & 0.000 & 1.000 & 1.000 & 1.414 & -- \\
 & 7 &  & 6606 & 0.000 & 1.000 & 1.000 & 1.414 & -- \\
 & 8 &  & 2760 & 0.999 & 0.019 & 0.000 & 0.027 & -- \\
 & 9 &  & 3460 & 0.000 & 0.999 & 0.999 & 1.413 & -- \\
 & 10 &  & 3214 & 0.000 & 0.999 & 1.000 & 1.413 & -- \\
\addlinespace
SnOH3-me S (32q), A5-46q & 1 &  & 395 & 0.000 & 1.000 & 1.000 & 1.414 & -- \\
 & 2 &  & 218 & 0.000 & 1.000 & 1.000 & 1.414 & -- \\
 & 3 & best & 31 & 0.998 & 0.035 & 0.001 & 0.049 & 1.69 \\
 & 4 &  & 37 & 0.998 & 0.035 & 0.001 & 0.049 & -- \\
 & 5 &  & 222 & 0.000 & 1.000 & 1.000 & 1.414 & -- \\
 & 6 &  & 48 & 0.998 & 0.035 & 0.001 & 0.049 & -- \\
 & 7 &  & 37 & 0.998 & 0.035 & 0.001 & 0.049 & -- \\
 & 8 &  & 41 & 0.998 & 0.035 & 0.001 & 0.049 & -- \\
 & 9 & 2nd & 30 & 0.998 & 0.035 & 0.001 & 0.049 & 1.96 \\
 & 10 &  & 69 & 0.998 & 0.035 & 0.001 & 0.049 & -- \\
\addlinespace
SnOH3-me S (32q), A5-70q & 1 &  & 7403 & 0.000 & 1.000 & 1.000 & 1.414 & -- \\
 & 2 &  & 6470 & 0.994 & 0.053 & 0.003 & 0.075 & -- \\
 & 3 &  & 83 & 0.998 & 0.035 & 0.001 & 0.049 & -- \\
 & 4 &  & 230 & 0.000 & 0.999 & 1.000 & 1.414 & -- \\
 & 5 &  & 224 & 0.000 & 1.000 & 1.000 & 1.414 & -- \\
 & 6 & 2nd & 64 & 0.993 & 0.059 & 0.004 & 0.084 & 1.72 \\
 & 7 &  & 2587 & 0.998 & 0.035 & 0.001 & 0.049 & -- \\
 & 8 & best & 69 & 0.954 & 0.152 & 0.024 & 0.216 & 1.58 \\
 & 9 &  & 272 & 0.000 & 1.000 & 1.000 & 1.414 & -- \\
 & 10 &  & 59 & 0.978 & 0.104 & 0.011 & 0.148 & -- \\
\addlinespace
SnOH3-me-h2o S (38q), A1-46q & 1 &  & 8003 & 0.000 & 1.000 & 1.000 & 1.414 & -- \\
 & 2 & best & 35 & 1.000 & 0.015 & 0.000 & 0.022 & -- \\
 & 3 &  & 455 & 0.000 & 1.000 & 1.000 & 1.414 & -- \\
 & 4 & 2nd & 52 & 0.999 & 0.022 & 0.001 & 0.031 & -- \\
 & 5 &  & 41 & 0.999 & 0.022 & 0.001 & 0.031 & -- \\
 & 6 &  & 251 & 0.000 & 1.000 & 1.000 & 1.414 & -- \\
 & 7 &  & 7236 & 0.000 & 1.000 & 1.000 & 1.414 & -- \\
 & 8 &  & 30 & 1.000 & 0.015 & 0.000 & 0.022 & -- \\
 & 9 &  & 234 & 0.000 & 1.000 & 1.000 & 1.414 & -- \\
 & 10 &  & 4016 & 1.000 & 0.006 & 0.000 & 0.008 & -- \\
\addlinespace
SnOH3-me-h2o S (38q), A1-70q & 1 &  & 5630 & 0.000 & 1.000 & 1.000 & 1.414 & -- \\
 & 2 & best & 57 & 1.000 & 0.015 & 0.000 & 0.022 & -- \\
 & 3 &  & 485 & 0.000 & 1.000 & 1.000 & 1.414 & -- \\
 & 4 & 2nd & 51 & 0.999 & 0.022 & 0.001 & 0.031 & -- \\
 & 5 &  & 72 & 0.991 & 0.069 & 0.005 & 0.098 & -- \\
 & 6 &  & 5581 & 0.000 & 1.000 & 1.000 & 1.414 & -- \\
 & 7 &  & 2835 & 0.000 & 1.000 & 1.000 & 1.414 & -- \\
 & 8 &  & 2805 & 1.000 & 0.006 & 0.000 & 0.008 & -- \\
 & 9 &  & 473 & 0.000 & 1.000 & 1.000 & 1.414 & -- \\
 & 10 &  & 5322 & 0.000 & 1.000 & 1.000 & 1.414 & -- \\
\addlinespace
SnOH3-me-h2o S (38q), A5-46q & 1 &  & 11431 & 0.995 & 0.052 & 0.003 & 0.074 & -- \\
 & 2 &  & 3284 & 1.000 & 0.015 & 0.000 & 0.022 & -- \\
 & 3 &  & 3113 & 0.979 & 0.103 & 0.011 & 0.145 & -- \\
 & 4 & best & 34 & 1.000 & 0.015 & 0.000 & 0.022 & -- \\
 & 5 &  & 3389 & 1.000 & 0.015 & 0.000 & 0.022 & -- \\
 & 6 &  & 3808 & 1.000 & 0.015 & 0.000 & 0.022 & -- \\
 & 7 &  & 8494 & 0.000 & 1.000 & 1.000 & 1.414 & -- \\
 & 8 &  & 11595 & 1.000 & 0.015 & 0.000 & 0.022 & -- \\
 & 9 &  & 4075 & 0.000 & 1.000 & 1.000 & 1.414 & -- \\
 & 10 & 2nd & 3867 & 1.000 & 0.006 & 0.000 & 0.008 & -- \\
\addlinespace
SnOH3-me-h2o S (38q), A5-70q & 1 &  & 7577 & 1.000 & 0.015 & 0.000 & 0.022 & -- \\
 & 2 &  & 145 & 1.000 & 0.015 & 0.000 & 0.022 & -- \\
 & 3 &  & 168 & 1.000 & 0.015 & 0.000 & 0.022 & -- \\
 & 4 &  & 130 & 1.000 & 0.015 & 0.000 & 0.022 & -- \\
 & 5 &  & 7292 & 1.000 & 0.015 & 0.000 & 0.022 & -- \\
 & 6 &  & 6542 & 1.000 & 0.015 & 0.000 & 0.022 & -- \\
 & 7 &  & 4397 & 0.000 & 1.000 & 1.000 & 1.414 & -- \\
 & 8 & best & 110 & 1.000 & 0.015 & 0.000 & 0.022 & -- \\
 & 9 & 2nd & 161 & 1.000 & 0.015 & 0.000 & 0.022 & -- \\
 & 10 &  & 3802 & 0.994 & 0.052 & 0.003 & 0.074 & -- \\
\addlinespace
SnOH3-me-2h2o S (44q), A1-46q & 1 &  & 273 & 0.000 & 1.000 & 1.000 & 1.414 & -- \\
 & 2 & 2nd & 59 & 0.969 & 0.126 & 0.016 & 0.178 & -- \\
 & 3 &  & 540 & 0.000 & 1.000 & 1.000 & 1.414 & -- \\
 & 4 &  & 4704 & 1.000 & 0.013 & 0.000 & 0.018 & -- \\
 & 5 & best & 30 & 0.982 & 0.095 & 0.009 & 0.134 & -- \\
 & 6 &  & 285 & 0.000 & 1.000 & 1.000 & 1.414 & -- \\
 & 7 &  & 6505 & 0.000 & 1.000 & 1.000 & 1.414 & -- \\
 & 8 &  & 5009 & 1.000 & 0.013 & 0.000 & 0.018 & -- \\
 & 9 &  & 284 & 0.000 & 1.000 & 1.000 & 1.414 & -- \\
 & 10 &  & 4913 & 1.000 & 0.005 & 0.000 & 0.007 & -- \\
\addlinespace
SnOH3-me-2h2o S (44q), A1-70q & 1 &  & 6251 & 0.000 & 1.000 & 1.000 & 1.414 & -- \\
 & 2 & 2nd & 59 & 0.998 & 0.034 & 0.001 & 0.048 & -- \\
 & 3 &  & 569 & 0.000 & 1.000 & 1.000 & 1.414 & -- \\
 & 4 &  & 56 & 1.000 & 0.013 & 0.000 & 0.018 & -- \\
 & 5 & best & 60 & 0.995 & 0.048 & 0.002 & 0.069 & -- \\
 & 6 &  & 7112 & 0.000 & 1.000 & 1.000 &  1.414 & -- \\
 & 7 &  & 4460 & 0.000 & 1.000 & 1.000 & 1.414 & -- \\
 & 8 &  & 9153 & 1.000 & 0.005 & 0.000 & 0.016 & -- \\
 & 9 &  & 6979 & 0.000 & 1.000 & 1.000 & 1.414 & -- \\
 & 10 &  & 6506 & 0.000 & 1.000 & 1.000 & 1.414 & -- \\
\addlinespace
SnOH3-me-2h2o S (44q), A5-46q & 1 &  & 44 & 0.999 & 0.020 & 0.000 & 0.028 & -- \\
 & 2 &  & 11122 & 0.000 & 1.000 & 1.000 & 1.414 & -- \\
 & 3 & best & 49 & 1.000 & 0.013 & 0.000 & 0.018 & -- \\
 & 4 &  & 37 & 1.000 & 0.013 & 0.000 & 0.018 & -- \\
 & 5 & 2nd & 62 & 1.000 & 0.013 & 0.000 & 0.018 & -- \\
 & 6 &  & 5757 & 0.000 & 1.000 & 1.000 & 1.414 & -- \\
 & 7 &  & 32 & 1.000 & 0.013 & 0.000 & 0.018 & -- \\
 & 8 &  & 5062 & 1.000 & 0.013 & 0.000 & 0.018 & -- \\
 & 9 &  & 4180 & 1.000 & 0.005 & 0.000 & 0.007 & -- \\
 & 10 &  & 4886 & 1.000 & 0.013 & 0.000 & 0.018 & -- \\
\addlinespace
SnOH3-me-2h2o S (44q), A5-70q & 1 & 2nd & 142 & 0.992 & 0.064 & 0.004 & 0.090 & -- \\
 & 2 &  & 10847 & 0.000 & 1.000 & 1.000 & 1.414 & -- \\
 & 3 & best & 168 & 0.992 & 0.064 & 0.004 & 0.090 & -- \\
 & 4 &  & 410 & 0.000 & 1.000 & 1.000 & 1.414 & -- \\
 & 5 &  & 5564 & 0.000 & 1.000 & 1.000 & 1.414 & -- \\
 & 6 &  & 6509 & 0.000 & 1.000 & 1.000 & 1.414 & -- \\
 & 7 &  & 4229 & 1.000 & 0.013 & 0.000 & 0.018 & -- \\
 & 8 &  & 4403 & 0.995 & 0.048 & 0.002 & 0.069 & -- \\
 & 9 &  & 5665 & 0.000 & 1.000 & 1.000 & 1.414 & -- \\
 & 10 &  & 4271 & 0.999 & 0.020 & 0.000 & 0.028 & -- \\
\addlinespace
\bottomrule
\end{longtable}

\begin{longtable}{@{}lllrrrrrr@{}}
\caption{Per-circuit ideal divergence from the LUCJ circuit, tuned reward set (Section~\ref{sec:divergence}). $F$, $H$, JS (bits) and $d_{\mathrm{HF}}$ as in Table~\ref{tab:divergence-summary}. ``op.\ norm*'' is a lower-bound estimate of $\|U_g-U_{\mathrm{LUCJ}}\|$ (maximum 2) computed without phase minimisation, and only for the best and second circuit of SnOH3-me S; 2.00 indicates a non-converged value (Lanczos stopped at the step cap). $F=0.000$ means $F<10^{-5}$, and $H$ or JS values above 0.998 are rounded to 1.000. Depth is as-submitted. ``sel.'': best or second circuit.}\label{tab:divergence-circuit_analysis_tuned}\\
\toprule
Run & \# & sel. & depth & $F$ & $H$ & JS & $d_{\mathrm{HF}}$ & op.\ norm* \\
\midrule
\endfirsthead
\toprule
Run & \# & sel. & depth & $F$ & $H$ & JS & $d_{\mathrm{HF}}$ & op.\ norm* \\
\midrule
\endhead
SnOH3-me S (32q), A1-46q & 1 &  & 5561 & 0.000 & 0.999 & 1.000 & 1.414 & -- \\
 & 2 &  & 2789 & 0.954 & 0.152 & 0.024 & 0.216 & -- \\
 & 3 &  & 3873 & 0.000 & 1.000 & 1.000 & 1.414 & -- \\
 & 4 & best & 27 & 0.998 & 0.035 & 0.001 & 0.049 & 1.36 \\
 & 5 &  & 219 & 0.000 & 0.999 & 1.000 & 1.414 & -- \\
 & 6 & 2nd & 4954 & 0.000 & 1.000 & 1.000 & 1.414 & 2.00 \\
 & 7 &  & 7665 & 0.000 & 0.999 & 1.000 & 1.413 & -- \\
 & 8 &  & 2564 & 0.999 & 0.017 & 0.000 & 0.024 & -- \\
 & 9 &  & 196 & 0.000 & 1.000 & 1.000 & 1.414 & -- \\
 & 10 &  & 6806 & 0.999 & 0.017 & 0.000 & 0.024 & -- \\
\addlinespace
SnOH3-me S (32q), A1-70q & 1 &  & 6932 & 0.000 & 1.000 & 1.000 & 1.414 & -- \\
 & 2 & 2nd & 63 & 0.921 & 0.201 & 0.042 & 0.284 & 1.61 \\
 & 3 &  & 8003 & 0.000 & 1.000 &  1.000 & 1.414 & -- \\
 & 4 & best & 34 & 0.954 & 0.152 & 0.024 & 0.216 & 1.79 \\
 & 5 &  & 237 & 0.000 & 1.000 & 1.000 & 1.414 & -- \\
 & 6 &  & 5057 & 0.000 & 1.000 & 1.000 & 1.414 & -- \\
 & 7 &  & 7134 & 0.000 & 1.000 & 1.000 & 1.414 & -- \\
 & 8 &  & 2283 & 0.000 & 0.999 & 1.000 & 1.414 & -- \\
 & 9 &  & 7522 & 0.000 & 0.999 & 0.999 & 1.413 & -- \\
 & 10 &  & 2877 & 0.000 & 0.999 & 1.000 & 1.413 & -- \\
\addlinespace
SnOH3-me S (32q), A5-46q & 1 &  & 763 & 0.000 & 1.000 & 1.000 & 1.414 & -- \\
 & 2 & 2nd & 5551 & 0.000 & 1.000 & 1.000 & 1.414 & 2.00 \\
 & 3 &  & 7077 & 0.000 & 1.000 & 1.000 & 1.414 & -- \\
 & 4 &  & 411 & 0.000 & 1.000 & 1.000 & 1.414 & -- \\
 & 5 &  & 418 & 0.000 & 1.000 & 1.000 & 1.414 & -- \\
 & 6 &  & 2888 & 0.000 & 0.999 & 1.000 & 1.414 & -- \\
 & 7 &  & 2713 & 0.993 & 0.059 & 0.004 & 0.084 & -- \\
 & 8 &  & 2757 & 0.000 & 1.000 & 1.000 & 1.414 & -- \\
 & 9 &  & 235 & 0.000 & 1.000 & 1.000 & 1.414 & -- \\
 & 10 & best & 2770 & 0.998 & 0.035 & 0.001 & 0.049 & 2.00 \\
\addlinespace
SnOH3-me S (32q), A5-70q & 1 &  & 15128 & 0.000 & 1.000 & 1.000 & 1.414 & -- \\
 & 2 &  & 6375 & 0.776 & 0.345 & 0.124 & 0.488 & -- \\
 & 3 &  & 3124 & 0.000 & 1.000 & 1.000 & 1.414 & -- \\
 & 4 & best & 92 & 0.776 & 0.345 & 0.124 & 0.488 & 1.99 \\
 & 5 &  & 6838 & 0.000 & 1.000 & 1.000 & 1.414 & -- \\
 & 6 &  & 2771 & 0.993 & 0.059 & 0.004 & 0.084 & -- \\
 & 7 &  & 3516 & 0.000 & 1.000 & 1.000 & 1.414 & -- \\
 & 8 &  & 3413 & 0.000 & 1.000 & 1.000 & 1.414 & -- \\
 & 9 &  & 3214 & 0.000 & 0.999 & 1.000 & 1.414 & -- \\
 & 10 & 2nd & 3041 & 0.000 & 1.000 & 1.000 & 1.414 & 2.00 \\
\addlinespace
SnOH3-me-h2o S (38q), A1-46q & 1 &  & 3627 & 0.000 & 1.000 & 1.000 & 1.414 & -- \\
 & 2 & best & 40 & 1.000 & 0.015 & 0.000 & 0.022 & -- \\
 & 3 &  & 453 & 0.000 & 1.000 & 1.000 & 1.414 & -- \\
 & 4 & 2nd & 31 & 0.999 & 0.022 & 0.001 & 0.031 & -- \\
 & 5 &  & 253 & 0.000 & 1.000 & 1.000 & 1.414 & -- \\
 & 6 &  & 470 & 0.000 & 1.000 & 1.000 & 1.414 & -- \\
 & 7 &  & 8879 & 0.000 & 1.000 & 1.000 & 1.414 & -- \\
 & 8 &  & 3381 & 1.000 & 0.006 & 0.000 & 0.008 & -- \\
 & 9 &  & 233 & 0.000 & 1.000 & 1.000 & 1.414 & -- \\
 & 10 &  & 2933 & 1.000 & 0.006 & 0.000 & 0.008 & -- \\
\addlinespace
SnOH3-me-h2o S (38q), A1-70q & 1 &  & 5592 & 0.000 & 1.000 & 1.000 & 1.414 & -- \\
 & 2 & 2nd & 51 & 0.999 & 0.022 & 0.001 & 0.031 & -- \\
 & 3 &  & 4446 & 0.000 & 1.000 & 1.000 & 1.414 & -- \\
 & 4 & best & 68 & 0.999 & 0.022 & 0.001 & 0.031 & -- \\
 & 5 &  & 292 & 0.000 & 1.000 & 1.000 & 1.414 & -- \\
 & 6 &  & 6360 & 0.000 & 1.000 & 1.000 & 1.414 & -- \\
 & 7 &  & 8075 & 0.000 & 1.000 & 1.000 & 1.414 & -- \\
 & 8 &  & 9337 & 0.000 & 1.000 & 1.000 & 1.414 & -- \\
 & 9 &  & 5594 & 0.000 & 1.000 & 1.000 & 1.414 & -- \\
 & 10 &  & 3326 & 0.000 & 1.000 & 1.000 & 1.414 & -- \\
\addlinespace
SnOH3-me-h2o S (38q), A5-46q & 1 &  & 7517 & 0.000 & 1.000 & 1.000 & 1.414 & -- \\
 & 2 &  & 35 & 1.000 & 0.015 & 0.000 & 0.022 & -- \\
 & 3 &  & 3428 & 1.000 & 0.015 & 0.000 & 0.022 & -- \\
 & 4 &  & 32 & 1.000 & 0.015 & 0.000 & 0.022 & -- \\
 & 5 &  & 4205 & 0.000 & 1.000 & 1.000 & 1.414 & -- \\
 & 6 &  & 267 & 0.000 & 1.000 & 1.000 & 1.414 & -- \\
 & 7 &  & 670 & 0.000 & 1.000 & 1.000 & 1.414 & -- \\
 & 8 & 2nd & 53 & 1.000 & 0.015 & 0.000 & 0.022 & -- \\
 & 9 &  & 253 & 0.000 & 1.000 & 1.000 & 1.414 & -- \\
 & 10 & best & 64 & 1.000 & 0.015 & 0.000 & 0.022 & -- \\
\addlinespace
SnOH3-me-h2o S (38q), A5-70q & 1 &  & 7960 & 1.000 & 0.015 & 0.000 & 0.022 & -- \\
 & 2 &  & 3786 & 0.995 & 0.052 & 0.003 & 0.074 & -- \\
 & 3 &  & 8351 & 0.000 & 1.000 & 1.000 & 1.414 & -- \\
 & 4 & best & 45 & 1.000 & 0.015 & 0.000 & 0.022 & -- \\
 & 5 &  & 4040 & 0.994 & 0.052 & 0.003 & 0.074 & -- \\
 & 6 &  & 3726 & 0.994 & 0.052 & 0.003 & 0.074 & -- \\
 & 7 &  & 492 & 0.000 & 1.000 & 1.000 & 1.414 & -- \\
 & 8 &  & 7283 & 0.995 & 0.052 & 0.003 & 0.074 & -- \\
 & 9 & 2nd & 77 & 0.979 & 0.103 & 0.011 & 0.145 & -- \\
 & 10 &  & 3189 & 0.994 & 0.052 & 0.003 & 0.074 & -- \\
\addlinespace
SnOH3-me-2h2o S (44q), A1-46q & 1 &  & 283 & 0.000 & 1.000 & 1.000 & 1.414 & -- \\
 & 2 & best & 59 & 0.987 & 0.079 & 0.006 & 0.112 & -- \\
 & 3 &  & 530 & 0.000 & 1.000 & 1.000 & 1.414 & -- \\
 & 4 & 2nd & 30 & 0.999 & 0.020 & 0.000 & 0.028 & -- \\
 & 5 &  & 279 & 0.000 & 1.000 & 1.000 & 1.414 & -- \\
 & 6 &  & 527 & 0.000 & 1.000 & 1.000 & 1.414 & -- \\
 & 7 &  & 5410 & 0.000 & 1.000 & 1.000 & 1.414 & -- \\
 & 8 &  & 11178 & 1.000 & 0.005 & 0.000 & 0.016 & -- \\
 & 9 &  & 277 & 0.000 & 1.000 & 1.000 & 1.414 & -- \\
 & 10 &  & 4300 & 1.000 & 0.005 & 0.000 & 0.007 & -- \\
\addlinespace
SnOH3-me-2h2o S (44q), A1-70q & 1 &  & 6947 & 0.000 & 1.000 & 1.000 & 1.414 & -- \\
 & 2 &  & 5317 & 0.988 & 0.079 & 0.006 & 0.112 & -- \\
 & 3 &  & 6533 & 0.000 & 1.000 & 1.000 & 1.414 & -- \\
 & 4 &  & 4338 & 1.000 & 0.013 & 0.000 & 0.018 & -- \\
 & 5 & best & 298 & 0.000 & 1.000 & 1.000 & 1.414 & -- \\
 & 6 & 2nd & 806 & 0.000 & 1.000 & 1.000 & 1.414 & -- \\
 & 7 &  & 5296 & 0.000 & 1.000 & 1.000 & 1.414 & -- \\
 & 8 &  & 9736 & 0.000 & 1.000 & 1.000 & 1.414 & -- \\
 & 9 &  & 6734 & 0.000 & 1.000 & 1.000 & 1.414 & -- \\
 & 10 &  & 6418 & 0.000 & 1.000 & 1.000 & 1.414 & -- \\
\addlinespace
SnOH3-me-2h2o S (44q), A5-46q & 1 &  & 4696 & 0.995 & 0.048 & 0.002 & 0.069 & -- \\
 & 2 &  & 11682 & 0.000 & 1.000 & 1.000 & 1.414 & -- \\
 & 3 & 2nd & 45 & 1.000 & 0.013 & 0.000 & 0.018 & -- \\
 & 4 &  & 66 & 1.000 & 0.013 & 0.000 & 0.018 & -- \\
 & 5 & best & 33 & 1.000 & 0.013 & 0.000 & 0.018 & -- \\
 & 6 &  & 11243 & 0.000 & 1.000 & 1.000 & 1.414 & -- \\
 & 7 &  & 4257 & 1.000 & 0.013 & 0.000 & 0.018 & -- \\
 & 8 &  & 4855 & 1.000 & 0.013 & 0.000 & 0.018 & -- \\
 & 9 &  & 6320 & 0.995 & 0.047 & 0.002 & 0.067 & -- \\
 & 10 &  & 288 & 0.000 & 1.000 & 1.000 & 1.414 & -- \\
\addlinespace
SnOH3-me-2h2o S (44q), A5-70q & 1 &  & 570 & 0.000 & 1.000 & 1.000 & 1.414 & -- \\
 & 2 &  & 9938 & 0.000 & 1.000 & 1.000 & 1.414 & -- \\
 & 3 &  & 29113 & 0.000 & 1.000 & 1.000 & 1.414 & -- \\
 & 4 &  & 294 & 0.000 & 1.000 & 1.000 & 1.414 & -- \\
 & 5 & best & 66 & 0.999 & 0.020 & 0.000 & 0.028 & -- \\
 & 6 &  & 329 & 0.000 & 1.000 & 1.000 & 1.414 & -- \\
 & 7 &  & 21016 & 0.996 & 0.047 & 0.002 & 0.067 & -- \\
 & 8 & 2nd & 58 & 0.995 & 0.048 & 0.002 & 0.069 & -- \\
 & 9 &  & 11617 & 0.000 & 1.000 & 1.000 & 1.414 & -- \\
 & 10 &  & 5285 & 1.000 & 0.013 & 0.000 & 0.018 & -- \\
\addlinespace
\bottomrule
\end{longtable}


\section{Zero-shot inference: full detail}
\label{sec:appendix-inference-detail}

\begin{figure}[p]
\centering
\includegraphics[width=0.86\textwidth]{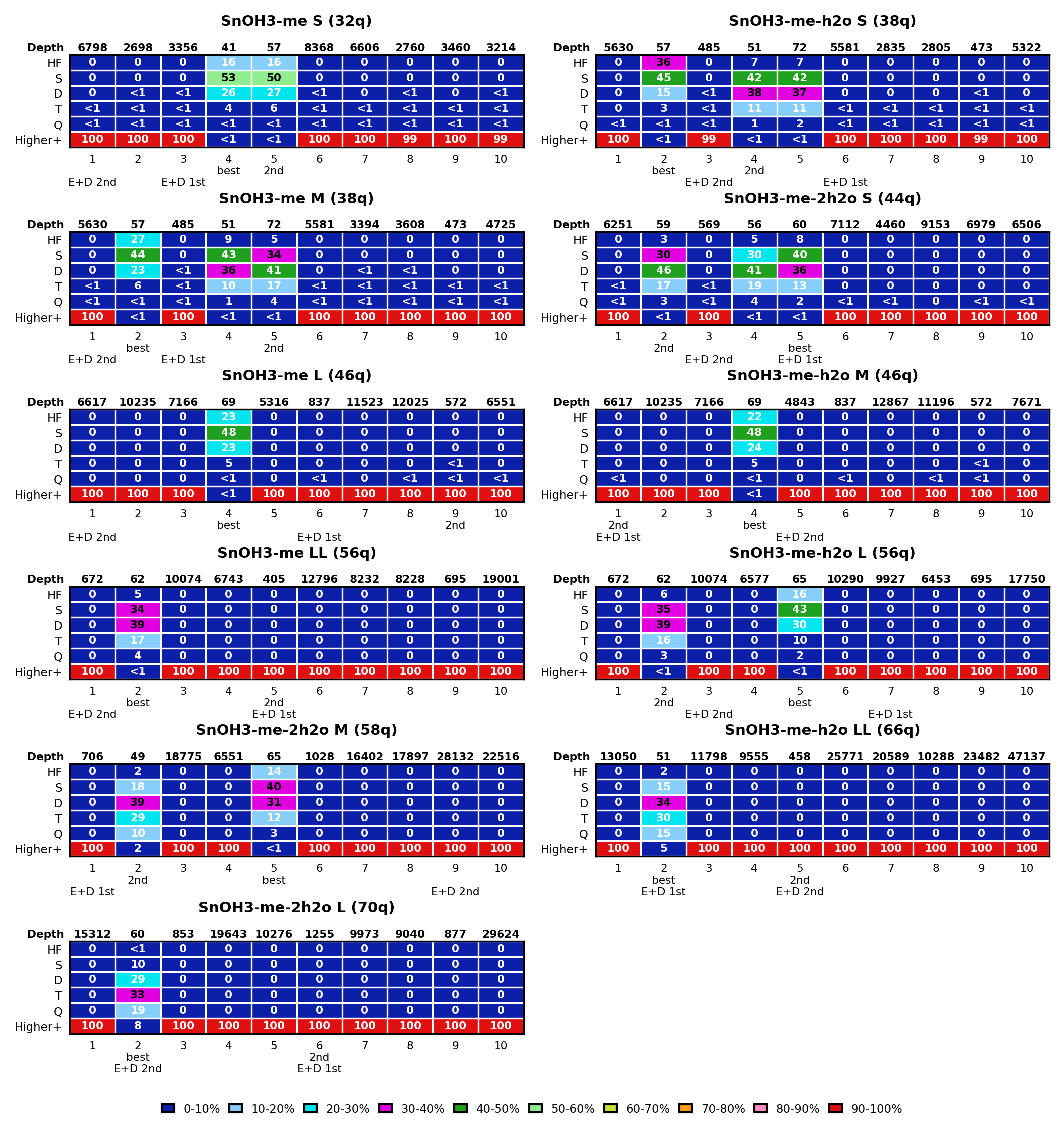}
\caption{As Fig.~\ref{fig:collage-a1-46q} for the A1 policy trained at 70 qubits. The best circuit is the shallowest of the ten in six of the eleven panels and among the three shallowest in all of them, and it has at most 7.9\% of its weight in ``higher'' (at most 0.6\% in nine of the panels). The second circuit has little weight in ``higher'' in six panels; in SnOH3-me L, SnOH3-me-h2o M, SnOH3-me LL, SnOH3-me-h2o LL and SnOH3-me-2h2o L it has essentially all its weight there. The other circuits are one to three orders of magnitude deeper with essentially all weight in ``higher''.}
\label{fig:collage-a1-70q}
\end{figure}

\begin{table}[ht]
\centering
\caption{Which (target, policy) runs have fresh-shot results. ``done'': sampled, ranked and recovered; ``not run'': not part of the design (untrained controls were run for a subset of targets); ``excluded$^{\dagger}$'': the batch of ten circuits had a single circuit with valid shots and a second draw gave a recovered error of about 200~mHa, so the run is not part of the statistics (Appendix~\ref{sec:appendix-repeat}). Columns are the lineage (A1/A5, trained at 46 or 70 qubits).}
\label{tab:run-matrix}
\scriptsize
\resizebox{\linewidth}{!}{%
\begin{tabular}{@{}l*{8}{c}@{}}
\toprule
 & \multicolumn{4}{c}{Trained} & \multicolumn{4}{c}{Untrained control} \\
Target & A1-46q & A1-70q & A5-46q & A5-70q & A1-46q & A1-70q & A5-46q & A5-70q \\
\midrule
SnOH3-me S (32q) & done & done & done & done & not run & not run & not run & not run \\
SnOH3-me-h2o S (38q) & done & done & done & done & not run & not run & not run & not run \\
SnOH3-me M (38q) & done & done & done & done & not run & not run & not run & not run \\
SnOH3-me-2h2o S (44q) & done & done & done & done & not run & not run & not run & not run \\
SnOH3-me L (46q) & done & done & done & done & done & done & done & done \\
SnOH3-me-h2o M (46q) & done & done & done & done & done & done & not run & done \\
SnOH3-me LL (56q) & done & done & excluded$^{\dagger}$ & done & not run & done & not run & not run \\
SnOH3-me-h2o L (56q) & done & done & done & done & not run & not run & not run & not run \\
SnOH3-me-2h2o M (58q) & done & done & done & done & not run & not run & not run & not run \\
SnOH3-me-h2o LL (66q) & done & done & done & done & not run & not run & not run & not run \\
SnOH3-me-2h2o L (70q) & done & done & done & done & not run & not run & not run & not run \\
\bottomrule
\end{tabular}}
\end{table}

\begin{table}[ht]
\centering
\caption{Zero-shot inference on the SnOH3-me L tier (46 qubits), per lineage and policy. Depth and 2q are the as-submitted (live) cost of the best/second circuit; ``valid shots'' is the number of the 100{,}000 shots with the correct electron numbers, i.e.\ the data behind each pick. ``SI'' (singles-injected) adds the single-excitation configurations to the best circuit's recovered determinant set and re-diagonalizes. DMRG reference $-479.314216$~E$_h$. Ranking, selection, shot counts and recovery settings are as in Table~\ref{tab:zeroshot-master}, except for the E+D picks of the untrained controls: these eight runs were sampled in the first campaign, and their E+D circuits were chosen with the energy rank of the earlier ranking, which used a cap of 2{,}000 determinants instead of 40{,}000, added to the rank by distinct bitstrings. Their best, second and merged results use the 40{,}000-determinant ranking.}
\label{tab:zeroshot-detail}
\small
\resizebox{\linewidth}{!}{%
\begin{tabular}{@{}lrrrrrr@{}}
\toprule
Lineage/Policy & Depth (best/2nd) & 2q (best/2nd) & Valid shots (best/2nd) & $\Delta E$ best/2nd/merged (mHa) & $\Delta E$ E+D 1/2 (mHa) & $\Delta E$ best-SI (mHa) \\
\midrule
\multicolumn{7}{l}{\textbf{Trained policies}} \\
A1-46q trained & 37/4624 & 64/9866 & 11097/3 & 14.3/192.7/55.4 & 192.7/198.1 & 14.3 \\
A1-70q trained & 69/572 & 100/2104 & 9034/355 & 7.8/197.9/41.7 & 193.1/198.6 & 7.8 \\
A5-46q trained & 55/61 & 80/88 & 2445/5055 & 12.5/7.6/29.8 & 199.6/12.5 & 12.5 \\
A5-70q trained & 144/142 & 124/120 & 4324/15345 & 5.6/16.9/29.6 & 199.1/5.6 & 5.6 \\
\midrule
\multicolumn{7}{l}{\textbf{Untrained controls}} \\
A1-46q untrained & 14298/334 & 25505/1120 & 21/1859 & 200.2/196.2/188.7 & 196.2/200.0 & 39.7 \\
A1-70q untrained & 340/7267 & 1168/13540 & 605/62 & 193.7/194.7/192.2 & 196.9/199.9 & 38.2 \\
A5-46q untrained & 12650/24025 & 24649/44951 & 48/3 & 192.9/193.2/194.3 & 199.2/185.6 & 38.8 \\
A5-70q untrained & 10282/24119 & 17269/45351 & 8/26 & 188.2/200.9/198.9 & 198.1/200.9 & 39.2 \\
\bottomrule
\end{tabular}}
\end{table}

\begin{table}[ht]
\centering
\caption{Measured breakdown of the first-campaign sample-job elapsed time for the A1-70q jobs (not repeated in
Table~\ref{tab:zeroshot-detail} for the fresh-shot campaign), queried directly from the IBM
Quantum Runtime job metadata (the job-metrics interface) via saved
job IDs, not estimated from local logs. Both batch jobs started at the moment of submission (confirmed from the scheduler accounting records), so none
of the batch elapsed time is scheduler queue wait; the principal breakdown is between
IBM's device queue, device execution, and local overhead (Python
startup, pipeline construction, the circuit-depth transpilation
transpile loop over all 10 sampled circuits). ``Billed QPU'' is the quantum time IBM billed for the job (561 and 624~s here); ``per circuit'' is this billed time divided by the 10 circuits in the batched Runtime job (56.1 and 62.4~s). The IBM-reported device-execution time is not the billed time.}
\label{tab:zeroshot-ibm-timing}
\small
\begin{tabular}{@{}lrrrrrr@{}}
\toprule
 & sbatch total & IBM queue & IBM device exec. & IBM total & Local overhead & Billed QPU / per circuit \\
\midrule
A1-70q trained   & 12m18s & 5.3s   & 10m04s & 10m09s & \phantom{0}2m09s & 561~s / 56.1~s \\
A1-70q untrained & 39m46s & 4m40s  & 32m33s & 37m13s & \phantom{0}2m34s & 624~s / 62.4~s \\
\bottomrule
\end{tabular}
\end{table}

{\footnotesize
{\footnotesize
{\footnotesize
{\footnotesize
{\footnotesize
{\footnotesize
{\footnotesize
{\footnotesize
{\footnotesize
{\footnotesize
{\footnotesize
{\footnotesize
{\footnotesize
{\footnotesize
{\footnotesize
{\footnotesize
{\footnotesize
{\footnotesize
{\footnotesize
{\footnotesize
{\footnotesize
{\footnotesize
{\footnotesize
{\footnotesize
{\footnotesize
{\footnotesize
{\footnotesize
{\footnotesize
{\footnotesize
{\footnotesize
{\footnotesize
{\footnotesize
{\footnotesize
{\footnotesize
{\footnotesize
\begin{longtable}{@{}llrrrrrrr@{}}
\caption{Recovered energy and excitation-level composition of the recovered determinant set for the SnOH3-me L tier, per lineage/policy and circuit type. \textbf{Caution:} S/D/T/Q/higher are counts of determinant \emph{pairs} $(a_i,b_j)$ in the Cartesian product of the recovered string list with itself (the same list serves the $\alpha$ and the $\beta$ sector of these closed-shell systems), not counts out of the 100{,}000 shots; the per-shot picture is Table~\ref{tab:raw-shot-excitation}. With $n$ recovered strings the product has $n^2$ pairs, and the five columns sum to $n^2{-}1$ in every row (the Hartree--Fock pair is not listed); $n=20{,}000$, i.e. $4\times10^8$ pairs, in the rows that reached the cap. ``Higher'' absorbs every order above quadruple. Ranking, selection, shot counts and recovery settings are as in Table~\ref{tab:zeroshot-master}.}\label{tab:zeroshot-excitation}\\
\toprule
Target/Lineage & Type & Energy (E$_h$) & $\Delta E$ (mHa) & S & D & T & Q & Higher \\
\midrule
\endfirsthead
\toprule
Target/Lineage & Type & Energy (E$_h$) & $\Delta E$ (mHa) & S & D & T & Q & Higher \\
\midrule
\endhead
\bottomrule
\endfoot
A1-46q trained & best & $-$479.299926 & 14.3 & 224 & 16206 & 414092 & 0 & 18885502 \\
A1-46q trained & second & $-$479.121483 & 192.7 & 28 & 616 & 9224 & 0 & 399990131 \\
A1-46q trained & merged & $-$479.258786 & 55.4 & 202 & 11075 & 91470 & 0 & 399897252 \\
\addlinespace
A1-46q untrained & best & $-$479.114005 & 200.2 & 16 & 392 & 5434 & 0 & 399994157 \\
A1-46q untrained & second & $-$479.117972 & 196.2 & 16 & 488 & 6624 & 0 & 399992871 \\
A1-46q untrained & merged & $-$479.125474 & 188.7 & 26 & 661 & 9786 & 0 & 399989526 \\
\addlinespace
A1-70q trained & best & $-$479.306409 & 7.8 & 224 & 17064 & 515550 & 0 & 93984445 \\
A1-70q trained & second & $-$479.116356 & 197.9 & 24 & 624 & 9540 & 0 & 399989811 \\
A1-70q trained & merged & $-$479.272504 & 41.7 & 216 & 12692 & 113924 & 0 & 399873167 \\
\addlinespace
A1-70q untrained & best & $-$479.120472 & 193.7 & 30 & 751 & 11662 & 0 & 399987556 \\
A1-70q untrained & second & $-$479.119544 & 194.7 & 24 & 528 & 7576 & 0 & 399991871 \\
A1-70q untrained & merged & $-$479.122028 & 192.2 & 18 & 497 & 6830 & 0 & 399992654 \\
\addlinespace
A5-46q trained & best & $-$479.301748 & 12.5 & 224 & 16092 & 402204 & 0 & 36706128 \\
A5-46q trained & second & $-$479.306572 & 7.6 & 224 & 16448 & 444030 & 0 & 43615618 \\
A5-46q trained & merged & $-$479.284378 & 29.8 & 222 & 13875 & 174854 & 0 & 399811048 \\
\addlinespace
A5-46q untrained & best & $-$479.121282 & 192.9 & 14 & 473 & 6342 & 0 & 399993170 \\
A5-46q untrained & second & $-$479.120983 & 193.2 & 20 & 516 & 7320 & 0 & 399992143 \\
A5-46q untrained & merged & $-$479.119918 & 194.3 & 18 & 495 & 7024 & 0 & 399992462 \\
\addlinespace
A5-70q trained & best & $-$479.308601 & 5.6 & 224 & 17250 & 537532 & 0 & 97573829 \\
A5-70q trained & second & $-$479.297326 & 16.9 & 224 & 16526 & 452264 & 0 & 44945106 \\
A5-70q trained & merged & $-$479.284631 & 29.6 & 224 & 14468 & 218526 & 0 & 399766781 \\
\addlinespace
A5-70q untrained & best & $-$479.126065 & 188.2 & 34 & 739 & 11126 & 0 & 399988100 \\
A5-70q untrained & second & $-$479.113307 & 200.9 & 16 & 416 & 5772 & 0 & 399993795 \\
A5-70q untrained & merged & $-$479.115344 & 198.9 & 12 & 416 & 5374 & 0 & 399994197 \\
\end{longtable}}}}}}}}}}}}}}}}}}}}}}}}}}}}}}}}}}}}

Several patterns are visible from the eight L-tier lineages (Table~\ref{tab:zeroshot-detail}).

First, trained policies separate from untrained controls. Every untrained lineage's best, second, merged and
E+D result lies between 186 and 201~mHa, whereas each trained lineage has at least one individual-circuit result
at 5.6--14.3~mHa (A1-46q: 14.3 best; A5-46q: 12.5 best / 7.6 second; A1-70q: 7.8 best; A5-70q: 5.6 best).

Second, the 70q-trained lineages are as good as the 46q-trained ones on this target even though it is a cross-norb target for them: A1-70q and A5-70q reach 7.8 and 5.6~mHa, against 14.3 and 12.5~mHa for the matched-norb A1-46q and A5-46q. With one
molecule and one fresh sample per lineage we do not rank the lineages.

Third, a second circuit with few valid shots recovers poorly. The A1-46q and A1-70q second circuits have
3 and 355 valid shots and recover to 192.7 and 197.9~mHa, while the A5-46q second circuit (5055 valid shots) recovers to 7.6~mHa. A
finite ranking energy from a few valid shots is not a reliable ranking signal.

Fourth, merging pools weaker circuits with better ones: the merged result (29.6--55.4~mHa for the trained lineages) is
worse than the best individual circuit in every trained lineage. Finally, ranking-energy order is not recovered-energy
order: A5-46q trained's second circuit (7.6~mHa) beats its best (12.5~mHa) here.

\begin{table}[ht]
\centering
\caption{Qubit coverage of the fresh-shot sampled circuits, per run, over all ten circuits of each of the 43 runs (430 circuits). Each generated circuit was translated to \{rz, sx, cx\} without a coupling map and the qubits carrying at least one gate (or at least one two-qubit gate) were counted; ``idle'' means no gate at all after translation. 101 circuits leave at least one qubit idle. The earlier \code{active\_qubits} field stored with each circuit counts the qubits named by any instruction of the pre-transpile circuit, including composite gates that translate to nothing on some qubits; it equals the register width for every circuit and so does not detect these idle qubits. Targets of a different norb than the lineage's training norb are the cross-norb ablation of Section~\ref{sec:crossnorb}.}
\label{tab:coverage-idle}
\small
\resizebox{\linewidth}{!}{%
\begin{tabular}{@{}llrrrrr@{}}
\toprule
Target & Lineage/Policy & Target qubits & Qubits with a gate (min / max) & Qubits in a 2q gate (min / max) & Max idle qubits & Circuits with idle qubits \\
\midrule
SnOH3-me S (32q) & A1-46q trained & 32 & 30 / 32 & 20 / 32 & 2 & 2/10 \\
SnOH3-me S (32q) & A1-70q trained & 32 & 32 / 32 & 28 / 32 & 0 & 0/10 \\
SnOH3-me S (32q) & A5-46q trained & 32 & 28 / 32 & 22 / 32 & 4 & 4/10 \\
SnOH3-me S (32q) & A5-70q trained & 32 & 28 / 32 & 22 / 32 & 4 & 1/10 \\
\\addlinespace
SnOH3-me-h2o S (38q) & A1-46q trained & 38 & 34 / 38 & 26 / 38 & 4 & 4/10 \\
SnOH3-me-h2o S (38q) & A1-70q trained & 38 & 36 / 38 & 32 / 38 & 2 & 1/10 \\
SnOH3-me-h2o S (38q) & A5-46q trained & 38 & 30 / 38 & 20 / 38 & 8 & 1/10 \\
SnOH3-me-h2o S (38q) & A5-70q trained & 38 & 30 / 38 & 26 / 38 & 8 & 3/10 \\
\\addlinespace
SnOH3-me M (38q) & A1-46q trained & 38 & 32 / 38 & 26 / 38 & 6 & 4/10 \\
SnOH3-me M (38q) & A1-70q trained & 38 & 36 / 38 & 32 / 38 & 2 & 1/10 \\
SnOH3-me M (38q) & A5-46q trained & 38 & 38 / 38 & 30 / 38 & 0 & 0/10 \\
SnOH3-me M (38q) & A5-70q trained & 38 & 38 / 38 & 30 / 38 & 0 & 0/10 \\
\\addlinespace
SnOH3-me-2h2o S (44q) & A1-46q trained & 44 & 42 / 44 & 40 / 44 & 2 & 1/10 \\
SnOH3-me-2h2o S (44q) & A1-70q trained & 44 & 42 / 44 & 40 / 44 & 2 & 2/10 \\
SnOH3-me-2h2o S (44q) & A5-46q trained & 44 & 38 / 44 & 32 / 44 & 6 & 2/10 \\
SnOH3-me-2h2o S (44q) & A5-70q trained & 44 & 42 / 44 & 34 / 44 & 2 & 1/10 \\
\\addlinespace
SnOH3-me L (46q) & A1-46q trained & 46 & 42 / 46 & 34 / 46 & 4 & 1/10 \\
SnOH3-me L (46q) & A1-70q trained & 46 & 44 / 46 & 42 / 46 & 2 & 1/10 \\
SnOH3-me L (46q) & A5-46q trained & 46 & 38 / 46 & 34 / 46 & 8 & 3/10 \\
SnOH3-me L (46q) & A5-70q trained & 46 & 42 / 46 & 40 / 46 & 4 & 2/10 \\
\\addlinespace
SnOH3-me-h2o M (46q) & A1-46q trained & 46 & 42 / 46 & 34 / 46 & 4 & 1/10 \\
SnOH3-me-h2o M (46q) & A1-70q trained & 46 & 44 / 46 & 42 / 46 & 2 & 1/10 \\
SnOH3-me-h2o M (46q) & A5-46q trained & 46 & 40 / 46 & 30 / 46 & 6 & 7/10 \\
SnOH3-me-h2o M (46q) & A5-70q trained & 46 & 38 / 46 & 32 / 46 & 8 & 7/10 \\
\\addlinespace
SnOH3-me LL (56q) & A1-46q trained & 56 & 52 / 56 & 46 / 56 & 4 & 3/10 \\
SnOH3-me LL (56q) & A1-70q trained & 56 & 56 / 56 & 52 / 56 & 0 & 0/10 \\
SnOH3-me LL (56q) & A5-70q trained & 56 & 54 / 56 & 42 / 56 & 2 & 1/10 \\
\\addlinespace
SnOH3-me-h2o L (56q) & A1-46q trained & 56 & 52 / 56 & 46 / 56 & 4 & 3/10 \\
SnOH3-me-h2o L (56q) & A1-70q trained & 56 & 56 / 56 & 52 / 56 & 0 & 0/10 \\
SnOH3-me-h2o L (56q) & A5-46q trained & 56 & 44 / 56 & 38 / 56 & 12 & 4/10 \\
SnOH3-me-h2o L (56q) & A5-70q trained & 56 & 50 / 56 & 36 / 56 & 6 & 3/10 \\
\\addlinespace
SnOH3-me-2h2o M (58q) & A1-46q trained & 58 & 54 / 58 & 48 / 58 & 4 & 1/10 \\
SnOH3-me-2h2o M (58q) & A1-70q trained & 58 & 54 / 58 & 52 / 58 & 4 & 2/10 \\
SnOH3-me-2h2o M (58q) & A5-46q trained & 58 & 54 / 58 & 44 / 58 & 4 & 2/10 \\
SnOH3-me-2h2o M (58q) & A5-70q trained & 58 & 52 / 58 & 32 / 58 & 6 & 5/10 \\
\\addlinespace
SnOH3-me-h2o LL (66q) & A1-46q trained & 66 & 54 / 66 & 50 / 66 & 12 & 3/10 \\
SnOH3-me-h2o LL (66q) & A1-70q trained & 66 & 66 / 66 & 64 / 66 & 0 & 0/10 \\
SnOH3-me-h2o LL (66q) & A5-46q trained & 66 & 60 / 66 & 48 / 66 & 6 & 3/10 \\
SnOH3-me-h2o LL (66q) & A5-70q trained & 66 & 56 / 66 & 42 / 66 & 10 & 6/10 \\
\\addlinespace
SnOH3-me-2h2o L (70q) & A1-46q trained & 70 & 62 / 70 & 52 / 70 & 8 & 3/10 \\
SnOH3-me-2h2o L (70q) & A1-70q trained & 70 & 64 / 70 & 60 / 70 & 6 & 1/10 \\
SnOH3-me-2h2o L (70q) & A5-46q trained & 70 & 62 / 70 & 50 / 70 & 8 & 6/10 \\
SnOH3-me-2h2o L (70q) & A5-70q trained & 70 & 56 / 70 & 42 / 70 & 14 & 5/10 \\
\bottomrule
\end{tabular}}
\end{table}

Across the 43 HF-penalty runs and all 430 individual sampled circuits (ten per run, not only the best and second), 329 circuits carry at least one gate on every one of the target molecule's qubits; 101 circuits in 37 runs leave between 2 and 14 qubits with no gate after translation to \{rz, sx, cx\} (Table~\ref{tab:coverage-idle}), and 128 circuits have at least one qubit that is in no two-qubit gate. Idle-qubit circuits include many of the selected ones: 30 of the 43 best circuits and 21 of the 43 second circuits leave qubits idle, for example the best circuit of SnOH3-me-2h2o L with A5-70q (14 idle of 70). A coarser count stored with each circuit counts qubits named by any instruction of the pre-transpile circuit, including composite gates that translate to nothing on some of their qubits, so it does not detect these cases. Whether the idle qubits matter for accuracy was not tested: SnOH3-me-h2o M with A1-70q, whose best circuit has 2 idle qubits, gives the lowest error of that target (0.71~mHa). Operators that act on the full register (an orbital rotation, a diagonal-Coulomb layer, the LUCJ layer) link all qubits in one block, whereas circuits built only from local gates split into two spin blocks (Section~\ref{sec:architecture}).


Representative circuit diagrams are not shown, since no single-page rendering stays legible at these qubit counts and
depths. The per-circuit depth and two-qubit-gate counts are in Table~\ref{tab:zeroshot-detail}: the best circuits of the four trained L-tier lineages have 64--124 two-qubit gates (as-submitted depth 37--144), against depth 334--24{,}119 and 1{,}120--45{,}351 two-qubit gates for the untrained lineages' best and second circuits.

The gap between the trained and untrained A1-70q sample jobs' total
elapsed time (12m18s vs.\ 39m46s) is not Slurm queue wait or local
overhead -- both are a few minutes at most and comparable between the two
jobs (Table~\ref{tab:zeroshot-ibm-timing}) -- it is almost entirely time
spent on IBM's own side: the 27.5-minute gap lies about 22.5~min in device execution (32m33s vs.\ 10m04s) and about 4.6~min in the IBM device queue (4m40s vs.\ 5.3s). Despite the first-campaign untrained policy's circuits being 3.0--3.2$\times$ deeper/costlier (those sample jobs' depths,
not repeated in Table~\ref{tab:zeroshot-detail}), the billed QPU time differs by only 11\% (624 vs.\ 561~s; Table~\ref{tab:zeroshot-ibm-timing}). The metadata do not show the cause of the extra device-execution time.

\subsection{Cross-norb vocab-mismatch ablation}
\label{sec:crossnorb}

A separate campaign ran the 70q-trained A1/A5 checkpoints (and, for
one target, the 46q-trained A5 checkpoint) against three further unseen
molecules whose active-space orbital count (norb) differs from that
lineage's own training norb -- confirmed in direct code inspection
before running (Section~\ref{sec:lucj-comparison}) to be a
vocab-mismatch condition: the operator pool's gate vocabulary is built
fresh per molecule from that molecule's own norb, so a token index that
meant one specific gate during training can mean a different gate (or an
identity pad) at the target's different norb. The numbering depends on the electron count as well. Table~\ref{tab:token-map} gives the token ordering. In particular, the two Givens angles on neighbouring orbitals $(p,p{+}1)$ occupy tokens $6{+}2p$ and $6{+}2p{+}1$, so the HOMO--LUMO pair (the only neighbouring pair that changes the Hartree--Fock determinant) appears at different token ids in different molecules: token 28 for SnOH3-me S (16 orbitals, 24 electrons), 34 for the hydrate M tier (23 orbitals, 30 electrons), 36 for SnOH3-me L (23 orbitals, 32 electrons) and 52 for the dihydrate L tier (35 orbitals, 48 electrons). Whether the policies learned such a preference was not checked. Results below are reported
as an ablation of the vocabulary mismatch, not as clean zero-shot transfer.
Targets: SnOH3-me S (norb=16, 32 qubits, trained checkpoints only --
DMRG reference $-479.193382$~E$_h$), SnOH3-me-h2o M (norb=23, 46
qubits, trained and untrained -- DMRG reference $-555.125158$~E$_h$),
and SnOH3-me LL (norb=28, 56 qubits -- DMRG reference
$-479.315057$~E$_h$, Table~\ref{tab:dmrg-reference}).

\begin{table}[ht]
\centering
\caption{Cross-norb ablation and remaining matched-norb targets, per target, lineage and policy: SnOH3-me S (32q), LL (56q), SnOH3-me-h2o S/M/LL (38/46/66q) and SnOH3-me-2h2o S/M/L (44/58/70q), with DMRG references from Table~\ref{tab:dmrg-reference}. Same columns as Table~\ref{tab:zeroshot-detail}. Where valid shots are few (tens to hundreds of 100{,}000) the ranking energy and the pick rest on very little data. Wall times of the first campaign are not repeated here. Ranking, selection, shot counts and recovery settings are as in Table~\ref{tab:zeroshot-master}.}
\label{tab:crossnorb-detail}
\small
\resizebox{\linewidth}{!}{%
\begin{tabular}{@{}llrrrrrr@{}}
\toprule
Target & Lineage/Policy & Depth (best/2nd) & 2q (best/2nd) & Valid shots (best/2nd) & $\Delta E$ best/2nd/merged (mHa) & $\Delta E$ E+D 1/2 (mHa) & $\Delta E$ best-SI (mHa) \\
\midrule
\multicolumn{8}{l}{\textbf{Trained policies}} \\
SnOH3-me S (32q) & A1-46q trained & 50/30 & 56/36 & 2652/34212 & 0.12/0.71/0.01 & 0.01/0.12 & 0.1 \\
SnOH3-me S (32q) & A1-70q trained & 41/57 & 68/80 & 21300/4078 & 0.22/0.11/0.01 & 0.01/0.01 & 0.2 \\
SnOH3-me S (32q) & A5-46q trained & 31/30 & 36/36 & 40942/11383 & 0.65/0.30/0.01 & 0.30/0.01 & 0.7 \\
SnOH3-me S (32q) & A5-70q trained & 69/64 & 60/76 & 33072/14648 & 0.46/0.25/0.01 & 0.01/0.46 & 0.5 \\
\addlinespace
SnOH3-me-h2o S (38q) & A1-46q trained & 35/52 & 52/48 & 8991/13563 & 0.29/0.15/0.01 & 0.29/0.01 & 0.3 \\
SnOH3-me-h2o S (38q) & A1-70q trained & 57/51 & 92/80 & 14708/16736 & 0.07/0.08/0.01 & 0.01/0.01 & 0.1 \\
SnOH3-me-h2o S (38q) & A5-46q trained & 34/3867 & 40/5856 & 40442/3 & 0.12/0.01/0.01 & 0.12/0.01 & 0.1 \\
SnOH3-me-h2o S (38q) & A5-70q trained & 110/161 & 84/104 & 12917/23131 & 0.09/0.07/0.01 & 0.09/0.19 & 0.1 \\
\addlinespace
SnOH3-me M (38q) & A1-46q trained & 35/41 & 52/60 & 12049/3485 & 9.6/5.7/15.8 & 152.0/9.6 & 9.6 \\
SnOH3-me M (38q) & A1-70q trained & 57/72 & 92/100 & 14809/2736 & 3.8/6.6/19.7 & 144.4/139.1 & 3.8 \\
SnOH3-me M (38q) & A5-46q trained & 32/51 & 52/68 & 20660/25340 & 17.2/19.5/30.9 & 143.9/130.6 & 17.2 \\
SnOH3-me M (38q) & A5-70q trained & 152/401 & 104/788 & 6989/1005 & 13.6/149.8/36.4 & 131.8/149.1 & 13.6 \\
\addlinespace
SnOH3-me-2h2o S (44q) & A1-46q trained & 30/59 & 60/76 & 1544/768 & 0.21/0.07/0.04 & 0.04/0.21 & 0.2 \\
SnOH3-me-2h2o S (44q) & A1-70q trained & 60/59 & 100/108 & 1898/1854 & 0.06/0.23/0.04 & 0.06/0.04 & 0.1 \\
SnOH3-me-2h2o S (44q) & A5-46q trained & 49/62 & 64/80 & 11891/10298 & 0.13/0.11/0.04 & 0.11/0.06 & 0.1 \\
SnOH3-me-2h2o S (44q) & A5-70q trained & 168/142 & 104/112 & 14875/19156 & 0.22/0.27/0.04 & 0.22/0.27 & 0.2 \\
\addlinespace
SnOH3-me-h2o M (46q) & A1-46q trained & 37/309 & 64/1068 & 10254/771 & 1.4/22.7/7.1 & 22.7/22.7 & 1.4 \\
SnOH3-me-h2o M (46q) & A1-70q trained & 69/6617 & 100/13041 & 10230/13 & 0.71/23.0/6.4 & 23.0/22.6 & 0.7 \\
SnOH3-me-h2o M (46q) & A5-46q trained & 58/51 & 64/72 & 23929/3130 & 1.5/2.0/2.0 & 1.4/2.0 & 1.5 \\
SnOH3-me-h2o M (46q) & A5-70q trained & 108/139 & 100/104 & 5346/6485 & 1.4/1.7/0.91 & 1.4/1.7 & 1.4 \\
\addlinespace
SnOH3-me LL (56q) & A1-46q trained & 52/42 & 80/76 & 3788/3000 & 14.9/15.8/28.2 & 203.8/14.9 & 14.9 \\
SnOH3-me LL (56q) & A1-70q trained & 62/405 & 124/1636 & 1525/14 & 10.4/202.2/35.2 & 202.2/202.1 & 10.4 \\
SnOH3-me LL (56q) & A5-70q trained & 79/363 & 104/1588 & 1770/52 & 13.2/203.3/53.8 & 203.3/204.3 & 13.2 \\
\addlinespace
SnOH3-me-h2o L (56q) & A1-46q trained & 47/42 & 80/76 & 5017/12757 & 2.8/2.6/5.5 & 33.1/2.8 & 2.8 \\
SnOH3-me-h2o L (56q) & A1-70q trained & 65/62 & 124/124 & 7270/1656 & 1.9/2.5/7.8 & 33.1/33.0 & 1.9 \\
SnOH3-me-h2o L (56q) & A5-46q trained & 53/41 & 64/84 & 8388/2750 & 3.6/2.4/5.3 & 33.1/3.6 & 3.6 \\
SnOH3-me-h2o L (56q) & A5-70q trained & 44/69 & 76/80 & 14182/2605 & 2.7/2.6/4.3 & 2.7/33.1 & 2.7 \\
\addlinespace
SnOH3-me-2h2o M (58q) & A1-46q trained & 47/358 & 88/1692 & 2263/1196 & 6.5/41.1/23.5 & 41.1/41.1 & 6.0 \\
SnOH3-me-2h2o M (58q) & A1-70q trained & 65/49 & 132/124 & 5605/1202 & 4.3/5.6/12.4 & 41.1/41.1 & 4.3 \\
SnOH3-me-2h2o M (58q) & A5-46q trained & 45/46 & 72/88 & 7066/4184 & 4.5/5.0/13.6 & 41.1/41.1 & 4.5 \\
SnOH3-me-2h2o M (58q) & A5-70q trained & 62/41 & 68/64 & 19811/3396 & 4.9/6.7/7.9 & 41.1/41.1 & 4.9 \\
\addlinespace
SnOH3-me-h2o LL (66q) & A1-46q trained & 60/51 & 92/96 & 12073/3357 & 2.1/3.9/6.1 & 35.4/2.1 & 2.1 \\
SnOH3-me-h2o LL (66q) & A1-70q trained & 51/458 & 156/2256 & 2232/3 & 2.6/35.4/5.8 & 2.6/35.4 & 2.6 \\
SnOH3-me-h2o LL (66q) & A5-46q trained & 44/32 & 84/84 & 9310/10483 & 3.5/3.0/6.5 & 35.4/3.5 & 3.5 \\
SnOH3-me-h2o LL (66q) & A5-70q trained & 57/58 & 84/80 & 7521/1020 & 2.2/3.1/3.6 & 3.1/2.2 & 2.2 \\
\addlinespace
SnOH3-me-2h2o L (70q) & A1-46q trained & 48/55 & 100/100 & 3740/1157 & 8.6/9.0/12.8 & 8.6/54.8 & 8.6 \\
SnOH3-me-2h2o L (70q) & A1-70q trained & 60/1255 & 160/7296 & 393/29 & 7.7/54.8/32.4 & 54.8/7.7 & 7.7 \\
SnOH3-me-2h2o L (70q) & A5-46q trained & 44/68 & 96/84 & 5651/2579 & 7.4/9.0/9.6 & 54.8/7.2 & 7.4 \\
SnOH3-me-2h2o L (70q) & A5-70q trained & 134/160 & 128/132 & 4941/11792 & 12.0/7.9/21.5 & 12.0/54.8 & 9.6 \\
\midrule
\multicolumn{8}{l}{\textbf{Untrained controls}} \\
SnOH3-me-h2o M (46q) & A1-46q untrained & 11979/10835 & 19458/17473 & 177/55 & 22.9/22.8/22.8 & 22.5/22.8 & 5.0 \\
SnOH3-me-h2o M (46q) & A1-70q untrained & 14901/6614 & 27111/12183 & 181/123 & 22.7/22.7/22.9 & 22.9/22.9 & 4.9 \\
SnOH3-me-h2o M (46q) & A5-70q untrained & 13777/349 & 24044/1180 & 148/263 & 22.6/22.7/23.0 & 23.0/22.8 & 5.1 \\
\addlinespace
SnOH3-me LL (56q) & A1-70q untrained & 89/12313 & 152/23965 & 12522/1 & 12.3/204.9/35.5 & 12.3/204.4 & 12.3 \\
\bottomrule
\end{tabular}}
\end{table}

{\footnotesize
{\footnotesize
{\footnotesize
{\footnotesize
{\footnotesize
{\footnotesize
{\footnotesize
{\footnotesize
{\footnotesize
{\footnotesize
{\footnotesize
{\footnotesize
{\footnotesize
{\footnotesize
{\footnotesize
{\footnotesize
{\footnotesize
{\footnotesize
{\footnotesize
{\footnotesize
{\footnotesize
{\footnotesize
{\footnotesize
{\footnotesize
{\footnotesize
{\footnotesize
{\footnotesize
{\footnotesize
{\footnotesize
{\footnotesize
{\footnotesize
{\footnotesize
{\footnotesize
{\footnotesize
{\footnotesize
\begin{longtable}{@{}llrrrrrrr@{}}
\caption{Recovered energy and pair-counted excitation composition for the targets of Table~\ref{tab:crossnorb-detail}; same pair-counting convention and caution as Table~\ref{tab:zeroshot-excitation}. Ranking, selection, shot counts and recovery settings are as in Table~\ref{tab:zeroshot-master}.}\label{tab:crossnorb-excitation}\\
\toprule
Target/Lineage & Type & Energy (E$_h$) & $\Delta E$ (mHa) & S & D & T & Q & Higher \\
\midrule
\endfirsthead
\toprule
Target/Lineage & Type & Energy (E$_h$) & $\Delta E$ (mHa) & S & D & T & Q & Higher \\
\midrule
\endhead
\bottomrule
\endfoot
SnOH3-me S (32q) A1-46q trained & best & $-$479.193266 & 0.12 & 96 & 3088 & 38790 & 0 & 1248521 \\
SnOH3-me S (32q) A1-46q trained & second & $-$479.192672 & 0.71 & 96 & 2996 & 33466 & 0 & 239066 \\
SnOH3-me S (32q) A1-46q trained & merged & $-$479.193374 & 0.01 & 96 & 3096 & 39776 & 0 & 3269431 \\
\addlinespace
SnOH3-me S (32q) A1-70q trained & best & $-$479.193164 & 0.22 & 96 & 3096 & 39054 & 0 & 1053962 \\
SnOH3-me S (32q) A1-70q trained & second & $-$479.193270 & 0.11 & 96 & 3086 & 38722 & 0 & 1310664 \\
SnOH3-me S (32q) A1-70q trained & merged & $-$479.193374 & 0.01 & 96 & 3096 & 39776 & 0 & 3269431 \\
\addlinespace
SnOH3-me S (32q) A5-46q trained & best & $-$479.192727 & 0.65 & 96 & 3040 & 35626 & 0 & 296478 \\
SnOH3-me S (32q) A5-46q trained & second & $-$479.193081 & 0.30 & 96 & 3092 & 38574 & 0 & 690973 \\
SnOH3-me S (32q) A5-46q trained & merged & $-$479.193374 & 0.01 & 96 & 3096 & 39776 & 0 & 3269431 \\
\addlinespace
SnOH3-me S (32q) A5-70q trained & best & $-$479.192924 & 0.46 & 96 & 3074 & 37530 & 0 & 527815 \\
SnOH3-me S (32q) A5-70q trained & second & $-$479.193134 & 0.25 & 96 & 3072 & 37644 & 0 & 709143 \\
SnOH3-me S (32q) A5-70q trained & merged & $-$479.193374 & 0.01 & 96 & 3096 & 39776 & 0 & 3269431 \\
\addlinespace
SnOH3-me-h2o S (38q) A1-46q trained & best & $-$555.108538 & 0.29 & 120 & 4658 & 64666 & 0 & 1992651 \\
SnOH3-me-h2o S (38q) A1-46q trained & second & $-$555.108684 & 0.15 & 120 & 4730 & 68696 & 0 & 1151902 \\
SnOH3-me-h2o S (38q) A1-46q trained & merged & $-$555.108826 & 0.01 & 120 & 4860 & 79268 & 0 & 18466000 \\
\addlinespace
SnOH3-me-h2o S (38q) A1-70q trained & best & $-$555.108766 & 0.07 & 120 & 4834 & 75660 & 0 & 3293954 \\
SnOH3-me-h2o S (38q) A1-70q trained & second & $-$555.108751 & 0.08 & 120 & 4808 & 74266 & 0 & 3546021 \\
SnOH3-me-h2o S (38q) A1-70q trained & merged & $-$555.108826 & 0.01 & 120 & 4860 & 79268 & 0 & 18466000 \\
\addlinespace
SnOH3-me-h2o S (38q) A5-46q trained & best & $-$555.108713 & 0.12 & 120 & 4720 & 67806 & 0 & 1029853 \\
SnOH3-me-h2o S (38q) A5-46q trained & second & $-$555.108826 & 0.01 & 120 & 4860 & 79278 & 0 & 18873057 \\
SnOH3-me-h2o S (38q) A5-46q trained & merged & $-$555.108826 & 0.01 & 120 & 4860 & 79268 & 0 & 18466000 \\
\addlinespace
SnOH3-me-h2o S (38q) A5-70q trained & best & $-$555.108740 & 0.09 & 120 & 4774 & 71788 & 0 & 2276473 \\
SnOH3-me-h2o S (38q) A5-70q trained & second & $-$555.108764 & 0.07 & 120 & 4834 & 75254 & 0 & 2130960 \\
SnOH3-me-h2o S (38q) A5-70q trained & merged & $-$555.108826 & 0.01 & 120 & 4860 & 79274 & 0 & 18777394 \\
\addlinespace
SnOH3-me M (38q) A1-46q trained & best & $-$479.292683 & 9.6 & 168 & 9310 & 193124 & 0 & 19930566 \\
SnOH3-me M (38q) A1-46q trained & second & $-$479.296591 & 5.7 & 168 & 9408 & 201334 & 0 & 18955973 \\
SnOH3-me M (38q) A1-46q trained & merged & $-$479.286434 & 15.8 & 168 & 8936 & 162134 & 0 & 399828761 \\
\addlinespace
SnOH3-me M (38q) A1-70q trained & best & $-$479.298463 & 3.8 & 168 & 9688 & 226642 & 0 & 41404710 \\
SnOH3-me M (38q) A1-70q trained & second & $-$479.295672 & 6.6 & 168 & 9488 & 210426 & 0 & 61355326 \\
SnOH3-me M (38q) A1-70q trained & merged & $-$479.282569 & 19.7 & 168 & 8712 & 142664 & 0 & 399848455 \\
\addlinespace
SnOH3-me M (38q) A5-46q trained & best & $-$479.285058 & 17.2 & 168 & 8972 & 163332 & 0 & 9920856 \\
SnOH3-me M (38q) A5-46q trained & second & $-$479.282828 & 19.5 & 168 & 8982 & 164246 & 0 & 11557228 \\
SnOH3-me M (38q) A5-46q trained & merged & $-$479.271416 & 30.9 & 168 & 7994 & 81576 & 0 & 399910261 \\
\addlinespace
SnOH3-me M (38q) A5-70q trained & best & $-$479.288699 & 13.6 & 168 & 9142 & 178184 & 0 & 8526809 \\
SnOH3-me M (38q) A5-70q trained & second & $-$479.152486 & 149.8 & 56 & 1688 & 30954 & 0 & 399967301 \\
SnOH3-me M (38q) A5-70q trained & merged & $-$479.265913 & 36.4 & 162 & 7809 & 106826 & 0 & 399885202 \\
\addlinespace
SnOH3-me-2h2o S (44q) A1-46q trained & best & $-$631.083103 & 0.21 & 144 & 6886 & 125446 & 0 & 7640467 \\
SnOH3-me-2h2o S (44q) A1-46q trained & second & $-$631.083239 & 0.07 & 144 & 6904 & 126698 & 0 & 6798942 \\
SnOH3-me-2h2o S (44q) A1-46q trained & merged & $-$631.083269 & 0.04 & 144 & 7020 & 138720 & 0 & 53363340 \\
\addlinespace
SnOH3-me-2h2o S (44q) A1-70q trained & best & $-$631.083257 & 0.06 & 144 & 6964 & 131694 & 0 & 10685297 \\
SnOH3-me-2h2o S (44q) A1-70q trained & second & $-$631.083084 & 0.23 & 144 & 6818 & 120528 & 0 & 7380109 \\
SnOH3-me-2h2o S (44q) A1-70q trained & merged & $-$631.083269 & 0.04 & 144 & 7020 & 138720 & 0 & 53363340 \\
\addlinespace
SnOH3-me-2h2o S (44q) A5-46q trained & best & $-$631.083185 & 0.13 & 144 & 6836 & 120756 & 0 & 3444363 \\
SnOH3-me-2h2o S (44q) A5-46q trained & second & $-$631.083207 & 0.11 & 144 & 6876 & 124062 & 0 & 4841817 \\
SnOH3-me-2h2o S (44q) A5-46q trained & merged & $-$631.083267 & 0.04 & 144 & 7020 & 137030 & 0 & 15292846 \\
\addlinespace
SnOH3-me-2h2o S (44q) A5-70q trained & best & $-$631.083092 & 0.22 & 144 & 6834 & 120950 & 0 & 4690096 \\
SnOH3-me-2h2o S (44q) A5-70q trained & second & $-$631.083044 & 0.27 & 144 & 6798 & 118334 & 0 & 4309959 \\
SnOH3-me-2h2o S (44q) A5-70q trained & merged & $-$631.083269 & 0.04 & 144 & 7020 & 138720 & 0 & 53348711 \\
\addlinespace
SnOH3-me-h2o M (46q) A1-46q trained & best & $-$555.123748 & 1.4 & 240 & 18500 & 496376 & 0 & 24205667 \\
SnOH3-me-h2o M (46q) A1-46q trained & second & $-$555.102442 & 22.7 & 14 & 411 & 5278 & 0 & 399994296 \\
SnOH3-me-h2o M (46q) A1-46q trained & merged & $-$555.118018 & 7.1 & 210 & 11639 & 66550 & 0 & 399921600 \\
\addlinespace
SnOH3-me-h2o M (46q) A1-46q untrained & best & $-$555.102302 & 22.9 & 8 & 294 & 3508 & 0 & 399996189 \\
SnOH3-me-h2o M (46q) A1-46q untrained & second & $-$555.102364 & 22.8 & 10 & 339 & 3992 & 0 & 399995658 \\
SnOH3-me-h2o M (46q) A1-46q untrained & merged & $-$555.102310 & 22.8 & 6 & 295 & 3022 & 0 & 399996676 \\
\addlinespace
SnOH3-me-h2o M (46q) A1-70q trained & best & $-$555.124450 & 0.71 & 240 & 19496 & 621246 & 0 & 105552042 \\
SnOH3-me-h2o M (46q) A1-70q trained & second & $-$555.102207 & 23.0 & 4 & 212 & 2308 & 0 & 399997475 \\
SnOH3-me-h2o M (46q) A1-70q trained & merged & $-$555.118773 & 6.4 & 230 & 14021 & 93558 & 0 & 399892190 \\
\addlinespace
SnOH3-me-h2o M (46q) A1-70q untrained & best & $-$555.102456 & 22.7 & 10 & 249 & 3314 & 0 & 399996426 \\
SnOH3-me-h2o M (46q) A1-70q untrained & second & $-$555.102468 & 22.7 & 14 & 181 & 1768 & 0 & 399998036 \\
SnOH3-me-h2o M (46q) A1-70q untrained & merged & $-$555.102255 & 22.9 & 14 & 289 & 3702 & 0 & 399995994 \\
\addlinespace
SnOH3-me-h2o M (46q) A5-46q trained & best & $-$555.123672 & 1.5 & 240 & 18354 & 478028 & 0 & 16700986 \\
SnOH3-me-h2o M (46q) A5-46q trained & second & $-$555.123151 & 2.0 & 240 & 17760 & 408178 & 0 & 27918797 \\
SnOH3-me-h2o M (46q) A5-46q trained & merged & $-$555.123131 & 2.0 & 240 & 18042 & 441442 & 0 & 399540275 \\
\addlinespace
SnOH3-me-h2o M (46q) A5-70q trained & best & $-$555.123764 & 1.4 & 240 & 19000 & 561904 & 0 & 96224776 \\
SnOH3-me-h2o M (46q) A5-70q trained & second & $-$555.123482 & 1.7 & 240 & 18406 & 488464 & 0 & 59137618 \\
SnOH3-me-h2o M (46q) A5-70q trained & merged & $-$555.124246 & 0.91 & 240 & 19488 & 619226 & 0 & 399361045 \\
\addlinespace
SnOH3-me-h2o M (46q) A5-70q untrained & best & $-$555.102543 & 22.6 & 8 & 242 & 2946 & 0 & 399996803 \\
SnOH3-me-h2o M (46q) A5-70q untrained & second & $-$555.102455 & 22.7 & 6 & 43 & 580 & 0 & 399999370 \\
SnOH3-me-h2o M (46q) A5-70q untrained & merged & $-$555.102186 & 23.0 & 8 & 238 & 2950 & 0 & 399996803 \\
\addlinespace
SnOH3-me LL (56q) A1-46q trained & best & $-$479.300183 & 14.9 & 294 & 27785 & 919854 & 0 & 237940002 \\
SnOH3-me LL (56q) A1-46q trained & second & $-$479.299292 & 15.8 & 294 & 27607 & 892998 & 0 & 218178304 \\
SnOH3-me LL (56q) A1-46q trained & merged & $-$479.286822 & 28.2 & 294 & 25185 & 528534 & 0 & 399445986 \\
\addlinespace
SnOH3-me LL (56q) A1-70q trained & best & $-$479.304625 & 10.4 & 294 & 28161 & 975858 & 0 & 398995686 \\
SnOH3-me LL (56q) A1-70q trained & second & $-$479.112860 & 202.2 & 10 & 173 & 2370 & 0 & 399997446 \\
SnOH3-me LL (56q) A1-70q trained & merged & $-$479.279821 & 35.2 & 294 & 22871 & 187598 & 0 & 399789236 \\
\addlinespace
SnOH3-me LL (56q) A1-70q untrained & best & $-$479.302734 & 12.3 & 294 & 29007 & 1103188 & 0 & 379663706 \\
SnOH3-me LL (56q) A1-70q untrained & second & $-$479.110195 & 204.9 & 0 & 16 & 160 & 0 & 399999823 \\
SnOH3-me LL (56q) A1-70q untrained & merged & $-$479.279604 & 35.5 & 294 & 23847 & 331504 & 0 & 399644354 \\
\addlinespace
SnOH3-me LL (56q) A5-70q trained & best & $-$479.301896 & 13.2 & 294 & 27359 & 859128 & 0 & 303022707 \\
SnOH3-me LL (56q) A5-70q trained & second & $-$479.111752 & 203.3 & 10 & 241 & 3240 & 0 & 399996508 \\
SnOH3-me LL (56q) A5-70q trained & merged & $-$479.261236 & 53.8 & 254 & 17777 & 211854 & 0 & 399770114 \\
\addlinespace
SnOH3-me-h2o L (56q) A1-46q trained & best & $-$555.132208 & 2.8 & 320 & 32232 & 1073160 & 0 & 288166351 \\
SnOH3-me-h2o L (56q) A1-46q trained & second & $-$555.132421 & 2.6 & 320 & 31772 & 999304 & 0 & 272871103 \\
SnOH3-me-h2o L (56q) A1-46q trained & merged & $-$555.129520 & 5.5 & 314 & 26771 & 334486 & 0 & 399638428 \\
\addlinespace
SnOH3-me-h2o L (56q) A1-70q trained & best & $-$555.133193 & 1.9 & 320 & 33296 & 1245064 & 0 & 398721319 \\
SnOH3-me-h2o L (56q) A1-70q trained & second & $-$555.132552 & 2.5 & 320 & 32806 & 1166154 & 0 & 398800719 \\
SnOH3-me-h2o L (56q) A1-70q trained & merged & $-$555.127221 & 7.8 & 314 & 26311 & 262464 & 0 & 399710910 \\
\addlinespace
SnOH3-me-h2o L (56q) A5-46q trained & best & $-$555.131417 & 3.6 & 320 & 31506 & 954976 & 0 & 204677478 \\
SnOH3-me-h2o L (56q) A5-46q trained & second & $-$555.132660 & 2.4 & 320 & 32916 & 1182632 & 0 & 391101380 \\
SnOH3-me-h2o L (56q) A5-46q trained & merged & $-$555.129768 & 5.3 & 320 & 29162 & 573050 & 0 & 399397467 \\
\addlinespace
SnOH3-me-h2o L (56q) A5-70q trained & best & $-$555.132405 & 2.7 & 320 & 31550 & 962282 & 0 & 206625128 \\
SnOH3-me-h2o L (56q) A5-70q trained & second & $-$555.132408 & 2.6 & 320 & 31702 & 986450 & 0 & 186370248 \\
SnOH3-me-h2o L (56q) A5-70q trained & merged & $-$555.130714 & 4.3 & 320 & 29542 & 634626 & 0 & 399335511 \\
\addlinespace
SnOH3-me-2h2o M (58q) A1-46q trained & best & $-$631.112084 & 6.5 & 394 & 45927 & 1418666 & 0 & 398535012 \\
SnOH3-me-2h2o M (58q) A1-46q trained & second & $-$631.077480 & 41.1 & 0 & 14 & 190 & 0 & 399999795 \\
SnOH3-me-2h2o M (58q) A1-46q trained & merged & $-$631.095105 & 23.5 & 232 & 14422 & 112804 & 0 & 399872541 \\
\addlinespace
SnOH3-me-2h2o M (58q) A1-70q trained & best & $-$631.114317 & 4.3 & 396 & 49464 & 2047290 & 0 & 397902849 \\
SnOH3-me-2h2o M (58q) A1-70q trained & second & $-$631.113058 & 5.6 & 396 & 47314 & 1623144 & 0 & 398329145 \\
SnOH3-me-2h2o M (58q) A1-70q trained & merged & $-$631.106177 & 12.4 & 374 & 36859 & 354898 & 0 & 399607868 \\
\addlinespace
SnOH3-me-2h2o M (58q) A5-46q trained & best & $-$631.114080 & 4.5 & 396 & 48848 & 1927832 & 0 & 358908932 \\
SnOH3-me-2h2o M (58q) A5-46q trained & second & $-$631.113620 & 5.0 & 396 & 48692 & 1893402 & 0 & 227398245 \\
SnOH3-me-2h2o M (58q) A5-46q trained & merged & $-$631.105052 & 13.6 & 346 & 31705 & 308132 & 0 & 399659816 \\
\addlinespace
SnOH3-me-2h2o M (58q) A5-70q trained & best & $-$631.113705 & 4.9 & 396 & 47538 & 1658058 & 0 & 86973896 \\
SnOH3-me-2h2o M (58q) A5-70q trained & second & $-$631.111954 & 6.7 & 396 & 45066 & 1170682 & 0 & 115812979 \\
SnOH3-me-2h2o M (58q) A5-70q trained & merged & $-$631.110745 & 7.9 & 394 & 42011 & 632994 & 0 & 399324600 \\
\addlinespace
SnOH3-me-h2o LL (66q) A1-46q trained & best & $-$555.135328 & 2.1 & 400 & 51898 & 2393474 & 0 & 284585591 \\
SnOH3-me-h2o LL (66q) A1-46q trained & second & $-$555.133503 & 3.9 & 400 & 48660 & 1745814 & 0 & 236845829 \\
SnOH3-me-h2o LL (66q) A1-46q trained & merged & $-$555.131333 & 6.1 & 400 & 44596 & 923320 & 0 & 399031683 \\
\addlinespace
SnOH3-me-h2o LL (66q) A1-70q trained & best & $-$555.134846 & 2.6 & 400 & 49602 & 1935652 & 0 & 398014345 \\
SnOH3-me-h2o LL (66q) A1-70q trained & second & $-$555.102002 & 35.4 & 0 & 36 & 544 & 0 & 399999419 \\
SnOH3-me-h2o LL (66q) A1-70q trained & merged & $-$555.131599 & 5.8 & 398 & 42647 & 608990 & 0 & 399347964 \\
\addlinespace
SnOH3-me-h2o LL (66q) A5-46q trained & best & $-$555.133877 & 3.5 & 400 & 49706 & 1953242 & 0 & 197173420 \\
SnOH3-me-h2o LL (66q) A5-46q trained & second & $-$555.134389 & 3.0 & 400 & 51528 & 2322426 & 0 & 380884574 \\
SnOH3-me-h2o LL (66q) A5-46q trained & merged & $-$555.130850 & 6.5 & 400 & 42928 & 587058 & 0 & 399369613 \\
\addlinespace
SnOH3-me-h2o LL (66q) A5-70q trained & best & $-$555.135231 & 2.2 & 400 & 51502 & 2316104 & 0 & 318256829 \\
SnOH3-me-h2o LL (66q) A5-70q trained & second & $-$555.134346 & 3.1 & 400 & 49142 & 1846532 & 0 & 398103925 \\
SnOH3-me-h2o LL (66q) A5-70q trained & merged & $-$555.133797 & 3.6 & 400 & 49072 & 1824750 & 0 & 398125777 \\
\addlinespace
SnOH3-me-2h2o L (70q) A1-46q trained & best & $-$631.123614 & 8.6 & 528 & 81784 & 3207164 & 0 & 396710523 \\
SnOH3-me-2h2o L (70q) A1-46q trained & second & $-$631.123272 & 9.0 & 528 & 79590 & 2629546 & 0 & 397290335 \\
SnOH3-me-2h2o L (70q) A1-46q trained & merged & $-$631.119423 & 12.8 & 516 & 68948 & 616726 & 0 & 399313809 \\
\addlinespace
SnOH3-me-2h2o L (70q) A1-70q trained & best & $-$631.124552 & 7.7 & 528 & 79440 & 2588922 & 0 & 397331109 \\
SnOH3-me-2h2o L (70q) A1-70q trained & second & $-$631.077480 & 54.8 & 0 & 0 & 16 & 0 & 399999983 \\
SnOH3-me-2h2o L (70q) A1-70q trained & merged & $-$631.099821 & 32.4 & 330 & 28945 & 285278 & 0 & 399685446 \\
\addlinespace
SnOH3-me-2h2o L (70q) A5-46q trained & best & $-$631.124857 & 7.4 & 528 & 83086 & 3552970 & 0 & 396363415 \\
SnOH3-me-2h2o L (70q) A5-46q trained & second & $-$631.123201 & 9.0 & 526 & 79153 & 2643128 & 0 & 397277192 \\
SnOH3-me-2h2o L (70q) A5-46q trained & merged & $-$631.122669 & 9.6 & 526 & 76467 & 1925512 & 0 & 397997494 \\
\addlinespace
SnOH3-me-2h2o L (70q) A5-70q trained & best & $-$631.120256 & 12.0 & 500 & 67066 & 1152464 & 0 & 398779969 \\
SnOH3-me-2h2o L (70q) A5-70q trained & second & $-$631.124322 & 7.9 & 526 & 77617 & 2239082 & 0 & 334368575 \\
SnOH3-me-2h2o L (70q) A5-70q trained & merged & $-$631.110701 & 21.5 & 410 & 44667 & 546210 & 0 & 399408712 \\
\end{longtable}}}}}}}}}}}}}}}}}}}}}}}}}}}}}}}}}}}}

{\footnotesize
{\footnotesize
{\footnotesize
{\footnotesize
{\footnotesize
{\footnotesize
{\footnotesize
{\footnotesize
{\footnotesize
{\footnotesize
{\footnotesize
{\footnotesize
{\footnotesize
{\footnotesize
{\footnotesize
{\footnotesize
{\footnotesize
{\footnotesize
{\footnotesize
{\footnotesize
{\footnotesize
{\footnotesize
{\footnotesize
{\footnotesize
{\footnotesize
{\footnotesize
{\footnotesize
{\footnotesize
{\footnotesize
{\footnotesize
{\footnotesize
{\footnotesize
{\footnotesize
{\footnotesize
{\footnotesize
\begin{longtable}{@{}llrrrrrrr@{}}
\caption{Raw-shot excitation-level composition (HF/S/D/T/Q/higher) of the sampled bitstrings, per lineage/policy, for the best and second circuit and for the merged pool. Unlike Tables~\ref{tab:zeroshot-excitation} and \ref{tab:crossnorb-excitation}, counts are over the measured shots themselves (\code{raw\_shot\_excitation\_histogram}, weighted by multiplicity), so each best/second row sums to 100{,}000 (``n''); the merged row sums to the pooled shots of all finite-energy circuits. Ranking, selection, shot counts and recovery settings are as in Table~\ref{tab:zeroshot-master}.}\label{tab:raw-shot-excitation}\\
\toprule
Target/Lineage & Type & HF & S & D & T & Q & Higher & n \\
\midrule
\endfirsthead
\toprule
Target/Lineage & Type & HF & S & D & T & Q & Higher & n \\
\midrule
\endhead
\bottomrule
\endfoot
SnOH3-me S (32q) A1-46q trained & best & 9563 & 47174 & 35302 & 7217 & 704 & 40 & 100000 \\
SnOH3-me S (32q) A1-46q trained & second & 73648 & 24464 & 1823 & 64 & 1 & 0 & 100000 \\
SnOH3-me S (32q) A1-46q trained & merged & 174718 & 158128 & 57496 & 10431 & 9313 & 589914 & 1000000 \\
\addlinespace
SnOH3-me S (32q) A1-70q trained & best & 15864 & 53088 & 26321 & 4322 & 395 & 10 & 100000 \\
SnOH3-me S (32q) A1-70q trained & second & 15619 & 50395 & 27497 & 5836 & 601 & 52 & 100000 \\
SnOH3-me S (32q) A1-70q trained & merged & 31483 & 103483 & 53831 & 10447 & 4126 & 796630 & 1000000 \\
\addlinespace
SnOH3-me S (32q) A5-46q trained & best & 75767 & 22037 & 2076 & 119 & 1 & 0 & 100000 \\
SnOH3-me S (32q) A5-46q trained & second & 37986 & 49858 & 11077 & 1010 & 68 & 1 & 100000 \\
SnOH3-me S (32q) A5-46q trained & merged & 343611 & 271049 & 76571 & 9812 & 9776 & 289181 & 1000000 \\
\addlinespace
SnOH3-me S (32q) A5-70q trained & best & 38407 & 50890 & 9743 & 917 & 41 & 2 & 100000 \\
SnOH3-me S (32q) A5-70q trained & second & 41440 & 47694 & 9839 & 961 & 64 & 2 & 100000 \\
SnOH3-me S (32q) A5-70q trained & merged & 233308 & 140523 & 24451 & 6476 & 19835 & 575407 & 1000000 \\
\addlinespace
SnOH3-me-h2o S (38q) A1-46q trained & best & 29168 & 50965 & 17380 & 2308 & 170 & 9 & 100000 \\
SnOH3-me-h2o S (38q) A1-46q trained & second & 11483 & 51150 & 32605 & 4475 & 279 & 8 & 100000 \\
SnOH3-me-h2o S (38q) A1-46q trained & merged & 68750 & 216647 & 100105 & 13754 & 2961 & 397783 & 800000 \\
\addlinespace
SnOH3-me-h2o S (38q) A1-70q trained & best & 36208 & 45204 & 15453 & 2803 & 297 & 35 & 100000 \\
SnOH3-me-h2o S (38q) A1-70q trained & second & 7494 & 42005 & 37942 & 10962 & 1496 & 101 & 100000 \\
SnOH3-me-h2o S (38q) A1-70q trained & merged & 50894 & 129503 & 90324 & 25433 & 5245 & 598601 & 900000 \\
\addlinespace
SnOH3-me-h2o S (38q) A5-46q trained & best & 71348 & 25611 & 2864 & 171 & 6 & 0 & 100000 \\
SnOH3-me-h2o S (38q) A5-46q trained & second & 0 & 0 & 0 & 5 & 29 & 99966 & 100000 \\
SnOH3-me-h2o S (38q) A5-46q trained & merged & 71348 & 25611 & 2864 & 176 & 46 & 199955 & 300000 \\
\addlinespace
SnOH3-me-h2o S (38q) A5-70q trained & best & 33180 & 49434 & 15023 & 2157 & 193 & 13 & 100000 \\
SnOH3-me-h2o S (38q) A5-70q trained & second & 47581 & 42541 & 8936 & 900 & 41 & 1 & 100000 \\
SnOH3-me-h2o S (38q) A5-70q trained & merged & 141386 & 251716 & 92217 & 13551 & 1081 & 100049 & 600000 \\
\addlinespace
SnOH3-me M (38q) A1-46q trained & best & 29986 & 48487 & 18233 & 2995 & 284 & 15 & 100000 \\
SnOH3-me M (38q) A1-46q trained & second & 15637 & 59066 & 21921 & 3089 & 270 & 17 & 100000 \\
SnOH3-me M (38q) A1-46q trained & merged & 75820 & 217967 & 92926 & 12312 & 1482 & 599493 & 1000000 \\
\addlinespace
SnOH3-me M (38q) A1-70q trained & best & 26574 & 43964 & 22656 & 5803 & 911 & 92 & 100000 \\
SnOH3-me M (38q) A1-70q trained & second & 4831 & 33697 & 40786 & 16732 & 3508 & 446 & 100000 \\
SnOH3-me M (38q) A1-70q trained & merged & 40414 & 120949 & 99670 & 32619 & 6377 & 699971 & 1000000 \\
\addlinespace
SnOH3-me M (38q) A5-46q trained & best & 5815 & 40511 & 44079 & 8934 & 649 & 12 & 100000 \\
SnOH3-me M (38q) A5-46q trained & second & 9128 & 53451 & 31943 & 5040 & 412 & 26 & 100000 \\
SnOH3-me M (38q) A5-46q trained & merged & 14943 & 93962 & 76022 & 13993 & 1320 & 799760 & 1000000 \\
\addlinespace
SnOH3-me M (38q) A5-70q trained & best & 15975 & 51746 & 27391 & 4502 & 370 & 16 & 100000 \\
SnOH3-me M (38q) A5-70q trained & second & 0 & 0 & 1 & 4 & 93 & 99902 & 100000 \\
SnOH3-me M (38q) A5-70q trained & merged & 15975 & 51746 & 27392 & 4522 & 768 & 899597 & 1000000 \\
\addlinespace
SnOH3-me-2h2o S (44q) A1-46q trained & best & 6777 & 39820 & 38991 & 12317 & 1883 & 212 & 100000 \\
SnOH3-me-2h2o S (44q) A1-46q trained & second & 5397 & 43261 & 39345 & 10603 & 1297 & 97 & 100000 \\
SnOH3-me-2h2o S (44q) A1-46q trained & merged & 12174 & 83081 & 78336 & 22952 & 3611 & 399846 & 600000 \\
\addlinespace
SnOH3-me-2h2o S (44q) A1-70q trained & best & 8424 & 40114 & 35955 & 12733 & 2458 & 316 & 100000 \\
SnOH3-me-2h2o S (44q) A1-70q trained & second & 2848 & 30352 & 46181 & 17456 & 2875 & 288 & 100000 \\
SnOH3-me-2h2o S (44q) A1-70q trained & merged & 16026 & 100280 & 123367 & 49556 & 9698 & 301073 & 600000 \\
\addlinespace
SnOH3-me-2h2o S (44q) A5-46q trained & best & 23489 & 53166 & 20330 & 2792 & 213 & 10 & 100000 \\
SnOH3-me-2h2o S (44q) A5-46q trained & second & 13090 & 47446 & 30633 & 7606 & 1112 & 113 & 100000 \\
SnOH3-me-2h2o S (44q) A5-46q trained & merged & 60325 & 226041 & 164735 & 42601 & 5784 & 514 & 500000 \\
\addlinespace
SnOH3-me-2h2o S (44q) A5-70q trained & best & 13935 & 49122 & 30029 & 6236 & 631 & 47 & 100000 \\
SnOH3-me-2h2o S (44q) A5-70q trained & second & 14446 & 49445 & 29194 & 6252 & 631 & 32 & 100000 \\
SnOH3-me-2h2o S (44q) A5-70q trained & merged & 28381 & 98567 & 59231 & 12577 & 2021 & 99223 & 300000 \\
\addlinespace
SnOH3-me L (46q) A1-46q trained & best & 27683 & 52820 & 17094 & 2250 & 145 & 8 & 100000 \\
SnOH3-me L (46q) A1-46q trained & second & 0 & 0 & 0 & 0 & 0 & 100000 & 100000 \\
SnOH3-me L (46q) A1-46q trained & merged & 27683 & 52820 & 17094 & 2256 & 229 & 799918 & 900000 \\
\addlinespace
SnOH3-me L (46q) A1-46q untrained & best & 0 & 0 & 0 & 0 & 0 & 100000 & 100000 \\
SnOH3-me L (46q) A1-46q untrained & second & 0 & 0 & 0 & 2 & 51 & 99947 & 100000 \\
SnOH3-me L (46q) A1-46q untrained & merged & 0 & 0 & 0 & 4 & 80 & 899916 & 900000 \\
\addlinespace
SnOH3-me L (46q) A1-70q trained & best & 22910 & 47962 & 23493 & 5013 & 575 & 47 & 100000 \\
SnOH3-me L (46q) A1-70q trained & second & 0 & 0 & 0 & 1 & 6 & 99993 & 100000 \\
SnOH3-me L (46q) A1-70q trained & merged & 22910 & 47962 & 23493 & 5014 & 590 & 700031 & 800000 \\
\addlinespace
SnOH3-me L (46q) A1-70q untrained & best & 0 & 0 & 0 & 1 & 37 & 99962 & 100000 \\
SnOH3-me L (46q) A1-70q untrained & second & 0 & 0 & 0 & 0 & 0 & 100000 & 100000 \\
SnOH3-me L (46q) A1-70q untrained & merged & 0 & 0 & 0 & 1 & 44 & 999955 & 1000000 \\
\addlinespace
SnOH3-me L (46q) A5-46q trained & best & 11558 & 47087 & 32743 & 7713 & 844 & 55 & 100000 \\
SnOH3-me L (46q) A5-46q trained & second & 14844 & 50116 & 28378 & 5962 & 647 & 53 & 100000 \\
SnOH3-me L (46q) A5-46q trained & merged & 56270 & 147149 & 78341 & 16361 & 1774 & 500105 & 800000 \\
\addlinespace
SnOH3-me L (46q) A5-46q untrained & best & 0 & 0 & 0 & 0 & 4 & 99996 & 100000 \\
SnOH3-me L (46q) A5-46q untrained & second & 0 & 0 & 0 & 0 & 1 & 99999 & 100000 \\
SnOH3-me L (46q) A5-46q untrained & merged & 0 & 0 & 0 & 1 & 22 & 999977 & 1000000 \\
\addlinespace
SnOH3-me L (46q) A5-70q trained & best & 10334 & 44027 & 33879 & 10033 & 1563 & 164 & 100000 \\
SnOH3-me L (46q) A5-70q trained & second & 3866 & 32527 & 44595 & 16051 & 2690 & 271 & 100000 \\
SnOH3-me L (46q) A5-70q trained & merged & 21220 & 116464 & 117086 & 38289 & 6359 & 700582 & 1000000 \\
\addlinespace
SnOH3-me L (46q) A5-70q untrained & best & 0 & 0 & 0 & 0 & 0 & 100000 & 100000 \\
SnOH3-me L (46q) A5-70q untrained & second & 0 & 0 & 0 & 0 & 1 & 99999 & 100000 \\
SnOH3-me L (46q) A5-70q untrained & merged & 0 & 0 & 0 & 1 & 11 & 999988 & 1000000 \\
\addlinespace
SnOH3-me-h2o M (46q) A1-46q trained & best & 26532 & 52848 & 17834 & 2599 & 178 & 9 & 100000 \\
SnOH3-me-h2o M (46q) A1-46q trained & second & 0 & 0 & 0 & 1 & 12 & 99987 & 100000 \\
SnOH3-me-h2o M (46q) A1-46q trained & merged & 26532 & 52848 & 17834 & 2605 & 234 & 899947 & 1000000 \\
\addlinespace
SnOH3-me-h2o M (46q) A1-46q untrained & best & 0 & 0 & 0 & 0 & 4 & 99996 & 100000 \\
SnOH3-me-h2o M (46q) A1-46q untrained & second & 0 & 0 & 0 & 1 & 5 & 99994 & 100000 \\
SnOH3-me-h2o M (46q) A1-46q untrained & merged & 0 & 0 & 0 & 3 & 49 & 999948 & 1000000 \\
\addlinespace
SnOH3-me-h2o M (46q) A1-70q trained & best & 21761 & 47739 & 24387 & 5377 & 682 & 54 & 100000 \\
SnOH3-me-h2o M (46q) A1-70q trained & second & 0 & 0 & 0 & 0 & 1 & 99999 & 100000 \\
SnOH3-me-h2o M (46q) A1-70q trained & merged & 21761 & 47739 & 24387 & 5378 & 694 & 900041 & 1000000 \\
\addlinespace
SnOH3-me-h2o M (46q) A1-70q untrained & best & 0 & 0 & 0 & 0 & 2 & 99998 & 100000 \\
SnOH3-me-h2o M (46q) A1-70q untrained & second & 0 & 0 & 0 & 0 & 1 & 99999 & 100000 \\
SnOH3-me-h2o M (46q) A1-70q untrained & merged & 0 & 0 & 0 & 1 & 8 & 999991 & 1000000 \\
\addlinespace
SnOH3-me-h2o M (46q) A5-46q trained & best & 26761 & 54384 & 16890 & 1849 & 112 & 4 & 100000 \\
SnOH3-me-h2o M (46q) A5-46q trained & second & 6148 & 42569 & 40210 & 9857 & 1127 & 89 & 100000 \\
SnOH3-me-h2o M (46q) A5-46q trained & merged & 83510 & 271220 & 240736 & 87508 & 15304 & 301722 & 1000000 \\
\addlinespace
SnOH3-me-h2o M (46q) A5-70q trained & best & 15641 & 46468 & 29123 & 7493 & 1163 & 112 & 100000 \\
SnOH3-me-h2o M (46q) A5-70q trained & second & 14142 & 49156 & 29314 & 6458 & 860 & 70 & 100000 \\
SnOH3-me-h2o M (46q) A5-70q trained & merged & 107078 & 398102 & 297743 & 83360 & 12527 & 101190 & 1000000 \\
\addlinespace
SnOH3-me-h2o M (46q) A5-70q untrained & best & 0 & 0 & 0 & 0 & 3 & 99997 & 100000 \\
SnOH3-me-h2o M (46q) A5-70q untrained & second & 0 & 0 & 0 & 0 & 2 & 99998 & 100000 \\
SnOH3-me-h2o M (46q) A5-70q untrained & merged & 0 & 0 & 0 & 0 & 11 & 999989 & 1000000 \\
\addlinespace
SnOH3-me LL (56q) A1-46q trained & best & 11109 & 46363 & 32695 & 8558 & 1155 & 120 & 100000 \\
SnOH3-me LL (56q) A1-46q trained & second & 6209 & 37216 & 39430 & 14250 & 2589 & 306 & 100000 \\
SnOH3-me LL (56q) A1-46q trained & merged & 23464 & 117381 & 112204 & 39228 & 6898 & 300825 & 600000 \\
\addlinespace
SnOH3-me LL (56q) A1-70q trained & best & 5291 & 34388 & 39206 & 16886 & 3640 & 589 & 100000 \\
SnOH3-me LL (56q) A1-70q trained & second & 0 & 0 & 0 & 0 & 0 & 100000 & 100000 \\
SnOH3-me LL (56q) A1-70q trained & merged & 5291 & 34388 & 39206 & 16886 & 3640 & 400589 & 500000 \\
\addlinespace
SnOH3-me LL (56q) A1-70q untrained & best & 8196 & 33661 & 35363 & 17227 & 4661 & 892 & 100000 \\
SnOH3-me LL (56q) A1-70q untrained & second & 0 & 0 & 0 & 0 & 0 & 100000 & 100000 \\
SnOH3-me LL (56q) A1-70q untrained & merged & 8196 & 33661 & 35363 & 17227 & 4661 & 200892 & 300000 \\
\addlinespace
SnOH3-me LL (56q) A5-70q trained & best & 3320 & 26316 & 41043 & 21969 & 6207 & 1145 & 100000 \\
SnOH3-me LL (56q) A5-70q trained & second & 0 & 0 & 0 & 0 & 1 & 99999 & 100000 \\
SnOH3-me LL (56q) A5-70q trained & merged & 3320 & 26316 & 41043 & 21969 & 6208 & 301144 & 400000 \\
\addlinespace
SnOH3-me-h2o L (56q) A1-46q trained & best & 5390 & 31433 & 40536 & 18063 & 4021 & 557 & 100000 \\
SnOH3-me-h2o L (56q) A1-46q trained & second & 3280 & 28637 & 43522 & 19745 & 4254 & 562 & 100000 \\
SnOH3-me-h2o L (56q) A1-46q trained & merged & 19279 & 105553 & 117429 & 46890 & 9615 & 401234 & 700000 \\
\addlinespace
SnOH3-me-h2o L (56q) A1-70q trained & best & 15894 & 42646 & 29701 & 9609 & 1862 & 288 & 100000 \\
SnOH3-me-h2o L (56q) A1-70q trained & second & 5851 & 35489 & 39026 & 15758 & 3374 & 502 & 100000 \\
SnOH3-me-h2o L (56q) A1-70q trained & merged & 21745 & 78135 & 68727 & 25367 & 5236 & 700790 & 900000 \\
\addlinespace
SnOH3-me-h2o L (56q) A5-46q trained & best & 25345 & 48690 & 21049 & 4359 & 517 & 40 & 100000 \\
SnOH3-me-h2o L (56q) A5-46q trained & second & 10940 & 42947 & 34403 & 10013 & 1560 & 137 & 100000 \\
SnOH3-me-h2o L (56q) A5-46q trained & merged & 49353 & 168718 & 133148 & 41605 & 6515 & 200661 & 600000 \\
\addlinespace
SnOH3-me-h2o L (56q) A5-70q trained & best & 5621 & 34344 & 41636 & 15295 & 2757 & 347 & 100000 \\
SnOH3-me-h2o L (56q) A5-70q trained & second & 9317 & 43133 & 35791 & 10169 & 1464 & 126 & 100000 \\
SnOH3-me-h2o L (56q) A5-70q trained & merged & 19891 & 112942 & 120891 & 39422 & 6202 & 200652 & 500000 \\
\addlinespace
SnOH3-me-2h2o M (58q) A1-46q trained & best & 3041 & 23236 & 40636 & 24716 & 7127 & 1244 & 100000 \\
SnOH3-me-2h2o M (58q) A1-46q trained & second & 0 & 0 & 0 & 0 & 0 & 100000 & 100000 \\
SnOH3-me-2h2o M (58q) A1-46q trained & merged & 3041 & 23236 & 40636 & 24716 & 7127 & 901244 & 1000000 \\
\addlinespace
SnOH3-me-2h2o M (58q) A1-70q trained & best & 13830 & 39660 & 31266 & 11948 & 2812 & 484 & 100000 \\
SnOH3-me-2h2o M (58q) A1-70q trained & second & 1644 & 18231 & 38792 & 28591 & 10270 & 2472 & 100000 \\
SnOH3-me-2h2o M (58q) A1-70q trained & merged & 15474 & 57891 & 70058 & 40539 & 13082 & 802956 & 1000000 \\
\addlinespace
SnOH3-me-2h2o M (58q) A5-46q trained & best & 4001 & 28326 & 41146 & 20746 & 4958 & 823 & 100000 \\
SnOH3-me-2h2o M (58q) A5-46q trained & second & 7976 & 44070 & 35710 & 10444 & 1616 & 184 & 100000 \\
SnOH3-me-2h2o M (58q) A5-46q trained & merged & 11977 & 72396 & 76856 & 31190 & 6574 & 801007 & 1000000 \\
\addlinespace
SnOH3-me-2h2o M (58q) A5-70q trained & best & 37557 & 44812 & 14856 & 2536 & 218 & 21 & 100000 \\
SnOH3-me-2h2o M (58q) A5-70q trained & second & 9683 & 45606 & 34012 & 9236 & 1327 & 136 & 100000 \\
SnOH3-me-2h2o M (58q) A5-70q trained & merged & 72528 & 197532 & 160181 & 56849 & 11329 & 501581 & 1000000 \\
\addlinespace
SnOH3-me-h2o LL (66q) A1-46q trained & best & 31411 & 45412 & 18803 & 3853 & 476 & 45 & 100000 \\
SnOH3-me-h2o LL (66q) A1-46q trained & second & 7197 & 37715 & 38763 & 13528 & 2531 & 266 & 100000 \\
SnOH3-me-h2o LL (66q) A1-46q trained & merged & 43401 & 126104 & 137371 & 69934 & 19321 & 303869 & 700000 \\
\addlinespace
SnOH3-me-h2o LL (66q) A1-70q trained & best & 1510 & 15119 & 33817 & 29825 & 14518 & 5211 & 100000 \\
SnOH3-me-h2o LL (66q) A1-70q trained & second & 0 & 0 & 0 & 0 & 0 & 100000 & 100000 \\
SnOH3-me-h2o LL (66q) A1-70q trained & merged & 1510 & 15119 & 33817 & 29825 & 14518 & 105211 & 200000 \\
\addlinespace
SnOH3-me-h2o LL (66q) A5-46q trained & best & 18906 & 46401 & 26915 & 6685 & 1000 & 93 & 100000 \\
SnOH3-me-h2o LL (66q) A5-46q trained & second & 9291 & 39559 & 35828 & 12677 & 2316 & 329 & 100000 \\
SnOH3-me-h2o LL (66q) A5-46q trained & merged & 32446 & 115285 & 103553 & 39417 & 8154 & 301145 & 600000 \\
\addlinespace
SnOH3-me-h2o LL (66q) A5-70q trained & best & 23092 & 45098 & 24376 & 6353 & 967 & 114 & 100000 \\
SnOH3-me-h2o LL (66q) A5-70q trained & second & 2779 & 24312 & 41636 & 23602 & 6452 & 1219 & 100000 \\
SnOH3-me-h2o LL (66q) A5-70q trained & merged & 41304 & 180903 & 226667 & 115109 & 30330 & 105687 & 700000 \\
\addlinespace
SnOH3-me-2h2o L (70q) A1-46q trained & best & 9888 & 36840 & 34706 & 14553 & 3453 & 560 & 100000 \\
SnOH3-me-2h2o L (70q) A1-46q trained & second & 3686 & 27231 & 40572 & 21692 & 5757 & 1062 & 100000 \\
SnOH3-me-2h2o L (70q) A1-46q trained & merged & 14444 & 80586 & 115765 & 66438 & 19022 & 403745 & 700000 \\
\addlinespace
SnOH3-me-2h2o L (70q) A1-70q trained & best & 648 & 9799 & 29489 & 33364 & 18814 & 7886 & 100000 \\
SnOH3-me-2h2o L (70q) A1-70q trained & second & 0 & 0 & 0 & 0 & 0 & 100000 & 100000 \\
SnOH3-me-2h2o L (70q) A1-70q trained & merged & 648 & 9799 & 29489 & 33364 & 18814 & 607886 & 700000 \\
\addlinespace
SnOH3-me-2h2o L (70q) A5-46q trained & best & 16872 & 44324 & 28844 & 8366 & 1404 & 190 & 100000 \\
SnOH3-me-2h2o L (70q) A5-46q trained & second & 7448 & 36091 & 38624 & 14732 & 2770 & 335 & 100000 \\
SnOH3-me-2h2o L (70q) A5-46q trained & merged & 38431 & 185481 & 223634 & 114076 & 32098 & 306280 & 900000 \\
\addlinespace
SnOH3-me-2h2o L (70q) A5-70q trained & best & 145 & 3260 & 22593 & 43300 & 24325 & 6377 & 100000 \\
SnOH3-me-2h2o L (70q) A5-70q trained & second & 612 & 8134 & 32663 & 39988 & 15446 & 3157 & 100000 \\
SnOH3-me-2h2o L (70q) A5-70q trained & merged & 944 & 16631 & 98517 & 194045 & 135955 & 553908 & 1000000 \\
\end{longtable}}}}}}}}}}}}}}}}}}}}}}}}}}}}}}}}}}}}

\textbf{Is the amount of training what matters?} A trained policy is not always better than a random one, and the separation depends on the target. Three patterns stand out. First, at the h2o-M target all four trained lineages (A1-46q, A1-70q, A5-46q, A5-70q) reach 0.71--1.5~mHa with the circuit chosen by lowest ranking energy, while for A1-46q and A1-70q the E+D 1/2 circuits (22.6--23.0~mHa) and every untrained control (22.6--22.9~mHa) stay at the level of the untrained policies (for the two A5 lineages an E+D pick coincides with the best or second circuit and is also at 1.4--2.0~mHa); the cause, e.g.\ valid-shot
count versus ranking energy versus depth, was not isolated. Merged recoveries are 0.91--7.1~mHa for the trained lineages
and 22.9--23.0~mHa for the untrained ones. The singles-injected error of the untrained lineages' best circuit is
4.9--5.1~mHa, while for the trained lineages' best it equals the plain best deviation. Second, at the 32q target
the merged result of all four trained checkpoints is exact to 0.01~mHa and the best circuit's is within
0.12--0.65~mHa; this active space (norb=16) is small enough that the result says little about whether the vocab mismatch
is harmless at larger norb. Third, at the largest tier where an untrained control was run (LL, norb=28, 56 qubits) the trained A1-70q best circuit reaches 10.4~mHa (as-submitted depth 62, 124 two-qubit gates, 1{,}525 valid shots) and A5-70q 13.2~mHa (depth 79), but the A1-70q untrained
control's best circuit reaches 12.3~mHa (as-submitted depth 89, 12{,}522 valid shots), so the trained-versus-untrained separation seen
at h2o-M and L is not present at this target; this is one pair of runs. Coverage of these circuits is in Table~\ref{tab:coverage-idle}.

\FloatBarrier
\section{LUCJ baseline: transpilation and hardware benchmark}
\label{sec:appendix-lucj}
Circuit cost and hardware results of the fixed LUCJ ansatz that serves as the classical-ansatz comparator of Section~\ref{sec:lucj-comparison}.

\begin{table}[ht]
\centering
\caption{LUCJ-ansatz hardware benchmark, \code{optimization\_level=1},
energy and pair-counted subspace excitation (same convention as
Table~\ref{tab:zeroshot-excitation}/Table~\ref{tab:crossnorb-excitation}:
counts of determinant pairs in the recovered $n\times n$ Cartesian
product, not out of the 100{,}000 physical shots---see
Table~\ref{tab:lucj-hw-raw} for that). ``SI'' = singles-injected error.
All 9 target molecules across the hardware benchmark campaign.}
\label{tab:lucj-hw}
\small
\begin{tabular}{@{}lrrrrrr@{}}
\toprule
Target & Energy (E$_h$) / $\Delta E$ (mHa) & $\Delta E$ SI (mHa) & S & D & T & Higher \\
\midrule
S (32q) & $-$479.193374 / 0.0076 & 0.0076 & 96 & 3096 & 39776 & 3269431 \\
h2o-S (38q) & $-$555.108826 / 0.0069 & 0.0069 & 120 & 4860 & 79278 & 18864350 \\
h2o-M (46q) & $-$555.102419 / 22.7 & 5.2 & 4 & 42 & 410 & 399999543 \\
2h2o-S (44q) & $-$631.083269 / 0.04 & 0.04 & 144 & 7020 & 138720 & 53363340 \\
LL (56q) & $-$479.110941 / 204.1 & 41.2 & 10 & 145 & 2152 & 399997692 \\
2h2o-M (58q) & $-$631.077480 / 41.1 & 9.6 & 2 & 21 & 220 & 399999756 \\
h2o-LL (66q) & $-$555.102002 / 35.4 & 8.0 & 2 & 25 & 388 & 399999584 \\
2h2o-L (70q) & $-$631.077480 / 54.8 & 12.6 & 0 & 0 & 26 & 399999973 \\
h2o-L (56q) & $-$555.102002 / 33.1 & 7.3 & 2 & 45 & 862 & 399999090 \\
\bottomrule
\end{tabular}
\end{table}

\begin{table}[ht]
\centering
\caption{LUCJ-ansatz hardware benchmark, \code{optimization\_level=1}:
raw-shot excitation distributions (evaluated directly out of the 100{,}000 physical sampled
shots from each circuit's measured counts, with the same convention as
Table~\ref{tab:raw-shot-excitation}) alongside the offline circuit-cost estimate (idealized heavy-hex model, \code{ibm\_heron\_r2} basis, optimisation level~1; the circuits submitted to \code{ibm\_kobe} were larger). ``n'' is the total
valid sampled-shot count attributed to a named excitation rank; S (32q) has
n$<$100{,}000 because 13 sampled shots were non-particle-number-
conserving readout errors, excluded from every named bucket (confirmed
directly in the raw histogram). All 9
target molecules across the benchmark campaign.}
\label{tab:lucj-hw-raw}
\small
\begin{tabular}{@{}lrrrrrrrr@{}}
\toprule
Target & HF & S & D & T & Q & Higher & Depth & 2q \\
\midrule
S (32q) & 6 & 97 & 672 & 2694 & 7922 & 88609 & 251 & 1368 \\
h2o-S (38q) & 0 & 0 & 3 & 5 & 124 & 99868 & 293 & 1872 \\
h2o-M (46q) & 0 & 0 & 0 & 0 & 1 & 99999 & 356 & 3040 \\
2h2o-S (44q) & 0 & 0 & 0 & 0 & 4 & 99996 & 337 & 2448 \\
LL (56q) & 0 & 0 & 0 & 0 & 0 & 100000 & 428 & 4248 \\
2h2o-M (58q) & 0 & 0 & 0 & 0 & 0 & 100000 & 446 & 4912 \\
h2o-LL (66q) & 0 & 0 & 0 & 0 & 0 & 100000 & 506 & 5880 \\
2h2o-L (70q) & 0 & 0 & 0 & 0 & 0 & 100000 & 536 & 6952 \\
h2o-L (56q) & 0 & 0 & 0 & 0 & 1 & 99999 & 430 & 4360 \\
\bottomrule
\end{tabular}
\end{table}

\section{Dataset-level numbers}
\label{sec:appendix-data}
The deposit accompanying this paper (Zenodo, Section~\ref{sec:appendix-availability}) consists of two self-contained HTML reports, one for each training configuration. For every run they give the ten sampled circuits as gate-token sequences with their depth and two-qubit-gate count, the valid-shot, Hartree--Fock, HOMO$\to$LUMO and spin-resolved shares of the measured shots, and the five recovery energies, together with the timing of the campaign. The raw counts and the transpiled circuits are not part of the deposit.

The tables of this appendix are the numbers most often needed to reuse the data: Tables~\ref{tab:zeroshot-detail}--\ref{tab:raw-shot-excitation} (energies, depth, valid shots, excitation), Table~\ref{tab:coverage-idle} (coverage), the cost tables of Section~\ref{sec:cost} and, once fetched, the \code{ibm\_kobe} calibration tables.
Circuits were generated and transpiled with Qiskit \cite{javadiabhari2024qiskit}.
Software versions (training, sampling, ranking, recovery and the LUCJ baseline): Python 3.11, Qiskit 2.5.1, Qiskit Runtime 0.49.0, qiskit-addon-sqd 0.12.1, ffsim 0.0.80 \cite{sung2026ffsim}, PySCF 2.11.0, PyTorch 2.11.0 and Transformers 4.57.3.


\subsection{\code{ibm\_kobe} calibration during the experiments}
\label{sec:kobe-calibration}
Median over the device's qubits (or couplers) of the calibration IBM reported as in force at the stated time (UTC), from the IBM Quantum
backend properties (fetched 2026-10-09 09:04 UTC). T$_1$/T$_2$ are medians; the last column is the worst two-qubit error on the device.
The first table samples the device every few hours over the last days of training and inference; the second gives the calibration in force when each
fresh-shot hardware sampling job was created. IBM recalibrates several times a day, so a job's actual calibration is the snapshot in force at its execution time. A worst two-qubit error of 100\% means that IBM lists some couplers with error 1 (unusable; 10 of the 352 listed two-qubit gate entries, both directions, in the first snapshot); the median is unaffected.

\begin{table}[ht]
\centering
\caption{\code{ibm\_kobe} calibration snapshots over the last days of training and inference.}
\label{tab:kobe-calibration-daily}
\scriptsize
\begin{tabular}{@{}lrrrrrr@{}}
\toprule
Snapshot (UTC) & T$_1$ ($\mu$s) & T$_2$ ($\mu$s) & readout error (\%) & 1q (\code{sx}) error (\%) & 2q (\code{cz}) error (\%) & worst 2q error (\%) \\
\midrule
2026-10-05 09:04 & 226 & 120 & 0.61 & 0.026 & 0.21 & 100.0 \\
2026-10-05 15:04 & 226 & 120 & 0.66 & 0.026 & 0.21 & 100.0 \\
2026-10-05 21:04 & 226 & 120 & 0.59 & 0.026 & 0.21 & 100.0 \\
2026-10-06 03:04 & 188 & 113 & 0.63 & 0.029 & 0.21 & 100.0 \\
2026-10-06 09:04 & 188 & 113 & 0.66 & 0.029 & 0.22 & 100.0 \\
2026-10-06 15:04 & 188 & 113 & 0.66 & 0.029 & 0.22 & 100.0 \\
2026-10-06 21:04 & 188 & 113 & 0.62 & 0.029 & 0.22 & 100.0 \\
2026-10-07 03:04 & 188 & 113 & 0.62 & 0.029 & 0.22 & 100.0 \\
2026-10-07 09:04 & 183 & 104 & 0.61 & 0.028 & 0.20 & 100.0 \\
2026-10-07 15:04 & 183 & 104 & 0.68 & 0.028 & 0.20 & 100.0 \\
2026-10-07 21:04 & 183 & 104 & 0.62 & 0.028 & 0.20 & 100.0 \\
2026-10-08 03:04 & 183 & 104 & 0.62 & 0.028 & 0.20 & 100.0 \\
2026-10-08 09:04 & 218 & 105 & 0.62 & 0.030 & 0.20 & 100.0 \\
2026-10-08 15:04 & 218 & 105 & 0.60 & 0.030 & 0.20 & 100.0 \\
2026-10-08 21:04 & 218 & 105 & 0.63 & 0.030 & 0.20 & 100.0 \\
2026-10-09 03:04 & 218 & 105 & 0.64 & 0.030 & 0.20 & 100.0 \\
2026-10-09 09:04 & 184 & 101 & 0.59 & 0.029 & 0.20 & 100.0 \\
\bottomrule
\end{tabular}
\end{table}

{\scriptsize
\begin{longtable}{@{}p{2.3in}rrrrrr@{}}
\caption{\code{ibm\_kobe} calibration in force at the creation of each fresh-shot sampling job (run, job creation time UTC). RO: readout error; 1q: single-qubit \code{sx} error; 2q: median two-qubit \code{cz} error; worst 2q: largest two-qubit error on the device.}\label{tab:kobe-calibration-jobs}\\
\toprule
Run, job created (UTC) & T$_1$ ($\mu$s) & T$_2$ ($\mu$s) & RO (\%) & 1q (\%) & 2q (\%) & worst 2q (\%) \\
\midrule
\endfirsthead
\toprule
Run, job created (UTC) & T$_1$ ($\mu$s) & T$_2$ ($\mu$s) & RO (\%) & 1q (\%) & 2q (\%) & worst 2q (\%) \\
\midrule
\endhead
\bottomrule
\endfoot
h2o m no23 a5-46q trained, 2026-10-08 11:01 & 218 & 105 & 0.62 & 0.030 & 0.20 & 100.0 \\
h2o m no23 a5-70q trained, 2026-10-08 11:02 & 218 & 105 & 0.62 & 0.030 & 0.20 & 100.0 \\
sno32 a5-46q trained, 2026-10-08 11:02 & 218 & 105 & 0.62 & 0.030 & 0.20 & 100.0 \\
sno32 a1-70q trained, 2026-10-08 11:02 & 218 & 105 & 0.62 & 0.030 & 0.20 & 100.0 \\
2h2o s22 a5-46q trained, 2026-10-08 11:02 & 218 & 105 & 0.62 & 0.030 & 0.20 & 100.0 \\
sno32 a5-70q trained, 2026-10-08 11:02 & 218 & 105 & 0.62 & 0.030 & 0.20 & 100.0 \\
h2o s19 a1-46q trained, 2026-10-08 11:02 & 218 & 105 & 0.62 & 0.030 & 0.20 & 100.0 \\
h2o m no23 a1-46q untrained, 2026-10-08 11:02 & 218 & 105 & 0.62 & 0.030 & 0.20 & 100.0 \\
2h2o s22 a1-46q trained, 2026-10-08 11:02 & 218 & 105 & 0.62 & 0.030 & 0.20 & 100.0 \\
h2o m no23 a1-46q trained, 2026-10-08 11:02 & 218 & 105 & 0.62 & 0.030 & 0.20 & 100.0 \\
h2o m no23 a1-70q trained, 2026-10-08 11:02 & 218 & 105 & 0.62 & 0.030 & 0.20 & 100.0 \\
h2o ll33 a5-70q trained, 2026-10-08 11:02 & 218 & 105 & 0.62 & 0.030 & 0.20 & 100.0 \\
2h2o l35 a5-70q trained, 2026-10-08 11:02 & 218 & 105 & 0.62 & 0.030 & 0.20 & 100.0 \\
h2o s19 a1-46q untrained, 2026-10-08 11:02 & 218 & 105 & 0.62 & 0.030 & 0.20 & 100.0 \\
h2o m no23 a5-70q untrained, 2026-10-08 11:02 & 218 & 105 & 0.62 & 0.030 & 0.20 & 100.0 \\
llno28 a5-70q trained, 2026-10-08 11:02 & 218 & 105 & 0.62 & 0.030 & 0.20 & 100.0 \\
h2o m no23 a1-70q untrained, 2026-10-08 11:02 & 218 & 105 & 0.62 & 0.030 & 0.20 & 100.0 \\
h2o ll33 a1-70q trained, 2026-10-08 11:02 & 218 & 105 & 0.62 & 0.030 & 0.20 & 100.0 \\
2h2o l35 a1-70q trained, 2026-10-08 11:02 & 218 & 105 & 0.62 & 0.030 & 0.20 & 100.0 \\
llno28 a1-70q untrained, 2026-10-08 11:02 & 218 & 105 & 0.62 & 0.030 & 0.20 & 100.0 \\
2h2o m no29 a1-70q trained, 2026-10-08 11:03 & 218 & 105 & 0.62 & 0.030 & 0.20 & 100.0 \\
L46 a1-46q untrained, 2026-10-08 12:34 & 218 & 105 & 0.62 & 0.030 & 0.20 & 100.0 \\
L46 a1-46q trained, 2026-10-08 12:34 & 218 & 105 & 0.62 & 0.030 & 0.20 & 100.0 \\
L46 a5-46q trained, 2026-10-08 12:34 & 218 & 105 & 0.62 & 0.030 & 0.20 & 100.0 \\
L46 a1-70q trained, 2026-10-08 12:34 & 218 & 105 & 0.62 & 0.030 & 0.20 & 100.0 \\
L46 a1-70q untrained, 2026-10-08 12:34 & 218 & 105 & 0.62 & 0.030 & 0.20 & 100.0 \\
L46 a5-70q trained, 2026-10-08 12:34 & 218 & 105 & 0.62 & 0.030 & 0.20 & 100.0 \\
L46 a5-46q untrained, 2026-10-08 12:34 & 218 & 105 & 0.62 & 0.030 & 0.20 & 100.0 \\
L46 a5-70q untrained, 2026-10-08 12:34 & 218 & 105 & 0.62 & 0.030 & 0.20 & 100.0 \\
\end{longtable}}

\section{Computational cost: complementary figures}
\label{sec:appendix-cost}
The main-text cost section (Section~\ref{sec:cost}, Table~\ref{tab:cost-summary}) gives the quantum resources. The training budget per lineage of the main (HF-penalty) recipe is collected here; the inference timing is in Table~\ref{tab:inference-timing}.

\begin{table}[ht]
\centering
\caption{Stage-2 training budget per lineage of the HF-penalty recipe (150 iterations each; shots = iterations $\times$ 20 circuits $\times$ 10{,}000 shots, one Runtime job per iteration; wall times per lineage are in Table~\ref{tab:t2-timing}, IBM-billed QPU time in Table~\ref{tab:t2-qpu}). The inference timing and cost of the two configurations is in Table~\ref{tab:inference-timing}.}
\label{tab:cost-appendix}
\footnotesize
\setlength{\tabcolsep}{5pt}
\begin{tabular}{@{}lrrrr@{}}
\toprule
Lineage & Stage-2 iterations & Circuits & Shots & Runtime jobs \\
\midrule
A1-46q & 150 & 3{,}000 & 30{,}000{,}000 & 150 \\
A5-46q & 150 & 3{,}000 & 30{,}000{,}000 & 150 \\
A1-70q & 150 & 3{,}000 & 30{,}000{,}000 & 150 \\
A5-70q & 150 & 3{,}000 & 30{,}000{,}000 & 150 \\
Total & 600 & 12{,}000 & 120{,}000{,}000 & 600 \\
\bottomrule
\end{tabular}
\end{table}

\FloatBarrier
\section{Repeated sampling of two test cases}
\label{sec:appendix-repeat}
In two of the 88 runs the ten sampled circuits contained a single circuit with valid shots, so the ranking found no best and second pair: the Hartree--Fock-penalty run of SnOH3-me LL (56 qubits) with A5-46q (one circuit, 5{,}772 valid shots) and the tuned-reward run of SnOH3-me-2h2o S (44 qubits) with A5-70q (one circuit, 28{,}055 valid shots). In both, the other nine circuits carried a diagonal-Coulomb layer and had a transpiled depth of several thousand; their valid-shot shares are near zero (Section~\ref{sec:analysis-selection}). The number of usable circuits depends on the draw of the policy: across the 88 runs, circuits without a diagonal-Coulomb operator are shallow in every case, and the share of such circuits differs between draws. We therefore repeated the draw of the ten gate sequences once for each of the two test cases with a different random seed (4321, 5678 and 8765 were drawn without hardware; the draw with the most circuits free of diagonal-Coulomb operators, seed 8765, was measured). Sampling, ranking and recovery are otherwise identical to the main campaign.

\begin{table}[ht]
\centering
\caption{Repeated draws for the two test cases with a single usable circuit. Errors are those of the recovered energy against DMRG (mHa). Usable: circuits with valid shots at the ranking step.}
\label{tab:repeat}
\small
\setlength{\tabcolsep}{5pt}
\begin{tabular}{@{}llrrrrr@{}}
\toprule
Test case & Draw & Usable & Best & Second & Merged & Reported \\
\midrule
Tuned reward, SnOH3-me-2h2o S, A5-70q & seed 1234 & 1 of 10 & -- & -- & -- & no \\
 & seed 8765 & 5 of 10 & 0.07 & 0.06 & 0.04 & yes \\
HF penalty, SnOH3-me LL, A5-46q & seed 1234 & 1 of 10 & -- & -- & -- & no \\
 & seed 8765 & 8 of 10 & 203 & 199 & 205 & no \\
\bottomrule
\end{tabular}
\end{table}

For the tuned-reward test case the second draw gives five usable circuits, with a best-circuit error of 0.07~mHa, and this draw is used in all tables of the main text. For the Hartree--Fock-penalty test case the second draw gives eight usable circuits but a recovered error of about 200~mHa for every selection (best, second, merged and the combined energy and distinct-bitstring picks), at no more than 188 valid shots per circuit. The same plateau of about 200~mHa appears at this molecule for the second circuits of the 70-qubit policies, and the 46-qubit A1 policy reaches 15~mHa. A policy trained at 46 qubits applied to a 56-qubit active space is a possible reason, but one further draw does not establish it, and the test case is excluded from the Hartree--Fock-penalty statistics (43 of 44 runs). Both outcomes show that the result is sensitive to the draw of the sequences. They do not establish that a policy has to be applied at the active-space size it was trained for (Section~\ref{sec:discussion} finds no clear size dependence).

\section{Noiseless matrix-product-state sampling of three earlier checkpoints}
\label{sec:appendix-noiseless-simulation}

This appendix reports a noiseless simulation of circuits generated by three 46-qubit checkpoints, labelled A1-v5, A5-v5 and A5-v6.
These checkpoints generate eight operator tokens per circuit, whereas the 46-qubit lineages of the main text use $L = 28$.
The simulated circuits are therefore different from those run on \code{ibm\_kobe}.
The results describe these checkpoints and are not a noiseless counterpart of the hardware runs.

\paragraph{Protocol.} We evaluated the three checkpoints on five active spaces:
the three S tiers (32, 38 and 44 qubits), Sn(OH)$_3$-Me~L (46 qubits) and Sn(OH)$_3$-Me$\cdot$H$_2$O~M (46 qubits).
For each checkpoint and active space we generated ten circuits with $\beta = 0.2$ and sampled each circuit
with $10^5$ shots.
The circuits were simulated as matrix product states with cuTensorNet, part of the cuQuantum SDK~\cite{cuquantum}.
We used a maximum bond dimension of $\chi = 512$, complex128 precision and a relative singular-value cutoff of $10^{-8}$.
No noise was added. The generation and recovery seed was 1234, and the measurement seeds were 1234--1243.
Ranking and recovery used the budgets of Table~\ref{tab:infer-hyperparams}: at most 200 strings per spin for ranking, and a target of 20,000 strings per spin with 25 iterations for recovery.
Unlike on the hardware, we recovered all ten circuits of each test case, so that the lowest error among the ten circuits after recovery is known.
Merged pools the shots of all ten circuits ($10^6$ shots).
 
\paragraph{Two kinds of circuits.}
Table~\ref{tab:mps} lists the errors against DMRG.
Of the 150 circuits, 30 contain a global orbital-rotation operator, two in every test case.
These returned the Hartree--Fock determinant at most once in $10^5$ shots and produced 73,000--100,000 distinct configurations.
The other 120 circuits returned the Hartree--Fock determinant in at least 90\% of their shots and produced at most 329 distinct configurations.
Of these, 4--7 per test case returned only the Hartree--Fock determinant.
 
\paragraph{Small active spaces.}
At the three S tiers, each orbital-rotation circuit alone covers 92--98\,\% of the $\alpha$-strings and recovers to 0.25--1.97~mHa.
The near-Hartree--Fock circuits recover only to 3.6--84~mHa.
Merging the ten circuits gives errors of 0.008--0.049~mHa and covers 99.58--99.95\,\% of the $\alpha$-strings.
This agrees with the counting in Section~\ref{sec:zero-shot}: at these tiers the 20,000-string cap exceeds the full $\alpha$-string dimension (1,820, 3,876 and 7,315 strings).
The merged energies also agree with the hardware merged energies in Table~\ref{tab:zeroshot-master} to within a few $\mu$Ha, although the circuits are different.
 
\paragraph{46-qubit active spaces.}
At the two 46-qubit tiers, almost every shot of an orbital-rotation circuit is a distinct configuration. The recovered subspaces reach the cap of 20,000 strings, which is 8.2\,\% of the $\alpha$-strings for Sn(OH)$_3$-Me~L and 4.1\,\% for Sn(OH)$_3$-Me$\cdot$H$_2$O~M.
The lowest error among the ten circuits is 73--173~mHa at Sn(OH)$_3$-Me~L and 14--22~mHa at Sn(OH)$_3$-Me$\cdot$H$_2$O~M.
The merged results are 43--149~mHa at Sn(OH)$_3$-Me~L and 13--21~mHa at Sn(OH)$_3$-Me$\cdot$H$_2$O~M. Merging improves on the lowest of the ten circuits in five of the six cases, but not for A5-v5 at Sn(OH)$_3$-Me$\cdot$H$_2$O~M (14.5 against 14.0~mHa), where the cap already binds.

\paragraph{Comparison with the hardware runs.}
On the same two active spaces the hardware runs of the main text are much more accurate: 5.6--14.3~mHa at Sn(OH)$_3$-Me~L and 0.71--1.5~mHa at Sn(OH)$_3$-Me$\cdot$H$_2$O~M (Table~\ref{tab:zeroshot-master}), against 73--173 and 14--22~mHa for the lowest of the ten noiseless circuits, and 30--41 and 4.9--5.2~mHa even after singles injection. The circuits differ (eight tokens against 28), so this is not a controlled comparison. It does show that sampling a comparable trained policy without noise does not give the accuracy of the hardware campaign. The noiseless circuits with a global orbital rotation spread their shots over so many configurations that the recovered subspace reaches only 4--8\,\% of the $\alpha$-strings, and the other circuits return the Hartree--Fock determinant. On the device the shots of the selected shallow circuits are spread by noise into single and double excitations (Table~\ref{tab:raw-shot-excitation}), which the recovery exploits. The accuracy of the hardware runs is therefore not an artefact of the ranking and recovery steps, which were identical here, and the device noise contributes to the result instead of only degrading it.
 
\paragraph{Ranking.}
The ranking placed the two orbital-rotation circuits last in 13 of the 15 test cases.
With at most 200 strings per spin, their ranking energies at the S tiers lie 1.0--2.3~$E_\mathrm{h}$ above DMRG, while those of the near-Hartree--Fock circuits lie only 3--84~mHa above it.
The ranking therefore favours circuits whose samples are concentrated on a few configurations.
As a result, the best circuit by ranking was not the lowest of the ten after recovery in 11 of the 15 test cases.
On the hardware, noise spreads the samples of every circuit, so this result may not carry over to the runs
in the main text, where the selected circuit recovered to within 15~mHa in 42 of 43 runs (Section~\ref{sec:zero-shot}).
 
\paragraph{Singles injection.}
Adding the single excitations of the Hartree--Fock determinant (SI) to the recovered set of the best circuit by ranking and diagonalizing again lowers its error, for example from 29.0 to 16.1~mHa at Sn(OH)$_3$-Me~S and from 73.4 to 38.5~mHa at Sn(OH)$_3$-Me~L for A1-v5.
No noise is present, so this gain comes from the larger basis and not from the removal of noise.
 
\paragraph{Limitations.}
Only one generation seed was used, and convergence beyond 25 recovery iterations was not checked.
The MPS truncation error was not recorded.
One of the $1.5\times10^7$ shots had a wrong number of alpha or beta electrons.
The circuits conserve both numbers, so this cannot occur in an exact simulation.
The shot came from an orbital-rotation circuit at 44 qubits and may reflect MPS truncation.

Ranking and recovery build their bases differently.
Ranking performs spin-symmetry completion.
In eight test cases, the recovered basis of the best circuit by ranking omitted strings added during ranking, and its energy was 0.6--12~mHa above the ranking energy.
Independent PySCF calculations~\cite{sun2018pyscf} on the same bases reproduced these increases.
In three of the eight affected test cases, all at 46 qubits, the affected circuit was also the lowest of the ten after recovery.

\begin{table*}[t]
\centering
\small
\caption{%
Deviation $\Delta E = E_\mathrm{SQD}-E_\mathrm{DMRG}$ (mHa) of the noiseless matrix-product-state sampling of the three earlier checkpoints.
Best and Second are the recovered deviations of the first- and second-ranked circuits by the pre-recovery ranking energy (one-shot diagonalization with at most 200 strings per spin sector, as in Section~\ref{sec:pipeline-overview}).
Where the two entries are equal (Sn(OH)$_3$-Me~S, A5-v5 and A5-v6), both circuits recovered to the same energy.
Lowest of ten is the lowest recovered deviation among all ten circuits, before SI; it cannot be known without recovering every circuit and is not a selection.
Merged pools all ten circuits.
SI augments the recovered $\alpha$ and $\beta$ string sets with strings associated with single excitations of the Hartree--Fock determinant and diagonalizes again.
The 20,000-string cap applies before SI.
HF-only is the number of circuits, out of ten, whose shots contain only the Hartree--Fock determinant.
Coverage is the share of the $\alpha$-strings in the merged recovered set.
The circuits differ from those run on \code{ibm\_kobe}.
}
\label{tab:mps}
\begin{tabular}{@{}lrrrrrrrr@{}}
\toprule
Checkpoint & Best & Second & Lowest & Merged & Best & Merged & HF-only & Coverage \\
 & & & of ten & & + SI & + SI & & (\%) \\
\midrule
\multicolumn{9}{@{}l}{Sn(OH)$_3$-Me S (32 qubits, 16o24e)} \\
A1-v5 & 28.956 & 78.702 & 0.505 & 0.008 & 16.050 & 0.008 & 4 & 99.89 \\
A5-v5 & 82.595 & 82.595 & 0.398 & 0.008 & 19.121 & 0.008 & 5 & 99.95 \\
A5-v6 & 82.595 & 82.595 & 0.501 & 0.008 & 19.121 & 0.008 & 6 & 99.89 \\
\addlinespace
\multicolumn{9}{@{}l}{Sn(OH)$_3$-Me$\cdot$H$_2$O S (38 qubits, 19o30e)} \\
A1-v5 & 3.629 & 6.425 & 0.248 & 0.008 & 1.226 & 0.008 & 6 & 99.85 \\
A5-v5 & 3.629 & 6.425 & 0.255 & 0.008 & 1.226 & 0.008 & 5 & 99.85 \\
A5-v6 & 6.357 & 6.425 & 0.259 & 0.008 & 1.476 & 0.008 & 6 & 99.85 \\
\addlinespace
\multicolumn{9}{@{}l}{Sn(OH)$_3$-Me$\cdot$2H$_2$O S (44 qubits, 22o36e)} \\
A1-v5 & 4.243 & 5.563 & 0.449 & 0.049 & 1.063 & 0.049 & 6 & 99.58 \\
A5-v5 & 4.243 & 5.563 & 0.380 & 0.045 & 1.063 & 0.045 & 6 & 99.60 \\
A5-v6 & 5.563 & 5.832 & 0.379 & 0.045 & 1.204 & 0.045 & 7 & 99.60 \\
\addlinespace
\multicolumn{9}{@{}l}{Sn(OH)$_3$-Me L (46 qubits, 23o32e)} \\
A1-v5 & 73.440 & 199.536 & 73.440 & 43.222 & 38.537 & 20.266 & 6 & 8.16 \\
A5-v5 & 199.536 & 203.430 & 173.048 & 148.614 & 41.332 & 26.471 & 6 & 8.16 \\
A5-v6 & 169.534 & 199.536 & 169.534 & 142.858 & 30.841 & 27.022 & 6 & 8.16 \\
\addlinespace
\multicolumn{9}{@{}l}{Sn(OH)$_3$-Me$\cdot$H$_2$O M (46 qubits, 23o30e)} \\
A1-v5 & 14.048 & 20.022 & 14.048 & 12.987 & 4.932 & 3.746 & 6 & 4.08 \\
A5-v5 & 14.048 & 22.750 & 14.048 & 14.488 & 4.932 & 4.147 & 5 & 4.08 \\
A5-v6 & 22.682 & 22.750 & 21.771 & 21.199 & 5.235 & 4.374 & 6 & 4.08 \\
\bottomrule
\end{tabular}
\end{table*}

\section{Data and code availability}
\label{sec:appendix-availability}

The data of this work are available on Zenodo at \url{https://zenodo.org/records/24069330}. The record contains two self-contained HTML reports, one for each training configuration: the Hartree--Fock penalty (the main configuration) and the tuned reward. Each report gives the training of the configuration, the table of every run, the chemistry of the generated circuits on unseen molecules (valid share, Hartree--Fock, HOMO$\to$LUMO and spin-resolved shares, operator-family statistics) set against the chemistry during training, the comparison of the two configurations, the timing and cost, and, for every run, the ten generated gate-token sequences with their depth, two-qubit-gate count and recovered energies. The code is available on GitHub at \url{https://github.com/kunalroy332/sqd-gqe}.

\clearpage

\clearpage
\bibliographystyle{apsrev4-2}
\bibliography{references}

\end{document}